# STERILE NEUTRINO SEARCHES AT THE ICECUBE NEUTRINO OBSERVATORY

by

Spencer Nicholas Gaelan Axani

Dip, Southern Alberta Institute of Technology (2008)
B.Sc. (Hons), University of Alberta (2014)
M.Sc, Massachusetts Institute of Technology (2019)

Submitted to the Department of Physics
in partial fulfillment of the requirements for the degree of

Doctor of Philosophy

at the

MASSACHUSETTS INSTITUTE OF TECHNOLOGY

Feb. 2020

Author....................................................................
Department of Physics
Oct. 31$^{\text{st}}$, 2019

Certified by..............................................................
Janet M. Conrad
Professor of Physics
Thesis Supervisor

Accepted by...............................................................
Nergis Mavalvala
Associate Department Head of Physics, MIT

# STERILE NEUTRINO SEARCHES AT THE ICECUBE NEUTRINO OBSERVATORY

by

Spencer Nicholas Gaelan Axani

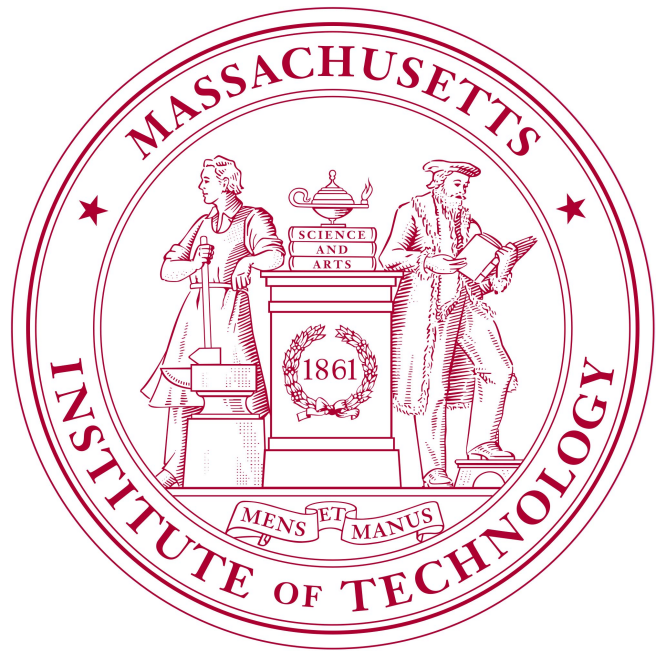

Submitted to the Department of Physics
on Oct. 31ˢᵗ, 2019, in partial fulfillment of the
requirements for the degree of
Doctor of Philosophy

# Abstract

The IceCube Neutrino Observatory is capable of performing a unique search for sterile neutrinos through the exploitation of a matter enhanced resonant neutrino oscillation phenomena. As atmospheric muon neutrinos pass the dense material within the Earth, neutral current elastic forward scattering is predicted to induce a transition into a sterile state.

This thesis presents two 3+1 sterile neutrino analyses by searching for spectral differences in the reconstructed energy and zenith direction of muon neutrino events, indicative of a transition into a sterile state. The first search probes the parameter space $\Delta m_{41}^2$ and $\sin^2(2\theta_{24})$ with relevant sensitivity to the global best fit region for a 3+1 sterile neutrino hypothesis. The second search performs a scan through $\sin^2(2\theta_{24})$ and $\sin^2(2\theta_{34})$ in the oscillation averaged out region of high-$\Delta m_{41}^2$ ($\Delta m_{41}^2 \gtrsim 10$ eV$^2$). The analyses are performed using an improved event selection, which was found to extract 305,891 well reconstructed muon neutrino events with a sample purity above 99.9%, from eight years of IceCube data. Novel simulation techniques, along with updated calibration, and a re-assessment of the systematic uncertainties are also discussed.

The first analysis finds a best fit sterile hypothesis point at $\Delta m_{41}^2 = 4.47\,\text{eV}^2$ and $\sin^2(\theta_{24}) = 0.10$, consistent with the no-sterile hypothesis at the 8% confidence level. The second analysis finds a best fit sterile hypothesis at $\sin^2(\theta_{34}) = 0.40$, $\sin^2(\theta_{24}) = 0.006$, consistent with the null hypothesis at the 19% confidence level.

Thesis Supervisor: Janet M. Conrad
Title: Professor of Physics

# Acknowledgments

I would like to express my profound gratitude to Professor Janet Conrad for bringing me into her research group and offering me an unfaltering supply of unique opportunities and research experience. Joining her group at MIT likely had one of the largest impacts of my life thus far. The career path that I pursue can largely be attributed to her mentorship. I wholeheartedly think that I could not have had a better or more influential PhD supervisor.

The work presented here would not have been possible without the effort of Carlos Argüelles, Ben Jones, and Marjon Moulai. They have each had their own significant impact on this thesis and stood by the analysis from start to finish. I am thankful to have them as friends and collaborators. I would also like to extend my sincere thanks to my good friend Daniel Winklehner, who I have worked closely with at MIT and has allowed me to crash on his couch more than a few times.

The IceCube collaboration is full of some of the best people I have ever met. I fully support and encourage any student curious about particle or astrophysics to pursue research opportunities with IceCube. The Oscillations Working Group provided a tremendous amount of thoughtful insight. Of particular importance to this analysis, I would like to thank Juan Pablo, Tom Stuttard and Philipp Eller for their roles in leading the working group, and Summer Blot and Claudio Kopper for doing an impeccable job reviewing this analysis. The calibration work that I performed would not have been possible without the guidance from the Calibration Working Group. In particular, I would like to thank Summer Blot, John Kelley, Keiichi Mase, Chris Wendt, and Dawn Williams. I'm also extremely grateful to the IceCube collaboration for sending me to the South Pole. It was a trip that I will never forget.

I would like to thank Professors Joe Formaggio and Scott A. Huges from MIT for being part of the doctoral review committee. Professor Hughes dedicated a immense amount of time in reading and adding input to both my master's and PhD thesis. Beyond their valuable contribution, I owe a great deal of gratitude to MIT for all the logistical support throughout my graduate work and giving me the opportunity to meet so many wonderful people.

There are many people that I worked closely with on various other projects who have had an important impact on my graduate work. I apologize in advance for not listing everyone individually but offer my thanks.

On a personal note, I would like to thank my parents, Bette and Dwayne Axani for guiding me to where I am now. I could not have made it without them. I would also like to thank my friends and brothers from Canada: Marshall, Grayson, and Jos Axani, and Neal McLaughlin, Mike Wilson, Brett Johnston, Chris Volkart, Steve Jorgenson, and Jayson Vavrek.

Lastly, the single greatest outcome of my PhD was being given the opportunity to meet my fiancé, Katarzyna Frankiewicz. I will never forget the love, support, patience, and friendship that she bestowed upon me throughout my masters and PhD theses. Traveling the world with her has been quite an adventure and I cannot wait for what comes next.

This doctoral thesis has been examined by a Committee of the Department of Physics as follows:

Janet M. Conrad . . . . . . . . . . . . . . . . . . . . . . . . . . . . . . . . . . . . . . . . . . . . . . . . . . . . . . . . . . . . . . . .
Thesis Supervisor
Professor of Physics

Joe Formaggio . . . . . . . . . . . . . . . . . . . . . . . . . . . . . . . . . . . . . . . . . . . . . . . . . . . . . . . . . . . . . . . . . .
Thesis Committee
Associate Professor of Physics

Scott A. Hughes . . . . . . . . . . . . . . . . . . . . . . . . . . . . . . . . . . . . . . . . . . . . . . . . . . . . . . . . . . . . . . . .
Thesis Committee
Professor of Physics

# Contents

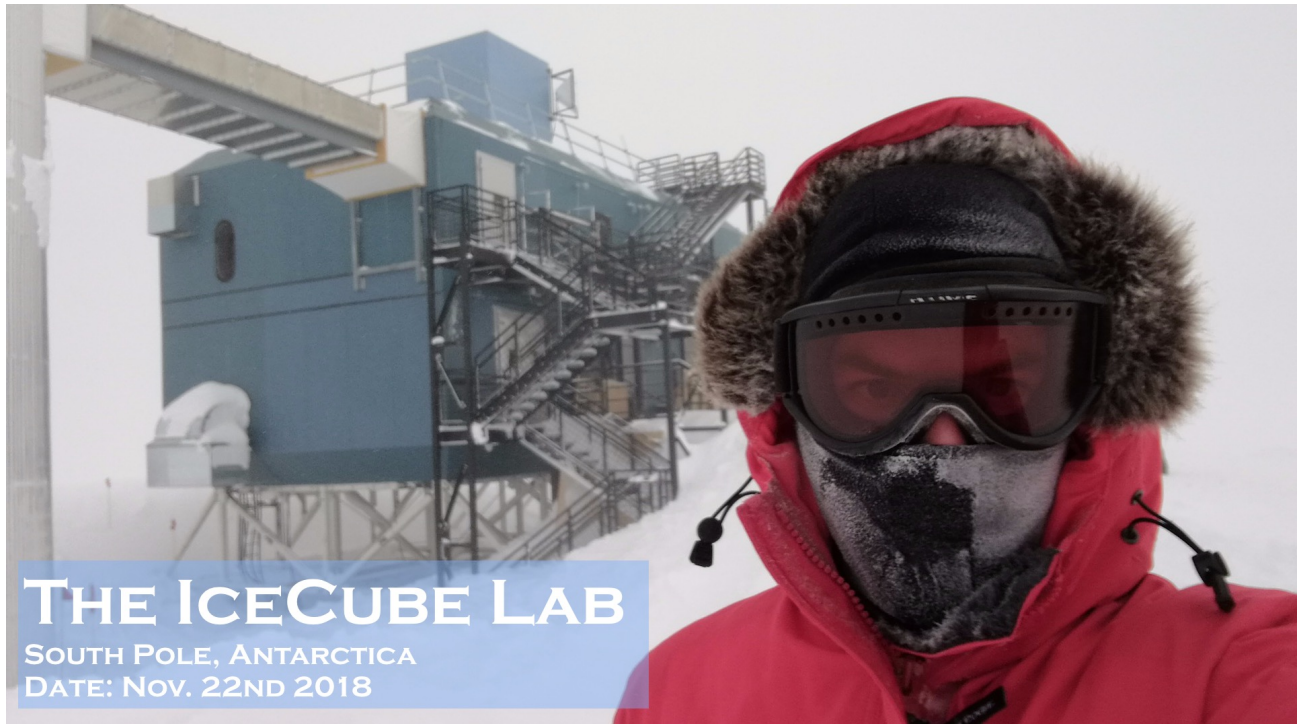

THE ICECUBE LAB
SOUTH POLE, ANTARCTICA
DATE: NOV. 22ND 2018

I have a strange hobby (to some) of measuring cosmic ray muons wherever I travel using custom designed pocket sized muon detectors [1–3]. It's a particularly interesting hobby when you're a physicist and able to travel to some rather dramatic and remote places on the planet. Since joining Prof. Conrad's group in 2014, we've performed measurements on top of mountains, kilometers underground and tens of kilometers above the surface of the Earth, inside an accelerator beamline, and on international flights. Perhaps, though, my favorite was the trip to the IceCube Neutrino Observatory at the geographical South Pole.

I left from Madison Wisconsin on November 20th 2018, a day after proposing to my now fiancé Kasia, for a three week stay at the Amundsen Scott south pole station in Antartica. Traveling there, I took six flights: four of which were operated by commercial airlines and the last two were by the US Air Force. Fig. 0-1 shows my measurement during these flights.

It's immediately obvious that there is a trend towards lower count rates heading towards the equator. This is most visible during the long flight from San Francisco United States (latitude

= +32°) heading south-west to Auckland New Zealand (latitude = -32°). The dip is due to the change in the Earth's magnetic field as a function of latitude. Near the equator, the magnetic field points parallel to the surface of the Earth, and since the cosmic ray muons flux peaks in the vertical direction, the force on the muon, $F = q\vec{v} \times B$, is maximal. Notice as well that although San Francisco and Auckland are at opposite latitudes, the observed count rate is not symmetric about the equator. This is due the magnetic latitudes being offset from the geographical latitudes (i.e. the geographical south pole is not at the same location as the magnetic south pole). This indicates that the ionizing radiation protection offered by the Earth's magnetic field is weaker at the high absolute latitudes.

The flight leaving from Christchurch New Zealand to McMurdo Antarctica was on a C-17 military jet operated by the US Air Force. Although flight information is not publicly available for military flights (I suppose for obvious reasons), I was able to calculate the altitude of the jet from the measured cosmic ray muon rate.

The final flight was aboard a C-130 military airplane. Myself and the other three scientists aboard landed at the geographical south pole approximately 6 hours later, on top of the 2700 m thick glacier. At this altitude, along with the minimal protection of the Earth's magnetic field, the measured cosmic ray muon rate was approximately 4.5 times larger than that at sea level near the equator.

Lastly, during my return flight, heading North through the equator at 35,000 ft, I setup two of my detectors and measured the cosmic ray muon flux coming from the east, then the west. I found a count rate coming from the east of 0.69 +/- 0.02 cps, while from the west 0.84 +/- 0.03 cps. This represents a 22.2±7.4% increase in the westward direction. The east-west asymmetry is produced by the cosmic-ray particles being predominately positively charged. The positively charged muons curve towards the east, meaning that the intensity from the west is stronger. From conservation laws, I can then deduce that the atmospheric neutrino flux is going to be larger than the antineutrino flux. Sadly, this means that the final result presented in this thesis is weaker than had nature realized the alternative.

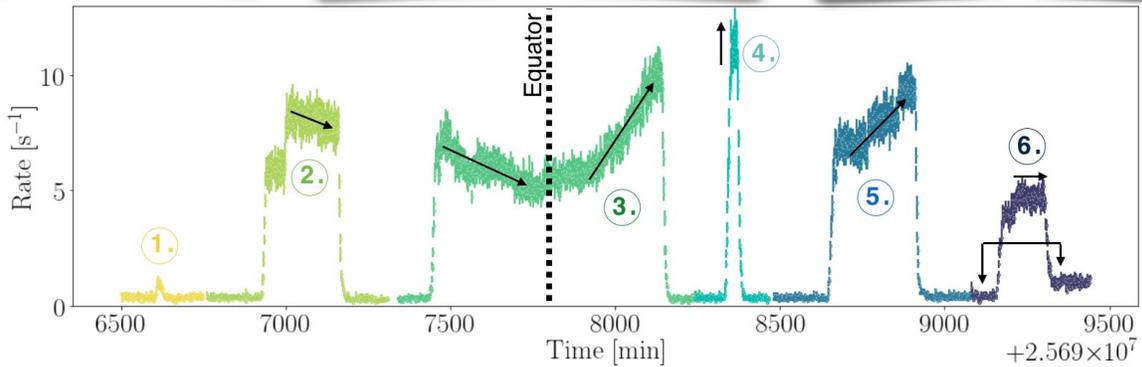

Figure 0-1: The cosmic ray muon measurements during each of the flights to the South Pole. From Ref. [4].

THIS PAGE INTENTIONALLY LEFT BLANK

# Chapter 1

# INTRODUCTION

The Standard Model (SM) of particle physics is currently the leading fundamental theory governing the laws of physics of our universe. It reduces everything we observe in nature down to a simple set of matter particles (the fermions) and force carrying particles (the bosons). While this theory has been tested to outstanding accuracy, the observation of neutrino oscillation was the first, and only, direct evidence indicating physics beyond the Standard Model. This thesis describes an analysis that searches for spectral distortions in the predicted atmospheric neutrino flux using the IceCube Neutrino Observatory, that would be indicative of a fourth neutrino state, often referred to as a "sterile" neutrino. Sterile neutrinos are a natural extension to the SM and could shed light on several anomalous neutrino oscillation measurements.

The thesis is organized as follows. Chapter 1 outlines a concise overview of the history of neutrino discoveries, highlighting the particularly important achievements in this field of research. Chapter 2 goes into detail about the formulation of neutrino oscillations and how they are described in the $\nu$Standard Model. Here, a set of anomalous neutrino oscillation measurements will be introduced that motivate the pursuit of a sterile neutrinos search, and hence this thesis. In Chapter 3, a description of the IceCube Neutrino Observatory will be presented in sufficient detail to understand the subtleties associated with the analysis. Chapter 4 describes the technical aspects of simulating neutrino events in IceCube. Chapter 5 describes the new event selection, that is, the method used to reduce all of the IceCube Neutrino Observatory data down to a high-purity sample of muon neutrinos/antineutrinos charged current events. Chapter 6 introduces the two sterile neutrino searches that will be performed along with the statistical framework used for analyzing the data. Chapter 7 goes through the systematic uncertainties associated with the detector, atmospheric neutrino flux, astrophysical neutrino flux, and interaction cross sections. Chapter 8 gives a concise overview of the methods used to characterize the sensitivity of IceCube to a sterile neutrino signal, the pre-unblinding tests designed to provide confidence in the result while preserving blindness, and finally, the result. A discussion follows the result and concludes with Chapter 9.

## 1.1 The Little Neutral One(s)

The earliest glimpse of the weak interaction came from H. Becquerel's discovery of radioactivity in 1896 [5]. Several years later it was shown that, unlike $\alpha$ and $\gamma$ rays, the beta-ray energy spectrum was unexpectedly continuous [6]. After eliminating the possibility of spectral broadening, there were two remaining explanations: either reject energy conservation or introduce a light neutral particle turning beta decay into a three body interaction. In an attempt to save the notion of conservation of energy, the second conjecture was championed by W. Pauli in 1930. Adhering to the conservation of electrical charge and Pauli exclusion principle (1945 Nobel Prize), the light neutral particle would have to be spin $1/2$ and a mass less than 0.01 times that of the proton.

The first milestone in the theory of weak interactions was established in 1934, when E. Fermi formulated a theory of beta-decay. However, this lead H. Bethe and R. Peierls to postulate that in the energy range of a characteristic beta-decay (MeV), the interaction cross section for this light neutral particle would have to be less than $10^{-44}$ cm$^2$, thus making it very difficult to measure. Fermi coined the term "neutrino" to describe this particle, which in Italian translates roughly to the "little neutral one."

Developments in liquid scintillator technology and photomultiplier tubes, coupled with the expansion of nuclear research due to the Manhattan Project, F. Reines and C. Cowan focused their research on experimentally measuring the neutrino via inverse beta decay, $\nu + p \rightarrow n + e^+$. To overcome the minute cross-section, Reines and Cowan required a large neutrino flux. This motivated them to place their detector near a nuclear reactor at Savannah River. In June of 1956 [7], they sent a telegram informing Pauli of the discovery of the neutrino [8]. This discovery was awarded the Nobel Prize in Physics nearly 40 years later.

Upon this measurement, R. Davis set out to observe $^{137}$C $\rightarrow$ $^{37}$Ar conversions in a tank containing a thousand gallons of carbon tetrachloride, also placed near the Savannah River reactor. The lack of neutrino observation elucidated the differences between neutrinos and their anti-particle partner, the antineutrino. The antineutrinos emanating from its fission reactions could not induce the transition above, while neutrinos should have done so.

In 1962, L. Lederman, M. Schwartz, and J. Steinberger discovered that more than one type of neutrino exists by detecting the secondary particles from the charged current interaction not seen by Reines and Cowan [9] (1988 Nobel Prize). Unlike the measurement at Savannah River, they observed a muon in the final state of the interaction, indicating that the neutrinos were divided into the "flavors" of the known leptons (at this point, the only leptons known were the electron and muon). When a third type of lepton, the tau, was discovered in 1975 at the Stanford Linear Accelerator [10], it too was expected to have an associated neutrino. Equipped also with the 1990 LEP measurements of the Z-boson invisible decay with suggesting that there are three active neutrino flavors interacting with the Z, the first observation of the tau neutrino was in 2000 by the DONUT collaboration at Fermilab [11].

In 1957, B. Pontecorvo formulated the idea of neutrino oscillation [12], which was subsequently refined, shortly after the discovery of muon neutrino, by K. Maki, M. Nakagawa, and S. Sakata [13]. The theory included the possibility of neutrino flavor transitions (which they called "virtual transmutations"). In 1978 [14], L. Wolfenstein formulated how neutrino oscillation would be modified if they propagated through matter. In 1985, S. Mikheyev and A. Smirnov illustrated that neutrino oscillations in a slowly varying medium can undergo a resonance enhancement (come to be known as the MSW resonance).

Neutrino oscillation would require the neutrinos to have a non-zero mass; this would be the first evidence for physics beyond the Standard Model of fundamental particle interactions.

During the formulation of neutrino oscillation, J. Bahcall was calculating the expected neutrino flux from the sun given the advances of nuclear physics during the $2^{\text{nd}}$ World War. Davis' Homestake experiment was designed specifically to measure the solar neutrinos flux, however after years of running, it was only able to account for approximately $1/3^{\text{rd}}$ of the expected flux (a measurement that was subsequently confirmed by the Kamiokande, GALLEX/GNO, and SAGE). This became known as the "solar neutrino anomaly" and the work from Davis earned him the Nobel Prize in 2002 "for the detection of cosmic neutrinos."

In 1992, the Kamiokande experiment confirmed [15] earlier reports from IMB-3 [16, 17] of a smaller than expected atmospheric $\nu_\mu/\nu_e$ ratio. They were comparing the number of $\nu_\mu$ events to the number of $\nu_e$ events from atmospheric neutrino and found a $4\sigma$ deficit compared to prediction. This became known as the "atmospheric neutrino anomaly."

At this stage there were two significant anomalies and it was understood that both anomalies might be resolved with neutrino oscillations. Driven also by the unexpected measurement of neutrinos from Supernova 1987A [18, 19], several large scale neutrino detectors were proposed. Perhaps most notably were the SNO experiment (proposed in1987 [20]) and the Super-Kamiokande experiment (introduced in 1984 [21, 22] and began construction 1991 [23]). The atmospheric neutrino anomaly was resolved by Super-Kamiokande in 1998 [24], favoring a mass of mass squared splitting of a few times $10^{-3}$ eV$^2$ (confirmed by the MACRO experiment [25]). The MSW resonance was found to be responsible for the solar neutrino anomaly, which was resolved by the SNO experiment in 2001 and found a preferred mass squared splitting just below $10^{-4}$ eV$^2$ [26]. These two experiments were jointly awarded the Noble Prize in 2015. The discovery of neutrino oscillation was arguably the largest paradigmatic shift in neutrino physics and still resonates to to this day. It is the first direct observation of physics beyond the Standard Model of particle physics and implies that at least two of the three neutrino states must have a non-zero mass.

However, this is not the end of the story. In fact, in parallel with the Super-Kamiokande and SNO discovery, a set of new anomalies began appearing. The first of which was by the LSND experiment. Originally proposed in 1989 [27], the LSND collaboration reported evidence for neutrino oscillations at the Los Alamos Meson Physics Facility in 1996 [28]. Unlike the SNO and Super-Kamiokande measurements, the signal preferred a mass squared splitting on the order of 1 eV$^2$. The preferred explanation for this observation was a fourth neutrino state and would come to be called the "eV-scale sterile neutrino." Many experiments were subsequently built to explore the LSND claim.

The MiniBooNE experiment was one of the first detectors to probe the LSND signal. It too found a significant excess partially compatible with the sterile neutrino hypothesis [29, 30].

Then, after revisiting the measured neutrino flux from the neutrino detectors placed at nuclear reactors, a deficit in the number of neutrinos observed in these measurements would also lend credence to this hypothesis [31]. And finally, during the calibration of the GALLEX [32,33] and SAGE [34,35] detectors, a deficit in the number of observed neutrino interactions was observed and could also be interpreted as a signal from a sterile neutrino.

While the above suggests consistency with the sterile neutrino hypothesis, the situation is much more complicated. In contrast to the anomalous measurements, there have been numerous experiments that have searched for the LSND signal and reported null (no sterile) results. Beyond this, the subset of experiments measuring signals in neutrinos tend to be incompatible with signals measured in antineutrinos. Further, the signals observed in $\nu_\mu$ disappearance experiments are in tension with the $\nu_e$ appearance and $\nu_e$ disappearance measurements.

Similar to the situation presented during the era of the solar neutrino anomaly and the atmospheric neutrino anomaly, the situation has become diluted with surprising and sometimes confusing results. These surprises have taken time to resolve because of the technical difficulty of measuring interactions with such low cross sections. At present, being able to resolve the remaining anomalous measurements through the introduction of a sterile neutrino is disputable. The aim of this thesis is to introduce a new result using data from the IceCube Neutrino Observatory and hopefully help clarify the situation.

# Chapter 2

# Neutrinos

This chapter provides an introduction to neutrino physics. We begin by outlining the neutrino as described by the Standard Model of Particle Physics, then expand on it to include the quantum mechanical effect of neutrino oscillation (which will be called the $\nu$Standard Model). At this point, several anomalous neutrino oscillation measurements will be introduced along with an interpretation of their result.

## 2.1 Standard Model neutrino properties

The Standard Model contains three spin 1/2, massless, uncharged leptons called neutrinos along with their three corresponding antiparticles, the antineutrinos. They do not carry the electromagnetic or color charge and therefore do not participate in either electromagnetic or strong interactions. They do, however, carry *weak isospin* which plays an analogous role for weak interactions. Similar to the conservation of charge in electromagnetic interactions, weak isospin is also conserved. The neutrino weak charged-current ($\nu$CC) interactions (Fig. 2-1 left and middle) produce a charged lepton and $W^{\pm}$ boson. In such a reaction, the charged lepton defines the *flavor* of the neutrino involved in the interaction: $\nu_e$, $\nu_\mu$, and $\nu_\tau$, where the subscript denotes the lepton flavor: electron, muon, and tau respectively. All flavors of neutrinos participate equally in the neutral-current (NC) interactions. NC interactions are defined by the interaction via the charge-less Z-boson, whose invisible decay width was measured by four experiments at the LEP collider and used to constrain the number of active neutrinos (with masses less than that of the Z-boson) to $2.9840 \pm 0.0082$ [36].

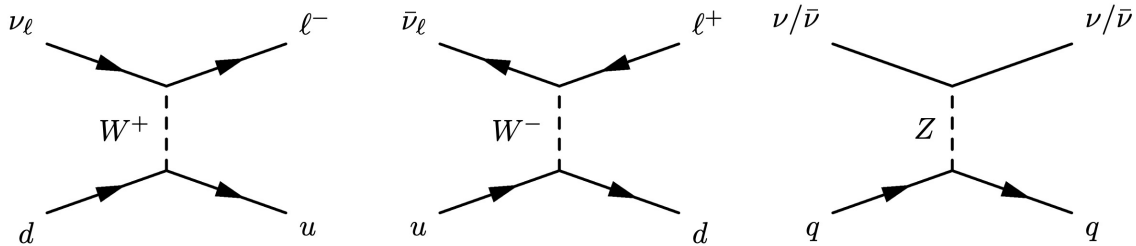

Figure 2-1: The Feynman diagrams for the weak interactions.

**Helicity, chirality, and parity** The *Helicity*, H, is a conserved quantity defined as the particle spin, $\sigma$, projected onto the direction of the momentum vector, $p$: $H = \sigma \cdot p/|p|$. It is also often referred to as *handedness*. When the spin direction is parallel to the momentum direction we define the helicity as positive, H= +1 (right-handed, RH); when the spin direction is opposite to the momentum direction, the helicity is negative, H= −1 (left-handed, LH). The first measurement of the neutrino helicity was performed in 1958 by M. Goldhaber, L. Grodzins and A. W. Sunyar [37]. Chirality is an intrinsic property of a particle and relates to both the helicity and how the particle transforms under parity. For massless particles helicity and chirality are equivalent, that is, a left-handed particle is said to be left-chiral. Early experiments by C. S. Wu [38] and theoretical predictions by T. D. Lee and C. N. Yang [39] tested the law of conservation of parity. In 1957, it was experimentally shown that weak interactions do not conserve parity, earning Lee and Yang the 1957 Nobel Prize in Physics [40]. To date, every neutrino ever observed is left-handed and every antineutrino is right-handed. The weak interaction only couples to the left-chiral particles and right-chiral antiparticles. Since neutrinos are produced in weak interactions, the neutrino is always emitted as a left-chiral particle (or right-chiral antiparticles).

## 2.2 $\nu$Standard Model neutrino properties

For many decades, the neutrino was considered a massless particle, and the SM treated them as such. It wasn't until July 1998 that the Super-Kamiokande experiment reported evidence of neutrino oscillation using atmospheric neutrinos [24]. This was later confirmed by the Sudbury Neutrino Observatory SNO [41–43] in 2002. Neutrino oscillations implied a non-zero neutrino mass and also indicated that lepton number is not separately conserved [44]; in contrast to the framework developed for the SM, resulting in a model that we can call the $\nu SM$.

With the inclusion of a non-zero neutrino mass, the helicity also needed to be expanded. The non-zero neutrino mass implies that the velocity of any neutrino must be less than the speed of light. Therefore, an observer would be able to transform their reference frame such that the momentum vector reverses, thus producing a right-handed neutrino (or left-handed antineutrino). If the neutrino is a *Dirac* particle, this state would be identifiable as a separate particle. While, there also exists the possibility of the neutrino being its own antiparticle (referred to as a *Majorana* fermion), in which the right-handed neutrino state is simply the antineutrino.

### 2.2.1 Theoretical framework of neutrino oscillation

The establishment of neutrino oscillation from solar, atmospheric, reactor, and man-made accelerator based neutrinos has firmly established the existence of three active flavors of neutrinos, which can be represented as a superposition of three mass eigenstates. Neutrinos are detected in the flavor basis, while they propagate in the mass basis (analogous to the Cabibbo-Kobayashi-Maskawa (CKM) matrix in the quark sector [45]). The unitary transformation relating the flavor to the mass eigenstate left-handed neutrino fields is the lepton mixing matrix, known as the PMNS Matrix, U (named after the primary authors of the theory B. Pontecorvo [46], Z. Maki, M. Nakagawa, and S. Sakata [13]):

$$|\nu_{\alpha L}\rangle = \sum_i U_{\alpha i} |\nu_{iL}\rangle. \tag{2.1}$$

Here, I denote flavor eigenstates by Greek indices ($\alpha$ = e, $\mu$, $\tau$) and mass eigenstates by Latin indices ($i$ = 1, 2, 3). I will subsequently drop the left-handed field notation, L. The orthogonality relationship for the flavor and mass eigenstates are specified as:

$$\langle \nu_\alpha | \nu_\beta \rangle = \delta_{\alpha\beta}, \qquad \langle \nu_i | \nu_j \rangle = \delta_{ij}. \tag{2.2}$$

We can expand Eq. 2.1 for the three active neutrino flavors:

$$\begin{pmatrix} \nu_e(t) \\ \nu_\mu(t) \\ \nu_\tau(t) \end{pmatrix} = U \begin{pmatrix} \nu_1(t) \\ \nu_2(t) \\ \nu_3(t) \end{pmatrix} = \begin{pmatrix} U_{e1} & U_{e2} & U_{e3} \\ U_{\mu 1} & U_{\mu 2} & U_{\mu 3} \\ U_{\tau 1} & U_{\tau 2} & U_{\tau 3} \end{pmatrix} \begin{pmatrix} \nu_1(t) \\ \nu_2(t) \\ \nu_3(t) \end{pmatrix}. \tag{2.3}$$

For example, the flavor state $|\nu_e\rangle$ is a superposition of the mass states $U_{e1} |\nu_1\rangle + U_{e2} |\nu_2\rangle + U_{e3} |\nu_3\rangle$. The probability of the neutrino being measured in a particular mass state, i, is therefore $|U_{ei}|^2$.

Being associated with a change of basis, the PMNS matrix is unitary. Like the CKM matrix, it satisfies unitary relations, derived from $UU^\dagger = U^\dagger U = \mathbb{1}$:

$$\sum_i U_{\alpha i} U^*_{\beta i} = \delta_{\alpha\beta}, \qquad \sum_\alpha U_{\alpha i} U_{\alpha j} = \delta_{ij}. \tag{2.4}$$

The simplest of the lepton unitary mixing matrix which relates the flavor and mass eigenstate (with mass $m_i$) neutrinos is given in terms of three mixing angles ($\theta_{12}, \theta_{13}, \theta_{23}$) and one (two) CP-violating Dirac (Majorana) phases ($\delta_{cp}$, $\alpha_{21}$, $\alpha_{31}$). This parametrization is analogous to three unitary rotation matrices, $U_{\theta_{23}}, U_{\theta_{13}}, U_{\theta_{12}}$, where the amount of "rotation" is determined through the mixing angles.

$$U = U_{\theta_{23}}U_{\theta_{13}}U_{\theta_{12}}P = \begin{pmatrix} 1 & 0 & 0 \\ 0 & c_{23} & s_{23} \\ 0 & -s_{23} & c_{23} \end{pmatrix} \begin{pmatrix} c_{13} & 0 & s_{13}e^{-i\delta_{cp}} \\ 0 & 1 & 0 \\ -s_{13}e^{i\delta_{cp}} & 0 & c_{13} \end{pmatrix} \begin{pmatrix} c_{12} & s_{12} & 0 \\ -s_{12} & c_{12} & 0 \\ 0 & 0 & 1 \end{pmatrix} P \quad (2.5)$$

Here, $U_{\theta_{23}}$ is called the atmospheric mixing, $U_{\theta_{13}}$ is called the reactor mixing, $U_{\theta_{12}}$ is called the solar mixing, whose names simply represent the method typically used to measure them. $P$ is a diagonal phase matrix. For simplicity, $c_{ij} = \cos(\theta_{ij})$, $s_{ij} = \sin(\theta_{ij})$, and $P$ can be either $\mathbb{1}$ in the Dirac case or $\text{diag}(e^{i\alpha_1}, e^{i\alpha_1}, 1)$ in the case of Majorana. We can multiply out these matrices to get to the standard PNMS matrix:

$$U_{PNMS} = \begin{pmatrix} c_{12}c_{13} & s_{12}c_{13} & s_{13}e^{-i\delta_{cp}} \\ -s_{12}c_{23} - c_{12}s_{13}s_{23}e^{i\delta_{cp}} & c_{12}c_{23} - s_{12}s_{13}s_{23}e^{i\delta_{cp}} & c_{13}s_{23} \\ s_{12}s_{23} - c_{12}s_{13}c_{23}e^{i\delta_{cp}} & -c_{12}s_{23} - s_{12}s_{13}c_{23}e^{i\delta_{cp}} & c_{13}c_{23} \end{pmatrix} P. \quad (2.6)$$

The current best fit values for the PNMS matrix along with the $3\sigma$ uncertainties are [47]:

$$U_{PNMS} = \begin{pmatrix} 0.824^{+0.022}_{-0.024} & 0.547^{+0.034}_{-0.032} & 0.147^{+0.008}_{-0.008} \\ 0.409^{+0.107}_{-0.180} & 0.634^{+0.065}_{-0.196} & 0.657^{+0.133}_{-0.043} \\ 0.392^{+0.136}_{-0.143} & 0.547^{+0.168}_{-0.085} & 0.740^{+0.036}_{-0.145} \end{pmatrix} \quad (2.7)$$

Unlike the CKM matrix, the PNMS matrix is far from diagonal, indicating large coupling between neutrino states. While unitary suggests that the sum of the squares of each row or column will add up to unity, the uncertainty of the global best fit values does not significantly constrain this condition.

With three neutrino mass states, there are three mass splittings defined as:

$$\Delta m_{21}^2 = m_2^2 - m_1^2, \quad \Delta m_{32}^2 = m_3^2 - m_2^2, \quad \Delta m_{31}^2 = m_3^2 - m_1^2, \quad (2.8)$$

however, only two of them are independent: $\Delta m_{31}^2 = \Delta m_{21}^2 + \Delta m_{32}^2$. Cosmology provides model-based upper bounds on the neutrino mass, both the neutrino mass ordering (often referred to as the neutrino mass hierarchy) and the absolute neutrino mass scale are still active areas of research. The current measured values for the mass splittings (assuming a normal hierarchy) are [47]:

$$\Delta m_{21}^2 = 7.50^{+0.59}_{-0.47} \times 10^{-5} \text{eV}^2, \quad \Delta m_{32}^2 = 2.524^{+0.119}_{-0.117} \times 10^{-3} \text{eV}^2, \quad (2.9)$$

At the stage of writing this thesis, there are no $3\sigma$ constraints on $\delta_{cp}/°$, but the current best fit with $1\sigma$ uncertainties is $261^{+51}_{-59}$ [47].

The mass eigenstates are equivalently eigenstates of the Hamiltonian, therefore the time dependence can be expanded to:

$$|\nu_\alpha(t)\rangle = \sum_i U_{\alpha\beta} |\nu_i\rangle e^{-itE} \quad (2.10)$$

For simplicity, assuming CP invariance ($U_{\alpha i}$ is real) and a neutrino mass much less than the momentum, $m_i \ll p_i$, the transition probability, P, starting with a neutrino flavor state $\alpha$ and being detected in state $\beta$ is:

$$P(\alpha \to \beta)(t) = \delta_{\alpha\beta} - 4 \sum_{j>i} U_{\alpha j} U_{\alpha j} U_{\beta i} U_{\beta j} \sin^2\left(\frac{\Delta m_{ij}^2}{4} \frac{L}{E}\right), \quad (2.11)$$

where L is baseline in km (m) and E is the neutrino energy in units of GeV (MeV).

Given the best fit values for the 3-neutrino oscillation model in Eq. 2.7, we can calculate the oscillation probability as a function of baseline per unit energy (L/E). As an example, Fig. 2-2 shows the survival probability of a beam that was initially $\nu_e$ in black. The red (blue) line shows the appearance probability of a neutrino being detected in a $\nu_\mu$ ($\nu_\tau$) state at a given L/E.

One could also ask at what baseline will the state undergo a complete oscillation and return to the original state. The oscillation length, L, is defined by the energy of the neutrino and mass

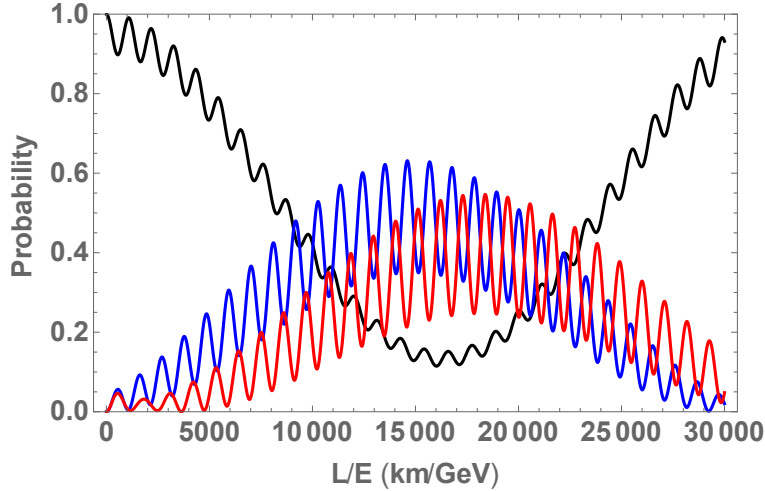

Figure 2-2: An example survival probability of beam of initially $\nu_e$ (black curve) as a function of distance travelled per unit energy. The appearance probability into the $\nu_\mu$ state is shown in red, while the appearance probability into the $\nu_\tau$ is shown in blue.

squared splitting:

$$L = \frac{4\pi E}{\Delta m^2} \qquad (2.12)$$

### 2.2.2 Neutrino oscillations in matter

The Mikheyev-Smirnov-Wolfenstein (MSW) effect was first published in 1978 [14, 48], and describes how neutrino oscillations are modified in the presence of matter by coherent forward scattering. It plays a very significant role in modern neutrino oscillation experiments [49]. All active neutrinos can interact with matter through the NC channel, however only the $\nu_e$ will interact via the CC channel with ordinary matter. This causes electron-flavor component of the neutrino state to have an additional potential (or negative potential in the case of antielectron neutrinos) in the Hamiltonian. This potential is related to the electron density in the real matter and becomes a non-negligible effect at high-energies and long baselines, such as in the Earth of Sun.

$$E \to E - V(x) \quad \text{where} \quad V(x) = \pm\sqrt{2}G_F \times \begin{cases} n_e(x) - \frac{1}{2}n_n(x) & \text{for} \quad \nu_e \\ -\frac{1}{2}n_n(x) & \text{for} \quad \nu_\mu, \nu_\tau \end{cases} \quad (2.13)$$

The upper sign is for neutrinos and the lower for antineutrinos, $G_F/(\hbar c)^3$ is Fermi's constant of $(1.166\,378\,7\pm6) \times 10^{-5}$ GeV$^{-2}$ [50], and $n_e(x)$ and $n_n(x)$ correspond to the electron and neutron number density respectively at a given location, $x$. In the flavor basis, we can write the Schrödinger equation :

$$i\frac{d}{dt}\begin{pmatrix}\nu_e \\ \nu_\mu \\ \nu_\tau\end{pmatrix} = \frac{1}{2E}U\begin{pmatrix}m_1^2 & 0 & 0 \\ 0 & m_2^2 & 0 \\ 0 & 0 & m_3^2\end{pmatrix}U^\dagger\begin{pmatrix}\nu_e \\ \nu_\mu \\ \nu_\tau\end{pmatrix} + \begin{pmatrix}n_e(x)-\frac{1}{2}n_n(x) & 0 & 0 \\ 0 & \frac{1}{2}n_n(x) & 0 \\ 0 & 0 & \frac{1}{2}n_n(x)\end{pmatrix}\begin{pmatrix}\nu_e \\ \nu_\mu \\ \nu_\tau\end{pmatrix}. \quad (2.14)$$

This can be simplified by re-phasing all of the neutrino flavor states by $\exp[-im_1^2 x/(2E)]$. Also, since the NC interactions is diagonal and does not distinguish between neutrino flavors, it just causes an overall phase shift of no physical importance as it drops out in Eq. 2.14:

$$i\frac{d}{dt}\begin{pmatrix}\nu_e \\ \nu_\mu \\ \nu_\tau\end{pmatrix} = \frac{1}{2E}U\begin{pmatrix}0 & 0 & 0 \\ 0 & \Delta m_{21}^2 & 0 \\ 0 & 0 & \Delta m_{31}^2\end{pmatrix}U^\dagger\begin{pmatrix}\nu_e \\ \nu_\mu \\ \nu_\tau\end{pmatrix} + \begin{pmatrix}n_e(x) & 0 & 0 \\ 0 & 0 & 0 \\ 0 & 0 & 0\end{pmatrix}\begin{pmatrix}\nu_e \\ \nu_\mu \\ \nu_\tau\end{pmatrix}. \quad (2.15)$$

For antineutrinos, the same equation holds with the change V(x)$\to$-V(x) and U$\to$U$^*$ (or equivalently $\delta \to -\delta$). The difference in $\nu_e$ and $\nu_\alpha$ interactions implies that the $\nu_i \to \nu_j$ transitions exist. This in its turn means that the eigenstates of propagation in matter $\nu_i^m$ and $\nu_j^m$, do not coincide with $\nu_i$ and $\nu_j$. Because the oscillation angles in matter, $\theta_{ij}^m$, is determined with respect to $\nu_i^m$ and $\nu_j^m$, it differs from $\theta_{ij}$. For illustration, in a two neutrino model, we find the oscillation amplitude in matter to be [51]:

$$\sin^2(2\theta^m) = \frac{\sin^2(2\theta)}{(2\sqrt{2}G_F E n_e(x)/\Delta m^2 - \cos 2\theta)^2 + \sin^2(2\theta)}. \quad (2.16)$$

Immediately, it is recognizable that when $2\sqrt{2}G_F E n_e(x)/\Delta m^2 = \cos 2\theta$, we run into the case where the mixing is maximal ($\sin^2(2\theta^m) = 1$). This occurs at the critical energy (resonance condition):

$$E_{\text{res}} = \frac{\Delta m^2 \cos(2\theta)}{2\sqrt{2}G_F n_e}, \qquad (2.17)$$

which can also be solved for the critical electron density, $n_e^{res}$ at a given neutrino energy.

$$n_{\text{res}} = \frac{\Delta m^2 \cos(2\theta)}{2\sqrt{2}G_F E}, \qquad (2.18)$$

The corresponding oscillation length is [52]:

$$L_{\text{res}} = \frac{4\pi E}{\Delta m^2 \sin 2\theta} \qquad (2.19)$$

The resonant condition can be satisfied only for neutrinos or antineutrinos separately, since they have opposite signs for the matter potentials.

The mass squared splitting in matter is also modified such that we define:

$$m^2_{1m,2m} = \frac{1}{2}[(2\sqrt{2}G_F n_e(x)E) \mp \sqrt{(\Delta m^2 \cos 2\theta - 2\sqrt{2}G_F n_e(x)E)^2 + (\Delta m^2 \sin 2\theta)^2}] \qquad (2.20)$$

It should be noted that neutrino oscillations in matter of constant density proceeds exactly as the oscillations in vacuum, the only difference being that the oscillation amplitude (Eq. 2.16) and mass squared splittings (Eq. 2.20) are different from those in vacuum.

It is also possible to have large flavor transitions through a *parametric oscillation enhancement*, when the neutrino travels through matter of changing density. It occurs if the variation of the matter density along the neutrino trajectory is correlated in a certain way with the change of the oscillation phase. Even if the mixing angles both in vacuum and in matter are small, this can lead to large probabilities of neutrino flavor transition in matter. For the parametric resonance to occur, the exact shape of the density profile is not very important; it is simply important that the change in the density be synchronized with the change of the oscillation phase [53].

## 2.3 Light sterile neutrinos

### 2.3.1 Anomalous neutrino oscillation measurements

The three active neutrino oscillation framework has been well established experimentally. However, several experiments have reported statistically significant observations that do not conform with the $\nu$SM. The measurements can be broken up into *appearance* and *disappearance* measurements. Appearance measurements aim to measure the appearance of a neutrino flavor composition some distance from the source. For example, this could be observing a $\nu_e$ flux at some baseline in a source that originated as a $\nu_\mu$ flux. The appearance probability from neutrino flavor $\alpha$ to neutrino flavor $\beta$ will be written as P($\nu_\alpha \to \nu_\beta$). Disappearance measurements look at the disappearance of a particular flavor at some distance from the source. As an example, the measurement of a $\nu_\mu$ flux in a source originating from a $\nu_\mu$ flux. This is often refered to as a survival probability and will be written as P($\nu_\alpha \to \nu_\alpha$).

**The LSND Anomaly**

The first anomalous short baseline oscillation measurement was reported by the Liquid Scintillator Neutrino Detector (LSND) collaboration. In 1996, they published evidence for neutrino oscillations from the observation of $\overline{\nu}_e$ appearance in a $\overline{\nu}_\mu$ beam, in a region inconsistent with the modern three active neutrino paradigm [28,54]. Their results appeared to be consistent with a fourth neutrino mass state $\Delta m^2 > 0.03$ eV$^2$. LSND recorded data from 1993 to 1998, always observing an above background (measured with beam-off data) excess of $\nu_e$-like events [55].

The detector was located at the Los Alamos Meson Physics Facility (LAMPF, previously known as LANSCE). Neutrinos were produced by impinging a high intensity (1 mA) 789 MeV proton beam on a water target (later updated to a high-Z target). This interaction produced a meson flux dominated by $\pi^+$ (the negatively charged pions are suppressed by the high probability of nuclear capture in the iron shielding and copper beam dump [56]). The majority of the $\pi^+$ would decay-at-rest (DAR), $\pi^+ \to \mu^+ + \nu_\mu$, followed by $\mu^+ \to e^+ + \nu_e + \overline{\nu}_\mu$. The $\overline{\nu}_e$ were detected through the weak charge interaction $\nu_e + p \to e^+ + n$, by measuring the delayed coincidence of the annihilation of the $e^+$ and the 2.2 MeV gamma ray emitted by the captured neutron on a free proton.

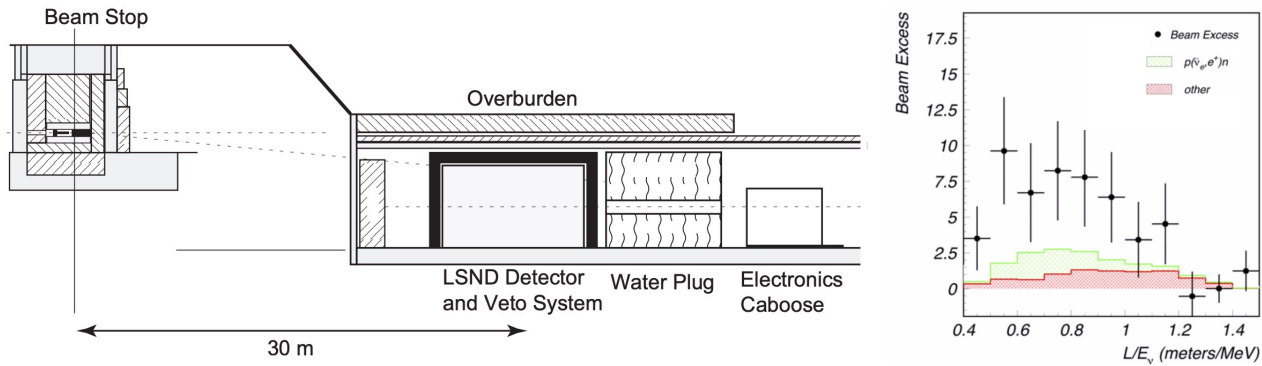

Figure 2-3: Left: The LSND detector layout, from [55]. Right: The observed excess in $\nu_e$ events in terms of the distance traveled by neutrinos (L) divided by the measured neutrino energy ($E_\nu$).

The observed data, shown in Fig. 2-3 (right), corresponds to a $> 3\sigma$ excess in $\nu_e$-like events (above the expected beam-off and neutrino background).

**MiniBooNE and the low-energy excess**

The Mini Booster Neutrino Experiment (MiniBooNE) [29, 30] was primarily designed to investigate the evidence of short baseline neutrino oscillations reported by LSND. It is based at Fermilab and has been reliably taking data since 2002. MiniBooNE is sensitive to the same part of the oscillation phase space as LSND (access to the same L/E), however, it operates at a distance and neutrino energy spectrum approximately 10 times larger than LSND. However, MiniBooNE has a larger range of L/E. The design MiniBooNE is also capable of operating in both neutrino and antineutrino mode and therefore is able to search for short baseline oscillations in both $\overline{\nu}_\mu \to \overline{\nu}_e$ and $\nu_\mu \to \nu_e$.

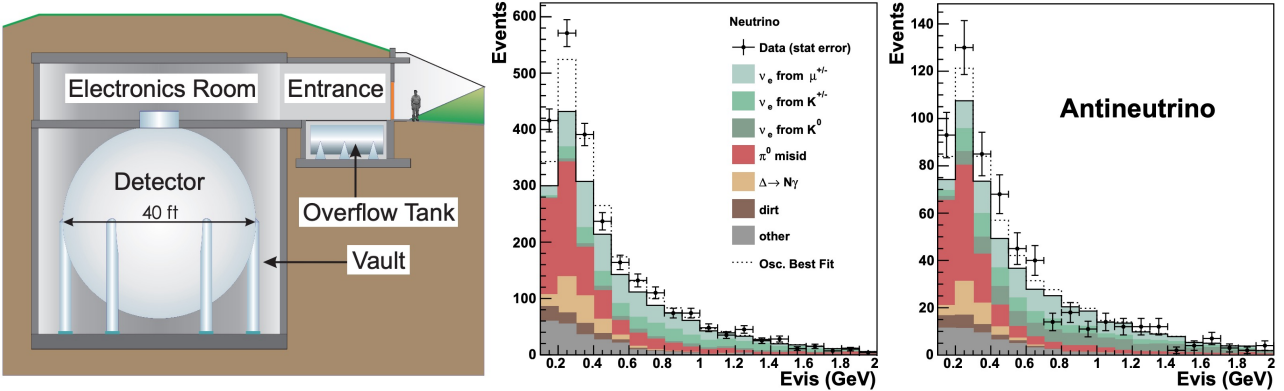

Figure 2-4: Left: The MiniBooNE detector, from [55]. Right: The observed data compared to the different stacked background predictions as a function of visible energy.

The source of neutrinos for MiniBooNE is the Booster Neutrino Beam (BNB) at Fermilab. Neutrinos are produced by an 8 GeV proton beam interacting with a beryllium target inside a magnetic focusing horn. Depending on the current in the horn, either the $\pi^+$ or $\pi^-$ produced in the interaction can be focused into a decay pipe, while the other pion species is de-focused. Using this method, The BNB is able to produce a relatively pure beam of $\bar{\nu}_\mu$ or $\nu_\mu$ [57].

MiniBooNE observed a significant low-energy excess in both the measurement of $\nu_e$ and $\bar{\nu}_e$ induced charged current quasi-elastic (CCQE) events. The total number of excess $\nu_e$ and $\bar{\nu}_e$ CCQE events observed was $460.5 \pm 99.0$, corresponding to a $4.7\sigma$ excess in the visible energy range of $200 < E_\nu^{QE} < 1250$ MeV. The MiniBooNE data is shown in Fig. 2-4 (right).

**The reactor antineutrino anomaly**

The Reactor Antineutrino Anomaly was originally described in 2011 [31], when a re-evaluation of the expected antineutrino flux from nuclear reactors brought the measured flux into disagreement with prediction. Reactor-based neutrino detectors located at short baselines to the reactor core were shown to measure a deficit number of $\bar{\nu}_e$ interactions compared to the standard 3-flavor neutrino model. The $\bar{\nu}_e$ energy spectrum from the reactors peaks approximately at 2 MeV and has a broad distribution that extends from sub-MeV to approximately 10 MeV. While this is a markedly different energy range than both LSND and MiniBooNE, the short baseline between the reactor core and the detectors provided access to a similar L/E, albeit with very different systematic uncertainties.

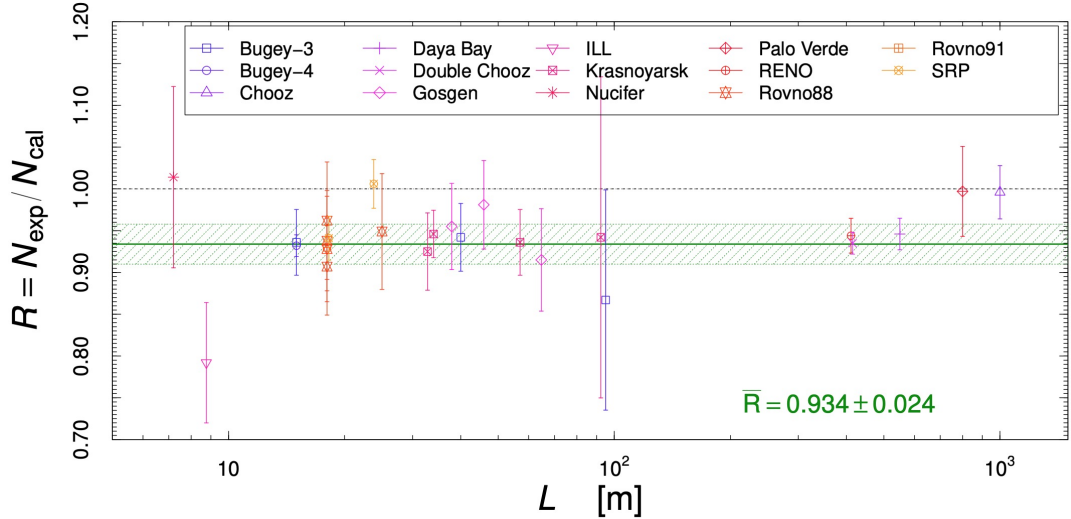

Figure 2-5: The measured $\bar{\nu}_e$ flux from various reactor experiments compared to prediction. From Ref. [58].

A recent nuclear reactor electron antineutrino flux evaluation [58] lead the global observed rate for reactor-detector distances >100 m of 0.934 ± 0.024. This represents a deviation from unity at the 98.6% CL, see Fig. 2-5. Beyond the $\bar{\nu}_e$ rate, the measured energy spectrum contains features that are not understood. In particular, a few percent excess in the reactor visible energy spectrum is observed at 5 MeV. At present, these two discrepancies between data and prediction are an outstanding issue and are not known whether or not they are related.

**The gallium anomalies**

In 1997-99, the GALLEX [32,33] and the Russian-American Gallium Experiment (SAGE) [34,35] were investigating the solar neutrino problem [59] using Gallium-based detectors. The detectors were originally calibrated using a $^{51}$Cr source of known activity. Upon electron capture, the source emits a mono-energetic $\nu_e$:

$$e^- + {}^{51}\text{Cr} \rightarrow {}^{51}\text{V} + \nu_e \quad \text{(half-life of 27.7 days)} \tag{2.21}$$

The outgoing $\nu_e$ might then interact with $^{71}$Ga via a CC interaction producing $^{71}$Ge + e$^-$. The $^{71}$Ge was then extracted and counted (analogous to Davis' radio-chemical solar neutrino experiments).

The measurements were repeated in 2004 through 2006 using a well understood $^{37}$Ar [60] source:

$$e^- + {}^{37}\text{Ar} \rightarrow {}^{37}\text{Cl} + \nu_e \quad \text{(half-life of 35.04 days)}. \tag{2.22}$$

Both GALLEX and SAGE concluded a ratio of observation to expectation of $0.88 \pm 0.05$ [61], corresponding to a >$2\sigma$ deficit. While the baselines for GALLEX and SAGE were approximately 1-2 m, the monoenergetic neutrino energies for $^{51}$Cr and $^{37}$Ar were 0.82 MeV and 0.90 MeV respectively. This also corresponds to a compatible L/E as LSND and MiniBooNE anomaly.

### 2.3.2 A fourth neutrino state in a 3+1 sterile neutrino model

The anomalous oscillation measurements observed by LSND, MiniBooNE, the global reactor measurements, and gallium radioactive source measurements may be explained with the introduction of a new eV-scale neutrino state. This idea is often invoked since many BSM models involve new neutrino states [62]. A new active neutrino state however contradicts measurement performed by four LEP experiments which constrained the number of active neutrinos states that interact with the Z-boson to $N_\nu = 2.9840 \pm 0.0082$ [63]. One can mitigate this restriction by simply assuming that the new neutrino state does not interact with the Z-boson, i.e. is in a "sterile" state. This state would still be able to interact via the mechanisms used for standard neutrino oscillation.

The introduction of a new state is commonly referred to as a 3+1 model, where there are 3

active neutrino states and a single, non-weakly (and non-strongly), interacting state with all active states being smaller than the sterile sates (a 1+3 model has the sterile state lighter than the active states). In terms of the mixing matrix described in Eq. 2.7 is then expanded to:

$$U_{3+1} = \begin{pmatrix} U_{e1} & U_{e2} & U_{e3} & U_{e4} \\ U_{\mu 1} & U_{\mu 2} & U_{\mu 3} & U_{\mu 4} \\ U_{\tau 1} & U_{\tau 2} & U_{\tau 3} & U_{\tau 4} \\ U_{s1} & U_{s2} & U_{s3} & U_{s4} \end{pmatrix}, \quad (2.23)$$

where "s" corresponds to the sterile state and the "4" is the equivalent mass state. Beyond this, we also require a new mass splitting, $\Delta m_{41}^2$, where the subscript represents the mass squared difference between the sterile mass state and the lowest active mass state. Since the sterile state is assumed to be eV-scale, this means that the other active states are all essentially degenerate ($\Delta m_{12}^2 \approx \Delta m_{23}^2 \approx \Delta m_{13}^2$). The 3+1 model also introduces two new phases, $\delta_{24}$ and $\delta_{14}$.

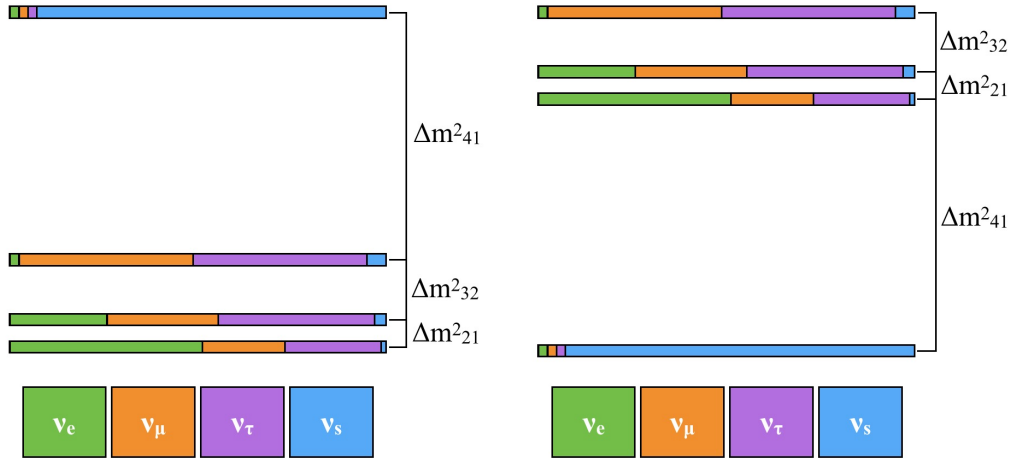

Figure 2-6: The mass splittings assuming a normal hierarchy for the active neutrinos. Left: A 3+1 model in which the sterile mass state is more massive than the active states. Right: A 1+3 model, where the sterile mass state is the least massive.

In a 3+1 model, the probability for finding a neutrino in flavor state $\beta$ after propagating a distance L in vacuum and being produced as a flavor state $\alpha$ is given by:

$$P_{\alpha \to \beta} = \delta_{\alpha\beta} - 4(\delta_{\alpha\beta} - U_{\alpha 4}U_{\beta 4}^*)U_{\alpha 4}^* U_{\beta 4} \sin^2(1.27 \frac{\text{GeV}}{\text{eV}^2\text{km}} \frac{\Delta m_{41}^2 L}{E}) \quad (2.24)$$

More generally we can include N sterile states, described as a 3 + N model. In this case, the oscillation probability is given [5] by:

$$P_{\alpha \to \beta} = \delta_{\alpha\beta} - 4 \sum_{ji} Re[U^*_{\alpha i} U_{\beta i} U_{\alpha j} U^*_{\beta j}] \sin^2(1.27 \frac{\text{GeV}}{\text{eV}^2 \text{km}} \frac{\Delta m^2_{ji} L}{E})$$
$$+ 2 \sum_{ji} Im[U^*_{\alpha i} U_{\beta i} U_{\alpha j} U^*_{\beta j}] \sin^2(2.54 \frac{\text{GeV}}{\text{eV}^2 \text{km}} \frac{\Delta m^2_{ji} L}{E}) \quad (2.25)$$

The degeneracy of the active mass states allows us to also write the 2-neutrino approximation:

$$P_{\alpha \to \beta} = \delta_{\alpha\beta} - \sin^2(2\theta) \sin^2(1.27 \frac{\text{GeV}}{\text{eV}^2 \text{km}} \frac{\Delta m^2_{ji} L}{E}) \quad (2.26)$$

Eq. 2.26 is important for pedagogical reasons. Notice that oscillation amplitude scales as $\sin^2(2\theta)$ and the oscillation frequency as $\Delta m^2_{ji} L/E$. This form is useful since at a given baseline and energy (L and E), the other two parameters will often be the parameters of interest in many sterile neutrino searches.

The anomalous oscillation measurements are found in $\nu_e$ disappearance, $P(\nu_e \to \nu_e)$, and $\nu_e$ appearance $P(\nu_\mu \to \nu_e)$ channels. There has yet to be anomalous measurement in $\nu_\mu$ disappearance, $P(\nu_\mu \to \nu_\mu)$. Assuming a two-flavor neutrino model, the $\nu_\mu \to \nu_e$ oscillation amplitude observed by LSND and MiniBooNE is $\sin^2(2\theta_{\mu e}) = 4|U_{e4}|^2 |U_{\mu 4}|^2$, indicating that both $U_{e4}$ and $U_{\mu 4} \neq 0$. The $\nu_e \to \nu_e$ observed by the reactor experiments and intense source measurements oscillation amplitude is $\sin^2(2\theta_{ee}) = 4 |U_{e4}|^2(1 - |U_{e4}|^2)$, which indicates that $U_{e4}$ must be sufficiently large. This in turn puts constraints on the allowed values of $U_{\mu 4}$. Finally, in a $\nu_\mu \to \nu_\mu$ experiment the oscillation amplitude is $\sin^2(2\theta_{\mu\mu}) = 4 |U_{\mu 4}|^2(1 - |U_{\mu 4}|^2)$ [64]. If we attribute the aforementioned anomalies to a 3+1 sterile neutrino model, $|U_{\mu 4}|^2 \neq 0$; there should be a measurable amount of $\nu_\mu \to \nu_\mu$, which has yet to be observed.

### 2.3.3 Null sterile oscillation measurements

It is important to recognize that although there have been several experimental anomalous measurements, there are many experiments that have performed sterile neutrino searches in the relevant parameter space and have reported no significant signal (called "null" experiments). These null measurements place strong constraints on the interpretation of the anomalies.

The experiment KARMEN [65] was the original experiment designed to test the $\nu_\mu \to \nu_e$ LSND signal. Similar to LSND, it used DAR pions as a neutrino source, however, at a shorter baseline. The experiment ruled out a large portion of the LSND allowed parameter space (particularly above $\Delta m_{41}^2 \approx 2$ eV$^2$), however at a lower confidence level because the intensity of the flux was lower than the LSND flux. The NOMAD [66] experiment was designed to search for $\nu_\mu \to \nu_\tau$ oscillations (through the $\nu_\tau$CC interaction, the final state contains an electron) and were also able to perform a $\nu_\mu \to \nu_e$ measurement. NOMAD found no signal and was particularly sensitive to $\Delta m_{41}^2 > 10$ eV$^2$. Similarly, the experiments OPERA [67] and ICARUS [68] has also reported limits on $\nu_\mu \to \nu_e$.

Bugey [69] was a $\overline{\nu}_e \to \overline{\nu}_e$ reactor experiment, consisting of three detectors positioned at different baselines from a 2800 Megawatt reactor. Bugey predated the reactor anomaly and observed no significant deviation from expectation at the 90% CL. At baselines shorter than 10 m, no significant deviation from expectation was also observed by PROSPECT [70], ILL [71], and STEREO [72]. It is interesting to note, that the reactor anomaly seems to affect experiments at >10 m baselines [73]. Recent results from DANSS [74] appear to favor a $\Delta m_{41}^2 \approx 1.4$ eV$^2$, however, they still assessing their systematic uncertainties. No evidence for $\overline{\nu}_e \to \overline{\nu}_e$ oscillations were reported by the recent NEOS/Daya Bay [75] measurement, and excludes $\sin^2(2\theta) < 0.1$ for 0.1 eV$^2 < \Delta m_{41}^2 < 2$ eV$^2$ at 90% CL.

Finally, experiments investigating $\nu_\mu \to \nu_\mu$ oscillations have all reported null results. Of particular interest for this analysis is the three previous searches performed by the IceCube collaboration. These will be described in detail in the Sec. 3.5. Measurements by Super-Kamiokande [76] using atmospheric neutrinos and MINOS [77] using neutrinos from accelerator decay-in-flight pions, have further restricted the allowed region of the parameter space. A combined analysis from MiniBooNE/SciBooNE [77,78] measuring $P(\nu_\mu \to \nu_\mu)$ also reported null results. As well as from early experiments, prior to the discovery of neutrino oscillation, that limit the allowed parameter space, such as from CDHSW [79] and CCFR [80].

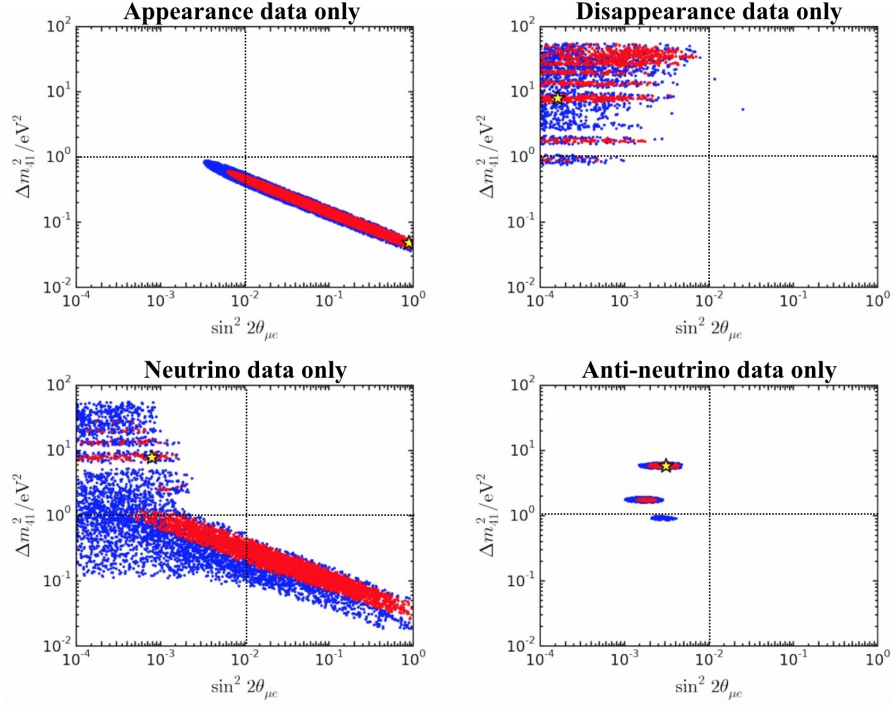

Figure 2-7: Frequentist confidence intervals for a $3+1$ model in terms of $\Delta m^2_{41}$ and $\sin^2(\theta_{e\mu})$. Red indicates the 90% CL and blue indicates the 99% CL allowed regions. Top left: Appearance data only. Top right: Disappearance data only. Bottom left: Neutrino data only. Bottom right: Antineutrino data only. Modified from Ref. [81].

### 2.3.4 Global fits to neutrino data

Data from many experiments can be used to constrain allowed regions for the 3+1 sterile hypothesis. An example of this is shown in the Fig. 2-7 [81]. Here, the allowed regions, i.e. not excluded at more than 90% C.L. and 99% CL, are shown in red and blue. Dashed lines are $\Delta m^2_{41} = 1\,\mathrm{eV}^2$ and $\sin^2(2\theta_{\mu e}) = 0.01$ are simply shown as a reference point between plots. One sees that there is tension between the allowed regions for appearance versus disappearance data

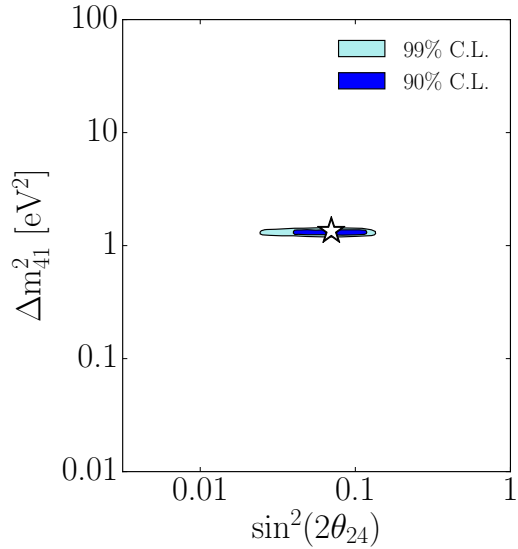

Figure 2-8: Frequentist confidence regions for a 3+1 model in terms of $\Delta m^2_{41}$ and $\sin^2(\theta_{24})$, for the 90% CL and 99% CL. The star indicates the best fit point at $\Delta m^2_{41} = 1.35\,\text{eV}^2$ and $\sin^2(\theta_{24}) = 0.07$. Modified from Ref. [81].

sets, and neutrino versus antineutrino data sets.

Of particular interest to this analysis is the global allowed region in terms of $\Delta m^2_{41}$ and $\sin^2(2\theta_{24})$. A recent analysis [73] including modern sterile neutrino measurements from DANSS, NEOS, PROSPECT, and an update from MiniBooNE, finds the global allowed region for a 3+1 sterile hypothesis shown in Fig. 2-8. The best fit sterile hypothesis point appears at $\Delta m^2_{41} = 1.35\,\text{eV}^2$ and $\sin^2(\theta_{24}) = 0.07$, and will be referred to throughout this thesis.

### 2.3.5 Neutrino oscillations in matter including a sterile state

Recall from Sec. 2.3.2 that sterile neutrinos do not interact weakly but do participate in the standard neutrino oscillation. Therefore, the active neutrinos have an additional matter potential due to NC interaction. This can be expressed in the Hamiltonian as an effective matter potential for the sterile neutrino states equal to the matter potential of NC interactions for active neutrinos with an opposite sign. Following the same logic presented in Sec. 2.2.2, the

Schrödinger equation is modified to accordingly:

$$i\frac{d}{dt}\begin{pmatrix}\nu_e\\\nu_\mu\\\nu_\tau\\\nu_s\end{pmatrix}=\frac{1}{2\text{E}}U\begin{pmatrix}0&0&0&0\\0&\Delta m_{21}^2&0&0\\0&0&\Delta m_{31}^2&0\\0&0&0&\Delta m_{41}^2\end{pmatrix}U^\dagger\begin{pmatrix}\nu_e\\\nu_\mu\\\nu_\tau\\\nu_s\end{pmatrix}+\sqrt{2}G_F\begin{pmatrix}n_e(x)&0&0&0\\0&0&0&0\\0&0&0&0\\0&0&0&-n_n/2\end{pmatrix}\begin{pmatrix}\nu_e\\\nu_\mu\\\nu_\tau\\\nu_s\end{pmatrix},$$

(2.27)

where $n_n$ is the neutron number density. For antineutrinos, we need to replace U→U* and $G_F \rightarrow -G_F$. Note that in electrically and isotopically neutral medium $n_n = n_p = n_e$, therefore it is common to see this expressed solely in terms of $n_e$.

The MSW resonance condition now reads:

$$E_{crit}^M = \frac{\Delta m^2 \cos(2\theta)}{\sqrt{2}G_F n_e} \approx \frac{\Delta m^2 \cos(2\theta)}{0.038(\rho[\text{g/cm}^3])}. \qquad (2.28)$$

If the resonance condition above is negative (either $\theta < \pi/4$ and a 3+1 sterile hierarchy, or $\theta > \pi/4$ and a 1+3 sterile hierarchy), the resonance occurs in the antineutrinos. Otherwise, the resonance condition applies to neutrinos. For neutrino flux passing through the Earth, this matter enhanced resonance can lead to a distortion of the observed energy and zenith angle distribution of atmospheric neutrino events [82], even for small mixing angles.

The density profile of the Earth has three main structures: the inner core ($\rho \approx 13$ g/cm$^3$), the outer core ($\rho \approx 11$ g/cm$^3$), and the mantle ($\rho \approx 4$g/cm$^3$). Density changes rapidly between the intersection of the three main structures. Neutrinos arriving at the detector with a cosine of the trajectory defined relative to the zenith angle, $\cos(\theta_z) > -0.45$, cross the mantle only, whereas for $\cos(\theta_z) < -0.8$, neutrinos cross six layers: mantle, outer core, inner core, and then back again. From Eq. 2.28, assuming an eV-scale sterile neutrino and small mixing angle, the matter effect induces an MSW active-sterile resonance flavor conversion for neutrino energies of $\sim 2$ TeV and $\sim 6$ TeV passing through the core and mantle respectively.

The change in Earth's density can also lead resonant like flavor transitions. Relating the periodic mantle-core-mantle density to a periodic potential, a parametric enhanced resonance [53, 83–85] or *Neutrino Oscillation Length Resonance-like* (NOLR) [86–89]) can occur. Similar to the MSW resonance condition, the energy corresponding to the resonance is proportional to the sterile mass-squared difference $\Delta m_{41}^2$. In Ref. [90] the parametric enhancement has been applied to $\nu_\mu \to \nu_s$ oscillations of atmospheric neutrinos. At $\cos(\theta_z) < -0.83$, neutrinos cross both the mantle and the core. The effect is different than the MSW effect and the enhancement happens in-between the MSW resonant energy found in the core and the mantle. The Earth mantle-core enhancement is due to the effect of maximal constructive interference between the amplitudes of the $\nu_\alpha \to \nu_\beta$, transitions in the Earth mantle and in the Earth core. It occurs when the neutrinos pass through a multi-layer medium of (non-periodic) constant density layers. This effect has a strong dependence on the neutrino energy.

We will refer to the above resonant effects as "matter enhanced oscillations." An example of this is shown in Fig. 2-9. This figure shows the $P(\overline{\nu}_\mu \to \overline{\nu}_\mu)$ (left) and $P(\nu_\mu \to \nu_\mu)$ (right) disappearance in terms of true neutrino energy and zenith angle for a sterile hypothesis of $\Delta m_{41}^2 = 1.3$ eV$^2$ and $\sin^2(2\theta_{24}) = 0.07$ (the global best fit point described in Sec. 2.3.4). Several features are present. Most notably, and relevant to this analysis, is the matter enhanced resonance observed in the $\overline{\nu}_\mu$ disappearance at approximately 2 TeV passing through the core of the Earth. This is due to the $\overline{\nu}_\mu$ state resonantly transitioning into a sterile state and results in the near to total depletion of the $\overline{\nu}_\mu$ flux in that region. Another feature illustrated by these plots is the high energy neutrinos attenuation due to the increase in the increased cross section at high energies. At low energies, the matter effects are less important and neutrinos exhibit vacuum-like oscillations.

It has been noted in Ref. [91] that the TeV-scale matter effects illustrated above reduce to vacuum oscillations at large $\theta_{34}$ values. However, constraints from the low energy atmospheric neutrino measurements [76, 92], eliminate this region of phase space for a 3+1 model.

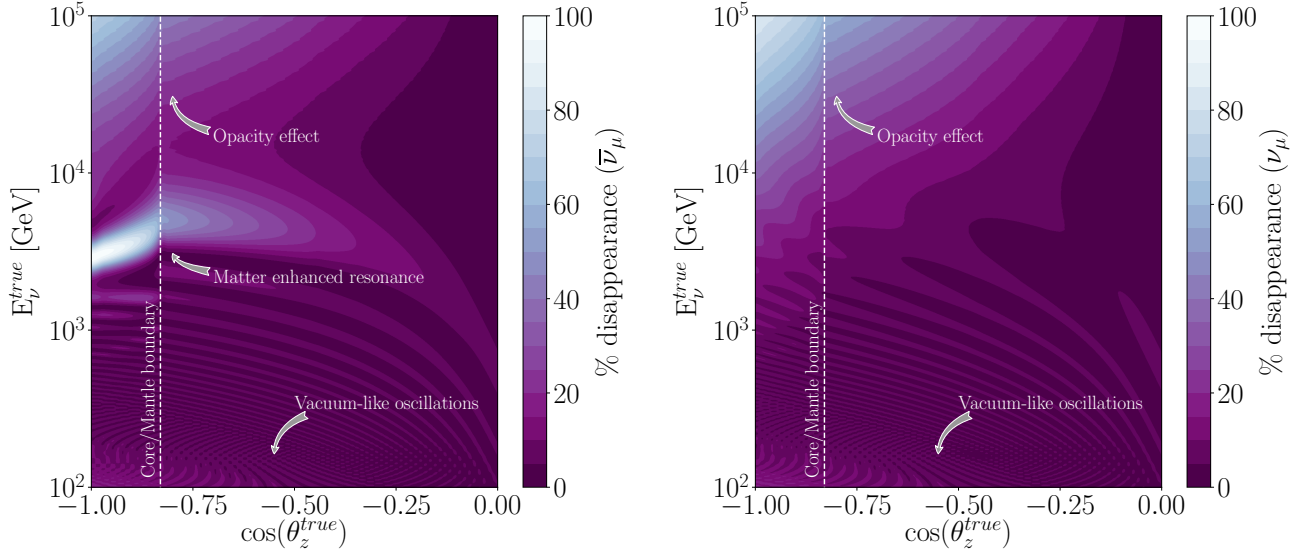

Figure 2-9: The change in the atmospheric spectrum for $\bar{\nu}_\mu$ (left) and $\nu_\mu$ (right) given a sterile neutrino 3+1 hypothesis of $\Delta m^2_{41}$ = 1.3 eV$^2$ and $\sin^2(2\theta_{24})$ = 0.07. Increasing the value of $\Delta m^2_{41}$ pushes the matter enhanced resonance to proportionally larger energies. Increasing the value of the mixing parameter $\sin^2(2\theta_{24})$, pushes the resonance away from the core and into the mantle.

## 2.4 High energy neutrino interactions with matter

Neutrinos interact exclusively via the weak nuclear force and it is only possible to detect them through the secondary particles produced in the interaction. The GeV and above scale neutrino-nucleon interactions are dominated by three processes: quasi-elastic (QE) scattering, resonant production, and deep inelastic scattering (DIS). Their relative contributions to the total cross section are shown in Fig. 2-10. In the energies related to this analysis, the DIS regime is particularly important and will be described in detail. Further information can be found in Refs. [50, 93].

The QE scattering is the dominant interaction for neutrino energy up to approximately 1 GeV and represents the coherent scattering off the nucleon as a whole rather than the individual partons (quarks and gluons). In light targets (H$_2$ or D$_2$), the CC interaction results in either a neutron converting to proton and lepton in the final state (e.g. $\nu_\mu$ + n → p + $\mu^-$), or a proton converting to a neutron and lepton (e.g. $\bar{\nu}_\mu$ + p → n + $\mu^+$). In the NC reaction, the target remains unchanged (e.g. $\nu + N \to \nu + N$). At energies between 1 and 10 GeV, resonant production dominates the total neutrino-nucleon cross section. This is the inelastic scattering off an individual nucleon via either a CC or NC interaction, producing an excited state, which subsequently decays often producing a single-pion final state. The final state often carries a large uncertainty due to secondary nuclear effects, such as pion re-scattering, charge exchange, and absorption.

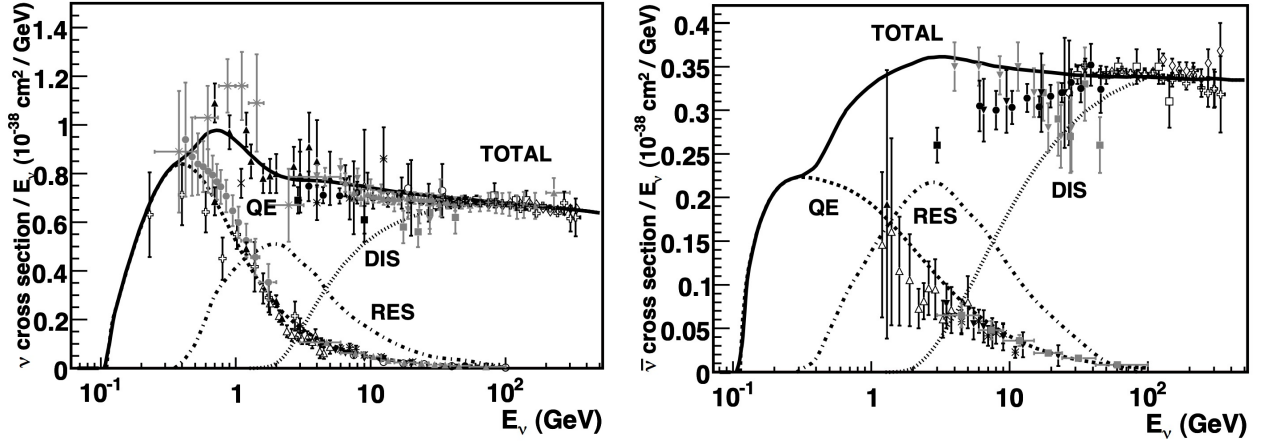

Figure 2-10: The total neutrino (left) and antineutrino (right) per nucleon CC cross sections, divided by the neutrino energy, as a function of neutrino energy. The total cross section is broken down into the three dominant interaction mechanisms relevant at these energies: quasi-elastic (QE) scattering, resonant production, and deep inelastic scattering. From Ref. [93].

Above 10 GeV, the dominant interaction is DIS and the neutrino momentum is large enough to resolve the internal structure of nucleons. DIS is the result of an incident neutrino scattering directly off a parton via the exchange of a virtual $W^{\pm}$ or $Z^0$ boson. In both the CC and NC interactions, the cross section increases with energy; rising linearly with energy up to approximately 3 TeV, then abating to $E_\nu^{0.3}$ due to the finite $W^{\pm}$ and $Z^0$ masses (see Fig. 2-11 (right)).

The generic form of the CC interaction is:

$$\nu_l(k) + N(P) \to l^-(k') + X(p'), \tag{2.29}$$

where the momenta of the different particles are shown in the parenthesis, N can either be a proton or neutron, and X represents set of final hadrons.

The cross section of DIS is commonly parametrized in terms of the dimensionless Lorentz invariant Bjorken scaling variables x and y, and $\nu$ defined by:

$$\text{x} \equiv \frac{Q^2}{2p \cdot q}, \qquad \text{y} \equiv \frac{p \cdot q}{p \cdot k}, \qquad \nu \equiv \frac{p \cdot q}{m_N}, \tag{2.30}$$

where $Q^2 \equiv -q^2$ ($Q^2 > 0$) is the momentum transfer, y is the fractional energy loss of the incoming particle: 1 - $E_{k'}/E+k$ and is often referred to as the inelasticity. The Bjorken scaling variables are therefore bound between [0,1). We are in the DIS regime when $Q^2 \gg m_N^2$ and $p \cdot q \gg m_N^2$. The Feynman diagram of the DIS interaction, along with the relevant variable definition is shown in Fig. 2-11 (left).

In the case of a CC interaction, DIS results in the production of a flavor conserving lepton and a jet of fragmented particles, consisting primarily of hadrons. On average, the secondary lepton carries 50% (for $E_\nu \approx 10\,\text{GeV}$) to 80% (at higher energies) of the neutrino energy [94]. The remainder of the energy is transferred to the nuclear target and released in the form of hadronic showers.

The neutrino and antineutrino DIS differential cross sections are given by [5]:

$$\frac{d^2\sigma_{CC}}{dxdy} = \sigma_{CC}^0 [xy^2 F_1^{W^\pm N} + (1-y) F_2^{W^\pm N} \pm xy(1-y/2) F_3^{W^\pm N}], \tag{2.31}$$

with

$$\sigma_{CC}^0 = \frac{G_F^2}{2\pi} s (1 - \frac{Q^2}{m_W^2})^{-2}, \tag{2.32}$$

where s is the Lorentz- invariant squared center-of-mass energy:

$$s = (k+P)^2 = m_N^2 + 2k \cdot P, \tag{2.33}$$

and $F_1^{W^\pm N}$, $F_2^{W^\pm N}$, and $F_3^{W^\pm N}$ are *structure functions*, which encode information about the structure of the nucleon, i.e. the momentum distribution of the quarks within the proton. The $\pm$ refers to the cross sections for $\nu$ and $\bar{\nu}$ scattering. The variables $m_W$ and $m_Z$ correspond to the mass of the mediating boson ($W^\pm$ or $Z^0$), and $G_F = 1.16632 \times 10^{-5}$ GeV$^{-2}$ is the Fermi constant.

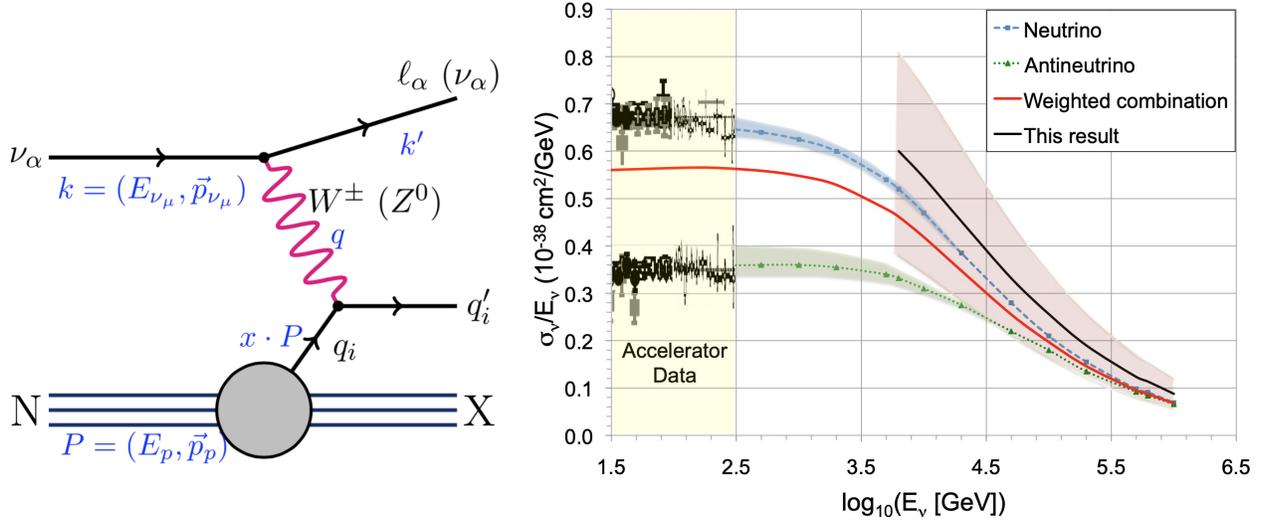

Figure 2-11: Left: The Feynman diagram for $\nu p$ interaction in the DIS regime. From Ref. [64]. Right: A compilation of neutrino charged current cross section measurements, divided by neutrino energy, from accelerator experiments in the sub-300 GeV range. The cross section grows linearly with energy up to approximately 1TeV and is approximately twice as small for antineutrinos as the neutrinos. The blue and green lines are the SM predictions for $\nu_\mu$ and $\bar{\nu}_\mu$ respectively, with the uncertainties on the deep inelastic cross sections shown by the shaded bands. The red line is for the expected mixture of $\nu_\mu$ and $\bar{\nu}_\mu$ in the IceCube sample. The black line shows the current result, assuming that the charged and neutral current cross sections vary in proportion, and that the ratio between the actual cross section and the SM prediction does not depend on energy. The pink band shows the total $1\sigma$ (statistical plus systematic) uncertainty. From Ref. [95].

In NC interactions, no charged lepton is present in the final state. Hence, the only way to observe them is through the shower from the nuclear remnant. The generic form of the NC interaction is:

$$\nu_l(k) + N(P) \rightarrow \nu_l(k') + X(p'). \qquad (2.34)$$

Nearly identical to the differential cross section in Eq. 2.31, except now with the exchange of a $Z^0$ boson:

$$\frac{d^2\sigma_{NC}}{dxdy} = \sigma_{NC}^0 [xy^2 F_1^{ZN} + (1-y)F_2^{ZN} \pm xy(1-y/2)F_3^{ZN}], \tag{2.35}$$

with

$$\sigma_{NC}^0 = \frac{G_F^2}{2\pi} s (1 - \frac{Q^2}{m_Z^2})^{-2}. \tag{2.36}$$

## 2.5 Sources of neutrinos

### 2.5.1 Atmospheric neutrinos: conventional and prompt

A continuous flux of high energy cosmic rays continuously bombards the Earth's atmosphere. While the production origin of the cosmic rays is still unknown, the bulk of the flux (below $E_k = 10^{15}$ eV) is thought to due to a Galactic source, from energetic arguments. At higher energies the flux is thought to transition to an extra-galactic composition above $E_k = 10^{19}$ eV. The majority of the cosmic ray flux (nearly 80% by mass) comes from ionized hydrogen (free protons) with the remainder primarily in form of helium nuclei (two protons and two neutrons) and trace amounts of heavier elements [96].

The differential energy spectrum of the cosmic ray flux follows a power law of the form

$$N(E)dE \propto E^{-\gamma}dE, \tag{2.37}$$

with $\gamma \approx 2.7$ below $10^6$ GeV. At around $10^6$ GeV, referred to as the "knee," the spectrum steepens to $\gamma \approx -3.1$ [97]. Recent high energy measurements indicate that the spectrum flattens at $10^9$ GeV, often referred to as the "ankle." Going beyond this, the flux extends up to $10^{11}$ GeV, at which point they loses energy from interactions with the cosmic microwave background (the GZK cutoff [98, 99]). This represents a theoretical upper limit on the energy of the cosmic ray protons.

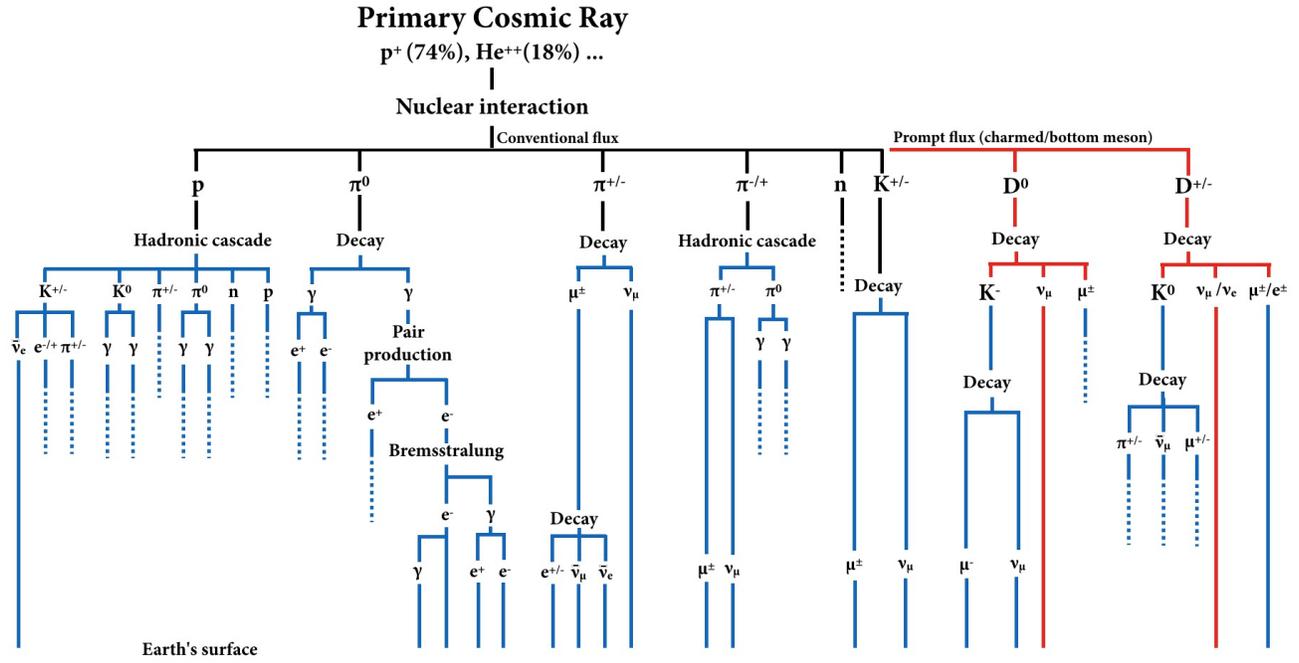

Figure 2-12: An illustration of the various decay and interaction channels through which the atmopheric flux is produced. The prompt neutrino chain is indicated in red. Modified from Ref. [1].

Various models are available to describe the cosmic ray flux. For example, HillasGaisser2012 (H3a) [100] is widely used. The Hillas model was formulated assuming the cosmic ray acceleration mechanism originates from three populations. The spectrum below the knee (E $\approx 10^{6.6}$ GeV) represents the particles accelerated by supernova remnants within the Milky Way. The higher energy components are composed of both a galactic component of unknown origin and an extragalactic component. T. Gaisser defined each population in terms of five groups of nuclei (p, He, CNO, Mg-Si, and Fe), which cut off exponentially at a characteristic rigidity [101]. Other models mentioned in this thesis are: Zatsepin-Sokolskaya/PAMELA [102, 103], Polygonato [104, 105], and Global Spline Fit (GSF) [106].

When a primary cosmic ray collides with a nucleus in the upper atmosphere, typically an oxygen or nitrogen molecule in the upper atmosphere (15-20 km [107]), the energies are often sufficient to break apart both or either of the primary particle or the target nucleus through a nuclear interaction. Much of the energy of the collision goes into producing short lived particles known as *mesons* [44, 108]. Simulation of the extensive air shower produced during the collision is often handled by a hadronic interaction model. An example of such a model is Sibyll2.3C [109]. This model was recently updated from the widely used Sibyll2.1 [110] to include data from fixed-target measurements from the LHC.

The most common mesons produced in the cosmic ray interactions are pions ($\pi^+$, $\pi^-$, $\pi^0$) and Kaon ($K^+$, $K^-$, $K^0$), and shorter lived mesons comprised of charged quarks (such as the D meson). An illustrative diagram of a cosmic ray interaction is shown in Fig. 2-12.

The charged pions have a mean life of $(2.6033 \pm 0.0005) \times 10^{-8}$s [50] producing a same charge muons and a $\nu_\mu$ (with the branching ratio of the decay in the parentheses):

$$\pi^\pm \to \mu^\pm + \nu_\mu(\bar{\nu}_\mu) \ ... \ (99.98770 \pm 0.00004\%) \tag{2.38}$$

Muons have a mass of 105.65 MeV and are also unstable particles with a half-life of $2.2 \times 10^{-6}$s. They decay to an electron and two neutrinos.

$$\mu^\pm \to e^\pm + \nu_e(\bar{\nu}_e) + \bar{\nu}_\mu(\nu_\mu) \ ... \ (100.0\%), \tag{2.39}$$

The neutral pions ($\pi^0$) have a much shorter mean lifetime of $(8.52 \pm 0.18) \times 10^{-17}$s, and have a much smaller chance of producing neutrinos ($\pi_0$ primarily decay to back-to-back gammas, with the neutrino decay channel being $\pi_0 \to \nu + \bar{\nu}$ having a branching fraction of $2.7 \times 10^{-7}\%$). Below approximately 80 GeV, pions are the dominant source of atmospheric neutrinos [111]. Above this, the kaons become the dominant neutrino source up to approximately $10^6$GeV. The relevant kaon decay channels through which a secondary neutrino flux is produced are [50]:

$$
\begin{aligned}
K^{\pm} &\to \mu^{\pm} + \nu_{\mu}(\bar{\nu}_{\mu}) \;...\; (63.56 \pm 0.11\%) \\
&\to \pi^{\pm} + \pi^{0} \;...\; (20.67 \pm 0.08\%) \\
&\to \pi^{\pm} + \pi^{+} + \pi^{-} \;...\; (5.583 \pm 0.024\%) \\
&\to \pi^{0} + \mu^{\pm} + \nu_{\mu}(\bar{\nu}_{\mu}) \;...\; (3.352 \pm 0.033\%) \\
&\to \pi^{0} + e^{\pm} + \nu_{e}(\bar{\nu}_{e}) \;...\; (5.07 \pm 0.04\%) \\
&\to \pi^{\pm} + \pi^{0} + \pi^{0} \;...\; (1.760 \pm 0.023\%) \\
K^{0}_{S} &\to \pi^{+} + \pi^{-} \;...\; ((69.20 \pm 0.05\%) \\
K^{0}_{L} &\to \pi^{+} + \pi^{-} + \pi^{0} \;...\; (12.54 \pm 0.05\%) \\
&\to \pi^{\pm} + \mu^{\mp} + \nu_{\mu}(\bar{\nu}_{\mu}) \;...\; (27.04 \pm 0.07\%) \\
&\to \pi^{\pm} + e^{\mp} + \nu_{e}(\bar{\nu}_{e}) \;...\; (40.55 \pm 0.11\%)
\end{aligned} \quad (2.40)
$$

The charged kaons, $K^{\pm}$, have a mean lifetime of $(1.2380 \pm 0.0020) \times 10^{-8}$s. They too preferentially decay to neutrinos as well (or to pions, which subsequently follow Eq. 2.40). The ratio of different flavored leptons at the surface, normalized to the muon neutrino flux is shown in Fig. 2-13. It is shown that the $\nu_{\mu}$ component dominates the neutrino flux up to approximately 1 PeV.

More massive mesons ($D^{\pm}$, $D_0$, $D_s$, $\Lambda_c$) may contain a charmed quark. Since these mesons are substantially heavier than the kaons and pions, they are short-lived (live times on the order $10^{-12}$ s) and decay "promptly," ergo giving the name: prompt neutrino flux. The critical energy for charm decay is $\sim 5 \times 10^{7}$ GeV, giving them the energy spectrum of the primary cosmic ray up to this energy. This makes their spectrum much harder spectrum than the conventional neutrino flux and only becomes dominant above $10^5$ to $10^{5.5}$ GeV [113]. The prompt neutrino flux has yet to be measured since their production is strongly suppressed, however various theoretical models exist accounting for modern understanding of charmed meson production. For example, the atmospheric prompt neutrino flux model, BERSS [114], is the successor to the standard prompt neutrino flux prediction from ERS [113], with updates to nuclear effects in the target, modern parton distribution functions, and new charm cross section measurements from the LHC and RHIC.

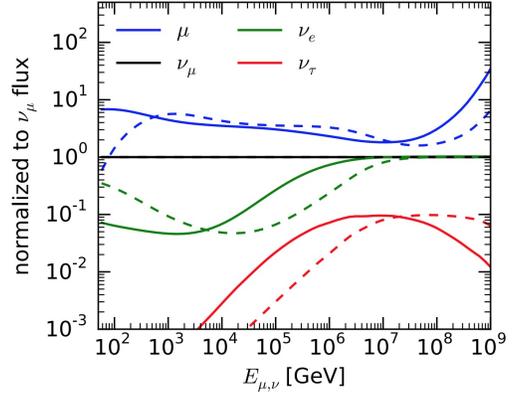

Figure 2-13: The lepton flavor ratio at the surface of the Earth, normalized to the muon neutrino flux. The calculation was performed using H3a primary flux and SIBYLL-2.3 RC1. The solid line indicates the flux for $\theta = 0°$ (perpendicular to the surface) and the dashed line is for $\theta = 90°$ (parallel to the surface). From Ref. [112].

### 2.5.2 Astrophysical neutrinos

Astrophysical neutrinos are thought to originate at or near the astrophysical accelerators that produce the high-energy cosmic rays. A recent observation suggests that one source of high-energy astrophysical neutrinos could be active blazars [115]. Other potential candidates are: supernovas, black holes, pulsars, active galactic nuclei and other extreme extra-galactic phenomena. Similar to the atmospheric neutrino production, astrophysical neutrinos are thought to be produced from high-energy cosmic rays interacting with local and interstellar particles creating charged mesons, which subsequently decay to neutrinos. The interaction may occur hadronically with protons or neutrons (following the description of the interaction producing atmospheric neutrinos) or more exotically with photons:

$$\begin{aligned} p + \gamma \rightarrow \Delta^+ &\rightarrow p + \pi^0 \\ &\rightarrow n + \pi^+ \end{aligned} \quad (2.41)$$

This interaction produces a short-lived Delta baryon, decaying to a proton of lower energy that the primary proton, or a neutron and charged pion. This is the resonant interaction which produces the GZK cuttoff mentioned in the previous subsection.

From the decay of the charged pions and kaons (Eq. 2.38 and Eq. 2.40), through the decay of the muon (Eq. 2.39), we expect the number of $\nu_\mu$ to be twice that of the $\nu_e$ (at the source). However, at the detection point, neutrino oscillations on astrophysical length scales is expected this to wash out the original flavor state, such that we expect equal numbers of each neutrino flavor. Based on the observation of the astrophysical neutrino flux [92], we expect the astrophysical neutrino flux to be the dominate neutrino source above approximately 100 TeV. As indicated in the discussion regarding high-energy neutrino interactions (Sec. 2.4), the neutrino CC interaction cross section grows with energy. This increases the interaction rate, particularly in the dense core of the Earth, of the high energy neutrinos, thus attenuating the flux. The astrophysical flux energy spectrum is often modeled as a single power law, with an isotropic distribution.

THIS PAGE INTENTIONALLY LEFT BLANK

# Chapter 3

# The IceCube experiment

In 1996, the Antarctic Muon And Neutrino Detector Array (AMANDA) collaboration demonstrated that one could use the natural ice at the US Amundsen-Scott South Pole station (located in Antarctica) as a Cherenkov medium for detecting high energy neutrino interactions. The AMANDA detector was the first neutrino telescope with an effective area greater than 10,000 square meters [51] and its successor, AMANDA-II proved the technology to be expandable and affordable. AMANDA-II introduced an order of magnitude more photomultiplier tubes (PMTs), and paved the way for the next order of magnitude expansion: the largest neutrino detector ever built, IceCube.

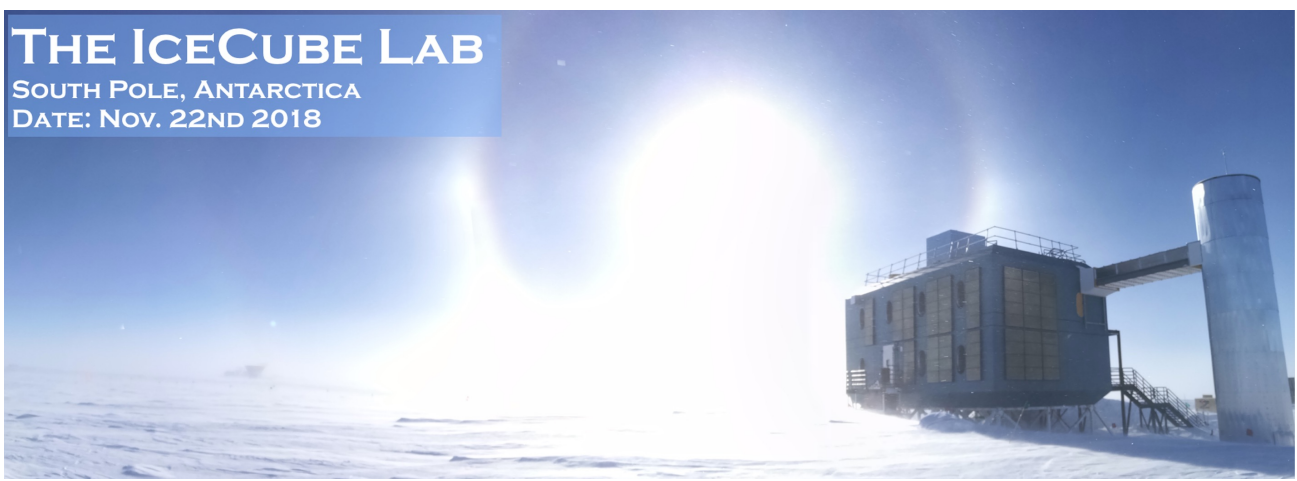

Figure 3-1: A photo of the the IceCube Lab taken during the 2018 austral summer. The IceCube detector extends radially outwards, approximately 2km below the surface.

Fully operational in 2011, the IceCube Neutrino Observatory has already made significant scientific discoveries and published results that span a broad range of scientific studies. These include the first evidence for astrophysical neutrinos from outside our local galactic group [92], the first astrophysics multi-messenger with neutrinos [115], world leading precision neutrino oscillation measurements [116–119], competitive limits on dark matter annihilation cross-sections [120], glaciology [121], and leading limits in Lorentz-violation and BSM searches [122]. Beyond this, IceCube is in a unique position to search for eV-scale sterile neutrinos. While the anomalous oscillation measurements (described in Sec. 2.3.1) hint at the potential existence of sterile neutrinos, all search for anomalous behaviors were performed at relatively low baselines and energies. IceCube is the only detector capable of exploring the same interesting area of parameter space using an entirely unique method, at vastly different energies and baselines, with completely different systematics. This has been previously been shown to provide world leading limits in the favored sterile neutrino hypothesis region [123].

## 3.1 The in-ice detectors

The IceCube Neutrino Observatory [126, 127] is a cubic-kilometer sized array of 5,160 PMTs buried in the Antarctic ice sheet designed to observe high-energy neutrinos interacting with the ice [128]. As of 2011, the IceCube collaboration completed the installation of the main IceCube detector consisting of 78 vertical strings of PMT modules, 125 m apart in a hexagonal grid, and the low-energy infill, DeepCore, consisting of a more densely arranged array of 8 strings [129] (see Figs. 3-2). Each string in the detector contains 60 digital optical modules (DOMs), each of which house a single down-facing PMT as well as all required electronics [130]. In strings 1 through 78, the DOMs extend from 1450 m to 2450 m, measured relative to the surface of the ice sheet, and have a vertical spacing of 17 m. The lower 50 DOMs on the DeepCore strings (79 to 86) are positioned in the clearest ice near the bottom of the detector and have a vertical spacing of 7 m and extend from 2100 m to 2450 m below the surface. The upper 10 DOMs on these strings form a cosmic ray muon veto and have a vertical spacing of 10 m (visible in the left side of Fig. 3-2). The veto is positioned just above the "dust layer," extending from 1900 m to 2000 m below the surface.

DOMs will be referred to as DOM (i, j), where i the string number and j is the optical module (OM) number. The OM numbers increase going deeper into the ice. Due to the ice properties where DeepCore is located, the denser instrumentation, and higher quantum efficiency PMTs, the energy threshold of DeepCore is approximately one order of magnitude lower ($\sim$10 GeV) than the IceCube detector ($\sim$100 GeV).

Beyond the in-ice detectors, there exists a surface array, IceTop, consisting of 81 stations located just above the in-ice IceCube strings. Each station consists of pairs of tanks separated by approximately 10 m. The tanks are filled with frozen water and each contain two down-facing DOMs. The primary goal of IceTop is to study the cosmic ray composition over a wide range of energies, however, it is also used as a surface veto for IceCube. IceTop may only be mentioned in passing, as it is not used directly in this analysis.

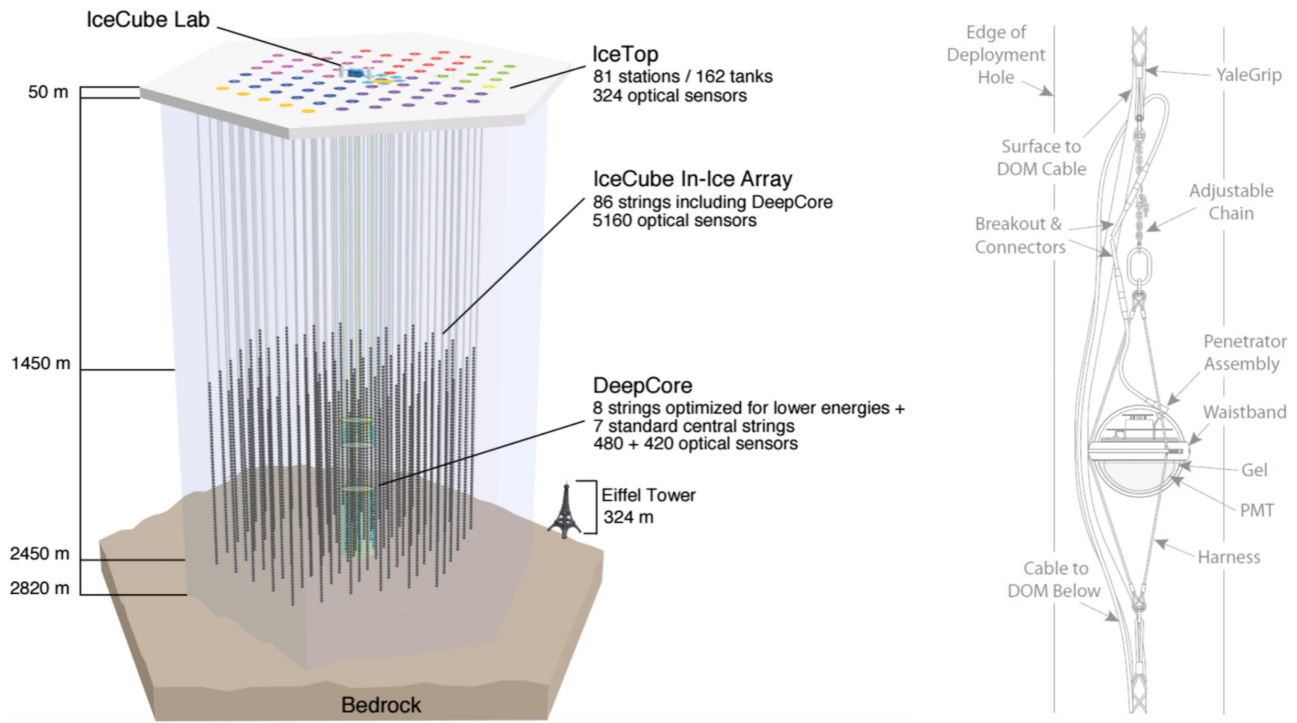

Figure 3-2: Left: A diagram of the IceCube Neutrino Observatory consisting of the two in-ice detector arrays: IceCube and DeepCore, as well as the surface array IceTop. The Super-Kamiokande [124] detector in Japan, and SNO [125] detector in Canada, are shown for scale. Right: The in-ice cable assembly for the DOMs.

### 3.1.1 The photomultiplier tubes

The majority of the DOMs consists of a 0.5"-thick spherical borosilicate glass pressure vessel that houses a single down-facing 10" R7081-02 PMT from Hamamatsu Photonics [131] and filled with dry nitrogen. The PMT is most sensitive to wavelengths ranging from 300 nm to 650 nm, with peak quantum efficiency of 25% near 390 nm [132]. Each PMT is optically coupled to the glass housing with optical gel and is surrounded by a wire mesh of mu-metal to reduce the effect of the ambient Earth's magnetic field. The glass housing is transparent to wavelengths 350 nm and above [133]. The effective photocathode area of an average PMT is approximately 550 cm$^2$.

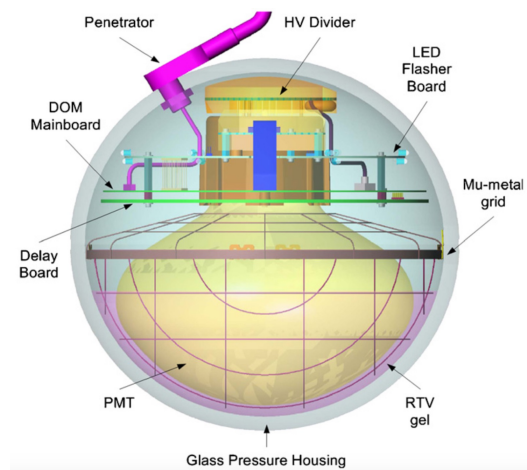

Figure 3-3: A schematic representation of the DOM including all of the components. From Ref. [130].

IceCube has also deployed 399 DOMs with Hamamatsu R7081-02MOD PMTs which are classified as high-quantum efficiency (HQE) DOMs [129]. They have a peak quantum efficiency of 34% near 390 nm, or equivalently, 36% higher efficiency than the Standard QE DOMs . These are primarily located in DeepCore and a few located on strings 36 and 43 as well, as shown in the left side of Fig. 3-4.

The R7081-02 and R7081-02MOD PMTs have 10 dynode stages and are operated with a gain of $1 \times 10^7$, corresponding to a high voltage of 1215 $\pm$ 83 V and 1309 $\pm$ 72 V respectively (see Fig. 3-5). The PMTs operate with the anodes at high voltage, therefore the signal is AC coupled to the front-end amplifiers. There are two versions of the AC coupling circuit, both of which use custom designed bifilar-wound 1:1 toroidal transformers. These are referred to as the *new* and *old* toroids, and their DOM specific positions are shown in the right side of Fig. F-1).

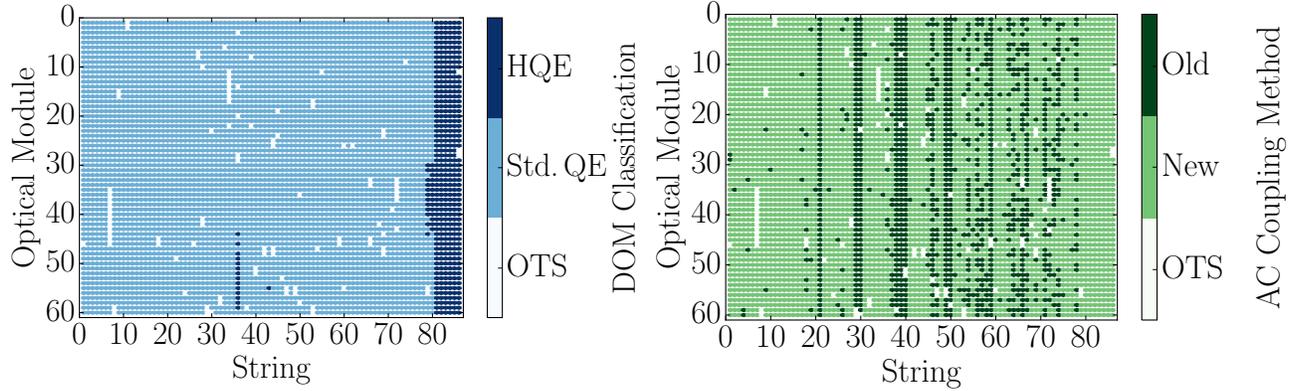

Figure 3-4: Left: The location in the detector containing the HQE (dark blue) and standard QE PMTs (light blue). Right: The DOMs instrumented with the old toroids (dark green) and new toroids (light green). Dead DOMs are shown in white.

The first 1129 DOMs were manufactured with 18 bifilar turns on a ferrite toroid core (the old toroid), which had a time constant of roughly $1.5\mu s$ at -30°C [130]. The time constant of the transformer pass the high-frequency components of the signals with negligible loss, but lead to a droop after large amplitude signals. The short time-constant transformer used in early DOM production was later replaced with one that produces less distortion and clipping. The updated DOMs, the later 4031 DOMs, were manufactured with a larger ferrite core and more windings (the new toroid), which reduced the recover time.

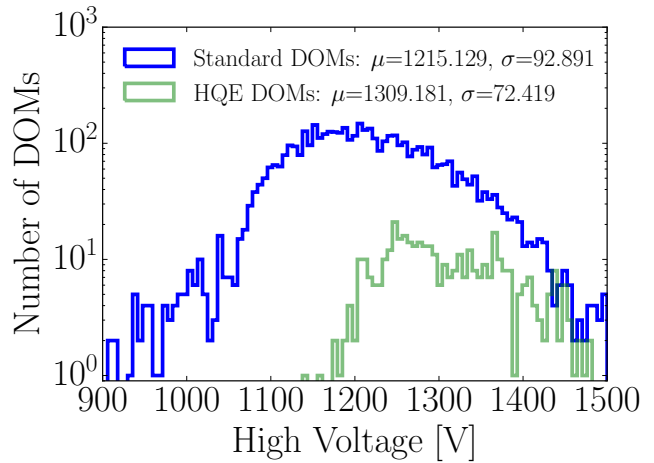

Figure 3-5: The distribution of the high-voltage in the detector separated between the HQE DOMs and the standard DOMs.

## 3.2 Data extraction

This section will describe the full data readout from the individual DOMs. This is schematically illustrated in Fig. 3-6.

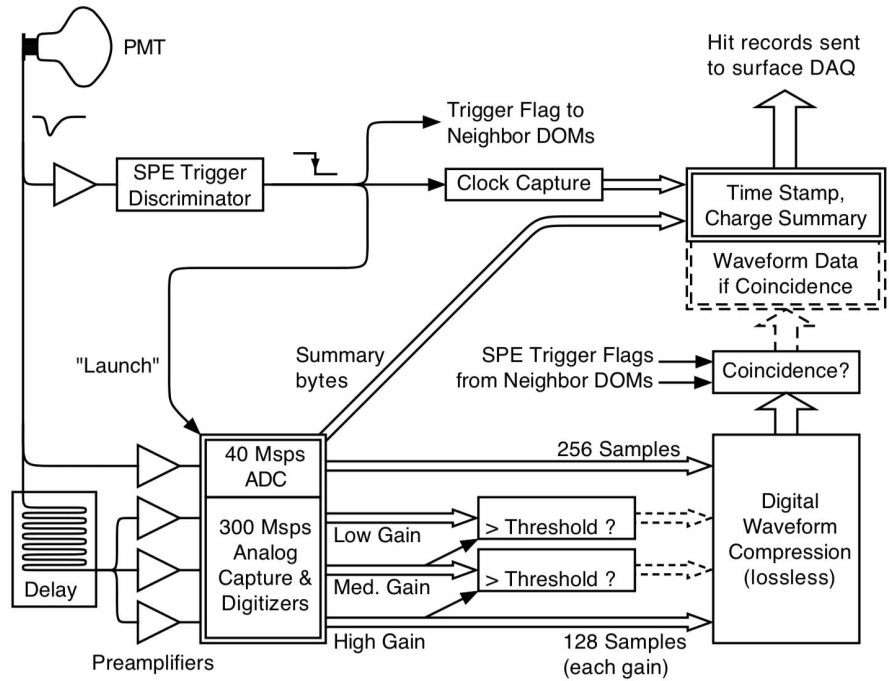

Figure 3-6: Data flow diagram for recording and processing of PMT waveforms in the DOM to form "Hit Record" that are sent to the surface data acquisition computers. From Ref. [134].

### 3.2.1 The main board readout electronics, signal digitization

When one or more photons produce a charge sufficient to trigger the on-board discriminator, the signal acquisition process is triggered. The DOM main board (MB) contains two built-in high-speed comparator circuits, each operating with a different threshold. The high-resolution comparator is used as a low charge trigger, with a resolution of 0.0024 PE/DAC count, and is currently set to trigger the data acquisition when the pulse peak voltage exceeds approximately 1.28 mV (or equivalently 0.2325 PE). This threshold was selected to avoid accidentally triggering on electronic noise. The low-resolution comparator is used for multi-PE (MPE) events and has a resolution that is coarser by a factor of 10 [130].

Once the DOM is triggered, this creates a "launch" condition and the signal is feed into four parallel input channels. Three of the channels first pass through a 75 ns delay loop in order to capture the leading edge of the pulse, then into three wide-band amplifiers each operating at a different gain: $15.7 \pm 0.6$, $1.79 \pm 0.06$, and $0.21 \pm 0.01$ [134]. The output of each amplifier feeds into separate inputs of a custom designed high-speed (300 MSPS for 128 samples, corresponding to a $\sim$427 ns readout) 10-bit Analog Transient Waveform Digitizer (ATWD). The different levels of amplification provide a large dynamic range to take into account single photons as well as very bright events. The highest gain channel is used for single photon detection and the channels with progressively lower gains are read out when the preceding channel has saturated. A second ATWD chips is included in parallel configuration with the first and operates in a ping-pong fashion to remove dead-time associated with the readout. The signal to the fourth channel is first shaped and amplified, then feed into a 10-bit fast Analog-To-Digital converter (fADC) operating at a sampling speed of 40 MSPS (for 256 samples, or correspondingly 6.4 $\mu$s). The amplification on this channel is similar to that of the high-gain ATWD channel.

Two example readout window are shown in Fig. 3-7, where the values from the ATWD channel with the highest amplification (ATWD0, blue) and fADC (green) are shown. At a nominal PMT gain of $1 \times 10^7$, a typical amplified single photoelectron will generate a $\sim$5.5 mV peak voltage with a full-width half max (FWHM) of $\sim$13 ns after pulse shaping.

### 3.2.2 Trigger conditions and information transmission

Individual DOMs are capable of determining local coincidence events through the cable network connecting the DOMs to the surface. When a DOM and its nearest or next-to-nearest neighbor observe a discriminator threshold crossing within a configurable time window (currently set to 1$\mu$s), it is referred to as a *Hard Local Coincidence* (HLC). When only a single DOM triggers, without seeing a local coincidence signal from one of its neighbors, the data is read out as a *Soft Local Coincidence* (SLC).

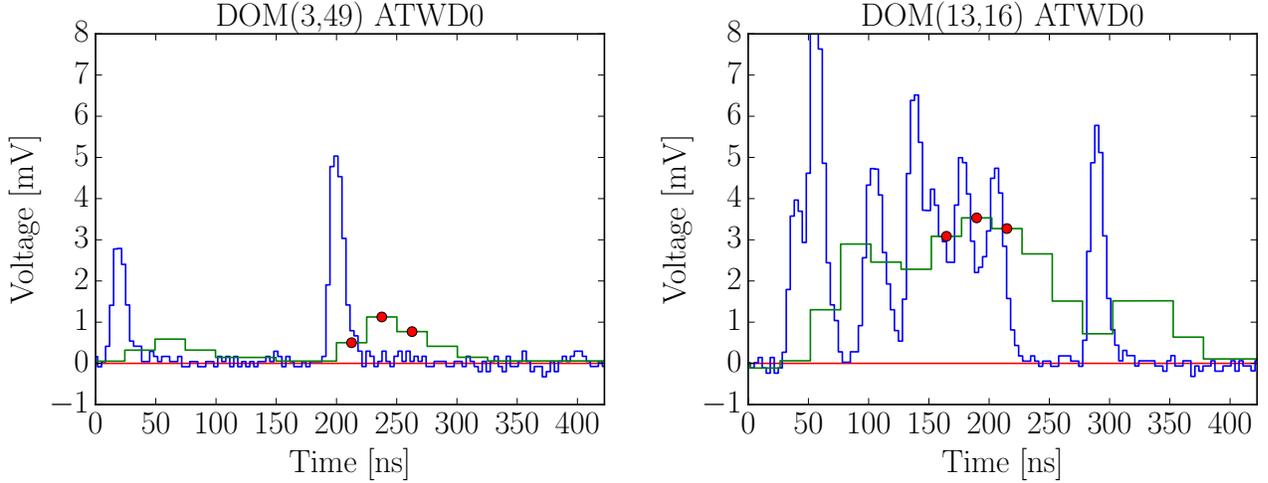

Figure 3-7: Two HLC waveforms showing the digitized signal from ATWD0 and the fADC. The three red dots represent what would have been read-out if these were SLC events.

Whenever either an HLC or SLC is satisfied, a *Hit* is generated. The Hit represents the fundamental unit of information transmitted from the in-ice detectors to the surface. It consists of a time stamp of the trigger, as well as the waveform readouts. Depending on the trigger condition, different information is transmitted. In the case of an SLC, only the highest fADC value and two adjacent sampling bins are readout with time stamps (see, for example, the bins with the red dots in Fig. 3-7). In the case of an HLC, the full waveform from the three ATWDs and the fADC are read out (blue and green waveform in Fig. 3-7, respectively).

The DOM stores approximately 1 second of data before transmitting the block of information to the surface, however, 16MB of onboard SDRam allows a single DOM to save up to 10 seconds of data in-case of an interruption.

The surface computers located in the ICL coordinate the received information and assemble the waveforms into events through the data acquisition software pDAQ. A variety of algorithms exist in the data acquisition software that group the waveforms into *triggers*. These are designed to collect and group periods of data of different levels of activity. An example is the Simple Majority Triggers (SMT). In particular, the SMT-8 trigger slides a 5 $\mu$s time window along the time ordered series of waveforms. If greater than 8 HLC hits are located in that time window, the trigger condition is satisfied and the data from -4$\mu$s to +6$\mu$s is sent to the high-level trigger filtering.

An event that passes through the trigger system is then subject to *filters* for classifying events. The different filters are designed to organize the triggers into different categories with various physics goals. The filters run on the Processing and Filtering (PnF) system and typically perform rudimentary reconstruction. When an event passes any filter, it is deemed to be important and worth further analysis. Since IceCube is located at the South Pole, real-time communication can only be provided by geosynchronous satellites, which have limited bandwidth available (100 GB/day). An event that passes one of the filters, is compressed and put in queue for satellite transmission to the IceCube Data Warehouse in Madison, WI. Filters are chosen to only run on a subset of the data. Each filter has a *prescale* which defines how often the filter should be run on triggers. A prescale of N means that the filter will run on 1/N randomly selected events. For example, the MinBias filter has a prescale of 1/1000, which means that 1/1000 randomly selected will pass through the MinBias filter. It should be noted, that all events in the detector are locally saved at the South Pole, and transported north for archival storage at the end of each austral winter.

An example of a filter used in this analysis is the online Muon Filter [135]. It is specifically designed to select muon-like events passing through the detector in any given direction, however, the selection varies as a function of angle from the zenith. It requires that down-going events trigger more than 10 DOMs, where as up-going events need to only trigger 8 DOMs. The filter preferentially extracts high-energy muons by requiring that the event deposited sufficient amount of charge in the detector.

Once the data is received in the North, it passes through the offline filters which perform much more sophisticated reconstruction, and is then sent on to analyzers to perform their independent event selections that are of interest to their specific analyses.

## 3.3 Calibration

The basis for all event reconstruction is derived from two main quantities recorded by IceCube: the pulse charge and time. Global time calibration across the detector is provided by the RAPCal procedure and runs during data-taking. The charge calibration is performed through the DOMCal routine, which is run at the beginning of each IceCube season. A description of these two calibration routines is found below.

### 3.3.1 Extracting timing information

The Reciprocal Active Pulsing Calibration (RAPCal) software represents a procedure for synchronizing the on-board DOM real time clocks with the ICL clocks. This system is able to accurately time stamp events down to nanosecond precision. The ICL is then able to convert the event time stamp to Coordinated Universal Time (UTC) for global comparison. The Master Clock keeps track of the absolute time of the ICL through the global positioning system (GPS). It then distributes 20 MHz bipolar square waves (and a pulse per second) through a fanout, which connects to each DOM Hub (a central computer used to power and communicate with the DOMs on a string). The DOM Hubs distribute packets of the bipolar signal to the individual DOMs. The packet is received and time stamped by the DOM according to its local time, then generates an identical 20 MHz packet which is sent back to the surface. The ICL receives the new packet and adds its own time stamp. The difference between the measured time stamp of the DOM compared to the ICL is then used to calculate the timing offset of each DOM. This method is suitable, since it does not rely on prior knowledge of the cable length. The calculated delay from the cable is used to monitor the timing stability and found to have an uncertainty of approximately 0.6 ns [134].

The time calibration through the entire data acquisition and software processing chain can be verified using the LED flashers. The LED flasher board is connected to the main board via a set of stand-offs, as shown in Fig. 3-3. This extension to the main board serves to provide an in-ice light source primarily used for calibration. The flasher board contains 12 independently operated LEDs arranged in pairs and space every 60°. One LED of each pair is positioned such that it is angled ~48° upwards, while the other is orientated to point radially outward from the DOM axis. The dimmest stable light pulse that can be achieved contains approximately $257 \times 10^6$ photons in a 30 ns long pulse [136]. They are used to stimulate and calibrate distant DOMs, simulate physical events, and to investigate optical properties of the ice.

### 3.3.2 Extracting charge information

The extracted waveforms from either the ATWDs or the fADC are sent through pulse deconvolution software called WaveDeform [137]. This software is used generate a *pulse series*. It operates by unfolding the waveform in terms of basis functions that represent the shape of a scalable average single photoelectron. The shape of the basis function depends on the front end electrons (i.e. the version of AC coupling). The basis functions are fit to the waveform through a minimized $\chi^2$ that extracts the best fit time and pulse height. The charge of the pulse is determined by finding the area of the fitted pulse, divided by the front-end impedance (43Ω for the original version of AC coupling and 50Ω for the new version of AC coupling), and further scaled by the PMT gain.

The gain on the PMT is determined through the extraction of the single photoelectron charge distribution probability distribution function (SPE charge template [138,139]). The SPE charge template is modeled as the sum of two exponentials and a Gaussian, explicitly:

$$f(q)_{\text{SPE}} = \text{E}_1 e^{-q/\text{w}_1} + \text{E}_2 e^{-q/\text{w}_2} + \text{N} e^{-\frac{(q-\mu)^2}{2\sigma^2}}, \qquad (3.1)$$

where $q$ represents the measured charge; $\text{E}_1$, $\text{E}_2$, and N are normalization factors; $\text{w}_1$ and $\text{w}_2$ are the exponential decay widths; and $\mu$, $\sigma$ are the Gaussian mean and width, respectively. The nonimal gain ($10^7$) on each PMT is defined such that 1 PE corresponds to the location of the Gaussian mean.

A significant effort was devoted in this analysis to re-calibrating the SPE charge templates for all DOMs and will be shown to dramatically improve the overall data/MC agreement in low-level variables (see Appendix F). A full description of the extraction of the SPE templates can be found in Refs. [139, 140], as well as a brief description in Sec. 4.2.3.

### 3.3.3 A description of the South Pole ice

**Bulk ice properties**

The optical properties of the ice affect photon propagation. Ref. [141] concludes that the in-ice scattering at the depths of the IceCube detector is almost entirely due to the presence in the ice of insoluble mineral dust grain, sea salt grains, and possibly liquid acid droplets deposits in snow as aerosols and subsequently compressed into the growing ice sheet. At depths shallower than 1400 m, the scattering is dominated by bubbles, however below this (in the region of the IceCube detector), the immense pressure of the ice causes the bubbles in the ice to collapse and form an air clathrate hydrate [142]. The clathrates have a similar index of refraction as the surrounding pure ice and do not increase the in-ice scattering. Since the ice was formed through stratification, the scattering and absorption is often described as a function of depth. Fig. 3-8 gives a graphical overview of our modern understanding of the ice, including the scattering (orange) and absorption (blue) coefficients at 400 nm along with the depth dependent age of the ice.

The undisturbed glacial ice, or "bulk ice," was originally studied using lasers [143] and later with a dust-loggers [144]. Modern studies on the optical properties of the ice involve using the built-in LED flashers (see Sec. 3.3.1). These have since replaced the earlier measurements, allowing us to probe the ice properties over the entire detector rather than near a single bore hole. Recent analyses using the LED flashers data have shown that the vertical stratification throughout the detector is not at a uniform depth (referred to as "tilt" [145]). Thus, the ice is instead defined in terms of ice layers of similar optical properties. The tilt direction (south-west) was found to the perpendicular to the ice flow direction (north-west). The flow of the ice was studied in Ref. [146]. It was shown that the ice at bedrock was essentially stationary and the ice approximately -2000m upwards had a horizontal velocity of approximately 9 m/year. The shear of the ice is thought to have produced an azimuthal anisotropy [147], in which the scattering length in the direction of flow is smaller than that of in the direction of the tilt (i.e. photons travel further before scattering in the direction of flow).

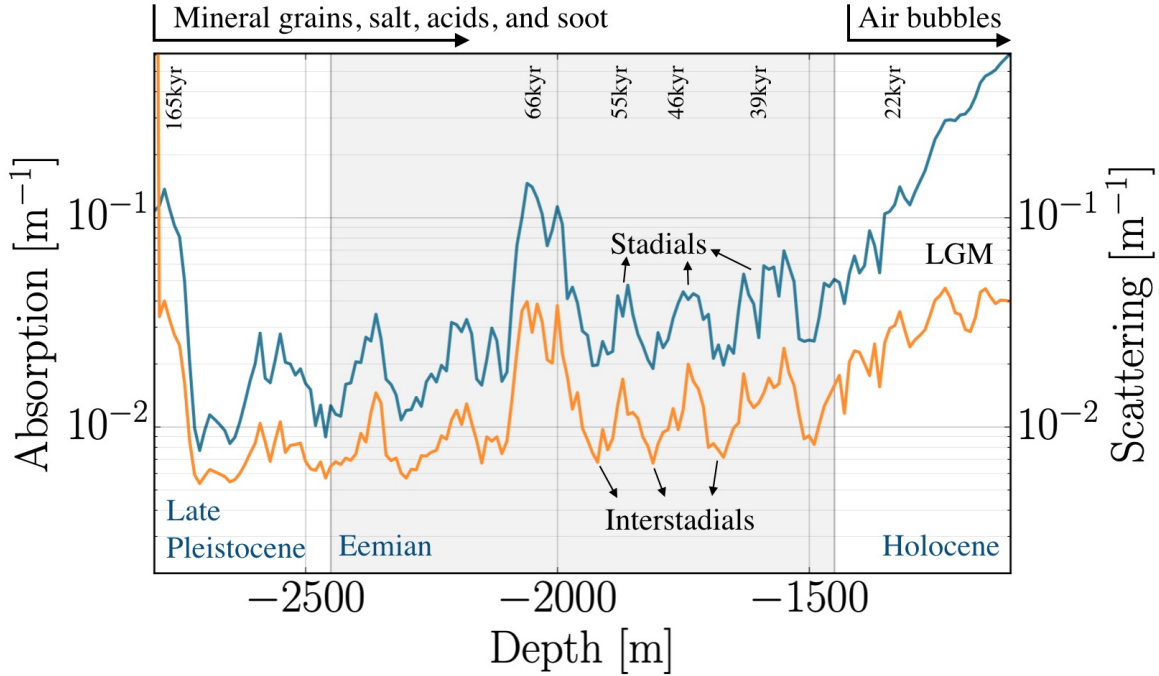

Figure 3-8: The absorption (blue) and scattering (orange) lengths at 400 nm as a function of depth. The values were extracted from the Spice3.2 ice model. The IceCube detector extend between the gray region. The large changes in both scattering and absorption have been correlated to periods of warmer (interstadials) and colder (stadials) climate. These known dates of this periods are shown at the top of the plot.

The effort of further refining the bulk ice model is ongoing as the calibration improves. The latest release of the bulk ice model used for simulation and event reconstruction is known as Spice3.2. This is the successor to the SpiceMie ice model described in Ref. [145]. It describes the ice in terms of wavelength dependent scattering and absorption coefficients as a function of layer (170 layers from across the extent of IceCube), along with a description of the tilt and anisotropy along a major and minor axis [148]. The wavelength dependent scattering and absorption coefficients used in Spice3.2 are shown in Fig. 3-9.

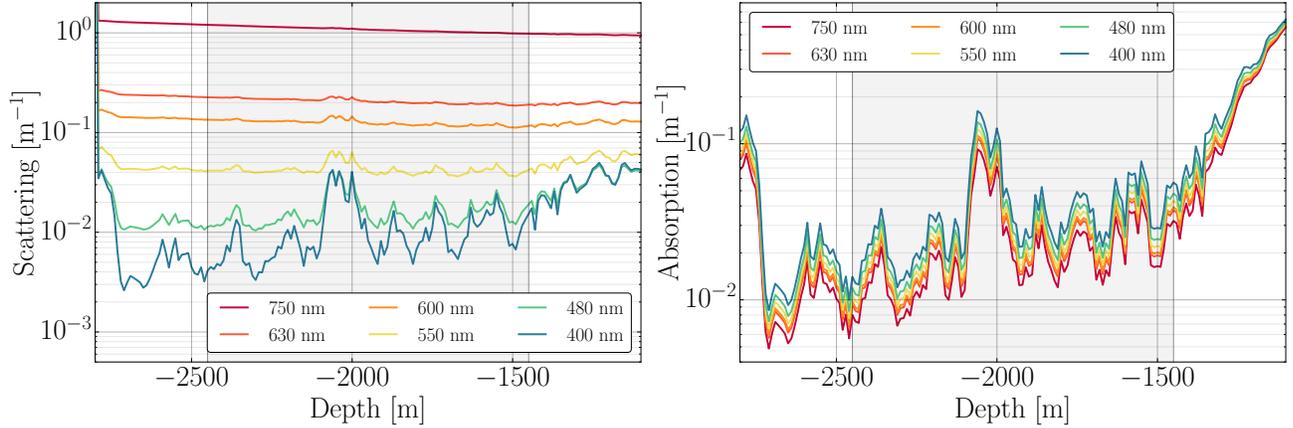

Figure 3-9: The wavelength depending scattering (left) and absorption (right), at string 63 in the IceCube detector, as a function of depth. The shaded region corresponds to the depth of IceCube. The large increase in both the scattering and absorption coefficients from -2000 m to -2200 m is internally known as the "dust layer."

**Hole ice properties**

The deployment of the IceCube strings required melting 60 cm diameter vertical columns of ice down to approximately 2500 m below the surface. These bore-holes remained water-filled while deploying the strings in order to sustain the pressure needed to avoid the melted column from collapsing. As the bore-holes refroze, the impurities and bubbles were forced into the center. A video recording from one of the in-ice DOMs of the hole ice refreezing processes indicated evidence for the formation of a narrow bubble column, 5-10 cm in diameter. It is this column of disturbed ice that is called the "hole ice." We model the effect of the hole ice as an angular sensitivity of the DOM. This will be discussed further in Sec. 7.1.3.

## 3.4 Particle interactions and event topologies in Ice

### 3.4.1 Muon energy loss in matter

The average stopping power for high-energy muons in matter can be described by:

$$-\langle \frac{dE}{dx} \rangle = \alpha(\text{E}) + \beta(\text{E})\text{E}, \tag{3.2}$$

where coefficients $\alpha(\text{E})$ is the electronic stopping power and $\beta(\text{E})$ is the radiative stopping power. Both $\alpha(\text{E})$ and $\beta(\text{E})$ depend weakly on the muon energy above 1 GeV. In ice, for energies between 20 and $10^{11}$ GeV, the values $\alpha = 0.260$ GeV m$^{-1}$ and $\beta = 0.357 \times 10^{-3}$ m$^{-1}$ agrees with measurements to within $\sim 6.6\%$ [149].

Below the critical energy – where electronic losses dominate over the radiative losses ($\sim$400 GeV for muons) – the energy loss is due to ionization (breaking electromagnetic bonds) and excitation (raise the electron to a higher-lying shell within the absorbed atom) of the incident particle and can be described through the Bethe Bloch formula [150–152]. This region is particularly interesting since the energy loss rate is nearly constant with an average energy loss rate of 2.2 MeV/cm in ice over many orders of magnitude in energy.

Above the critical energy the stopping power scales linearly with the muon energy. The losses in this regime can be highly stochastic causing the muon to potentially lose a significant fraction of its energy in a single interaction. The average fractional energy loss rate, however, of a muon is quite small, allowing high energy muons to travel large distances through a material. The fraction of muons able to penetrate a given distance in ice is shown in Fig. 3-10. Given these large penetration ranges, many of the muon events in IceCube are reconstructed as through-going.

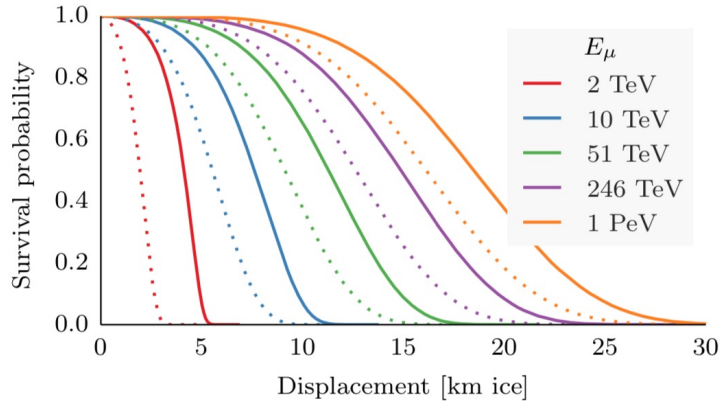

Figure 3-10: The survival probability of of muons penetrating a given distance through the ice. The solid lines show the total fraction of muons that survive and the dashed lines show the fraction that survive with greater an energy greater than 1 TeV. From Ref. [153].

### 3.4.2 Cherenkov radiation

A charged particle traveling through a medium will polarize the electron clouds of the atoms that form the medium. When the atom de-excites, it will emit electromagnetic radiation. If the incident particle is traveling above the phase velocity of light, $v > c/n$, the emission undergoes constructive interference. This coherent radiation emanates in a cone around the trajectory of the charged particle with an opening angle given by $\theta = \cos^{-1}(1/n\beta)$. For relativistic particles in ice, $n = 1.31$, this corresponds to approximately 41°. Cherenkov photons are emitted with a characteristic wavelength dependence of $1/\lambda^2$ in the wavelength range from 300 to 600 nm.

The critical energy of a charged particle to emit Cherenkov radiation is $E_c = mc^2 n/\sqrt{n^2-1}$. For a muon traveling through ice, the critical energy is 163 MeV.

### 3.4.3 Event topologies

IceCube does not directly detect neutrinos, rather, it measures the radiation emitted by the secondary charged leptons produced by the neutrino interaction, primarily in the form of Cherenkov radiation and electromagnetic cascades along the track at higher energies. There are two com-mon types of event topologies observed in IceCube. These are the "tracks" and "cascades".

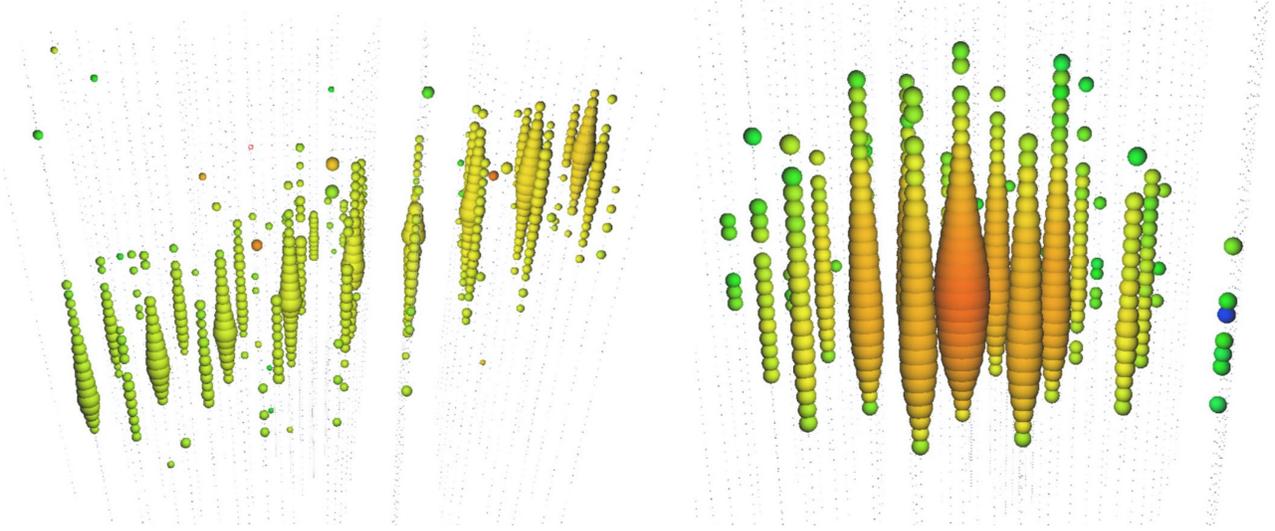

Figure 3-11: The IceCube Event Viewer illustrating the two common event topologies observed in IceCube. The color scheme shows the arrival time of the light; where blue is early and red is late. Left: A simulated through-going track of a 100 TeV muon in IceCube. Right: A simulated PeV cascade event. From Ref. [154].

Tracks are produced from relativistic muons traveling through the ice and are identifiable by the elongated photon emission pattern. The muon may have originated in a $\nu_\mu$CC interaction, from a cosmic ray muon, or from the decay of a $\tau$ lepton. In the case of a $\nu_\mu$CC interaction, the majority ($\sim$50-80%) of the energy of the incoming neutrino is carried away in the outgoing charged muon in the forward direction. The remaining energy of the interaction is transferred to the nuclear target and produces a hadronic shower near the interaction point. The large lever arm of track events allows for accurate trajectory reconstruction of the muon, often to sub $1°$. The track length may be larger than the detector, as shown in Fig. 3-11 and 3-10, therefore the start and end point might be unknown. This greatly limits the energy resolution of these types of events.

NC interactions of all neutrino flavors, $\nu_e$CC and $\nu_\tau$CC induce cascades. In the case of a NC interaction, the energy transfer from the neutrino goes primarily into the nucleus producing a outgoing hadronic shower. Because of the density of ice the shower dies out within several meters and has a signature inside of IceCube best described as a spherical emission with a slight asymmetry in the direction of motion. The $\nu_e$CC interactions produce a similar event topology in the detector. The resulting electron from the interaction begins a shower of gamma ray, positron, and electron production via bremsstrahlung and pair production respectively. It is often the case that a cascade is fully contained within the detector. This allows for a calorimetric measure of the energy.

## 3.5 Previous IceCube sterile neutrino searches

IceCube has published the result of three independent sterile neutrino searches using up-going atmospheric neutrinos [123, 155] that show no evidence for anomalous $\nu_\mu \to \nu_\mu$ oscillations.

The IC86 IceCube high-energy sterile neutrino search looked for matter enhanced oscillations in a sample consisting of 20,145 $\nu_\mu/\overline{\nu}_\mu$ events with reconstructed energies ranging from approximately 400 GeV to 20 TeV. By setting $|U_{\tau 4}|$ to zero, IceCube set a conservative limit on the mixing angle $\sin^2(2\theta_{24})$. The 90% C.L. limit is shown in the left side of Fig. 3-12 compared to other limits. The best fit point was found at to be at $\Delta m_{41}^2 = 10$ eV$^2$ and $\sin^2(2\theta_{24}) = 0.56$, and compatible with the no sterile hypothesis at the 15% level. The mixing element $U_{e4}$ as well as the sterile phases, $\delta_{14}$ and $\delta_{24}$, were also found to have no impact on the sensitivity and were set to zero. Alongside the IC86 result, an independent IC59 (59 strings) analysis was published. A description of this can be found in Ref. 3-12.

At low energies ($< 100$ GeV) IceCube is sensitive to standard atmospheric neutrino oscillations. The oscillation probability in this region is modified with the inclusion of sterile neutrino mixing, whose effect scales proportionally to the matter density traversed. The event selection included events solely from DeepCore, with reconstructed energies between 6.3 and 56 GeV. The final sample contained 5,118 events (tracks and cascades), but was not background free. The result places a limit on $U_{\mu 4}$ and $U_{\tau 4}$ assuming a $\Delta m_{41}^2 > 1$ eV$^2$. This result is shown in Fig. 3-12 right. The best fit point was found to be compatible with the no sterile hypothesis at the 30% level.

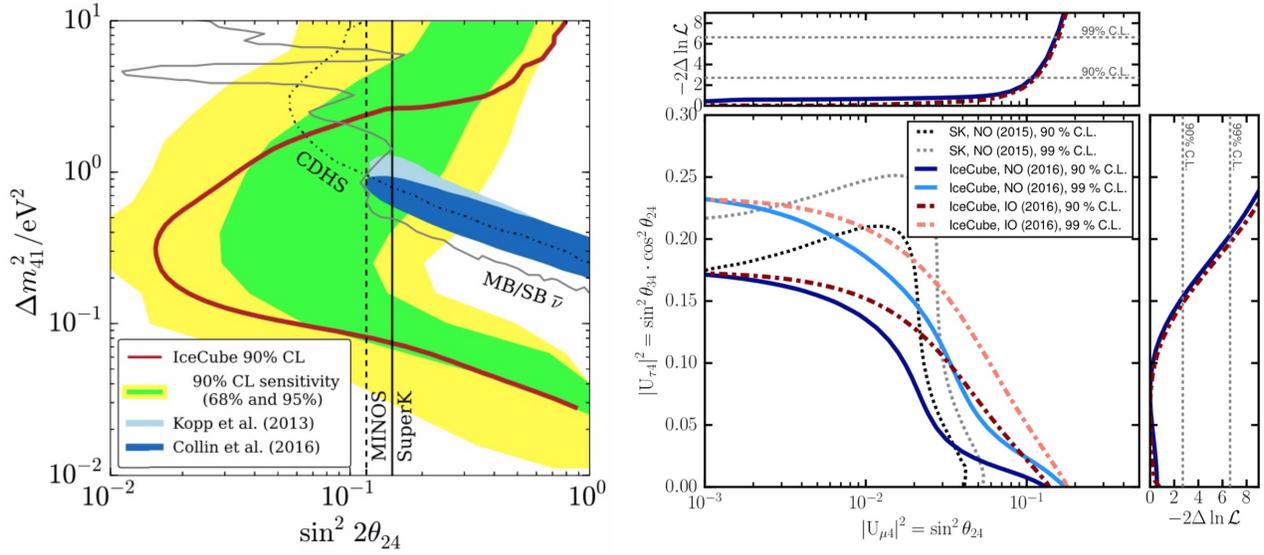

Figure 3-12: Left: The IceCube 90% C.L. limit from the high-energy sterile neutrino analysis compared to the allowed regions from appearance experiments (blue) and limits at the time of publication (grey/black), as well as subsequent $\nu_\mu$ disappearance results published since (purple). Right: Results from the IceCube low-energy sterile neutrino search assuming a normal mass ordering (NO) and an inverted mass ordering (IO), compared to similar results from Super-Kamiokande. From Ref. [156].

THIS PAGE INTENTIONALLY LEFT BLANK

# Chapter 4

# Simulating events in IceCube

## 4.1 Overview

The Monte Carlo (MC) simulation of the IceCube detector is a collaborative effort between many individuals. Over the years, the simulation has been continuously improving to incorporate new ideas and measurements. This particular analysis is systematics limited in a large portion of the physics parameter space and it is therefore particularly important to use the state-of-the-art simulation along with accurate descriptions of all pertinent uncertainties. At the start of this analysis, no up-to-date MC set existed with sufficient statistics to perform this analysis, so a tremendous amount of effort and resources went into its production. This happened over the course of approximately two years using both the resources available on the NPX cluster and the Open Science Grid (OSG) [157, 158].

A reference overview of the steps involved in the analysis-related processing chain is shown in Fig. 4-1. Relevant to this chapter are the steps contained within the dashed black line. The event filtering will be covered in the next chapter.

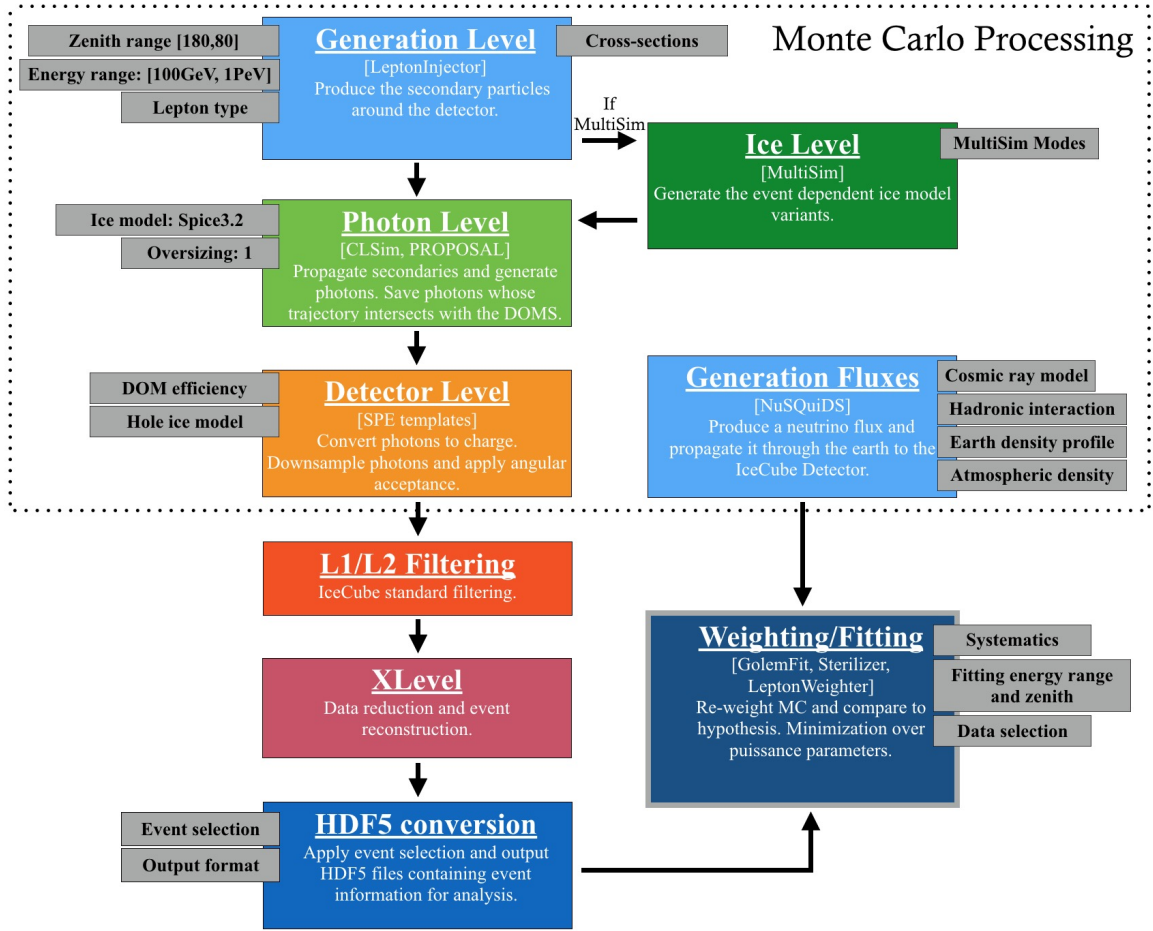

Figure 4-1: A graphical overview of the full analysis processing chain.

## 4.2 Monte Carlo generation processing chain

An overview of the physical process which we are simulating is shown in Fig. 4-2. Here, a cosmic ray is shown interacting with the upper atmosphere producing a shower of particles, including charged mesons which subsequently decay into the conventional and prompt atmospheric neutrino flux. The neutrinos propagate through the Earth and interact near the IceCube detector producing the final state charged particles that can trigger the IceCube detector. The simulation of this physical chain of events is broken up into discrete processing steps described below.

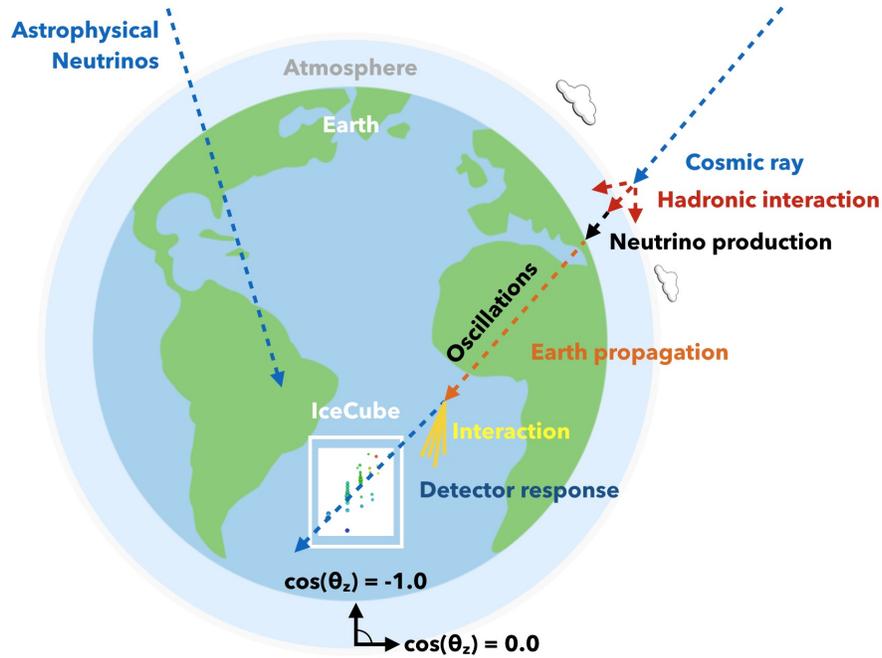

Figure 4-2: The physical process of the simulation. The coordinate system used in IceCube is also shown in this figure, where the trajectory is defined relative to the zenith angle in the southern sky.

### 4.2.1 Event generation

Prior to this analysis, the majority of the simulation production in IceCube used the neutrino generator, NuGen [159]. Conceptually, NuGen is simpler and follows the physical process rather closely: the simulation begins with an atmospheric neutrino flux hypothesis, propagates the neutrinos through the earth, then forces the neutrino interaction near the detector. This, however, means that the full simulation chain needs to be re-run for each flux hypothesis, propagation model, and cross-section model. This is unsuitable for analyses in which these variables are not constant, such as in the case for many BSM searches.

This analysis adopts a new lepton generation and weighting method called LeptonInjector [160] and LeptonWeighter [161]. These projects were primarily developed by C. Weaver and C. Argüelles, and expand on the MuonInjector project [162]. The philosophy of LeptonInjector/LeptonWeighter is slightly different than NuGen; rather than beginning at neutrino production, it begins with the simulation of the neutrino interaction point near the detector and produces the final state secondaries associated with the desired process. In our case, the signal component is the $\mu^{\pm}$ and hadrons from the $\nu_\mu$CC interactions. The final state is then connected to a neutrino flux at the re-weighting, via LeptonWeighter, in the processing chain. This method allows us to re-weight the MC (rather than reprocesses) on an event-to-event basis to any cosmic ray fluxes, atmospheric properties, cross-sections, Earth density models, and neutrino oscillation hypotheses without having to rerun the full simulation.

While developing LeptonInjector, it was important to check the produced leptons against NuGen. In doing so, we discovered several undesired features of NuGen that, once modified, reproduced the LeptonInjector distributions. A list of the differences found between event generators is shown in Table 4.1.

| **Difference** | NuGen | LeptonInjector/LeptonWeighter |
|---|---|---|
| Final state lepton | The kinematics assume the lepton is massless | Proper kinematics |
| Final state hadrons | The hadronic jet appears co-linear with the incident neutrinos. Also assumed to be massless. | Proper kinematics |
| Low energy leptons | The kinetic energy is forced to be the total lepton energy | Proper kinematics |
| Neutrino-atm. interactions | Does not allow for the neutrino to interact with the atmosphere | Allows neutrino-atm. interactions |
| Matter effects | Uses fixed cross-sections from tables | Density profile decoupled from event generation. Allows re-weighting. |
| Neutrino oscillations | Rerun simulatoin for each hypothesis | Oscillations are decoupled from event generation. Allows re-weighting. |
| Neutrino-nucleon cross sections | Coarsely sample from tables (111 bins) | Uses high fidelity splines. Allows re-weighting. |

Table 4.1: An overview of the differences between event generators: NuGen and LeptonInjector/LeptonWeighter. NuGen was subsequently updated to include the effects outlined here (internally referred to as *Detector Mode*).

LeptonInjector begins by injecting the final states secondaries in the desired volume encompassing the detector according to a reference energy spectrum and a continuous doubly differential cross-section. We consider a range of injected primary $\nu_\mu$ energies from 100 GeV to 1 PeV over a zenith angle of 78.5° (11.5° above the horizon) to 180° (up-going). The injected energies are sampled by an $\text{E}^{-2}$ power-law energy spectrum and the arrival directions are distributed isotrop-ically in azimuth and $\cos(\theta_z)$. The interaction point is assigned by randomly selecting a point in a cylinder whose axis is centered on the IceCube, with a radius (injection radius) of 800 m. The cylinder length is set to be the 99.9% muon range in ice plus two additional "endcaps" on either end, each with a length of 1200 m. An example sampling volume for a given muon energy is illustrated in Fig. 4-3.

For each event, the total injected energy (primary neutrino energy), final state lepton energy and zenith, Bjorken x and y interaction variables, inelasticity probability, and the properties associated with the injected point (total column depth and impact parameter) are recorded. A full simulation set in this analysis contains $2\times 10^9$ such events each generated with an independent seeds, yielding an effective live time of approximately 500 years.

### 4.2.2 Lepton and photon propagation

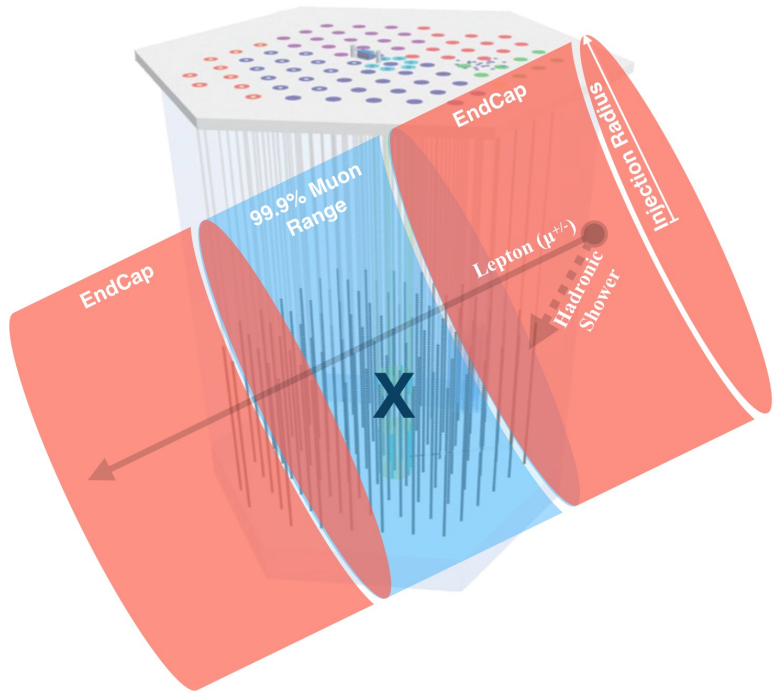

Figure 4-3: An illustration of the injection volume used in LeptonInjector. The blue region corresponds to the 99.9% muon range centered on the detector (indicated by the X). The endcaps are shown in red. This image is not to scale, but the endcaps are chosen to be large enough to extend well beyond the limits of the instrumented volume.s An example final state lepton trajectory and hadronic shower are also shown.

The final state secondaries are propagated through the ice according to the expected ionization energy loss and other stochastic losses using PROPOSAL [163]. Along the track, photons are generated randomly according to the parameterization of the Cherenkov radiative emission from tracks (muons) or cascades (electromagnetic interactions and hadronic showers) using CLSim [164]. Each photon is tracked as it propagates through ice until it is either absorbed by the ice or interacts with a DOM.

The photon propagation accounts for random scatters according to the ice model, described as depth and wavelength dependent scattering and absorption, and anisotropy along a major and minor axis (see Sec. 3.3.3). At each photon scattering point, the algorithm randomizes the new photon direction based on the scattering angle distribution, parametrized by the mean scattering angle and scattering coefficient [165].

We also introduce new method to parametrize the scattering and absorption uncertainty in the bulk ice, subsequently referred to as "SnowStorm". SnowStorm requires an extra processing step (labeled as "Ice Level" in Fig. 4-1), in which the scattering and absorption coefficients of the ice model are perturbed a predefined uncertainty for every 100 events. Details expanding on the implementation on of SnowStorm will be presented in Sec. 7.1.2.

A typical muon event in IceCube creates in excess of $10^7$ Cherenkov photons in the sensitive wavelength range of the PMTs [166]. This presents a considerable computational challenge when billions of such events are required to be simulated. GPUs are designed to perform the same computational operation in parallel across multiple threads. Each thread is used to propagate its own photon for as long as the photon exists. When the photon is absorbed or hits a DOM the thread receives the new photon from a pool of photons for as long as that pool is not empty. Although a single thread runs slower than a typical modern computer CPU core, running thousands of them in parallel results in the much faster processing of photons from the same pool on the GPU.

We overproduce the number of photons generated per event by a factor of approximately 30% such that we can later randomly reject (down-sample) photons to produce the *DOM efficiency* systematic data sets. This is further described in the following subsection as well as in Sec. 7.1.1. The photons that intersect a DOM have their angle relative to the DOM surface, position, and timing information recorded as a *photon series map* and sent on to the next processing step: the detector simulation. Photons that do not intersect a DOM are eliminated.

### 4.2.3 Detector simulation

At this stage, we have a photon series map for an event described by the intersection positional and directional information, as well as the time of intersection. The photons are then down-sampled by randomly rejecting photons to the desired DOM efficiency. The central MC set, for example, is down-sampled from a generation efficiency of 1.10 to a DOM efficiency of 0.97.[1] The photons that survive the down-sampling are then assigned a probability of acceptance depending on the interaction direction. This is determined by sampling from an angular acceptance curve (further information provided in Sec. 7.1.3). The set of remaining photons generate photoelectrons (MCPE) at the surface of the PMT photocathode.

The output charge at the PMT anode for each MCPE is determined by randomly sampling from the single photoelectron distribution (SPE charge template) based on in-situ low occupancy PMT measurements. An example of the measured charge distribution for DOM (1,1) is shown in Fig. 4-4 along with the extracted SPE charge template. The in-situ determination of the SPE charge templates is a new addition to the IceCube simulation production. Previously, a single distribution based on a lab measurement was used to describe all the DOMs in IceCube. A complete overview of the procedure used to extract the SPE charge templates for all DOMs is given in Appendix F (SPE Templates).

The detector simulation also includes PMT effects that contribute to the observed charge. These effects include pre-pusles, late pulses, after pulses, thermionic emission, and are all described in detail in Appendix F.1.1. Beyond this, the simulation accounts for timing difference due to non-uniformities in the photomultiplier itself [167], non-linearities associated with PMT amplification process [168], and the noise timing distribution [169]. Scaled single photoelectron pulse shapes (that take into account the version of the AC coupling) are then fit to the simulated waveforms using software referred to as WaveDeform [137, 170] (Waveform unfolding process), which determines the individual pulse time stamp and charge, and populates a pulse series.

---

[1] In the internal CLSim algorithm, the generation efficiency and DOM efficiencies are actually increased by a factor of 1.35 to account for higher quantum efficiency of the HQE DOMs, then again by another factor of 1.30 to account for the SPE charge templates, and finally by 1% as a safety margin.

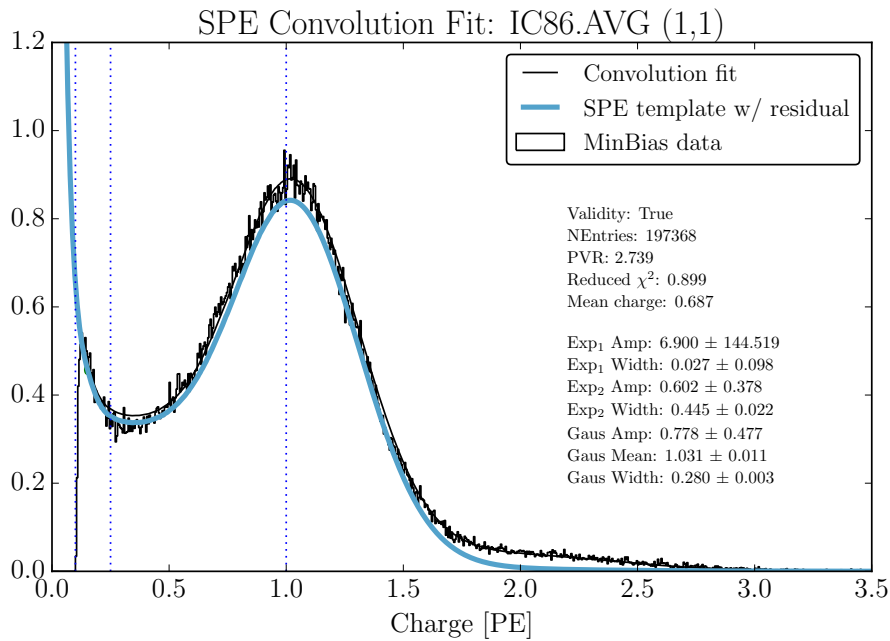

Figure 4-4: The SPE charge template for DOM (1,1) using the cumulative charge distribution from IC86.2011 to IC86.2016. The extracted charge per pulse is shown in the black histogram, the fit to the charge data is shown in black, and the SPE charge template is shown in teal. Two vertical dotted lines are shown at 0.25 PE and 1.0 PE for reference.

After the stage of running WaveDeform, the MC is treated identically to data (filtering and event reconstruction).

### 4.2.4 Neutrino flux generation and propagation through the Earth

The lepton generation discussed in Sec. 4.2.1 must now be tied to an incident neutrino flux at the IceCube detector. We begin by generating an atmospheric neutrino flux using the Matrix Cascade Equation (MCeq) program [112, 171], which solves the atmospheric shower coupled cascade equations numerically. MCEq allows us to specify the desired cosmic ray model, hadronic interaction model, and atmospheric temperature profile. The atmospheric temperature profile is extracted from AIRS satellite data [172] (further details provided in Sec. 7.2.4). We generate a separate flux for each month in the year to account for seasonal variations, then average over the fluxes accounting for monthly livetime differences in the data. Other atmospheric temperature profiles exist, for example the US std from 1976 [173], however this is disfavored in this thesis since it does not describe the time dependence of the temperature variations, nor is it an up-to-date measurement. Further, we require some knowledge of the uncertainty in the temperature profile, which is not reported in this model.

The atmospheric neutrinos are then propagated through the Earth using the neutrino Simple Quantum Integro-Differential Solver (nuSQuIDS) [174]. nuSQuIDS is a C++ code based on SQuIDS [175], that propagates an ensemble of neutrinos through a given media, e.g. the Sun, Earth, or in vacuum. It accounts for neutrino oscillations including coherent matter interactions and non-coherent interactions. In our case, the output of nuSQuIDS is an at-detector neutrino flux. For each point in the physics parameter space (the sterile neutrino hypothesis) a separate neutrino flux file is generated and stored for hypothesis testing. The Earth's density as a function of radius is described using the widely used PREM [176] model. Other models are available [177–179], however the differences between these are completely negligible for this analysis.

The astrophysical neutrino flux follows the same procedure outlined in the previous paragraph, however the initial flux fed in to nuSQuIDS is of the form of a single power law, with a normalization $\Phi_{\text{astro}}$ at 100 TeV and a spectral index $\gamma_{\text{astro}}$:

$$\Phi_{\nu_\mu + \bar{\nu}_\mu} = \Phi_{\text{astro}}\left(\frac{E}{100 TeV}\right)^{\gamma_{\text{astro}}}. \tag{4.1}$$

### 4.2.5 Event weighting

Following the description found in Ref. [64], the generation specifications of an IceCube event can be represented by:

1. **True quantities**:
    - Injected energy ($E_\nu$): The sum of the muon and hadronic shower energies (equivalent to the primary neutrino energy) [GeV].

- Interaction x: The Bjorken x of the interaction [dimensionless].

- Interaction y: The Bjorken y of the interaction [dimensionless].

- Injected muon energy ($E_\mu$): The initial energy of the muon [GeV].

- Injected muon zenith ($\theta_z$): The initial zenith angle of the muon [rad].

- Inelasticity probability (I): The probability which corresponds to having selected a given (x,y), namely:

$$I = \frac{1}{\sigma} \frac{\mathrm{d}^2\sigma}{\mathrm{dxdy}}(x,y)\Delta x \Delta y, \quad (4.2)$$

where $\Delta x$ and $\Delta y$ correspond to the MC kinematic discretized phase space size: $\Delta x = \Delta y = 10^{-10}$. [dimensionless]

- Total column depth ($h_{\mathrm{depth}}$): The column depth where the interaction was forced. [g/cm$^2$]

2. **Reconstructed quantities**:

    - Muon energy proxy ($E_\mu^{proxy}$): The reconstructed muon energy proxy. [dimensionless]

    - Reconstructed muon zenith ($\theta_z^{proxy}$): The reconstructed muon zenith angle. [rad]

3. **Flux weight**:

    - Event flux-less weight: The event weight without the flux. [GeV cm$^2$ sr]

The neutrino flux follows approximately a power law and therefore we generate MC according to an $E^{-2}$ spectrum. The azimuth and zenith change linearly, in which case it makes sense to sample uniformly in $\cos(\theta_z)$ and azimuth. A MC event is assigned a weight according to:

$$w_{\mathrm{event}}[s^{-1}] = w_{\mathrm{gen}} h_{\mathrm{depth}} N_a \frac{\mathrm{d}^2\sigma}{\mathrm{dxdy}} \times \phi_\nu, \quad (4.3)$$

where $N_a$ is Avogadro's number, $\phi_\nu$ is the neutrino flux in the center of the detector given in units of GeV$^{-1}$ cm$^{-1}$sr$^{-1}$s$^{-1}$, $\frac{d^2\sigma}{dxdy}$ is the charged current cross section, $h_{\mathrm{depth}}$ is the total collumn depth, and $w_{\text{gen}}$ is the MC generation weight given by:

$$w_{\text{gen}} = \frac{1}{0.5}\frac{\Delta x \Delta y}{I}\frac{1}{\text{N}_{MC}}\frac{\Omega_{\text{gen}}\pi R_{\text{inj}}^2}{E_\nu^{-\gamma}}\frac{E_{\text{max}}^{1-\gamma} - E_{\text{min}}^{1-\gamma}}{(1-\gamma)}, \tag{4.4}$$

where $E_\nu$ is the event injected energy, $N_{MC}$ is the total number of MC events, 0.5 arises from the fact that we generate equal amounts of neutrinos and antineutrinos, $\Omega = \Delta\phi_a \Delta\cos\theta_z$ is the generation solid angle, and $R_{\text{inj}}^2$ is the injection radius.

THIS PAGE INTENTIONALLY LEFT BLANK

# Chapter 5

# Selecting a pure sample of $\nu_\mu/\overline{\nu}_\mu$ events

This analysis aims to search for a spectral difference in the reconstructed muon energy and zenith distributions from $\nu_\mu/\overline{\nu}_\mu$ CC interactions. Muons are identifiable in IceCube by the track-like nature of the emitted Cherenkov light as they propagate through ice. The event selection defines the set of criteria used to reduce the background events (air shower cosmic ray muons, NC events, CC electron neutrino interactions, and CC tau neutrino interactions) passing through the standard IceCube filters, while maintaining a high efficiency selection of atmospheric muon neutrino events. The term "cut" is used to describe how the events are separated into a sample of signal-like events and background-like events. The background-like events are rejected from the final sample.

Despite the 1.5 km of overburden directly above IceCube, the detector is triggered at a rate of approximately 3 kHz [180] by downward-going cosmic ray muons produced in high-energy air showers. The simulation of cosmic ray air showers is handled by the widely used software package CORSIKA (COsmic Ray SImulations for KAscade ) [181, 182]. The CORSIKA simulation used in this analysis was produced by the IceCube Simulation Production Group (SimProd). Eight independent CORSIKA simulation sets were used to quantify the amount of cosmic ray muon contamination in the event selection, covering a primary cosmic ray (protons, photons, and other nuclei) energy from $6 \times 10^2$ GeV to $1 \times 10^{11}$ GeV. CORSIKA simulates the air showers to ground level, propagating the cosmic ray muons through the firn and ice using PROPOSAL to a sampling surface around the detector. The cosmic ray muons are then weighted to an initial cosmic ray flux, in this case HillasGaisser2012 H3a [100]. Verification the CORSIKA event rate prediction will be discussed in Appendix C.2.

At the Earth's surface, the conventional $\nu_\mu$ flux dominates the neutrino flavor composition. The sub-dominant neutrino flavors ($\nu_e$ and $\nu_\tau$) represent a much less significant component of the background compared to the cosmic ray muons. Nevertheless, we ensure that they are reduced to an insignificant level. The topological signature of cascades (primarily $\nu_e$ events and $\nu$NC events) in our energy range is sufficiently different from the track-like events that they are efficiently rejected. The $\nu_\tau$ can interact via a CC interaction, producing a $\tau$ lepton and cascade-like shower. These are also efficiently rejected. However, the $\tau$ can subsequently decay to a muon and flavor conserving neutrinos with a branching ratio of $17.4 \pm 0.03\%$ [50]. While the signature of these events are obviously track-like in nature, the $\nu_\tau$-appearance probability is small at TeV energies considering the first $\nu_\mu \to \nu_\tau$ oscillation maximum occurs at approximately 25 GeV for upward-going neutrinos. The $\nu_e$ and $\nu_\tau$ backgrounds were accounted for in dedicated simulation sets produced by LeptonInjector, each with an effective livetime of approximately 250 years.

The event selection contains two custom designed event filters. If an event passes either one of these filters, it is included in our final sample: the Platinum event selection. Each filter can be thought of as its own event selection, however to simplify terminology, they will be referred to as the Golden filter and the Diamond filter. For both filters, we first require that every event passes the online Muonfilter, has valid hit cleaning (PMT noise removal algorithm), and passes a set of early precuts used to reduce the data and MC to a manageable level. The list of precuts are:

1. If the reconstructed direction is above the horizon ($\cos(\theta_z) > 0.0$), we require that the total event charge (Qtot) is greater than 100 PE and the Average Charge Weighted Distance (AQWD) is less than 200 m/PE. The AQWD is defined as the average distance of the pulses, weighted by the total charge of the event charge, from the track hypothesis.

2. Reject all events with a reconstructed zenith angle above a $\cos(\theta_z) > 0.2$. The vast majority of which are muons produced in atmospheric showers.

3. Neglecting the DOMs in the DeepCore detector (NoDC), we require at least 20 triggered DOMs (NChan > 20) and Qtot > 20PE.

4. We also require that at least 12 DOMs triggered on direct light. Direct light refers to the Cherenkov photons which arrive at the optical modules prior to scattering.

5. Finally, the reconstructed track length using direct light in the detector must be greater than $200\,\mathrm{m}$ (DirL > 200 m), and the absolute value of the smoothness factor is smaller than 0.6 (|DirS| < 0.6). The smoothness factor is a measure that defines how smooth the distribution of triggered DOMs is around the reconstructed track.

For every event that passes the precuts, we apply the following reconstruction methodology:

1. The event passes through an *event splitter* (in our case, an updated version of the Topological Splitter [183]) to separate coincident events into multiple independent sub-events. A coincident event is defined as an event in which a uncorrelated cosmic ray muon entered the detector during the readout. Approximately 10% of neutrino events have an accompanying coincident muon in the time window.

2. Reconstruct the trajectory of each sub-event using the following algorithms: LineFit, SPEFit iterated five times, and then MPEFit, using each sequential fit to seed the following fit. These reconstructions rely primarily on timing information. LineFit provides the most basic reconstruction algorithm typically used for muons. It uses a simple least squares linear regression to fit the timing distribution of the first PE observed on each DOM [94]. The single photoelectron (SPEFit) and multi-photoelectron (MPEFit) algorithms are likelihood based and account for the Cherenkov emission profile as well as the ice scattering and absorption. SPEFit also only uses the first hit on each DOM, whereas MPEFit incorporates an arrival time PDFs to include information from the number of recorded photons as well (as well as a description of the PMT timing jitter width). We require that each fit succeeds, however only MPEFit is used as a final description of the trajectory. The trajectory information included in the reconstruction is:

(a) $x_0$, $y_0$, $z_0$: an arbitrary point along the track.

(b) $\theta$, $\phi$: the zenith and azimuthal direction.

(c) $t_0$, $E_0$: the time and energy at point ($x_0$, $y_0$, $z_0$).

However, we omit the energy calculation since it is performed later with a more sophisticated algorithm.

3. Compare the unconstrained reconstruction with one that has a Bayesian prior. The prior was defined in previous analyses [184] based the fact that the majority of observed events are truly down-going and should be reconstructed as such.

4. Include an appropriate variable to quantify the uncertainty in the reconstructed trajectory [185]. Internally this is referred to as "paraboloid sigma." It assesses the uncertainty on the trajectory reconstruction based on the likelihood profile around the best fit reconstructed track hypothesis. The resulting uncertainty is calculated as:

$$\sigma = \sqrt{\frac{\sigma_\phi^2 + \sigma_\theta^2}{2}}, \tag{5.1}$$

where $\sigma_\phi^2$ and $\sigma_\theta^2$ correspond to the $1\sigma$ confidence interval derived from the likelihood profile. A small paraboloid sigma value indicates higher precision in the reconstructed trajectory. A second variable, called the "reduced Log likelihood" (RLogL), uses the best fit likelihood value as a global measure of the success of the fit. Essentially, RLogL is a metric which quantifies how successful the reconstruction of an event was given other events.

5. Reconstruct the event energy with MuEX. Unlike the trajectory reconstructions, energy reconstruction relies heavily on the intensity of the light on each DOM. Given the output of MPEFit, an analytical approximation for the observed light distributions is used, which accounts for the gemoetry between the emitter and receiver, the ice absorption, and detector noise. Stochastic losses from high-energy interactions means that there will be points along the track with bursts of comparably more intense light. This is averaged out in MuEx by broadening the PDF that describes the energy loss expectation. Further information can be found in Refs. [170, 184].

The total rate in both signal and background after the precuts is approximately 1280 mHz. This is almost entirely composed of cosmic ray muons.

## 5.1 The Golden filter

The Golden filter was originally designed as the event selection for the IC79.2010 diffuse neutrino analysis [186]. It was optimized to accept high-energy muon neutrinos and was subsequently used in the IC86.2011 high-energy sterile neutrino search [123]. A detailed description of the cuts can be found in Refs. [184, 187].

The event selection for the IC79.2010 diffuse neutrino analysis was determined to have a greater than 99.9% muon neutrino purity. However, an event-by event hand-can through 1000 events revealed evidence for an approximate 1% contamination due to coincident cosmic ray muons. All these events were reported to have an AQWD greater than 100 m/PE. This level of contamination did not affect the result of the either analysis, but we include several supplementary safety cuts to reduce it further.

A coincident cosmic ray muon is likely to pass through the detector in a different volume than the track of interest. This causes an excess amount of charge to be located far away from the reconstructed trajectory. Such an event would have a larger AQWD. A cut on the AQWD was introduced into the Golden filter at a value of $90\,\mathrm{m/PE}$. It was also found that the events above the horizon contributed insignificantly to the sensitivity of the analysis and were subsequently cut. The total event expectation for signal and background passing through the Golden filter

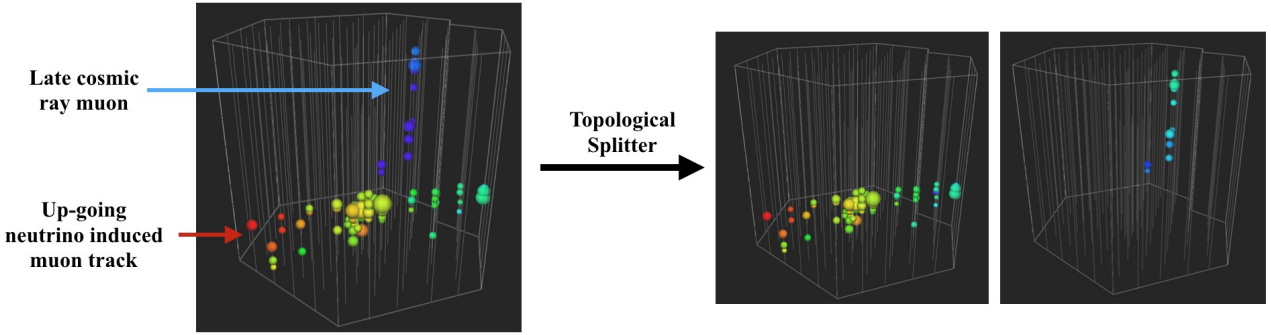

Figure 5-1: An example event of an up-going track from a moun neutrino CC interaction with a late coincident cosmic ray muon. This particular event was not split using the older version of the Topological Splitter but was successfully split with the updated version used in this analysis.

after these updates are shown in Table 5.1.

| Selection | $\nu_\mu$ | $\nu_\tau$ | $\nu_e$ | CR muons | Selection purity |
|---|---|---|---|---|---|
| Golden Filter | 154,970±393 | 16±4 | 1±1 | 16±4 | >99.9% |

Table 5.1: The expected number of events that pass the Golden filter. The uncertainties are statistical only.

## 5.2 The Diamond filter

The strategy used for the Diamond filter was to reduce the amount of data by first cutting away high background areas that are likely to be uninteresting (low-energy or poorly reconstructed), then maximize the analysis sensitivity. We sought to investigate three different strategies on what sort of events should be included in the final sample. It was found that optimizing the Diamond filter to include high-quality events (thus improving the overall resolution on reconstructed quantities) or strictly optimizing for high-energy events did not outperform the sensitivity of the Golden filter alone. In fact, the Golden filter was specifically designed to have a large active area at high energies, and was therefore difficult to improve on. The third strategy yielded useful results; the idea was to simply maximize the total number of $\nu_\mu$ events while keeping the event selection background free.

During the development of the Diamond filter, we also attempted to improve the sensitivity using another reconstruction algorithms, namely Truncated Energy [188], however MuEx was still found to provide the best sensitivity.

The following subsections describe the cuts introduced in the Diamond filter.

### 5.2.1 Data reduction

We begin with a second data reduction step in order to tightening up the precuts defined in Sec. 5, specifically:

1. The total NoDC charge of the event must be greater than 20PE (Qtot NoDC < 20 PE).

2. We require the event to have more than 15 triggered DOMs, excluding DeepCore (NChan NoDC < 15).

3. At least 12 DOMs must have seen direct light (DirN DOMs < 12).

4. The reconstructed trajectory cannot extend much above the horizon ($\cos(\theta_z) > 0.05$).

These cuts reduced the total rate to approximately 20 mHz, each of which are illustrated in Fig. 5-2.

### 5.2.2 Cosmic ray reduction

The overburden provides the greatest natural handle on the atmospheric muon contamination. For consideration, the horizontal overburden (trajectories with $\cos(\theta_z) > 0$) have approximately 157 kmwe (kilometer water equivalent) of ice shielding. As described in Sec. 3-10, even PeV muons will not be able to penetrate this amount of matter. Therefore, any atmospheric muons reconstructed with a trajectory originating below the horizon will likely have a poor reconstruction, or equivalently, large value of paraboloid sigma.

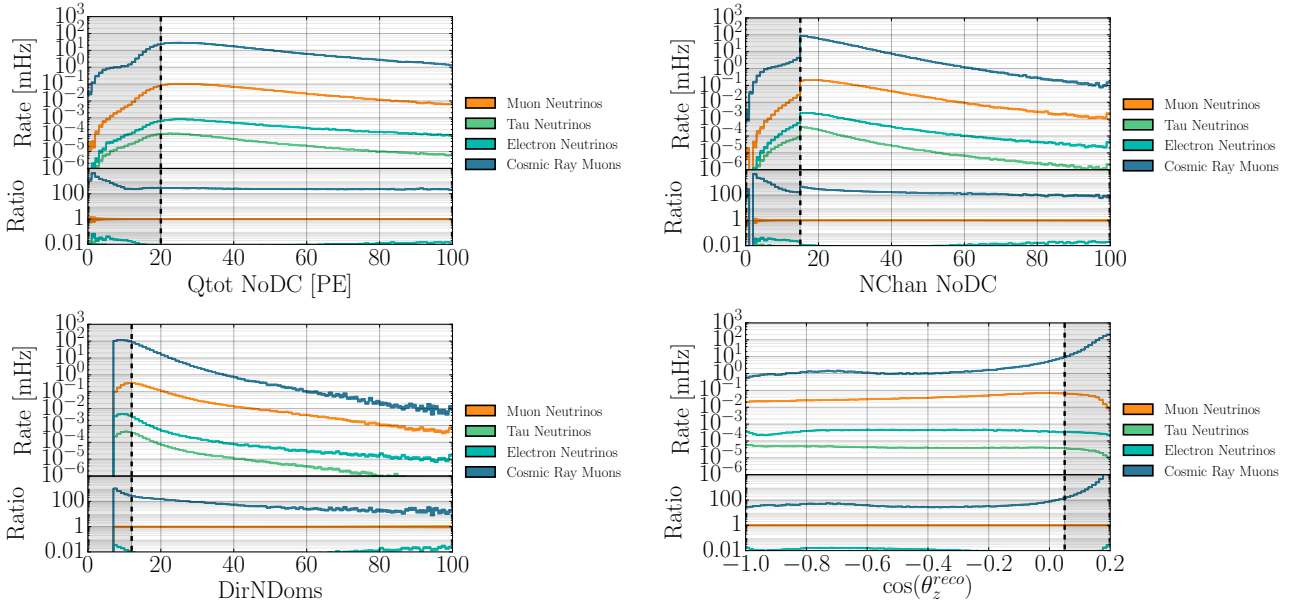

Figure 5-2: The MC event distribution of the cut variables used in the data reduction step of the Diamond filter. The signal (conventional muon neutrinos) is shown in orange, while the backgrounds are shown in blue, teal, and green (cosmic ray muons, electron neutrinos, and tau neutrinos respectively). The vertical-dashed line in each plot shows the location of the cut, and the shaded region is rejected.

The first two cuts introduced are performed in 2-dimensions and are shown in the bottom panels of Fig. 5-3. At small overburdens (events near the horizon), we require a smaller uncertainty in the track reconstruction (i.e. smaller values of paraboloid sigma). The Bayesian likelihood ratio was introduced specifically to reduce the cosmic ray muon backgrounds. We include a cut on the Bayesian likelihood ratio as a function of overburden.

A series of straight cuts were then introduced on the center of gravity of the charge in both the vertical direction (COGZ) and the radial direction (COGR). These cuts reduce the contamination of event near the edge of the detector (corner clippers) which have a higher probability of being mis-reconstructed cosmic ray muons. We also introduce the same updated AQWD cut found in the Golden filter. Fig. 5-3 shows these cuts, and are listed as:

1. If the value paraboloid sigma is greater than 0.03, cut event if $\text{Log}_{10}(\text{Overburden}) < 0.6 \times \text{Log}_{10}(\text{ParaboloidSigma-0.03}) + 7.5$.

2. If the Bayesian likelihood ratio is less than 33 units (BasyesLLH $< 33$).

3. The average charge weighted distance greater than 90 m (AQWD > 90 m).

4. The center of gravity of the charge in the vertical direction is above 450 m from the center of IceCube (COGZ > 450 m).

5. The center of gravity in the radial direction is greater than 650 m (COGR > 650 m).

6. The $\text{Log}_{10}$ of the overburden $< 10.0/(\text{BayesLLHR} - 30) + 4$.

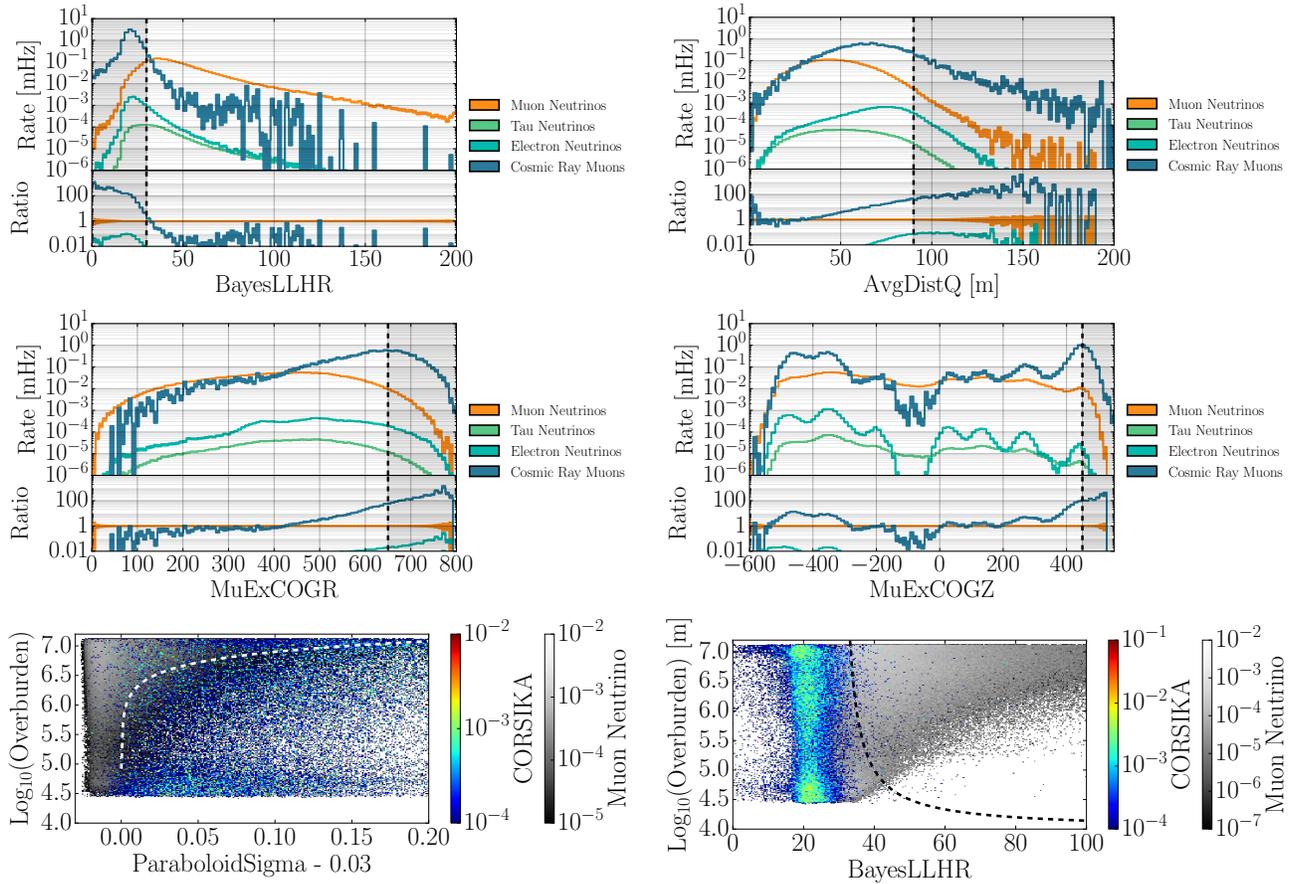

Figure 5-3: The top four plots show the 1D cuts on variables used to reduce the atmospheric shower background. The signal (muon neutrinos) is shown in orange, while the backgrounds are shown in blue, teal, and green (cosmic ray muons, electron neutrinos, and tau neutrinos respectively). The vertical-dashed line in each plot shows the location of the cut, and the shaded region is rejected. The bottom two plots show the 2D cuts on the overburden.

## 5.2.3 Background clean-up

At this stage, we attempt to simply remove the remaining background with some simple safety cuts. These are shown in Fig. 5-4. The 2D RLogL and DirNDoms cuts below were used in the Golden Filter and found to be useful without affecting neutrino data. After the clean-up, the final event rates for signal and background are shown in Table 5.2.

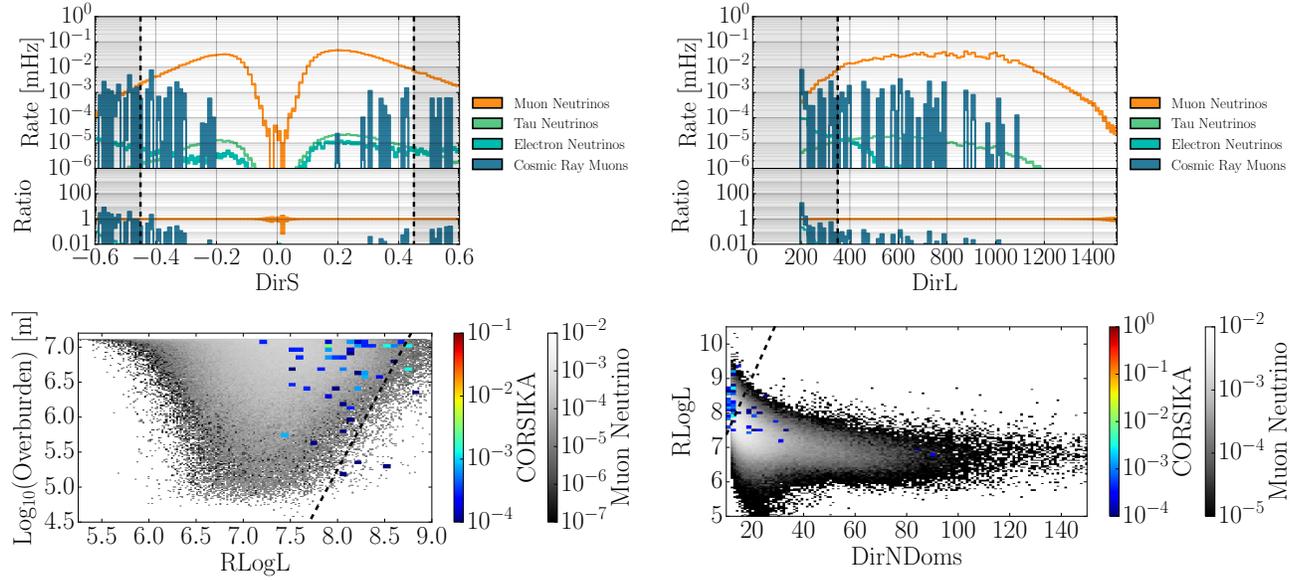

Figure 5-4: Top two plots show the 1D cuts used in the clean-up step. The signal (muon neutrinos) is shown in orange, while the backgrounds are shown in blue, teal, and green (cosmic ray muons, electron neutrinos, and tau neutrinos respectively). The vertical-dashed line in each plot shows the location of the cut, and the shaded region is rejected. The bottom two plots show the two dimensional cuts.

1. BayesLLHR < 33 m

2. AQWD > 90 m

3. COGZ > 450 m

4. COGR > 650 m

5. Remove event if RLogL > (3./18.)×(DirN DOMs) + 5.7 for all events where $\text{Log}_{10}$(Overburden) < 3/1.2 × (RLogL-7.1) +3.

| Selection | $\nu_\mu$ | $\nu_\tau$ | $\nu_e$ | CR muons | Selection purity |
|---|---|---|---|---|---|
| Diamond Filter | 295,416±543 | 22±5 | 1±1 | 4±2 | >99.9% |

Table 5.2: The expected number of events that pass the Diamond filter. The uncertainties are statistical only.

## 5.3 The Platinum event selection results

The Platinum event selection incorporates all events that pass either the Golden or Diamond filter. It was found that the sub-500 GeV reconstructed energy events contributed minimally to the sensitivity of the analysis and were therefore rejected. Similarly, we place a final safety cut at the horizon ($\cos(\theta_z) = 0$) where it is most likely to have atmospheric backgrounds leakage. Fig. 5-5 shows the Platinum event selection reconstructed energy (MuEx) and $\cos(\theta_z)$ distribution (from MPEFit). The numbers for this plot corrected for the livetime of the data is shown in Table 5.3. Although not explicitly used during the formulation of the event selection, this table also includes the predicted astrophysical and prompt neutrino flux contained in the sample. We also show the true neutrino energy of the conventional atmospheric neutrinos in the sample in Fig. 5-6. We find that greater than 90% of our events originate from a neutrino with an energy between 200 GeV and 10 TeV.

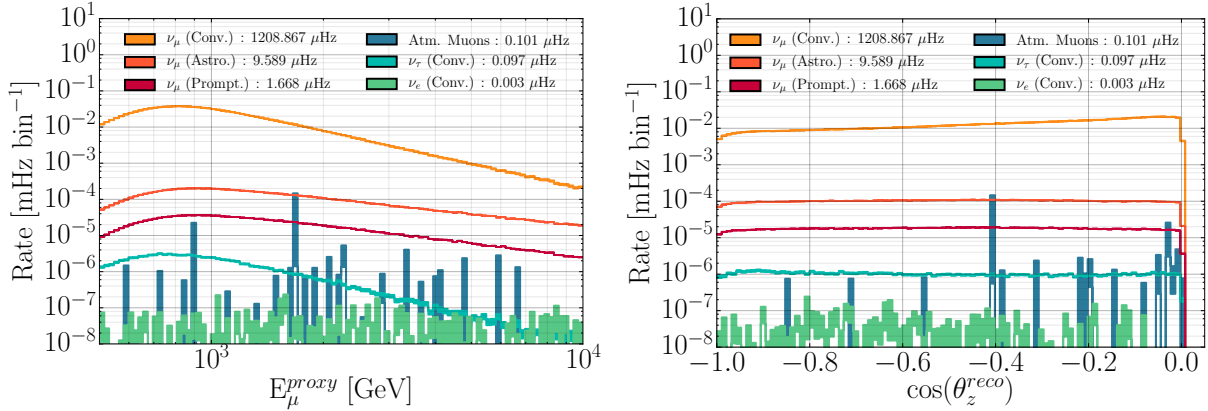

Figure 5-5: Left: The reconstructed energy distribution using MuEx for signal (conventional, prompt, and astrophysical $\nu_\mu$ flux) and the relevant backgrounds (atmospheric muons, $\nu_\tau$, and $\nu_e$). Right: The corresponding reconstructed zenith direction from MPEFit.

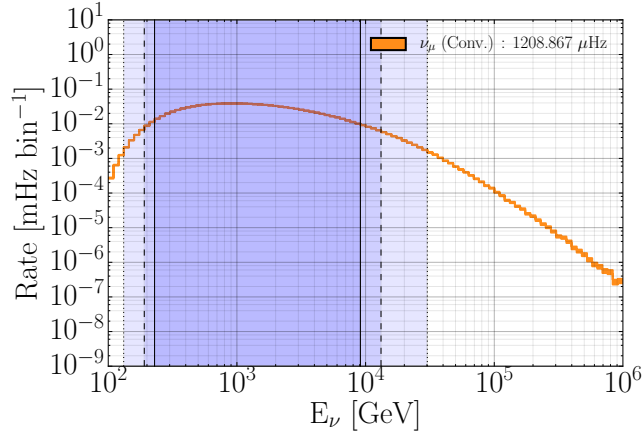

Figure 5-6: The predicted true conventional neutrino energy distribution of the Platinum event selection. The opaque regions show the regions containing 90% (solid lines), 95% (dashed), and 99% (dotted) of the data.

| Selection | Conv. $\nu_\mu$ | Astro $\nu_\mu$ | Prompt. $\nu_\mu$ | $\nu_\tau$ | $\nu_e$ | CR muons | Selection purity |
|---|---|---|---|---|---|---|---|
| Platinum | 315,214±561 | 2,350±48 | 481±22 | 23±5 | 1±1 | 18±4 | >99.9% |

Table 5.3: The final predicted signal (conventional, astrophysical, and prompt muon neutrinos) and background (cosmic ray muons, tau and electron neutrinos) rates for the Platinum event selection, assuming the central MC generation parameters.

## 5.4 IceCube data selection

The data used in this analysis spans from the original full configuration of the detector (the beginning of IC86) on May 13[th], 2011 to May 19[th], 2019. In 2016 through 2018, the IceCube collaboration reprocessed all the raw data to unify the event filtering procedure for all seasons. This reprocessing, referred internally as "Pass2," also re-calibrated the charge scaling for each PMT in the detector. This analysis uses only Pass2 data. The filtering and event selection for the Pass2 data follows the same processing chain as the MC described in the previous section.

IceCube data is typically broken up into 8-hour runs, which are vetted by the collaboration as having "good" in-ice data. For every good run, we require that all 86 strings are active, as well as at least 5,000 active in-ice DOMs. This helps normalize the data throughout the years with a minimal impact on the total livetime (∼0.4% reduction in data). Fig. 5-7 shows the number of active in-ice DOMs throughout the seasons under consideration. We find that we have an average of $5048 \pm 4$ active DOMs in the detector throughout the seasons investigated. We also find no significant deviation in the average event rate throughout the years. The data rates, broken up between seasons, is shown in Table 5.4.

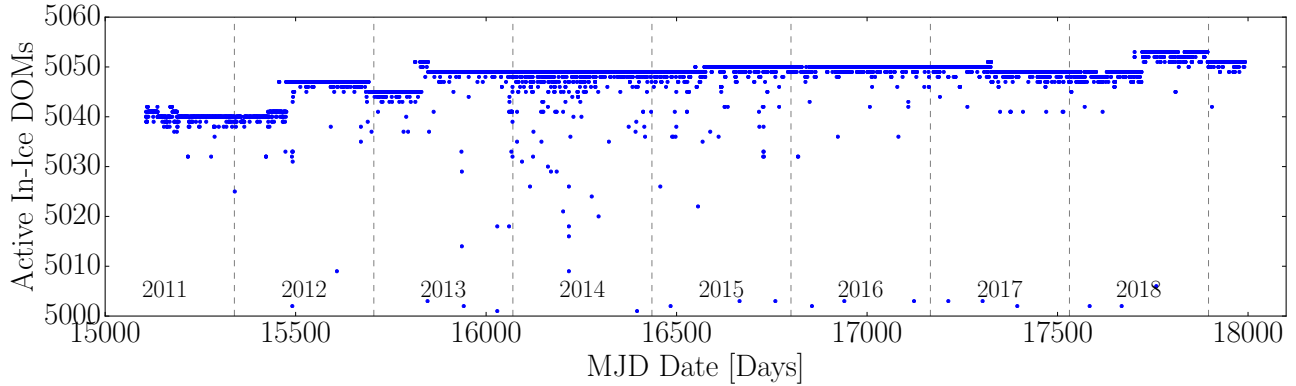

Figure 5-7: The number of active in-ice DOMs for all the considered IceCube runs used in this analysis. Each blue data point represents an individual run, while the vertical dashed lines indicate the start of a new year.

| **IceCube Season** | Start date | Number of Events | Livetime [s] | Rate [mHz] |
|---|---|---|---|---|
| IC86.2011 | 2011/05/13 | 36,293 | 28,748,982 | $1.262 \pm 0.007$ |
| IC86.2012 | 2012/05/15 | 35,728 | 27,950,931 | $1.278 \pm 0.007$ |
| IC86.2013 | 2013/05/02 | 37,823 | 29,844,500 | $1.267 \pm 0.007$ |
| IC86.2014 | 2014/05/06 | 38,926 | 30,874,229 | $1.261 \pm 0.006$ |
| IC86.2015 | 2015/05/18 | 39,930 | 31,325,562 | $1.275 \pm 0.006$ |
| IC86.2016 | 2016/05/25 | 38,765 | 30,549,531 | $1.269 \pm 0.006$ |
| IC86.2017 | 2017/05/25 | 44,403 | 34,733,434 | $1.278 \pm 0.006$ |
| IC86.2018 | 2018/06/19 | 33,867 | 26,720,667 | $1.267 \pm 0.007$ |
| **Total** | | 305,735 | 240,747,841 | $1.270 \pm 0.002$ |

Table 5.4: The total number of $\nu_\mu$ events, livetimes, and rates from each IceCube season considered in this analysis.

The gain in the DOMs is known to increase with time and therefore at the beginning of every season the high-voltage on the DOMs is adjusted accordingly to maintain a gain of $10^7$. To verify the stability of the extracted charge as a function of time, an analysis into the time variation of the single photoelectron charge distribution was performed. The variables used to describe the single photoelectron charge distribution (an exponential and Gaussian) were found to have no systematic variation as a function of time greater than that observed by randomly scrambling of the years, in agreement with the stability checks performed in Ref. [134].

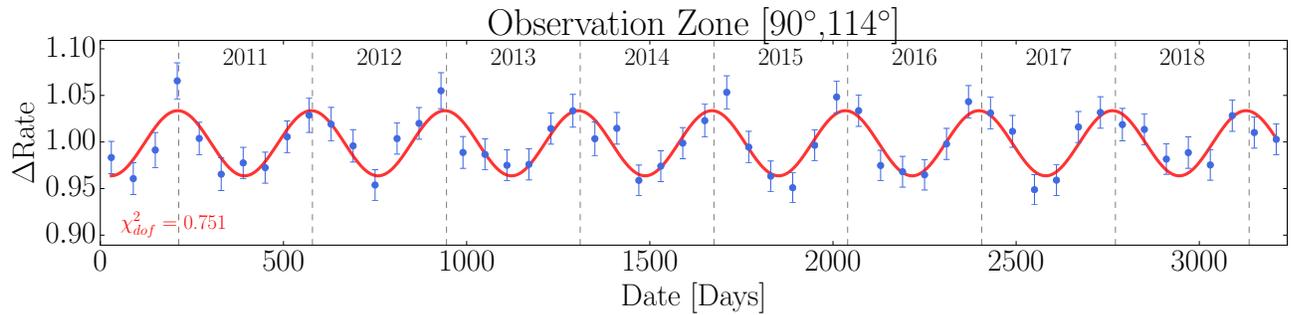

Figure 5-8: The seasonal variation in atmospheric neutrinos.

Finally, IceCube has previously shown that the atmospheric conditions presented to the cosmic ray flux affect the atmospheric neutrino spectrum [189]. This is observable as seasonal spectrum variations to the neutrino flux. When looking at the neutrino originating in the Antarctic atmosphere (24° below the horizon), we find the expected periodicity.

# Chapter 6

# ANALYSIS DESCRIPTION

This chapter will provide a complete overview of the full analysis including the statistical treatment of the hypothesis testing as well as a description of the expected signal shape in terms of reconstructed muon energy and $\cos(\theta_z)$.

## 6.1 Overview

We will be performing two searches for the signature of a 3+1 sterile neutrino using IceCube data. They will be referred to as "Analysis I" and "Analysis II" throughout the remainder of this thesis. Analysis I scans through the mass squared splitting $\Delta m^2_{41}$ over the mixing amplitude $\sin^2(\theta_{24})$. In this search, the 3+1 extended PNMS mixing elements $|U_{\tau 4}|^2$ is set to zero. This is a conservative choice and provides the weakest sensitivity. We also do not have sensitivity to $|U_{e4}|^2$ and it is therefore set to zero in both searches. Analysis II performs a scan in the oscillation averaged out regime, $\Delta m^2_{41} \gtrsim 20\text{eV}^2$ over the two mixing amplitudes $\sin^2(\theta_{34})$ and $\sin^2(\theta_{24})$. Specifically, we set $\Delta m^2_{41} = 50$ eV$^2$ for this particular search.

At $E_\nu > 100$ GeV, oscillations due to the active neutrino mass splittings have wavelengths larger than the diameter of the Earth and can be neglected, nevertheless the oscillation hypotheses are calculated using the full 3+1 flavor neutrino oscillation formalism outlined in Sec. 2.2.1. The active neutrino mixing parameters are all set to the current global best fit values described in Sec. 2.7. We are insensitive to the active neutrino mass ordering and assume a normal mass hierarchy. Similarly, we are also insensitive to all CP violating phases and therefore set them all to zero (i.e. $\delta_{cp} = \delta_{41} = \delta_{42} = 0.0$).

The at-detector neutrino flux is calculated at each hypothesis point in the physics parameter space. The granularity of the physics parameter space explored is:

- $\Delta m_{41}^2$: sampled from 0.01 eV$^2$ to 100 eV$^2$ logarithmically in steps of 0.05.

- $\sin^2(2\theta_{24})$: sampled from $10^{-2.6}$ to 1.0 logarithmically in steps of 0.05.

- $\sin^2(2\theta_{34})$: sampled from $10^{-3.1}$ to 1.0 logarithmically in steps of 0.05.

The neutrino flux is assumed to be a composite of atmospheric and astrophysical neutrinos (see Sec. 2.5). The conventional atmospheric neutrino flux is calculated using the HillasGaisser2012 H3a [100] cosmic ray model and the extensive air showers (EAS) initiated by high energy cosmic ray is modeled using Sibyll2.3c [109]. The prompt atmospheric neutrino flux is set to be the BERSS [114] model. The astrophysical neutrino flux is assumed to be isotropic, following a single power law energy spectrum.

The atmospheric temperature profile is modeled using the monthly livetime averaged temperature data from the Atmospheric Infrared Sounder, AIRS [172]. The atmospheric neutrino flux is calculated using MCEq [112, 171], and propagated through the Earth using nuSQuIDS [190] according to the sterile hypothesis. We use the CSMS [191] cross-section for both the neutrino-nucleon interaction during propagation and the interaction near the IceCube detector. The Earth density is assumed to be spherically symmetric and the radial density profile is given by the PREM [176] model.

The kinematics of the neutrino-neuclon interaction near the IceCube detector is simulated using LeptonInjector [160] and weighted according to the flux hypothesis and cross section model using

LeptonWeighter [161]. The bulk ice is described in terms of 170 layer dependent absorption and scattering coefficients, and the magnitude/direction of anisotropy. The DOMs are simulated with a DOM oversizing of 1.0 (no oversizing) and photons are propagated using CLSim [164]. This analysis uses the latest version of the SPE charge templates [138, 139] to describe the in-ice DOM dependent single-photoelectron charge distributions.

The data is divided into 260 bins in reconstructed muon energy and the cosine of the zenith angle ($\cos(\theta_z)$). The reconstructed energy is logarithmically binned in steps of 0.10, from 500 GeV to 9976 GeV (13 bins). The $\cos(\theta_z)$ is binned linearly in steps of 0.05, from -1.0 to 0.0 (20 bins). The energy proxy is calculated using the internal IceCube software MuEX [170] and directional reconstruction uses MPEFit [192].

Events are selected using the background free Platinum event selection described in Sec. 5. Data that has been reprocessed with the updated calibration (Pass2, see Sec. 5.4) and included all 86 active IceCube strings in the detector was strictly used in this analysis. The total data set contains 305,891 muon neutrinos collected over a livetime of 240,747,841 seconds (7.634 years), starting on May 13[th] 2011 and ending on May 19[th] 2019.

Along with the 3 physics parameters, the model includes 18 continuous nuisance parameters used to account for the systematic uncertainties related to the neutrino flux, cross section, and IceCube detector. Each nuisance parameter is continuous, and will be detailed in Sec. 7 in terms of its impact on the bin-wise rate in reconstructed energy and $\cos(\theta_z)$. The event re-weighting, systematic uncertainties implementation, and minimization is handled in the SterilizeSuperOptimized branch of the GolemFit project [193]. The prior on each nuisance parameter is defined in terms of a Gaussian with a central value and width.

## 6.2 Statistical treatment, hypothesis testing

The sterile neutrino hypotheses are tested using the binned maximum likelihood method [194]. The log-likelihood at a given set of physics parameters, $\vec{\Theta} = \{\Delta m_{41}^2, \theta_{24}, \theta_{34}\}$, is given by:

$$-LLH_{\vec{\Theta}} = \min_{\vec{\eta}} \left( \sum_{i=1}^{i=N_{\text{bins}}} [x_i \ln \lambda_i - \lambda_i] + \sum_{\alpha}^{\alpha=N_{\text{nuis.}}} \frac{(\eta_\alpha - \mu_\alpha)^2}{2\sigma_\alpha^2} \right) \quad \ldots \quad \lambda_i = \lambda_i(\vec{\Theta}, \vec{\eta}), \qquad (6.1)$$

where the first term represents the Poisson statistical probability at bin $i$ given the observed number of events $x_i$ with an expected number of events $\lambda_i(\vec{\Theta}, \vec{\eta})$. The set of nuisance parameters is given the symbol $\vec{\eta}$.

The second term is a penalizing factor that assumes a Gaussian prior uncertainty for each nuisance parameter $\alpha$. The prior on nuisance parameter $\alpha$ has the central value $\mu_\alpha$ and standard deviation $\sigma_\alpha$. The $LLH$ value for sterile hypothesis, $\vec{\Theta}$, is reported after being minimized over all the nuisance parameters. The minimization is handled by the Limited-memory BFGS-B algorithm [195], which includes box constraints. We further expand Eq. 6.1 to incorporate the statistical treatment of having finite MC, however a full description of this procedure is beyond the scope of the thesis and can be found in Ref. [196].

The test statistic (TS) used to compare different points in the physics parameter space is defined as two times the difference in $LLH$ between the point of interest, $\vec{\Theta}$, and the minimum $LLH$ value in the space, referred to as the "best fit point":

$$TS_{\vec{\Theta}} := 2\Delta LLH = 2(LLH_{\vec{\Theta}} - LLH_{\min}). \qquad (6.2)$$

According to Wilks' theorem [197], the TS distribution follows a chi-squared distribution with N degrees of freedom (DOF). A point in the parameter space with a TS greater than the critical value (given in Table 6.1) is rejected at the corresponding confidence level. Wilks' theorem assumes an ellipsoidal likelihood surface. This condition may not necessarily be assured, such as in the case of a boundary. The coverage (the value which defines the critical value) can be verified through simulated pseudo-experiments. In this analysis, the Wilks' contours will be used to describe the final result and several points along the reported contours will be checked to ensure proper coverage. Analysis I and II are found to have approximately 2 DOF and 1 DOF respectively.

| DOF | $0.25\sigma$ | $0.5\sigma$ | $1.0\sigma$ | $1.64\sigma$ | $1.96\sigma$ | $2.0\sigma$ | $2.58\sigma$ | $3.0\sigma$ |
|---|---|---|---|---|---|---|---|---|
|  | 19.7% CL | 38.3% CL | 68.3% CL | 90% CL | 95% CL | 95.5% CL | 99% C.L. | 99.7% CL |
| 1 | 0.06 | 0.25 | 1.00 | 2.71 | 3.84 | 4.00 | 6.63 | 9.00 |
| 2 | 0.44 | 0.97 | 2.30 | 4.61 | 5.99 | 6.18 | 9.21 | 11.83 |

Table 6.1: The critical values at a given level of confidence, in terms of -2ΔLLH for a specified number of degrees of freedom.

## 6.3  A 3+1 sterile neutrino signal expectation in IceCube

In a 3+1 model, the matter enhanced resonance effect described in Sec. 2.3.5 is observable as a depletion in the $\bar{\nu}_\mu$ flux at eV-scale sterile mass states. It is this resonant depletion that we will be searching for in the first analysis. An illustration of this is shown in Fig. 6-1 for a 3+1 sterile hypothesis of $\Delta m_{41}^2 = 1.35$ eV$^2$ and $\sin^2(\theta_{24}) = 0.07$ (the current global best fit point for a 3+1 sterile neutrino model, see Ref. [73]). Here, the left figure shows the disappearance, shape-only, signal in reconstructed energy and $\cos(\theta_z)$. The plot on the right shows the location in the parameter space where the signal was generated.

The IceCube detector cannot distinguish between a $\bar{\nu}_\mu$ and $\nu_\mu$ CC interaction. The signal manifests itself primarily in the $\bar{\nu}_\mu$ flux, therefore the dominant $\nu_\mu$ flux will wash out the signal in reconstructed quantities considerably. Beyond this, IceCube reconstructs the final state muon energy in the CC $\nu_\mu$ interaction with an average angular resolution below 1° and energy resolution of $\sim \text{Log}(\sigma_E) = 0.5$. While the angular resolution is smaller than the chosen $\cos(\theta_z)$ bin width, the energy resolution is rather poor and broadens the signal shape in energy. This can be seen comparing the true signal shape at this location (Fig. 2-9) compared to the reconstructed signal shape in Fig. 6-1.

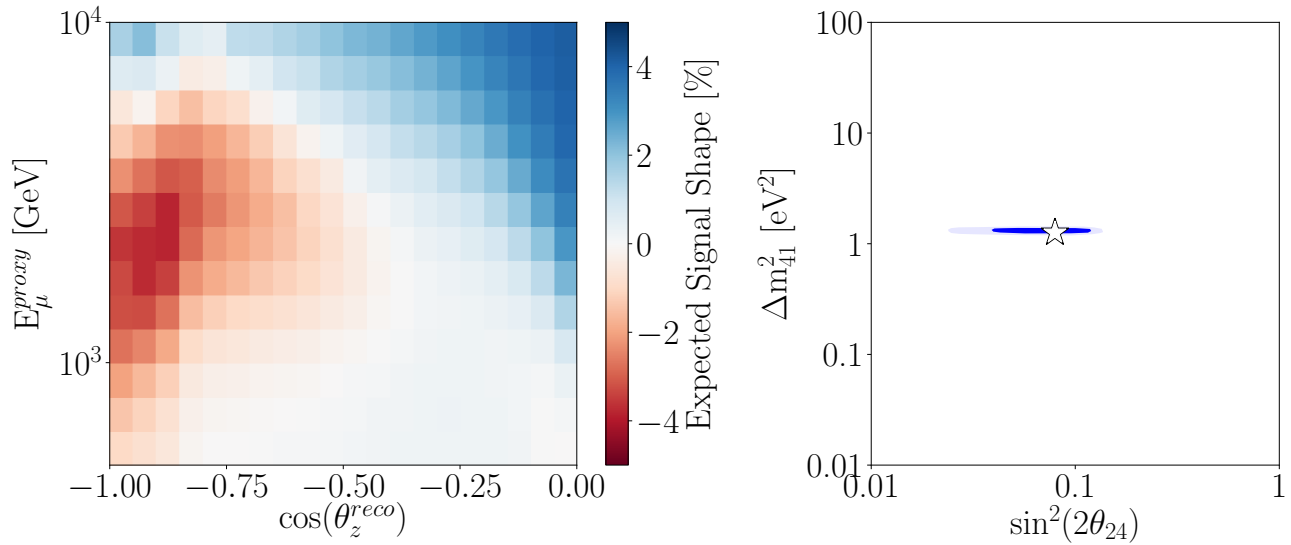

Figure 6-1: Left: The expected signal shape in IceCube (left) of a sterile neutrino with the physics parameters $\Delta m^2_{41} = 1.3$ eV$^2$ and $\sin^2(2\theta_{24}) = 0.07$. The signal shape is shown in terms of reconstructed energy and $\cos(\theta_z)$. The normalization of the signal has been removed in order to show the scale of the distortion. Right: The global best fit value (at the star) from Ref. [73], along with the 90% and 99% allowed regions in blue.

At larger values of $\Delta m^2_{41}$ the resonance occurs at larger energies (Eq. 2.28). At larger mixing angles, $\sin^2(2\theta_{24})$, the required density for the resonance decreases, pushing the signal away from the core and into the mantle. As the resonance moves beyond the upper energy limit of Analysis I, the higher frequency oscillations average out into an overall normalization shift dependent on the density profile from which they propagated through. This causes a zenith dependent effect as shown in Fig. 6-2 for a point in terms of Analysis II. Here, we show the signal shape along the 90% C.L. contour of the three year DeepCore analysis [155].

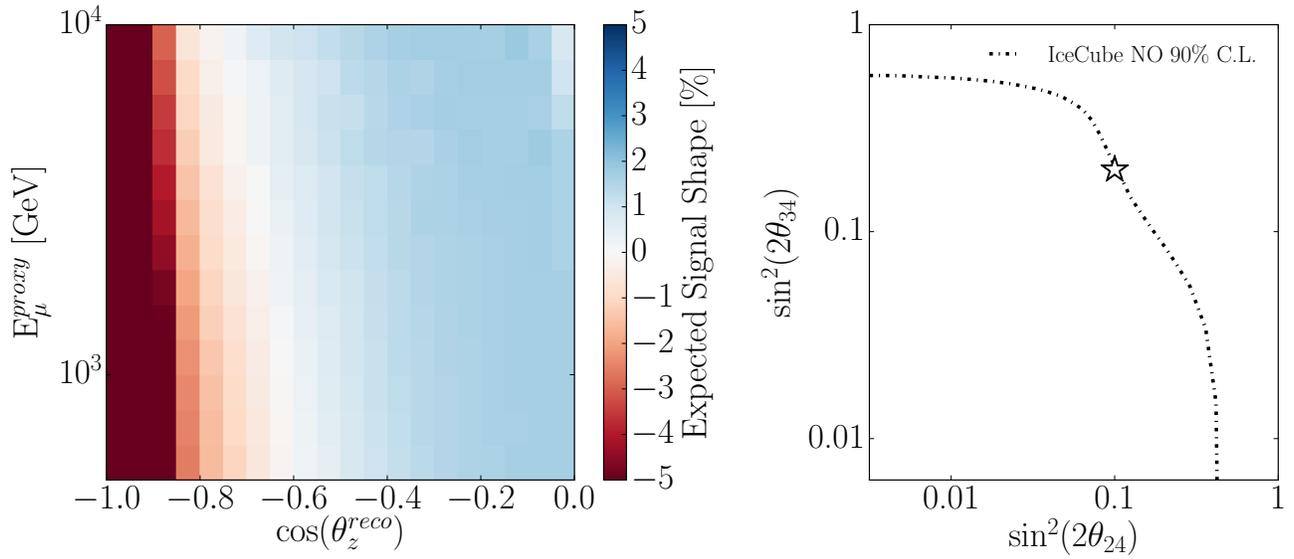

Figure 6-2: The expected signal shape in IceCube (left) of a sterile neutrino with the mixing parameters indicated in by the white star (right). The signal shape is shown in terms of reconstructed energy $\cos(\theta_z)$. The normalization of the signal has been removed in order to show the scale of the distortion. The dot-dashed contour on the right comes from the three year DeepCore low energy sterile analysis result [155].

THIS PAGE INTENTIONALLY LEFT BLANK

# Chapter 7

# Systematic uncertainties

Compared to the IC86.2011 high energy sterile neutrino search by IceCube [123], the number of $\nu_\mu$ events used in this analysis is nearly a factor of 15 larger. This corresponds to a rather dramatic decrease in the bin-wise statistical uncertainty. In many areas of the reconstructed energy-$\cos(\theta_z)$ space, this analysis is systematically limited and therefore a considerable amount of effort was devoted to properly modeling and understanding the systematic uncertainties. Each uncertainty reported in this chapter will be described in terms of the shape it generates on the reconstructed energy-$\cos(\theta_z)$ plane as it is perturbed within the limits of its prior range.

## 7.1 Detector uncertainties

### 7.1.1 DOM efficiency

The term *DOM efficiency* is used internally to describe the absolute photon detection efficiency of the full detector. While it is commonly associate specifically with DOMs, it also encompasses any physical property that changes the percentage of photons that deposit a measurable charge in the detector. This includes effects not only specific to the DOMs, such as photocathode efficiency, collection efficiency, angular sensitivity, wavelength acceptance, photocathode shadowing, and DOM glass transparency; but also external properties, such as the cable shadow, hole ice properties, and bulk ice properties. As the standard IceCube MC evolves, there is a corresponding DOM efficiency shift to maintain agreement with measured quantities, such as properties associated with minimum ionizing cosmic ray muons [170].

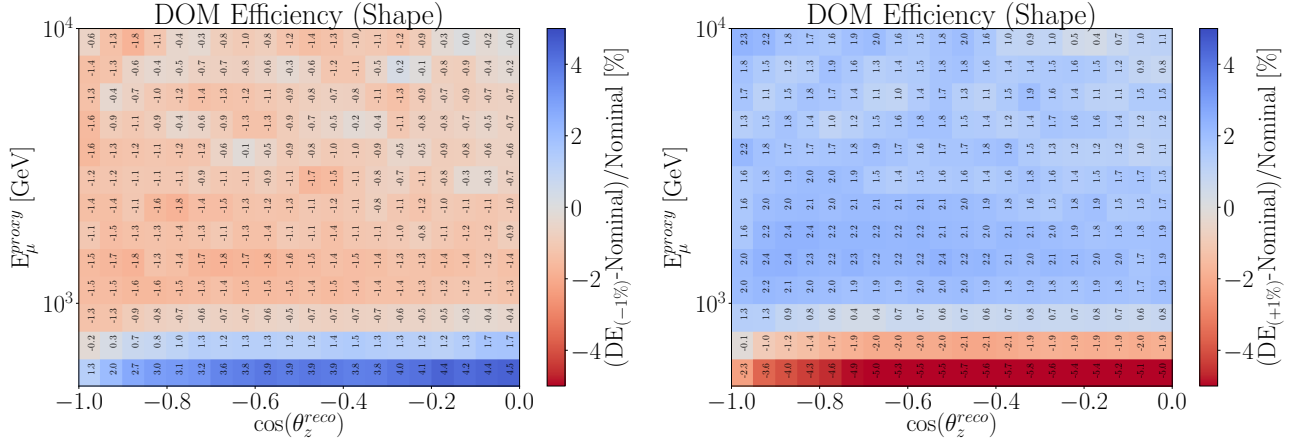

Figure 7-1: The shape-only rate change generated in the energy-$\cos(\theta_z)$ plane when comparing a DOM efficiency (DE) of -1% (left) and +1%(right) to the nominal MC set (at a DOM efficiency of 0.97).

The secondary particles in the simulated neutrino-nucleon interaction are propagated through the ice with an overabundance of photons produced along their track. During the detector level simulation (see Sec. 4.2.3), photons are down-sampled (i.e. a percentage of the propagated photons are randomly destroyed) to the desired DOM efficiency. In our case, the photons were generated at a DOM efficiency of 1.10, then down-sampled to the central value of 0.97. Five systematic data sets were generated relative to the central value at +6.3%, +4.7%, +2.4%, −1.6%, and −3.1%, which allow us to probe DOM efficiency values between approximately 0.93 and 1.03.

Each systematic data set was processed through the online filters to the final analysis level. They were then weighted to the central MC hypothesis and penalized splines [198] were generated of the 2D reconstructed energy and $\cos(\theta_z)$ distributions. We then linear interpolate between the splines and normalize the final 3D distribution by the central MC set. This procedure is used for various systematic data sets (the hole ice and cross-sections) and allows us to re-weight each event to any systematic value within the range of the splined region.

An example of the shape-only (normalization removed) effect of perturbing the DOM efficiency by ±1% relative to the central MC set is shown in Fig. 7-1. As expected from a change in the average observed charge, the shape manifests itself primarily in terms of a shift in reconstructed energy, with lower DOM efficiencies pulling the mean reconstructed energy to lower values. The prior is chosen to have a wide width, ±10%, in order to encompass all the modern simulation updates used in this analysis.

### 7.1.2 Bulk ice

The bulk ice corresponds to the undisturbed ice in-between the strings and is described in detail in Sec. 3.3.3. This ice is partially characterized by the scattering and absorption coefficients in a given layer of ice. Previous IceCube analyses quantified the bulk ice uncertainty by scaling all scattering and/or absorption coefficients by a fixed amount (typically ±5-10% depending on the ice model used in the analysis). This uncertainty, however, does not account for potential systematic differences in the depth dependent layers. Therefore, a new approach was investigated.

The ice model was first redefined in terms of a Fourier decomposition of $\log_{10}(\text{Abs}\times\text{Sca})$ as a function of ice layer number:

$$\frac{1}{2}\text{Log}_{10}(\text{Abs} \times \text{Sca}) = \frac{A_0}{2} + \sum_{n=1}^{N} A_n \sin\left(\frac{2n\pi x}{L} + \phi_n\right), \tag{7.1}$$

where x is the coordinate of the layer and L is the extent of the layers. This transition moves the parametrization from a depth dependent description of the ice (with 170 layers), to 170 Fourier modes, whose relative amplitudes are shown in Fig. 7-2. Low frequency modes reflects macroscopic changes over large regions of the IceCube detector. For example, Mode 1 describes the scattering and absorption at the upper part of the detector compared to the lower part of the detector. The high frequency modes, on the other hand, have a minimal effect in analysis space, since the effect of the fast-varying ice properties effectively cancels when integrating over the detector.

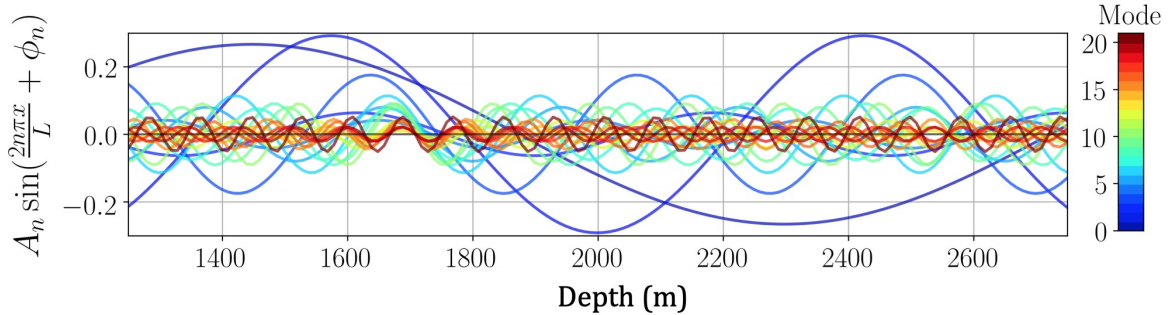

Figure 7-2: The first 20 Fourier modes of the decomposed ice model, neglecting mode zero (the DC mode). From Ref. [199].

Each Fourier mode can be treated as a separate systematic, where its amplitude and phase move continuously within a defined prior. However, it was found that only the first few modes dominate the uncertainty. The priors were determined by perturbing the central ice model by one of the amplitudes or phases. The likelihood profile, as a function of the size of the perturbation, was then calculated by comparing the perturbed simulation to flasher data. From this, we can define the prior width.

A full systematic data set, called SnowStorm, was generated where every group of 100 neutrino events was assigned a different ice model determined by perturbing each amplitude and phase within its prior by a randomly z-score. At analysis level, we asses the impact of each mode by splitting the MC into two sets: one set corresponds to the chosen amplitude/phase being positively perturbed and the other, negatively perturbed. We define the fractional gradient for a given amplitude or phase as:

$$G_i = \sqrt{\frac{\pi}{2}} \frac{[\psi(E_{\text{reco}})_{\eta_i > 0} - \psi(E_{\text{reco}})_{\eta_i < 0}]}{\psi(E_{\text{reco}})}, \qquad (7.2)$$

where $\psi(E_{\text{reco}})$ corresponds to the reconstructed energy distribution, $\eta_i > 0$ represents the positively perturbed mode $i$, and $\eta_i < 0$ represents the negatively perturbed mode. The fractional gradient is shown for both the amplitude and phase of the first 6 Fourier modes (the modes shown to have an impact at analysis level) in Fig. 7-3. The distributions shown in this figure can be further redefined in terms of two correlated basis functions. A linear combination of these functions was found to be able to reproduce the first 6 fractional gradients for the amplitudes and phases. These basis functions are what we refer to as the Ice Gradient 0 and Ice Gradient 1, and correspond to the actual shape of the systematic uncertainty used in this analysis. While they do not vary independently, Fig. 7-4 shows the shape of perturbing a single gradient by one sigma relative to the central value with the correlation removed. Further details regarding the implementation of this systematic uncertainty can be found in Ref. [199].

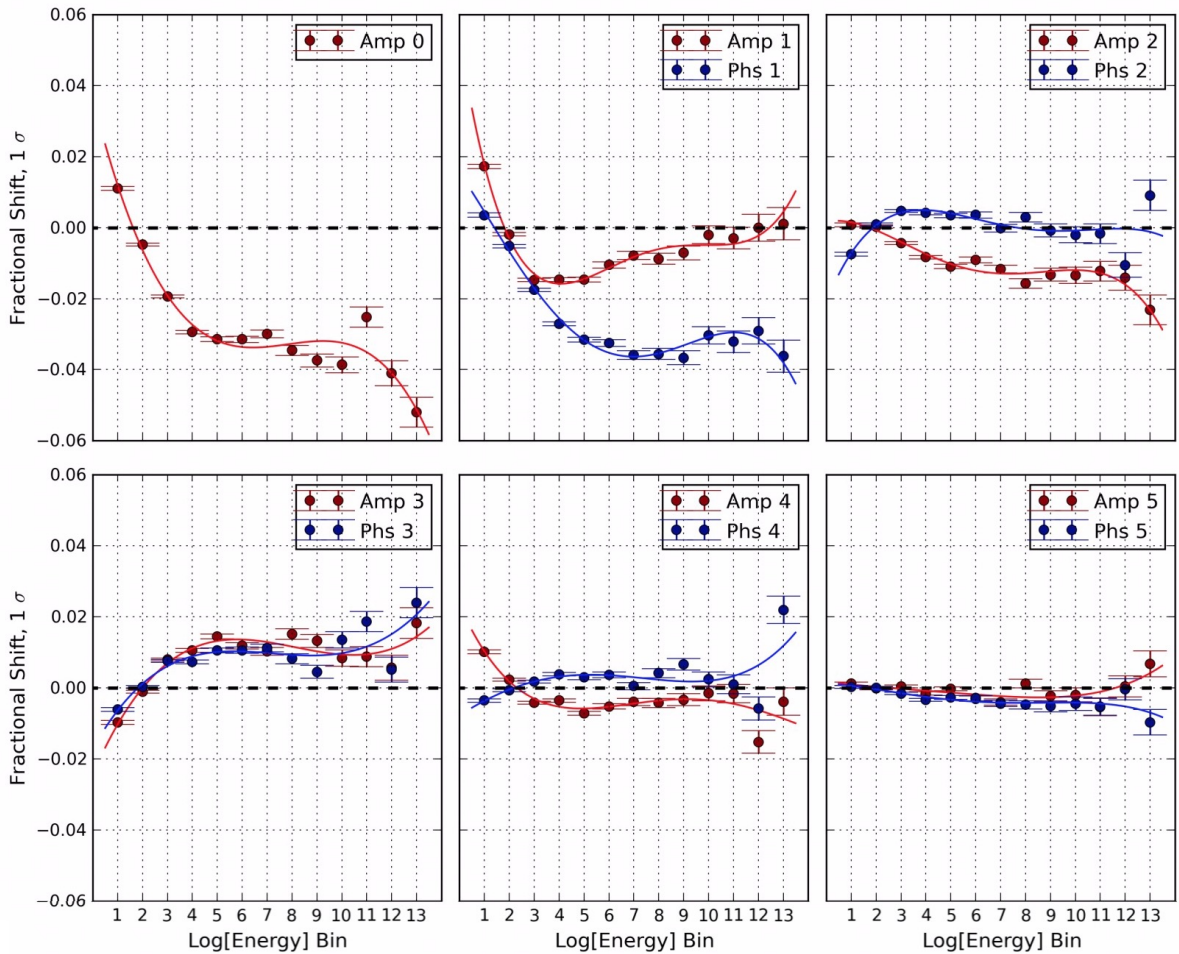

Figure 7-3: The fractional energy gradients split along Fourier mode amplitudes 0 to 5 (red), and phases 1 to 5 (blue). Mode 0 corresponds to the DC shift in the scattering and absorption and therefore has no associated phase. Modified from Ref. [199].

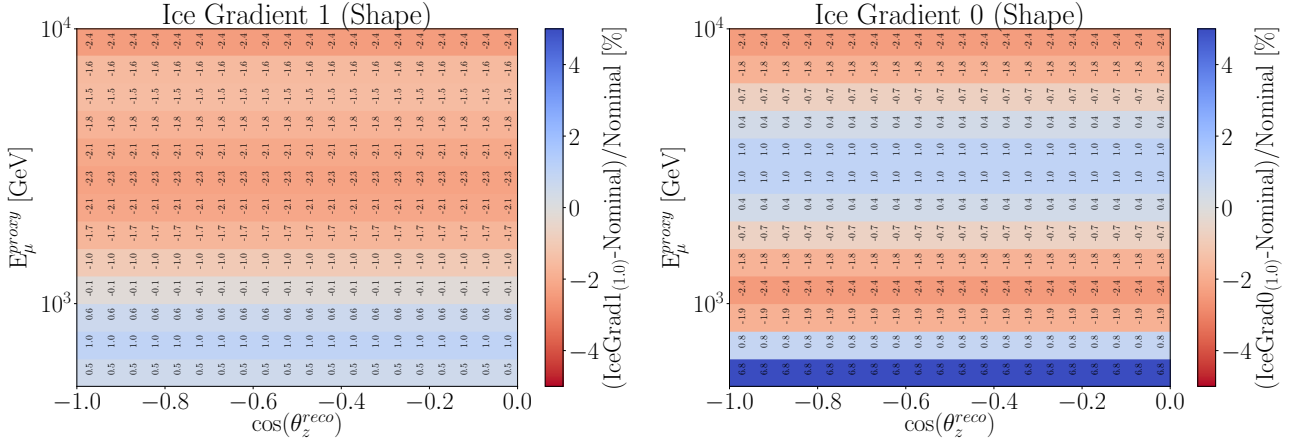

Figure 7-4: The energy and zenith shape difference comparing the central MC to the different Ice Gradients (with the correlation set to zero). Left: the shape generated by scaling the Ice Gradient 1 parameter by $1\sigma$. Right: The equivalent plot for Ice Gradient 0.

### 7.1.3 Hole ice

While the dominant part of the Cherenkov photon propagation occurs in the bulk ice, each photon detected by a DOM must also propagate through the refrozen ice in the boreholes (the hole ice [200]). Evidence from recorded images of the refreezing process [136] lead to the modeling hole ice as a transparent component the extends from the edge of the hole inwards and a column of bubbles/impurities (roughly 8 to 10 cm in diameter) in the center of the hole. We account for the change in optical properties due to the hole ice as change in the angular acceptance curve for all DOMs.

The angular acceptance is parameterized as:

$$A(\eta) = 0.34(1 + 1.5\eta^3/2) + p_1\eta(\eta^2 - 1)^3 + p_2 e^{(10(\eta-1.2))}, \qquad (7.3)$$

where $\eta$ is the angle of the incoming photon (as indicated in the left side of Fig. 7-5) and $p_1$, $p_2$ are free parameters. The $p_2$ parameter primarily varies the up-going photons ($\cos(\eta) = 1.0$), which are more likely to be affected by the bubble column. This parameter is often referred to as the "forward hole ice" and will be included as a systematic uncertainty in this analysis. The $p_1$ parameter, on the other hand, was found to have much less of an impact and was therefore

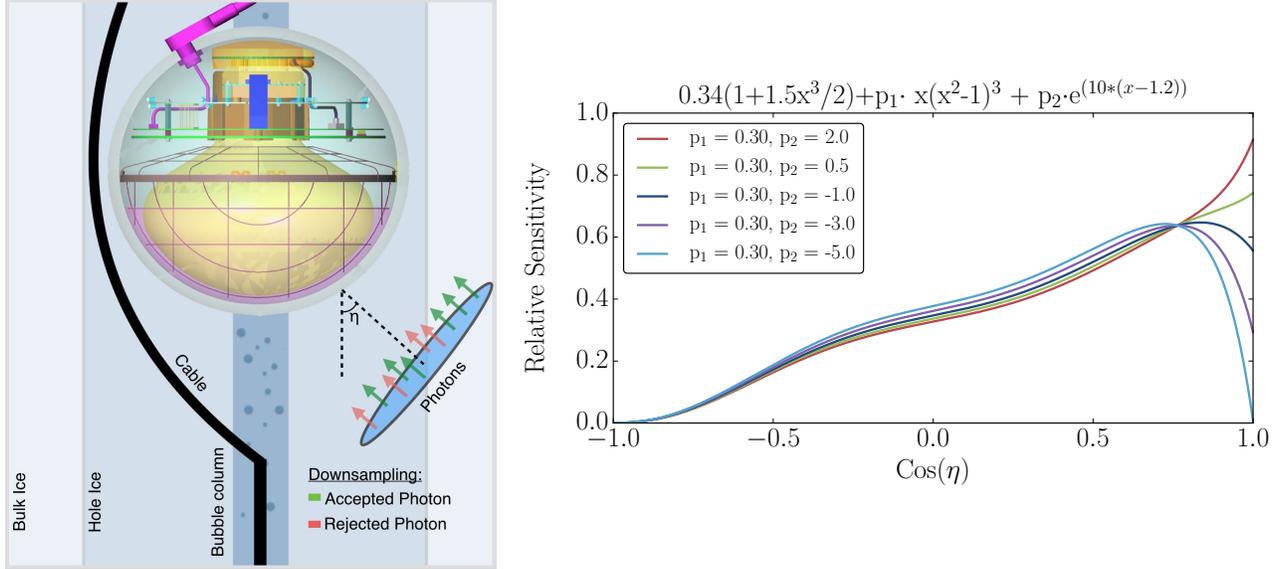

Figure 7-5: Left: The structure of the ice surrounding each DOM, illustrating the location of the bubble column, data transmission cable, bulk ice, and hole ice. Right: Various descriptions used in this analysis for the angular acceptance curves as a function of incoming photon angle relative to the DOM.

not included.

Five identical sets of MC were generated with the only difference being the description of the angular acceptance forward hole ice parameter ($p_2$ = [-5, -3, -1, 1,3] and $p_1$ = 0.3), these are shown in Fig. 7-5 (right). Each of these curves is normalized to 0.68 to maintain a constant overall efficiency factor. Following the description for the DOM efficiency, splines were generated which allow us to re-weight each event to any continuous value for $p_2$ between -5 and 3. The central MC set was chosen to be $p_2$ = -1.0 and $p_1$ = 0.3 and we assign a prior width of $p_2 = \pm 10$ (essentially flat). The shape generated by perturbing the forward hole ice to -3 and +1 relative to the central set is shown in Fig. 7-6.

## 7.2 Atmospheric neutrino flux uncertainties

This section outlines the central models used to define our our atmospheric neutrino flux and then breaks down how the uncertainties are implemented. Unlike the IC86.2011 high energy sterile search, we have moved from discrete variants of the cosmic ray and hadronic interaction models to continuous variables that describe the actual neutrino flux production. A comparison between the continuous parameterization of the uncertainty and the discrete models will be highlighted here and compared further in Sec. E.0.1.

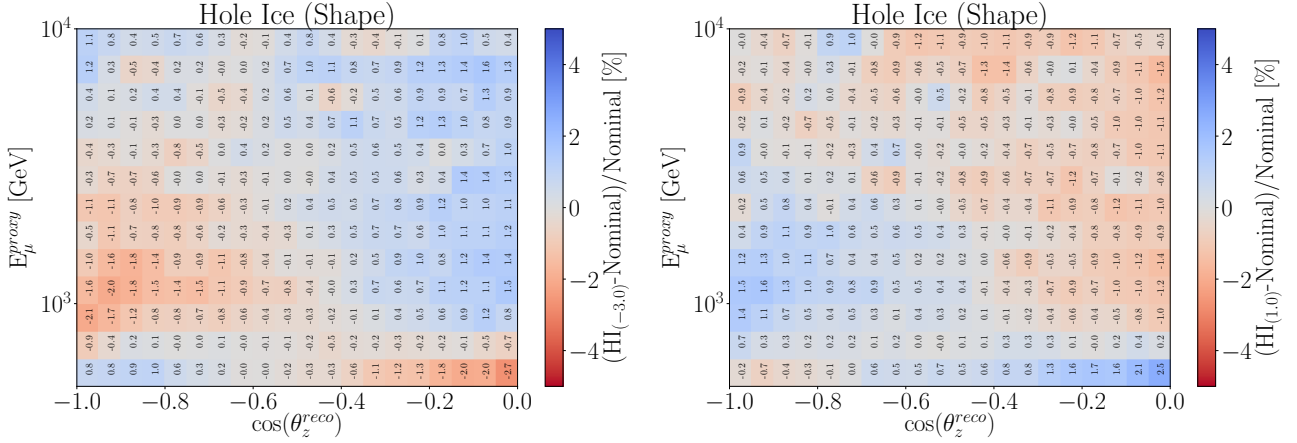

Figure 7-6: A comparison of the shapes generated relative to the central MC of two different forward hole ice parameter settings: p$_2$ = -3 (left) and +1 (right).

The nominal cosmic ray model used in this analysis is Hillas-Gaisser2012 with the H3a modification [201]. We use the hadronic interaction model Sibyll2.3C [109] to describe the extensive air showers initiated by high energy cosmic rays. The atmospheric prompt neutrino flux is described by the BERSS model [114], the relative contribution of which is shown in Fig. 7-7. The Earth's atmospheric temperature profile is provided by NASA's Atmospheric InfraRed Sounder (AIRS) satellite [172]. Monthly temperature variations and live time differences (see Fig. 7-9) are accounted for in the atmospheric neutrino flux prediction.

Fig. 7-8 shows the predicted atmospheric muon neutrino flux (relative to the total muon neutrino flux based on the central flux model) at the South Pole, coming from the horizon (left) and vertically up-going (right) for various models. In this figure, the color combination represents the neutrino flux from a given progenitor labeled at the top of the figure. The four subplots further reduce the flux components into different categories. The top subplot separates the flux into the neutrinos and anti-neutrinos components. The second from the top plot shows the span of three independent cosmic ray models. The third from the top shows two different hadronic

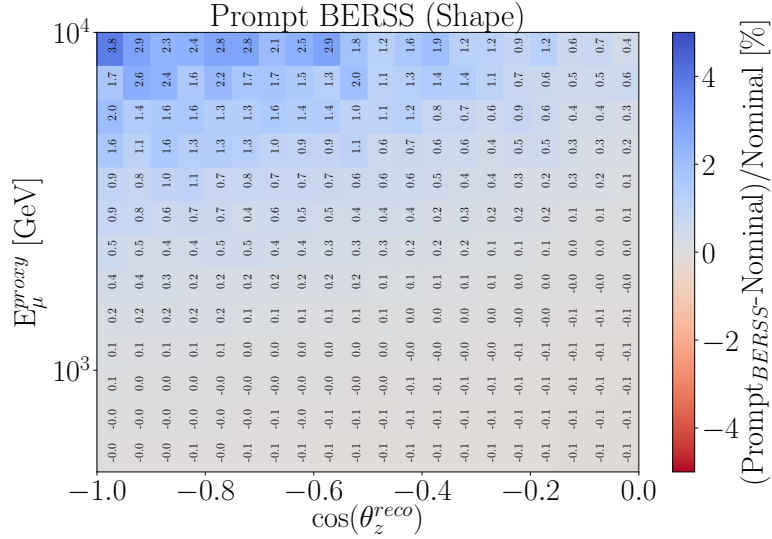

Figure 7-7: The shape generated by including the BERSS prompt neutrino flux prediction. Here, Nominal corresponds to the central neutrino flux without the prompt contribution.

interaction models. And the final plot shows the effect of taking two different atmospheric temperature profile models.

The uncertainty in the conventional neutrino spectrum is broken down into the uncertainty in the meson production in the atmosphere (Barr parameters), the overall normalization, the cosmic ray spectral index, and atmospheric density. This set of uncertainties is sufficient to span the space allowed by the models described above (see Sec. E.0.1). The following subsections give an overview of how each of these were implemented.

The monthly livetime throughout the seasons considered in this analysis compared to the total livetime of each month is shown in Fig. 7-9.

### 7.2.1 Barr parametrization

The Barr parameterization [203] describes the uncertainty associated with the production of pions and kaons in hadronic interactions based on accelerator data. The uncertainties are calculated for a given total incident parent energy, $E_i$, secondary total energy, $E_s$ (equivalently, the term $x_{lab} = E_i/E_s$), and separately for the charge of the produced meson. In the energy range of interest to this analysis (100 GeV to 10 TeV), the neutrino flux is dominated by kaon decay. The Barr parameters responsible for describing the uncertainties associated the pion production ($A^\pm$ to $I^\pm$) were found to be negligible at analysis level (see Sec. E.1.3), therefore we restrict ourselves to only those that impact the kaon production above 30 GeV: $W^\pm$, $Y^\pm$, and $Z^{\pm}$. The relevant phase space for each parameter in terms of $x_{lab}$ and primary energy are shown in the second and third column of Table 7.1.

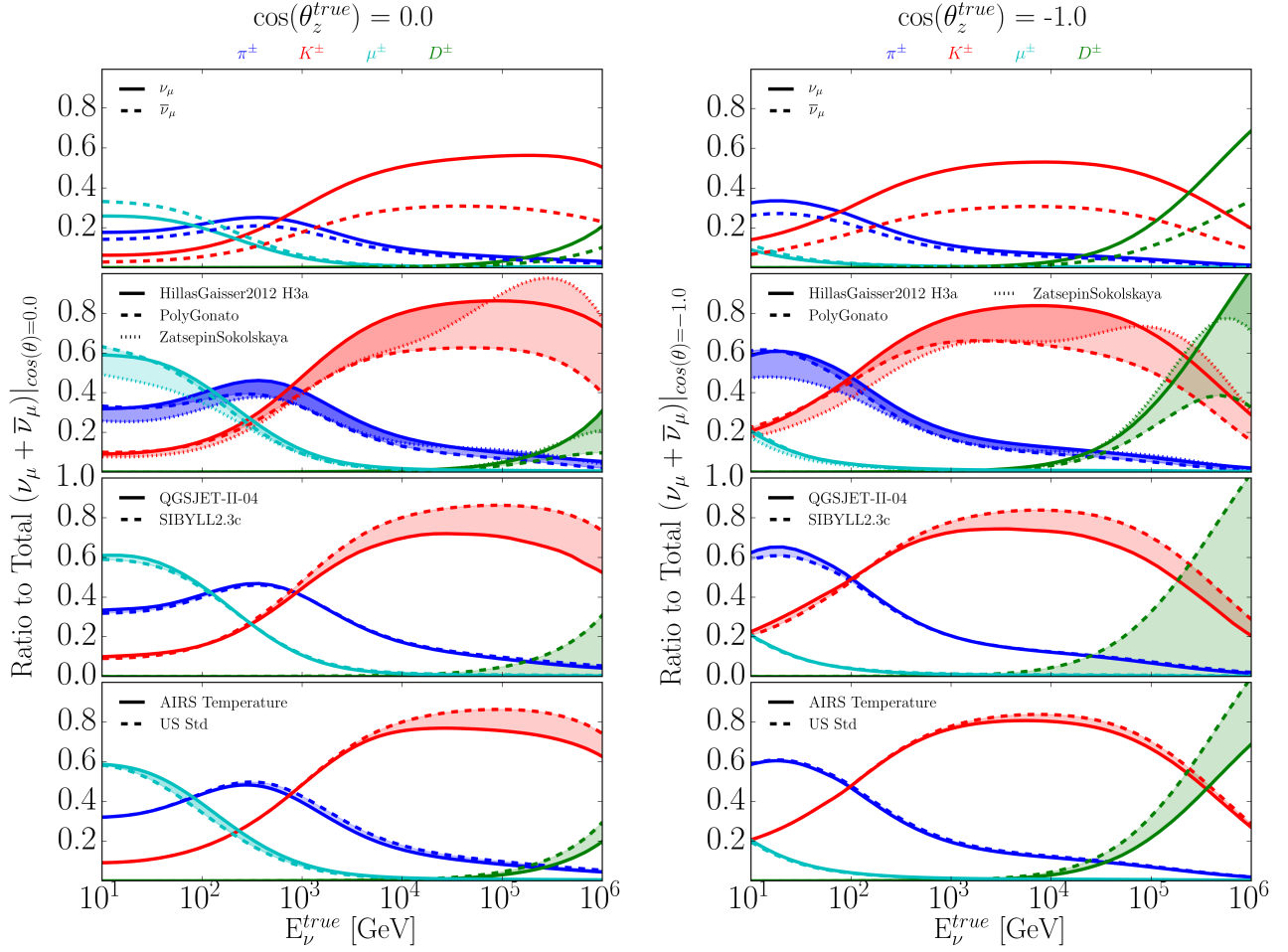

Figure 7-8: The MCEq calcualted flux broken down into its various progenitors (color component) compared to the total $\nu_\mu + \overline{\nu}_\mu$ flux at the IceCube detector, assuming a central model comprizing of HillasGaisser2012, Sibyll2.3c, and AIRS temperature profile. The top plots show the neutrino and antineutrino flux as a function of energy broken up into the individual progenitors. The second from the top shows the total $\nu_\mu + \overline{\nu}_\mu$ from the individual components and the spread given a different cosmic ray model (HillasGaisser2012 [201], Polygonato [104], and Zatespkin Sokolskaya [102]). The third plot from the top shows the change in the energy spectrum given two different hadronic flux models (QGSJET [202] and Sibyl 2.3c [110]). The bottom plots shows the flux hypothesis assuming two different earth temperature profiles: AIRS and the 1976 US Standard [173].

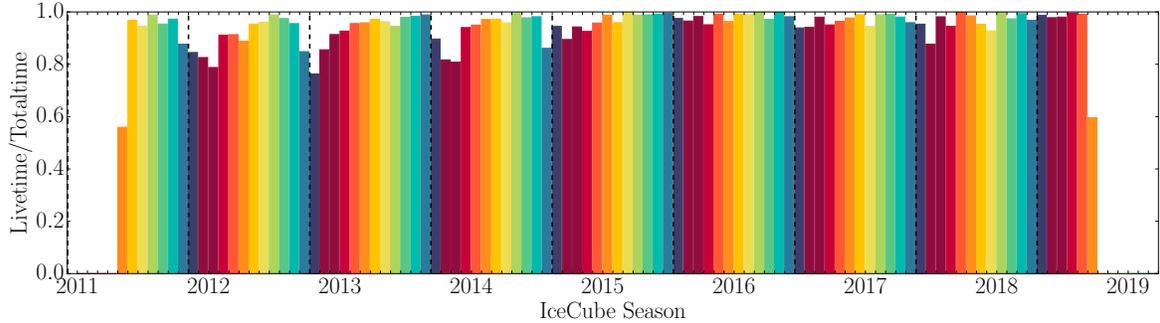

Figure 7-9: The ratio of total monthly livetime to the total length of each month. The atmospheric neutrino flux is calculated separately for each month, weighted by the livetime ratio shown here, then averaged over all months. This accounts for seasonal effects due to monthly livetime differences. The vertical dashed black lines indicate the beginning of a new year.

| Parameter | $x_{lab}$ | Energy [GeV] | Meson | Uncertainty |
|---|---|---|---|---|
| $W^{\pm}$ | 0.0-0.1 | $30 - 1\times10^{11}$ | $K^{\pm}$ | 40% |
| $Y^{\pm}$ | 0.1-1.0 | $30 - 1\times10^{11}$ | $K^{\pm}$ | 30% |
| $Z^{\pm}$ | 0.1-1.0 | $500 - 1\times10^{11}$ | $K^{\pm}$ | 12.2% $\log_{10}(E/500\text{GeV})$ |

Table 7.1: The uncertainties associated with the three relevant Barr parameters, along with the description of the phase space in which they are valid.

Fluxes are generated at the extremes of these uncertainties in order to derive gradients which can be used to translate the $1\sigma$ uncertainties into effective shapes in the reconstructed energy - $\cos(\theta_z)$ plane. Fig. 7-10 shows the shape of each of the Barr parameters perturbed to $+1\sigma$. We use the nomenclature "P" and "M" on the Barr parameters to denote whether they are used for the positively or negatively charged mesons.

### 7.2.2 Conventional neutrino flux normalization

At large values of $\Delta m^2_{41}$ there are areas in the physics parameter space with small signal shape and large normalization shifts due to the fast neutrino oscillations averaging out. The small signal shape can be rather localized, in which case statistical fluctuations in the data may tend to prefer that region, albeit for the normalization pulling to match the oscillations averaging out. It is therefore important to include an uncertainty on the conventional neutrino flux normalization, $\Phi_{\text{conv.}}$. The uncertainty remains the same as that used in the IC86.2011 high energy sterile analysis [204]. It was primarily derived from the theoretical uncertainty reported in Ref. [171] and an extrapolation from the uncertainties quoted in the HKKM calculation [205].

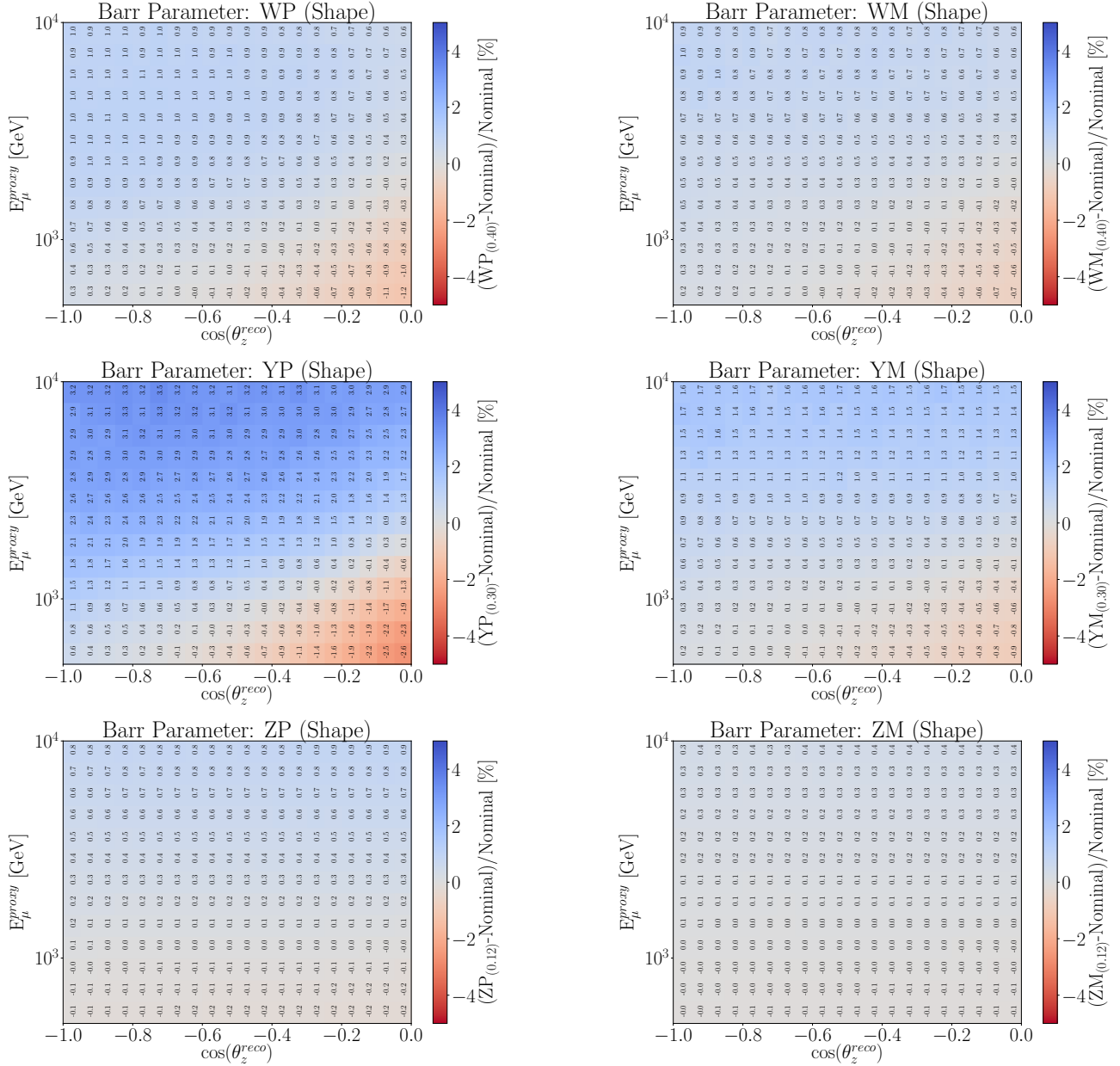

Figure 7-10: The shape-only change in reconstructed energy and zenith generated by perturbing the 6 relevant Barr parameters (WP, WM, YP, YM, ZP, and ZM) by $1\sigma$ from their central value.

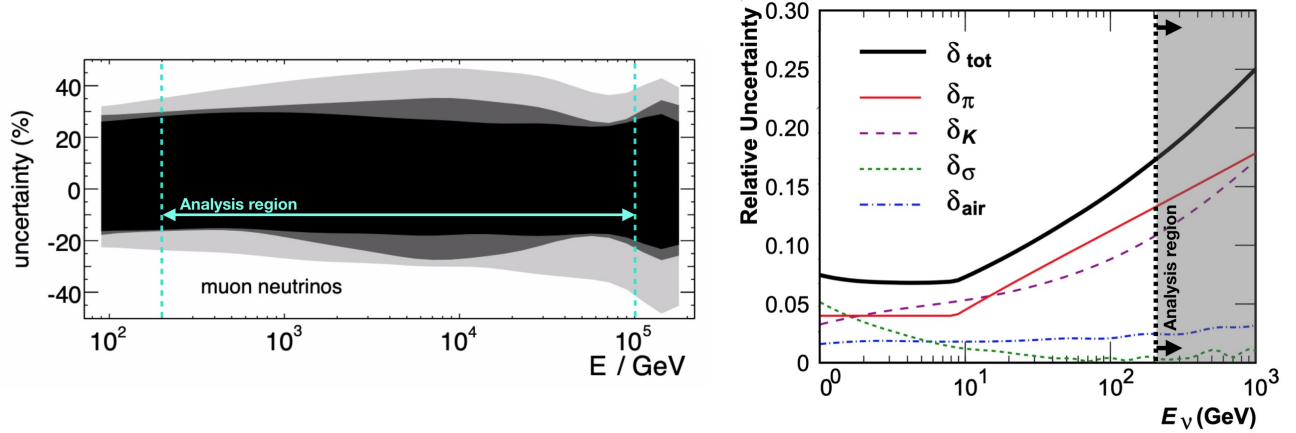

Figure 7-11: Left: The uncertainty derived from the variation of the interaction model (inner solid band) and from combinations of both the interaction model along with various cosmic ray model (shaded bands) as a function of muon neutrino energy. Modified from Ref. [171]. Right: The total neutrino production uncertainty (solid black line) broken down into the individual contributors. Here, $\delta_\pi$ ($\delta_K$) is the uncertainty due to the uncertainty of pion and kaon production, $\delta_\sigma$ is due to the hadronic interaction cross-sections, and $\delta_{\text{air}}$ is due to the atmospheric density profile. Modified from Ref. [205].

The theoretical uncertainty reported in Ref. [171], shown in Fig. 7-11 (left), accounts for both the cosmic ray and hadronic interaction model in the energy range of interest for this analysis. Up to approximately 1 TeV, the hadronic interaction model represents the majority of the uncertainty since the cosmic ray models in this regime are relatively well established. Above this energy, the uncertainty in the cosmic ray knee dominates the total uncertainty. The sub-TeV uncertainty is in agreement with the calculated total uncertainty found in the HKKM calculation [205], Fig. 7-11 (right). At 1 TeV, the uncertainty is reported as 25% and consists of the uncertainties associated with the pion ($\delta_\pi$) and kaon ($\delta_K$) production, hadronic interaction cross-section ($\delta_\sigma$), and atmospheric temperature profile ($\delta_{\text{air}}$).

Based on the findings described above, we include a 40% uncertainty on the conventional atmospheric neutrino normalization. The shape and normalization developed when perturbing the conventional atmospheric normalization by $\pm 1\sigma$ is shown in Fig. 7-12. A small shape emerges as the relative contribution of the astrophysical and prompt neutrino flux normalization varies.

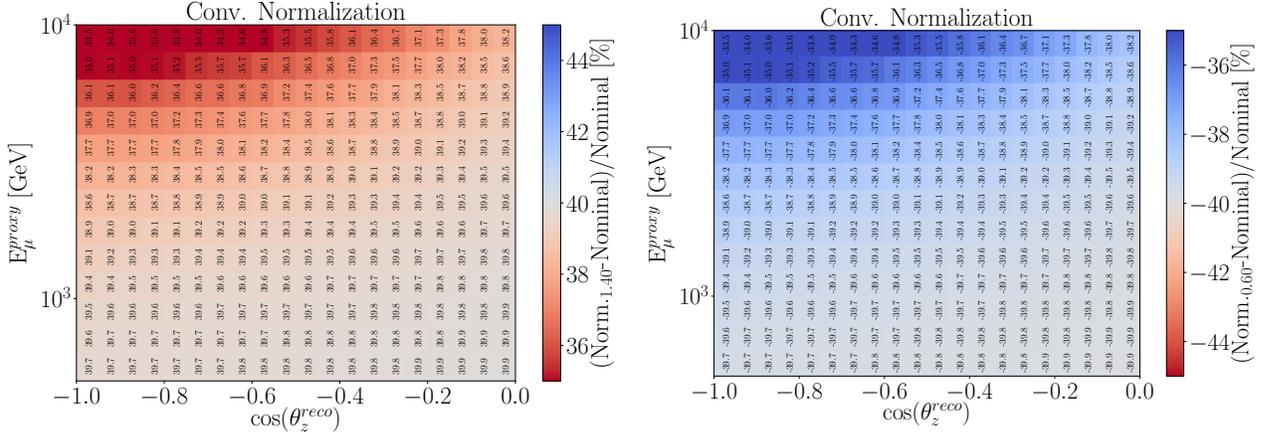

Figure 7-12: The shape generated when perturbing the normalization of the conventional neutrino flux by $\pm 1\sigma$.

### 7.2.3 Cosmic ray spectral slope

In the energy range of interest for this analysis, the cosmic ray spectrum responsible for producing the atmospheric neutrinos follows approximately an $E^{-2.65}$ energy dependence. We attribute a spectral shift, $\Delta\gamma$, to the energy dependence as:

$$\phi(E) = \phi(E)\left(\frac{E}{E_0}\right)^{-\Delta\gamma}, \tag{7.4}$$

where $E_0$ has been chosen to be 2.2 TeV in order to preserve the normalization.

The measured cosmic ray spectral index from the recent measurements is shown in Table 7.2. Based on these measurements, we assign an prior width on the cosmic ray spectral shift of $\Delta\gamma = 0.03$. The shape of the cosmic ray spectral shift at $\pm 1\sigma$ is shown in Fig. 7-13.

| Experiment | Year | Energy Range | C.R. Slope |
|---|---|---|---|
| CREAM-III [206] | 2017 | 1TeV - 200TeV | $-2.65 \pm 0.03$ |
| HAWC [207] | 2017 | 10TeV - 500TeV | $-2.63 \pm 0.01$ |
| Argo-YBJ [208] | 2016 | 3TeV - 300TeV | $-2.64 \pm 0.01$ |
| PAMELA [209] | 2011 | 50TeV - 15TeV | $-2.70 \pm 0.05$ |

Table 7.2: The measured cosmic ray spectral slope and uncertainty for several experiments.

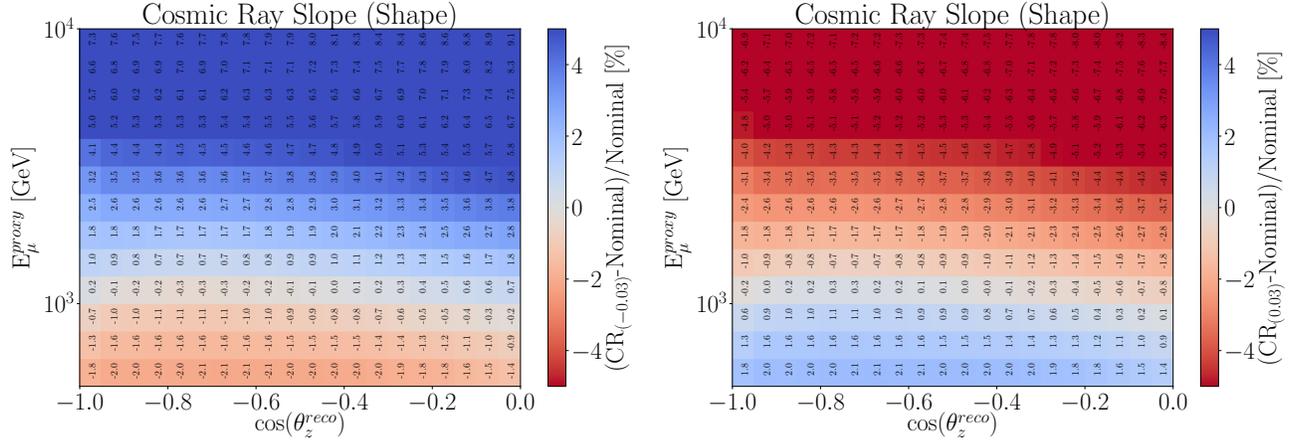

Figure 7-13: The shape of the a change in the cosmic ray spectral slope relative to the central value of $\pm 1\sigma$.

### 7.2.4 Atmospheric density

The pions and kaons produced in the hadronic showers of cosmic rays can either interact or decay producing the conventional neutrino flux. The competition between the two processes depends on the local atmospheric density. IceCube has previously shown experimentally that the atmospheric conditions presented to the cosmic ray flux can affect the atmospheric neutrino spectrum [189]. We have also observed this phenomena, as shown in Fig. 5-8.

We ascribe an uncertainty to the atmospheric density by perturbing the Earth's atmospheric temperature within a prior range given by the NASA Atmospheric InfraRed Sounder (AIRS) satellite [172] temperature data. The satellite provides open source atmospheric data for weather forecasting and climate science and reports the temperature profile as a function of atmospheric depth and location. Using monthly averaged temperature data arranged on a $180 \times 360$ grid (each element representing a $1° \times 1°$ area on the surface of the Earth), we calculate the density at 24 discrete altitudes assuming the ideal gas law, from which we can linearly interpolate between to describe the atmospheric density profile. A random z-score is chosen and all data points are shifted according to reported systematic error on AIRS measurement. The resulting atmospheric profile is injected into MCEq to generate a neutrino flux. This is performed independently for a variety of cosmic ray and hadronic interaction models:

1. The hadronic atmospheric shower model.

    - QGSJET-II-04 [210]
    - SIBYLL 2.3 RC1 [211]

1. The cosmic ray flux model.

    - Zatsepin-Sokolskaya/PAMELA [102, 103]
    - Hillas-Gaisser/Gaisser-Honda [201, 211, 212]
    - Poly-gonato [104, 105]

For a given model and neutrino energy, we average over all months and longitudinal variations to determine the change in the zenith distribution associated with the temperature profile perturbation. Fig. 7-14 shows an example of this for an true $\nu_\mu$ energy of 8.9 TeV. The standard deviation at every zenith angle is calculated (shown as the dotted red line) and is assigned as the atmospheric density uncertainty. Note: we force a crossing (such that the uncertainty goes negative) near $\cos(\theta_z) = $ -0.7 in order to account for the 180° temperature offset between the northern and southern hemispheres. The shape generated when perturbing the atmospheric density to $\pm 1\sigma$ is shown in Fig. 7-15. It appears primarily as a zenith dependent effect.

In total, 4450 different combinations of temperature shifts (z-score perturbations), hadronic interaction models, cosmic ray models, monthly variations, and sampling longitudes are used to assess the spread attributed to the temperature uncertainty.

## 7.3 Astrophysical neutrino flux uncertainties

The astrophysical neutrino flux is modeled as having a "single power law" (SBL) energy spectrum, equal component $\nu_\mu$ to $\bar{\nu}_\mu$, and isotropic distribution. The astrophysical neutrinos are thought to share a common origin as the primary cosmic rays, which is known to follow a broken power law. The energy spectrum is defined by the $\nu_\mu$ and $\bar\nu_\mu$ normalization, $\Phi_{\text{astro}}$, at 100 TeV and the change in the astrophysical spectral index, $\Delta\gamma_{\text{astro}}$, relative to a central value of $\gamma_{\text{astro}} = -2.5$:

$$\frac{dE_\nu}{dE} = \Phi_{\text{astro}} \left(\frac{E_\nu}{100\ TeV}\right)^{-2.5+\Delta\gamma_{astro}}. \tag{7.5}$$

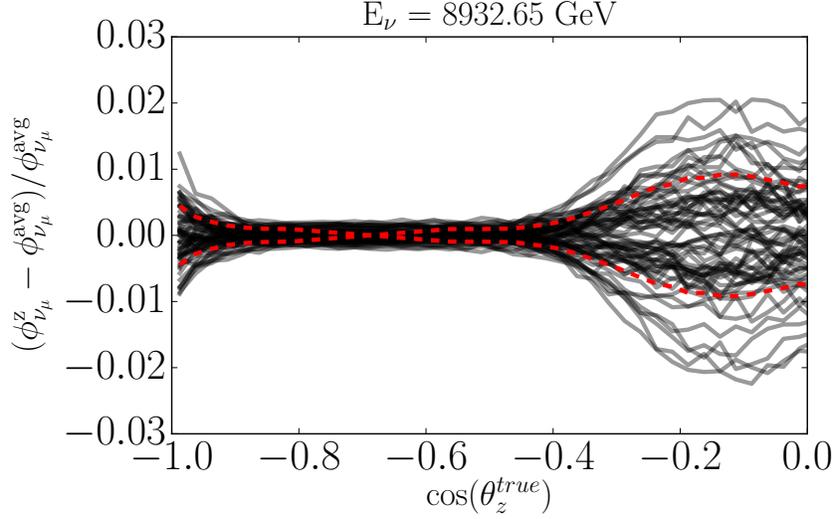

Figure 7-14: The change in neutrino flux relative to the average flux at 8.932 TeV given temperature variations from the AIRS satellite data (black). The standard deviation of the distribution of temperature fluctuations is shown as a dashed red line. A crossing is forced at $\cos\theta_z = -0.7$ in order to account for the temperature offset between Earth's hemispheres (i.e. we assume that when the southern hemisphere increase in temperature, the northern hemisphere decreases in temperature, analogous to seasonal effects).

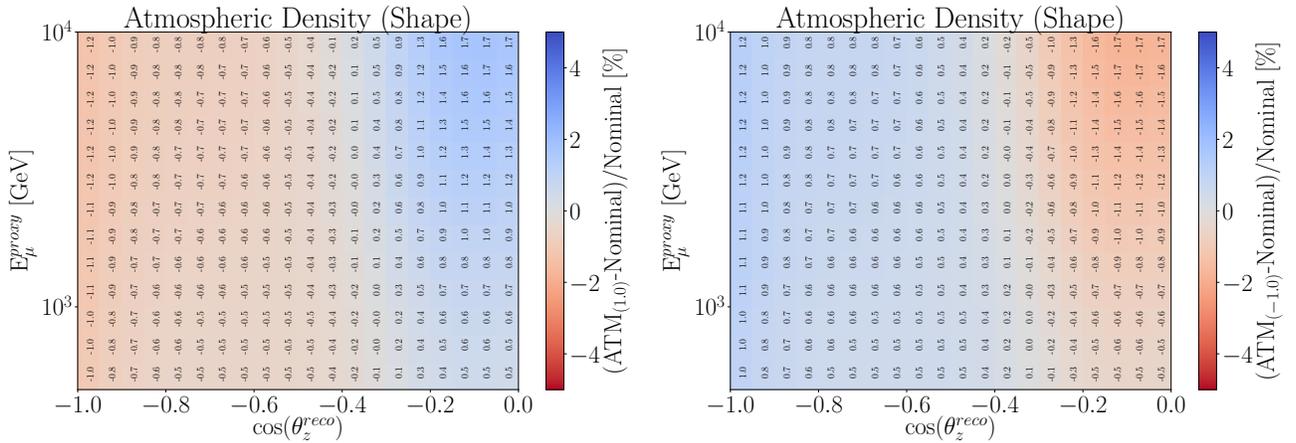

Figure 7-15: The effect at analysis level on reconstructed energy and $\cos(\theta_z)$ due to the atmospheric density uncertainty systematic at $\pm 1\sigma$.

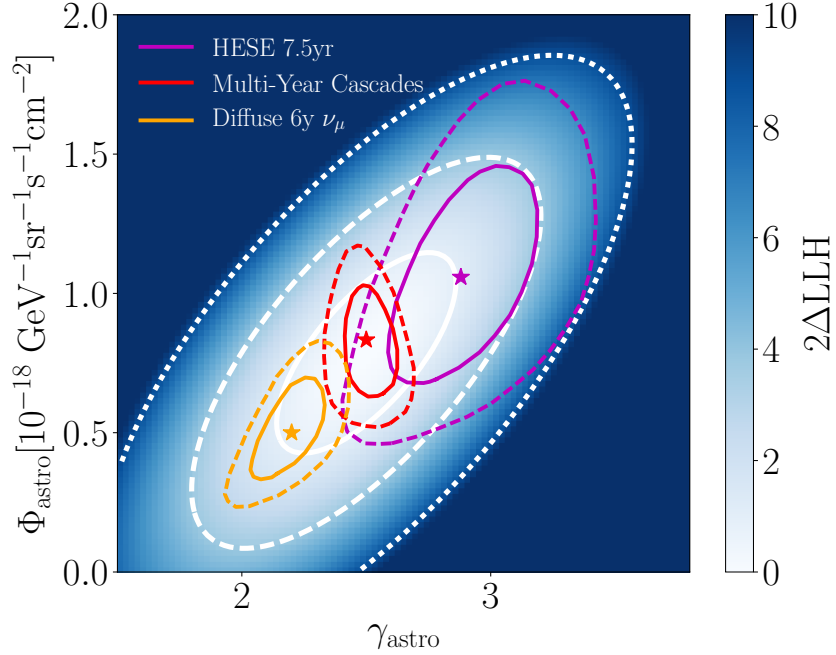

Figure 7-16: The results from three independent astrophysical neutrino flux measurements performed by IceCube [213]. The vertical axis shows the overall six-neutrino (assuming an equal component flux) normalization at 100 TeV. The horizontal axis shows the fitted spectral index. The stars correspond to the location of the best fit point of each measurement, and the solid (dashed) lines correspond to the 68.3% (95.4%) confidence regions. The z-axis shows the shape of the correlated prior at the 68.3% (white solid), 95.4% (white dashed), and 99.7% (white dotted) confidence level.

As with the atmospheric neutrino flux, the astrophysical neutrino flux is propagated through the Earth using nuSQuIDS accounting for the sterile hypothesis as well as high-energy attenuation within the Earth.

The central astrophysical neutrino flux has a astrophysical normalization at 100 TeV of $\Phi_{\text{astro}} = 0.787 \times 10^{-18}\,\text{GeV}^{-1}\text{sr}^{-1}\text{s}^{-1}\text{cm}^{-2}$ and $\Delta\gamma_{\text{astro}} = 0.0$. Both of these parameters are included as nuisance parameters in this analysis. Their prior uncertainty is define in term of two correlated Gaussian. The correlated uncertainty included on the priors are shown in white for the 68.3% (solid), 95.4% (dashed), and 99.7% (dotted) confidence level in Fig. 7-16. This figure also shows three previous single power-law fits to the astrophysical neutrino flux performed by IceCube [213–216]. The measured 68.3% and 95.4% confidence levels for each measurement are shown in the solid and dashed colored lines. The priors used for the astrophysical flux were set such that the best fit points of the previous IceCube astrophysical measurements lie within 1 standard deviation of our prior width.

The shape generated when comparing no astrophysical neutrino flux to one which includes our central model is shown in Fig. 7-17 (left). From the central model, a perturbation $\Delta\gamma_{\text{astro}}$ of slightly less than $1\sigma$ is shown on the right.

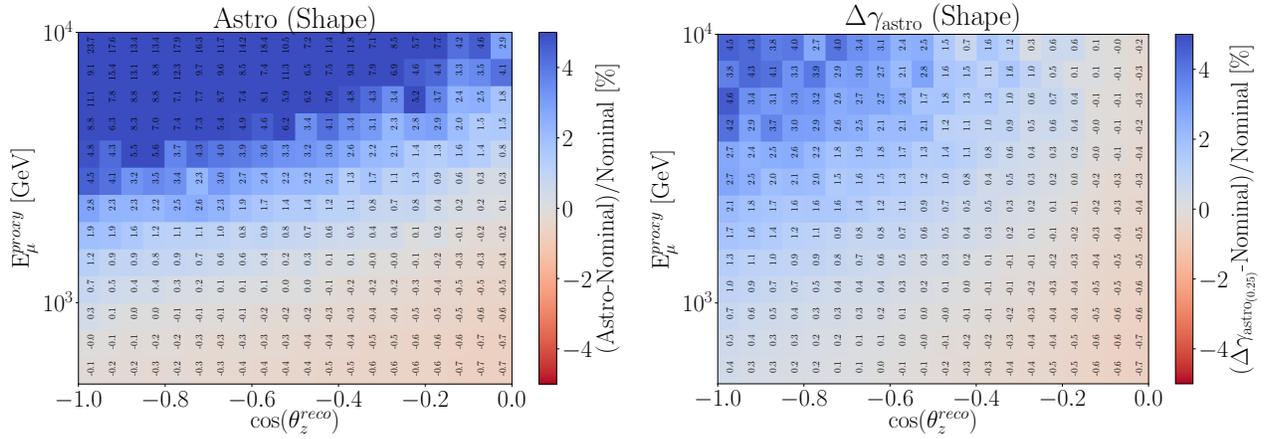

Figure 7-17: Left: The shape in reconstructed quantities comparing the nominal neutrino flux with and without the astrophysical flux component at its central values. Right: The expected change in spectrum due to a shift in the astrophysical spectral index of 0.25 compared to the nominal values.

## 7.4 Cross-section uncertainties

### 7.4.1 Kaon-nuclei total cross-section

Concerning the neutrino production via the decay of the charged mesons, we must account for the uncertainty in extensive air shower development as well as the energy losses in air. The shower development is modeled using Sybill2.3c with uncertainties in the meson production given by the Barr parameters (Sec. 7.2.1). Of the mesons responsible for the muon neutrino flux in our energy range, we are particularly interested in the uncertainty associated with the kaon re-interaction with oxygen (O) and nitrogen (N) nuclei within the atmosphere. This is investigated through the KO(N) total interaction cross section, the uncertainty of which will be described here.

The total cross-section for K$^{\pm}$-nucleon has not been measured above approximately 310 GeV [217], the lower end of our energy spectrum. From proton-proton (pp) cross section measurementsthis approach also one can theoretically derive the kaon-nucleus cross section through a Glauber [218, 219] and Gribov-Regge [220] multiple scattering formalism. This approach has been experimentally verified across a wide range of energies and projectile-target nuclear composition: $\sqrt{s} = 5.02$ TeV for proton-lead (pPb) collisions [221], $\sqrt{s} = 2.76$ TeV for PbPb collisions [222], and $\sqrt{s} = 57$ TeV for pAir [223]. However, verification that this approach also holds for pO (and thus KO(N)) interactions has yet to be realized and is currently the subject of a planned LHC run in 2021-2023 [224].

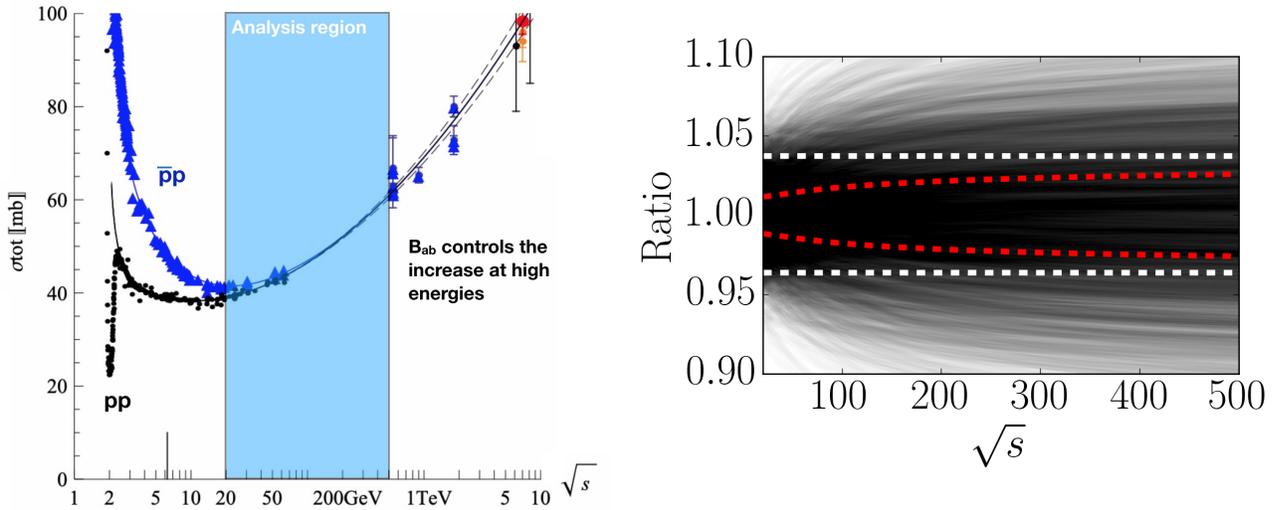

Figure 7-18: The measured total cross-section for $\bar{p}$p (blue triangles) and pp (black circles) interactions. The best fit described in the text is shown in the solid line, along with the one standard deviation (dashed). The analysis region lies between approximately $\sqrt{s} = 20$ GeV to $\sqrt{s} = 500$ GeV. Modified from Ref. [225]. Left: The calculated uncertainty of the kaon interaction cross-section from $\sqrt{s} = 20$ GeV to $\sqrt{s} = 500$ GeV. The standard deviation at a given energy is illustrated in the dashed red line, and the 7.5% uncertainty used in this analysis is illustrated in dashed white. Right: The shape derived from a 7.5% perturbation in the total hadronic interaction cross-section.

At high energies (above $\sqrt{s} \gtrsim 50\,\text{GeV}$), the total hadron-hadron cross-section as a function of center of mass energy, $\sqrt{s}$, is [225]:

$$\sigma_{\text{tot}} \approx Z_{ab} + B_{ab} Log(\frac{s}{s_0^{ab}}), \quad (7.6)$$

where $B_{ab}$ describes the shape and is universal for all hadron-hadron interactions ($B_{pp}$ = $B_{\pi p}$ = $B_{Kp}$ = $B_{pn}$ ≡ B) at high energies, $Z_{ab}$ is a normalization factor dependent on the projectile, and $s_0^{ab}$ is a scale factor for the collision. High energy $\pi$p (up to $\sqrt{s}$ = 600 GeV) and pp (up to $\sqrt{s}$ =50 TeV) data exists (see the left plot in Fig. 7-18) and is available to constrain the universality constant $B$, as well as the scaling of $Z_{ab}$ between projectiles. Ref. [225] finds $B_{Kp}$ = 0.293±0.026$_{\text{sys}}$±0.04$_{\text{stat}}$ mb and $Z_{Kp}$ = 17.76±0.43 mb. At energies above $\sqrt{s}$ = 40 GeV, the total uncertainty becomes dominated by the uncertainty in the $B$ parameter. By perturbing the total cross-section within the uncertainties of $B$ and $Z_{ab}$, we determine that the uncertainty over the range of interest for this analysis ($\sqrt{s} \approx$ 20 GeV to 500 GeV) is between 2.5% and 5.0%, as shown in Fig. 7-18 (right) as the dotted red line. Recent measurements indicate that the high energy pp total cross section uncertainty is known to ∼3.7% [226] and pPb to within ∼3.4% [221], in agreement with the Glauber and Gribov-Regge predictions. We include a conservative estimate on the total kaon-nuclei total cross section of ±7.5%. The shape generated when perturbing the kaon-nuclei total cross section terms by ±1$\sigma$ is shown in Fig. 7-19.

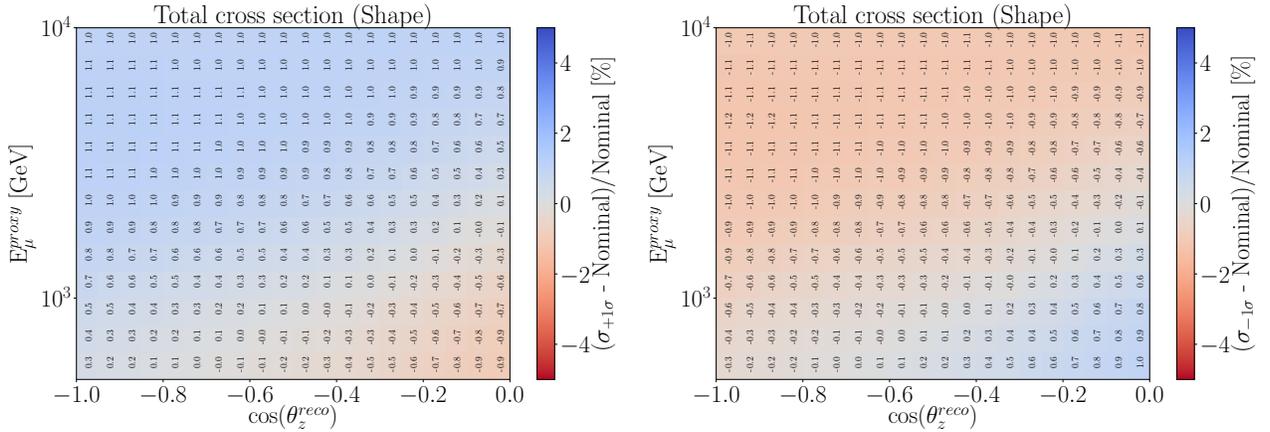

Figure 7-19: The shape generated when perturbing the kaon-nuclei total cross section terms by ±1$\sigma$.

## 7.4.2 Neutrino-nucleon interaction

The interaction between the neutrino and matter is dominated by Deep Inelastic Scattering (described in Sec. 2.4). The neutrino-nucleon cross section enters the analysis in two different parts of the simulation: during the neutrino propagation through the Earth and at the interaction location next to the IceCube detector. The later was previously investigated in Ref. [204] and found to have a minimal impact on the final event distribution. The effect of the propagation through the Earth required further investigation.

As noted in Sec. 2.4, the neutrino-nucleon DIS cross section increases with neutrino energy. This causes an attenuation of the high energy neutrino flux passing through the Earth. The attenuation in terms of true quantities was noted in Fig. 2-9 and shown to effect both the $\nu_\mu$ and $\overline{\nu}_\mu$ flux.

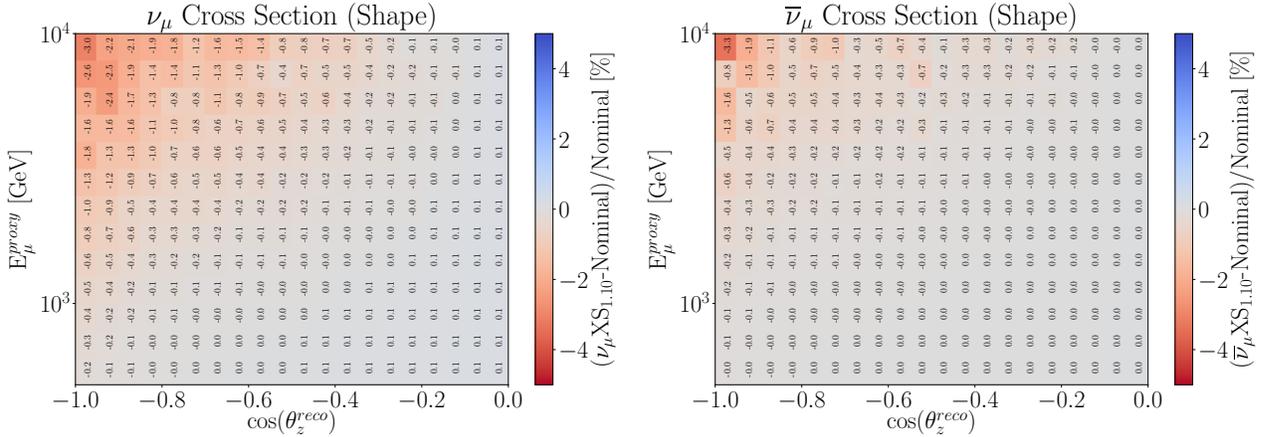

Figure 7-20: The shape in reconstructed quantities comparing a perturbation of the neutrino (left) and antinetrino (right) cross section by $+10\%$.

We use the CSMS cross sections described in Ref. [191] for both the neutrino-nucleon interaction during propagation and the interaction near the IceCube detector. Uncertainties are provide for both NC and CC interaction channels from 50 GeV to $5\times10^{20}$ GeV. From approximately 10 TeV upwards, the $\nu$CC and $\nu_\mu$CC uncertainties tend to be below 2% and 5%, respectively. Below this, the neutrino attenuation in the Earth is negligible and can therefore be neglected from this discussion. We include separate systematic uncertainties for the neutrinos and antineutrinos. The prior width on the neutrinos was chosen to be 3.0% and correspondingly 7.5% for the antineutrinos. The uncertainties are implemented via a 30-point penalized spline which spans cross section values from 50% to 150%. The shape in reconstructed energy and $\cos(\theta_z)$ when perturbing the cross sections by $+10\%$ is shown in Fig. 7-20. As expected the shape is primarily located in the upper energies, going through the center of the Earth. The shape is localized to the region of the distribution where we have minimal statistics, therefore, we expect this systematic to have a near negligible effect but will include it in the minimization nevertheless.

THIS PAGE INTENTIONALLY LEFT BLANK

# Chapter 8

# RESULTS

This chapter gives (1) a concise overview of the methods used to characterize the sensitivity of IceCube to a sterile neutrino signal, (2) the pre-unblinding tests designed to provide confidence in the result while preserving blindness, and finally, (3) the result and following discussion.

A summary of the full model hypothesis discussed prior to this chapter can be found in Table 8.1.

## 8.1 Expected sensitivity

The sensitivity is defined by the values of the sterile neutrino mixing parameters that can be excluded in case of the *null hypothesis*, or $H_0$ (no sterile neutrino mixing), at various confidence levels (CL). The final sensitivity is reported as the median value of the "Brazil bands" generated through an ensemble of pseudo-experiments. Various other tests were performed using an Asimov dataset, in which the expected distribution without statistical fluctuations is used as a representative average description of the signal hypothesis.

| Parameter | Central Value | Prior | Constraints |
|---|---|---|---|
| **Physics Mixing Parameters** | | | |
| $\Delta m_{41}^2$ | none | no prior | [0.01 eV$^2$, 100 eV$^2$] |
| $\sin^2(\theta_{24})$ | none | no prior | [$10^{-2.6}$, 1.0] |
| $\sin^2(\theta_{34})$ | none | no prior | [$10^{-3.1}$, 1.0] |
| **Detector parameters** | | | |
| DOM efficiency | 0.97 | 0.97 ± 0.10 | [0.94, 1.03] |
| Bulk Ice Gradient 0 | 0.0 | 0 ± 1.0* | NA |
| Bulk Ice Gradient 1 | 0.0 | 0 ± 1.0* | NA |
| Forward Hole Ice ($p_2$) | -1.0 | -1.0 ± 10.0 | [-5, 3] |
| **Conventional Flux parameters** | | | |
| Normalization ($\Phi_{\text{conv.}}$) | 1.0 | 1.0 ± 0.4 | NA |
| Spectral shift ($\Delta\gamma_{\text{conv.}}$) | 0.00 | 0.00 ± 0.03 | NA |
| Atm. Density | 0.0 | 0.0 ± 1.0 | NA |
| Barr WM | 0.0 | 0.0 ± 0.40 | [-0.5, 0.5] |
| Barr WP | 0.0 | 0.0 ±0.40 | [-0.5, 0.5] |
| Barr YM | 0.0 | 0.0 ± 0.30 | [-0.5, 0.5] |
| Barr YP | 0.0 | 0.0 ±0.30 | [-0.5, 0.5] |
| Barr ZM | 0.0 | 0.0 ± 0.12 | [-0.25, 0.5] |
| Barr ZP | 0.0 | 0.0 ± 0.12 | [-0.2, 0.5] |
| **Astrophysical Flux parameters** | | | |
| Normalization ($\Phi_{\text{astro.}}$) | 0.787 | 0.0 ± 0.36* | NA |
| Spectral shift ($\Delta\gamma_{\text{astro.}}$) | 0 | 0.0 ± 0.36* | NA |
| **Cross sections** | | | |
| Cross section $\sigma_{\nu_\mu}$ | 1.00 | 1.00 ± 0.03 | [0.5, 1.5] |
| Cross section $\sigma_{\bar{\nu}_\mu}$ | 1.000 | 1.000 ± 0.075 | [0.5, 1.5] |
| Kaon energy loss $\sigma_{KA}$ | 0.0 | 0.0 ± 1.0 | NA |

Table 8.1: A list of the physics and nuisance parameters along with a description of their priors and central values. The Bulk Ice Gradients 0 and 1 have correlated priors, along with $\Phi_{\text{astro.}}$ and $\Delta\gamma_{\text{astro.}}$.

### 8.1.1 Asimov dataset

An Asimov data set was used to access the impact of each set of systematic uncertainties, testing the implementation of the nuisance parameters (Sec. C.1.2), testing the minimizer (Sec. C.1.1), and assessing the impact of various model variations and how they would impact the result (Secs. E.0.1 - E.0.4 ).

Figs. 8-1 shows the Asimov sensitivity given a subset of the systematic uncertainties from Table 8.1 for both analyses. For each subset of systematic uncertainties, the conventional normalization nuisance parameter is always included. The conventional neutrino flux systematic

uncertainties are found to be the dominant source uncertainty along the 90% contour for Analysis I. The cross section uncertainties are found to have a negligible impact on on the sensitivity.

Analysis II shows that the sensitivity is nearly independent of the systematic uncertainties, albeit for the conventional normalization. This is a result of the conventional normalization pulling away from its central value to accommodate the shift due to the oscillations averaging out.

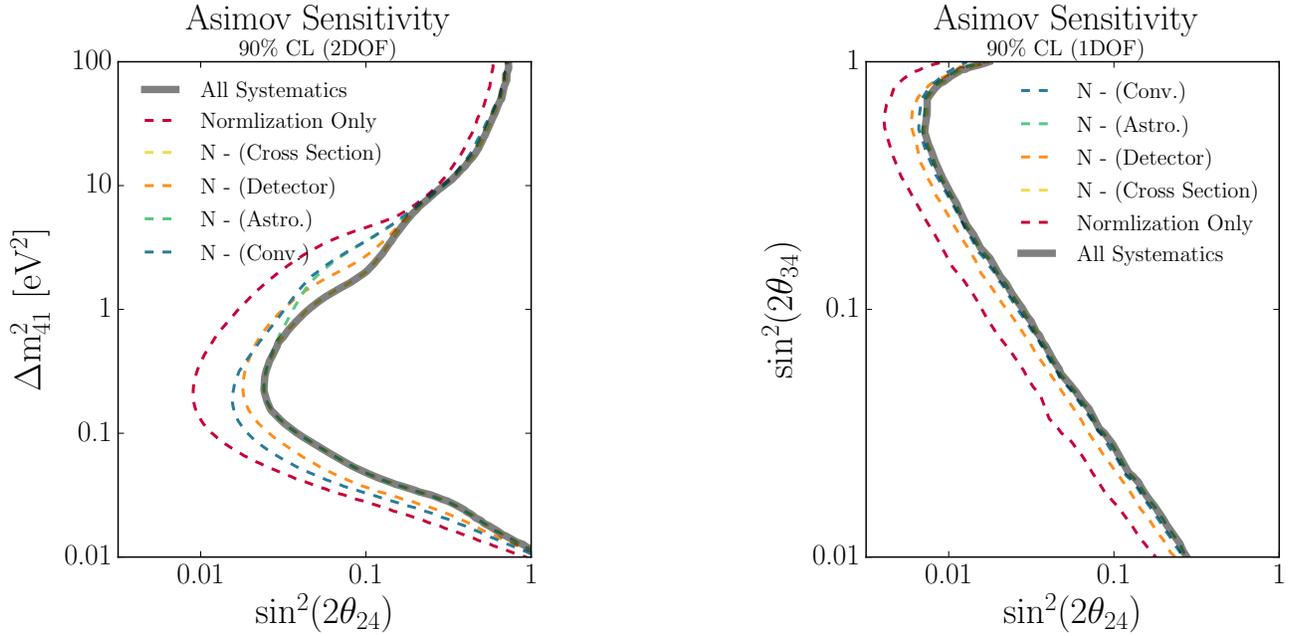

Figure 8-1: The expected 90%CL Asimov sensitivity given a subset of systematic uncertainties for Analysis I (left) and Analysis II (right). The sensitivity with all nuisance parameters included during the fit is shown in solid gray.
.

### 8.1.2 Ensemble dataset

The reported sensitivity of this analysis is calculated through an ensemble of 2,000 simulated pseudo-experiments, or "data realizations," each of which was generated by Poisson fluctuating the expected distribution at the null hypothesis with the nuisance parameters at their central value. For a given realization, we evaluate the Wilk's contour at the 90% CL and 99% CL. For every value of $\Delta m^2_{41}$ (or $\sin^2(\theta_{34})$ in the case of Analysis II), the coordinate of the contour in terms of $\sin^2(\theta_{24})$ is recorded. The distribution of the crossing values for $\sin^2(\theta_{24})$ are then used to define the 68.3% ($1\sigma$) and 95.4% ($2\sigma$) confidence intervals (CI). If the contour crosses more than once, we take the maximum $\sin^2(\theta_{24})$ value of the crossing. A graphical example of this procedure for $\Delta m^2_{41} = 10\,\mathrm{eV}^2$ is shown in Fig. 8-2. For this particular realization, we find that the 90% CL crosses at $\sin^2(\theta_{24}) = 0.2$. A histogram of 1,000 such realizations is shown on the right along with the 68.3% ($1\sigma$) CI and 95.4% ($2\sigma$) CI in green and yellow. This procedure is performed for each value in the y-axis, for both the 90% CL and 99% CL. The result is the "Brazil bands" and the median values of each histogram define the final analysis sensitivity. The resulting Brazil bands for both analyses at the 90% CL and 99% CL are shown in Fig. 8-3 and 8-4.

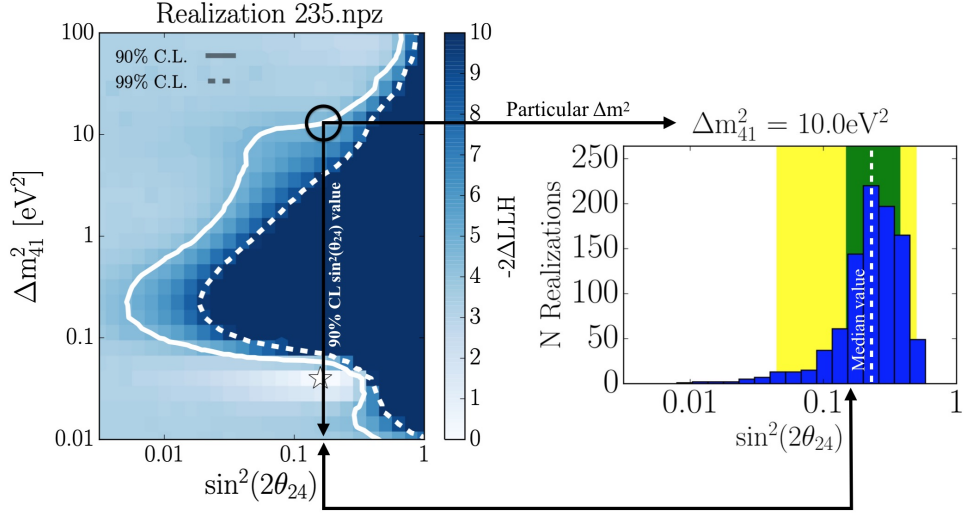

Figure 8-2: An example generation of the Brazil Bands at $10\,\mathrm{eV}^2$ for Analysis I. Left: The 90% CL and 99% Wilks' contour for Realization number 235. At $10\,\mathrm{eV}^2$, the 90% CL value is found at $\sin^2(\theta_{24}) = 0.2$. Right: The distribution of the 90% CL values for $\Delta m^2_{41} = 10\,\mathrm{eV}^2$, using 1000 realizations.

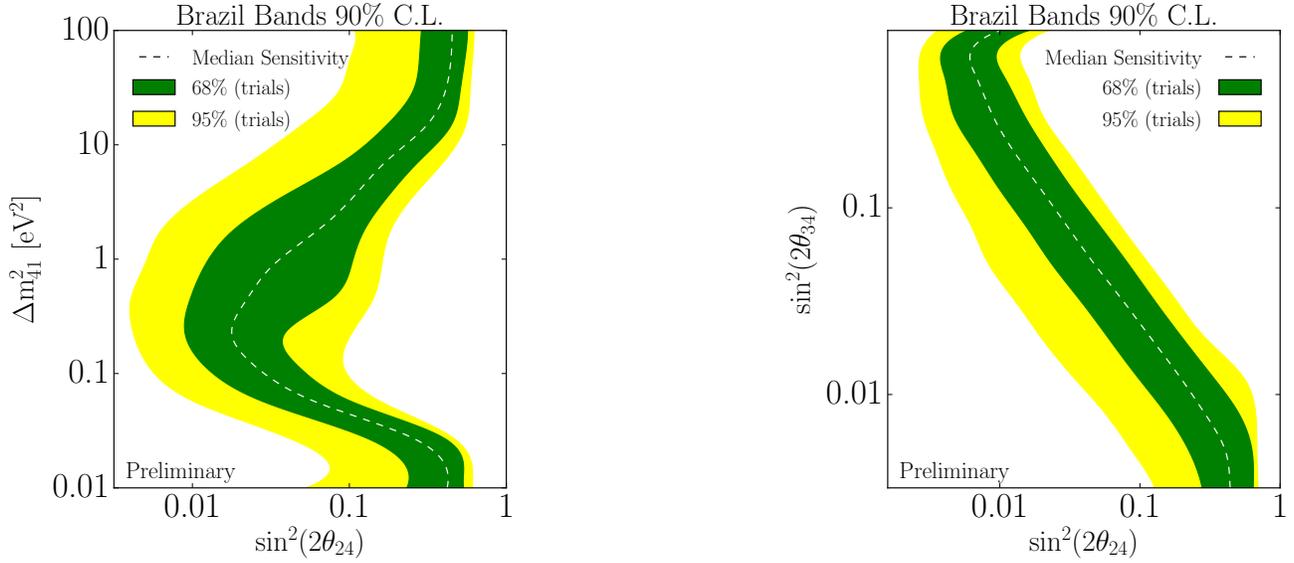

Figure 8-3: The 90% CL Brazil Bands for Analysis I (left) and Analysis II (right). The yellow band corresponds to the 95% spread, while the green to the 68%. The median sensitivity is shown as a dashed white line.
.

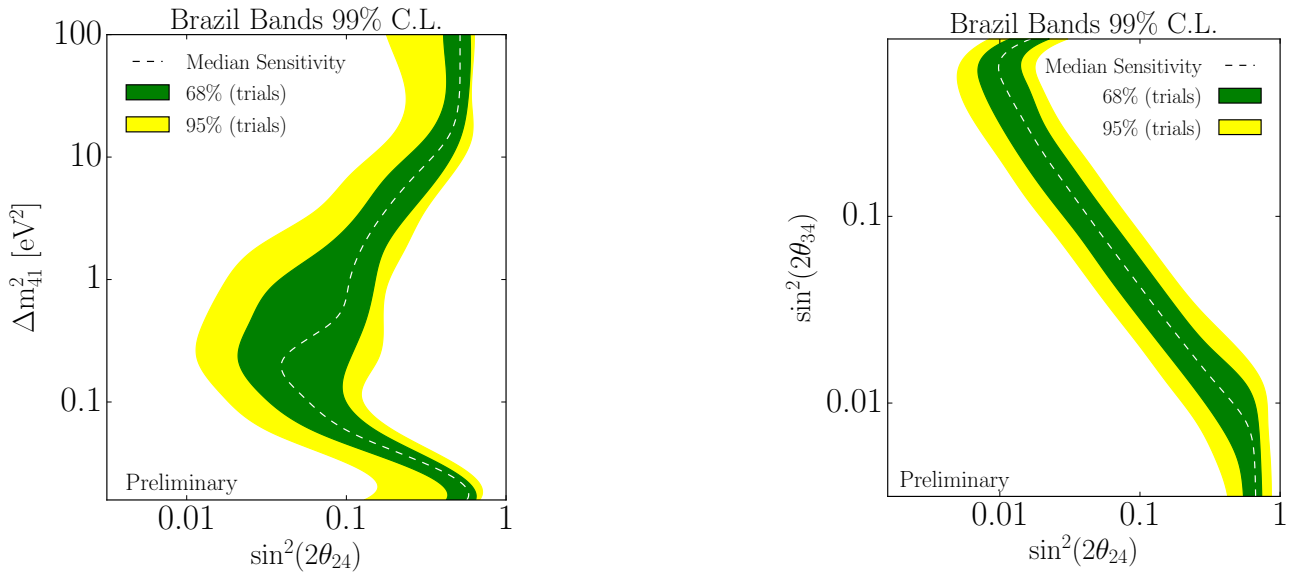

Figure 8-4: The 99% CL Brazil Bands for Analysis I (left) and Analysis II (right). The yellow band corresponds to the 95% spread, while the green to the 68%. The median sensitivity is shown as a dashed white line.
.

## 8.2 Pre-unblinding tests

We performed a set of successive steps using 100% of the data[1] to assess whether we have indeed found a good fit somewhere in the physics parameter space, without revealing where it is. Each test was designed such that it would protect against potential biases.

The first test examined the nuisance parameter pulls at the best fit point for both analyses (Fig. 8-5). The systematic pull was defined as:

$$\text{SysPull} = \frac{(\text{FitValue} - \text{PriorCenter})}{\text{PriorWidth}}. \tag{8.1}$$

None of the nuisance parameters, for either analysis, are in tension with the set priors (defined beforehand as a pull greater than $\pm 3\sigma$). It was first found that the largest systematic pull for both analyses was in the cosmic ray change in spectral index (see Fig. 8-5), pulling to a value of $\Delta\gamma_{\text{conv.}} = 0.067$ ($2.2\sigma$) for both analyses.

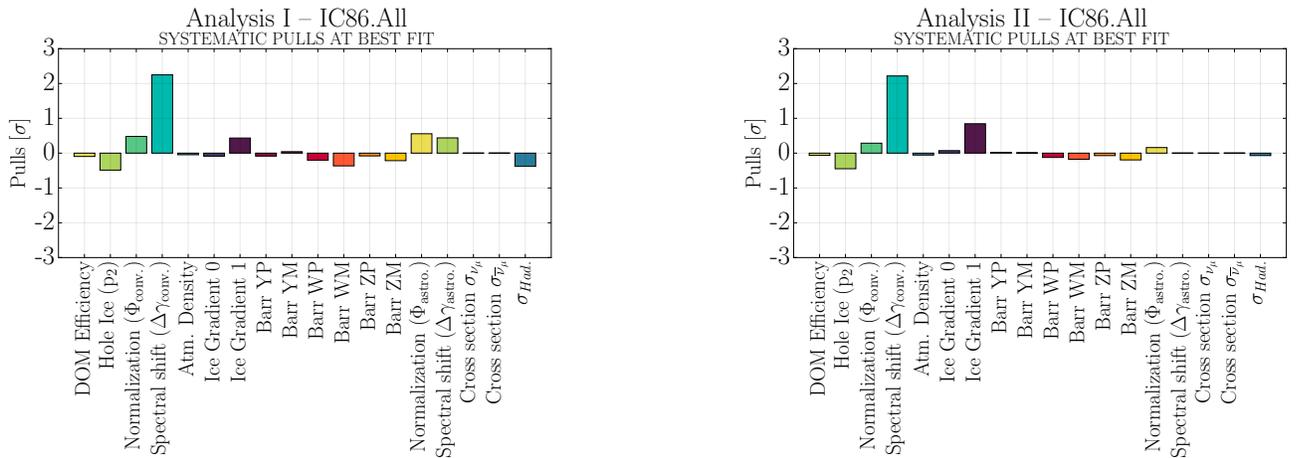

Figure 8-5: Systematic pulls for both analyses at their respective best fit points. Left: Analysis I. Right: Analysis II.
.

The statistical distribution of the data pulls at the best fit point was then examined. The data

---

[1]The first pre-unblinding was preformed on 5% of the total data, however since these tests all passed our criteria, we will restrict the discussion to the 100% pre-unblinding results.

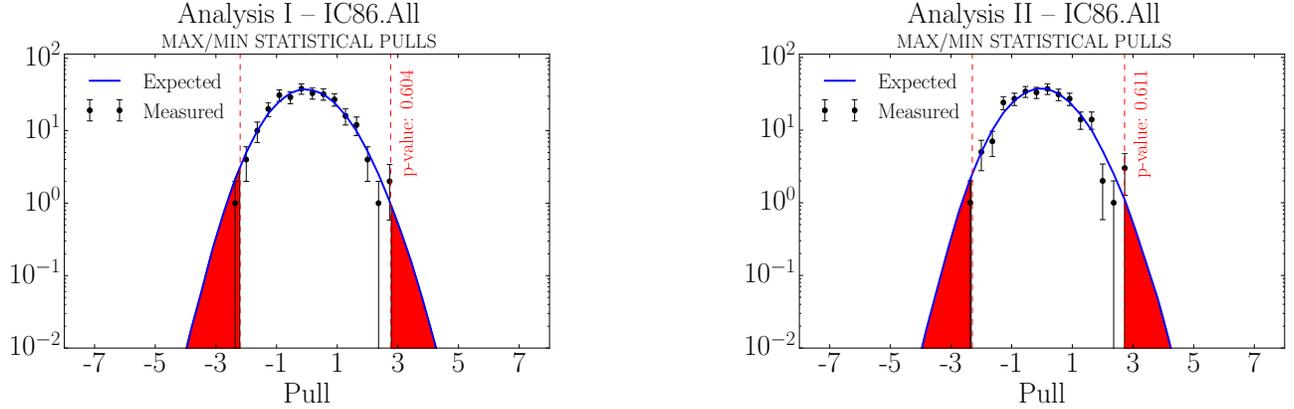

Figure 8-6: The observed statistical pull distribution (black) compared to the expectation (blue). The area in red represents the probability of observing a pull beyond the maximum and minimum observed. The p-value is calculated accounting for the 260 bins used in the analysis space. This test will be referred to as the Max/Min Data Pull in Table 8.2.

pull in bin $i$ was calculated as:

$$\text{Pull}_i = \frac{(\text{Data}_i - \text{Expectation}_i)}{\sqrt{\text{Expectation}_i}}. \tag{8.2}$$

The probability of observing at least one statistical pull larger, P(A), than the largest pull and at least one pull smaller, P(B), than the most negative pull is calculated as:

$$P(A \cap B) = P(A) \times P(B), \tag{8.3}$$

where both P(A) and P(B) are calculated accounting for the 260 bins (trials). The expected statistical pull distribution, in blue, is calculated by averaging over the best fit expectation Poisson fluctuated 10,000 times. The probability of observing at least one bin pulling beyond the limits of the observed pulls is shown in red along with the reported probability of observing a pull larger than this. In this case, we observe the maximum statistical pull to be $+2.7\sigma$ and the minimum pull to be $-2.2\sigma$. $P(A \cap B)$ is calculated to be 60.4% for Analysis I and 61.1% for Analysis II.

A $\chi^2$ was calculated, comparing the data (black) to the expectation (blue) in Fig. 8-6, where the uncertainty was defined as the square root of the expected number of events. The reported p-value from $\chi^2$ per degree of freedom is shown in the top right of both figures in Fig. 8-6.

Finally, we looked at the projections of the reconstructed energy and $\cos(\theta_z)$ distributions. The signal described in Analysis II (shown in Fig. 6-2) can be observed as a few percent change in the data distribution for events with $\cos(\theta_z) < -0.80$, therefore during this stage of the unblinding process, this region was not revealed until after unblinding.

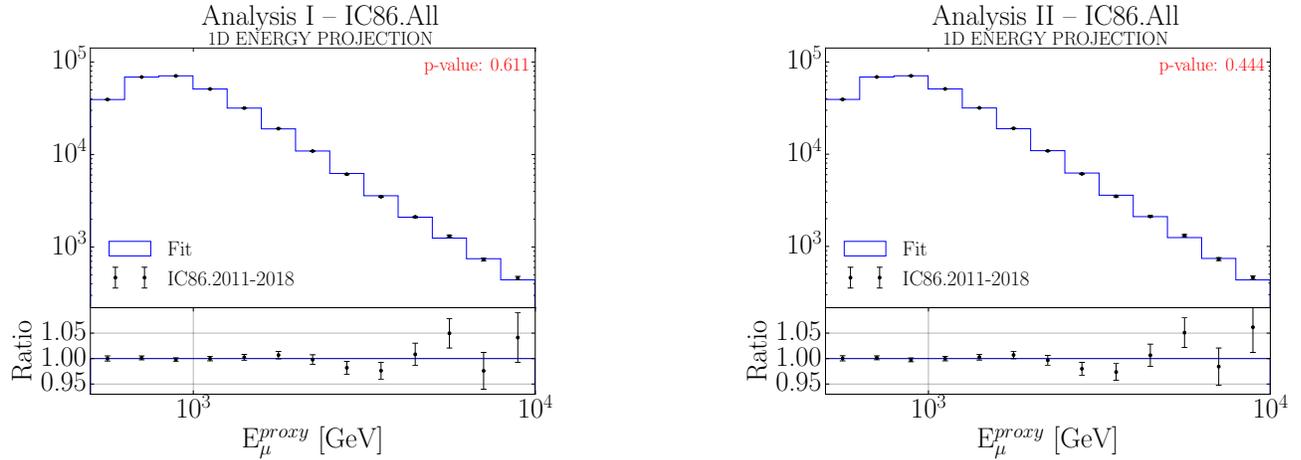

Figure 8-7: The 1D projections in reconstructed energy.

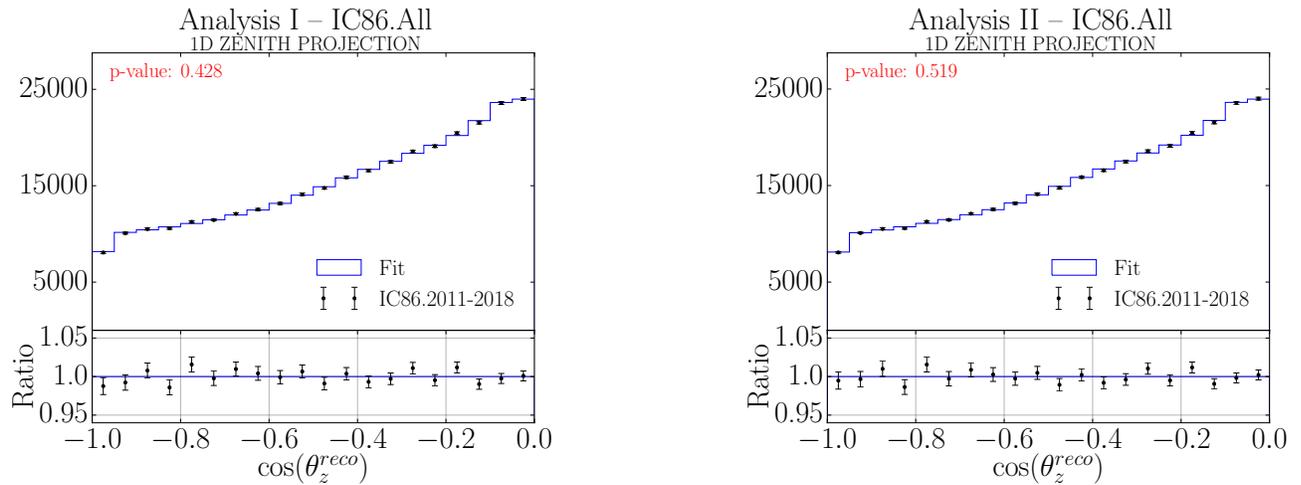

Figure 8-8: The 1D projections in reconstructed $\cos(\theta_z)$.

The final pre-unblinding test performed was to check to see if the best fit found when fitting to the cumulative data agrees with each IceCube season independently. This was checked by re-scaling the best fit distribution to the appropriate livetime for each season, and performing all previous tests described in this section (accounting for the statistical differences). The p-value of each test is reported in Table 8.2.

|  | Season | Max/Min Data Pull [%] | Data Stat. Pull Dist. [%] | 1D Energy Projection [%] | 1D Zenith Projection [%] |
|---|---|---|---|---|---|
| Analysis I | IC86.All | 60.4 | 96.7 | 61.1 | 42.8 |
|  | IC86.2011 | 46.4 | 89.3 | 87.9 | 32.3 |
|  | IC86.2012 | 22.6 | 91.2 | 207 | 61.7 |
|  | IC86.2013 | 3.0 | 29.2 | 78.0 | 53.8 |
|  | IC86.2014 | 14.3 | 44.8 | 42.7 | 97.7 |
|  | IC86.2015 | 24.7 | 69.8 | 64.1 | 21.2 |
|  | IC86.2016 | 0.8 | 0.2 | 29.8 | 1.3 |
|  | IC86.2017 | 19.3 | 18.0 | 11.8 | 84.2 |
|  | IC86.2018 | 39.2 | 16.1 | 51.2 | 2.5 |
| Analysis II | IC86.All | 61.1 | 51.3 | 44.4 | 51.9 |
|  | IC86.2011 | 47.6 | 79.7 | 86.6 | 30.7 |
|  | IC86.2012 | 23.5 | 52.1 | 15.7 | 57.9 |
|  | IC86.2013 | 2.6 | 38.9 | 78.8 | 57.1 |
|  | IC86.2014 | 14.1 | 53.8 | 42.9 | 98.5 |
|  | IC86.2015 | 25.7 | 73.5 | 60.9 | 19.4 |
|  | IC86.2016 | 0.8 | 0.1 | 29.4 | 1.9 |
|  | IC86.2017 | 17.6 | 9.2 | 11.0 | 84.9 |
|  | IC86.2018 | 48.9 | 10.5 | 49.8 | 2.5 |

Table 8.2: The p-values of each test comparing the best fit distribution using all 7.6 years of data to the individual IceCube seasons. The tests are described in the text.

## 8.3 Unblinding the result

The 8-year fully blind data was opened on August 1$^{\text{st}}$ 2019, though predefined post-unblinding checks led to the discovery of a significant bug and reassessment of several assumptions. This brought about several updates, the most notable being the implementation of a systematic uncertainty for the astrophysical neutrino flux and the bug fix. A detailed description of the unblinding and subsequent updates to the analysis is given in Appendix D, while this chapter will focus on the final results of this thesis.

On September 30$^{\text{th}}$ 2019, the IceCube collaboration granted permission to re-open the data. The observed event count in the reconstructed energy and $\cos(\theta_z)$ plane is shown in Fig. 8-9.

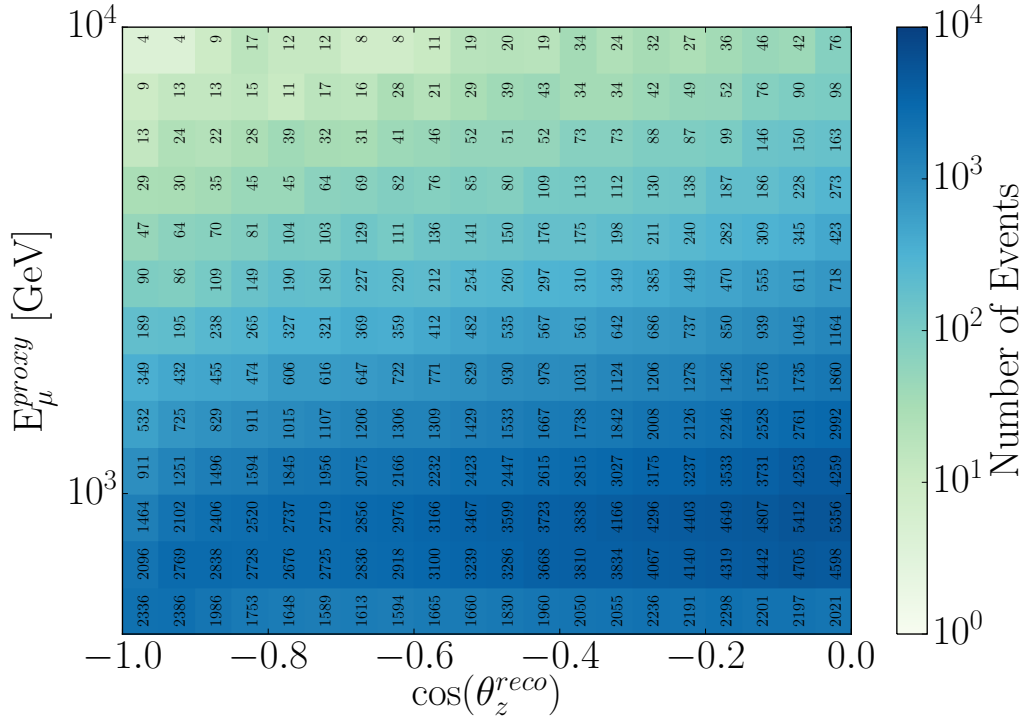

Figure 8-9: The number of observed events per bin.

The results of the analyses are shown in Fig. 8-10 in terms of the 90%, 95%, and 99% CL contours assuming Wilks' theorem. No evidence for a 3+1 sterile neutrino was observed. Analysis I was found to have a best fit point at $\Delta m_{41}^2 = 4.47\,\text{eV}^2$ and $\sin^2(\theta_{24}) = 0.10$. The TS compared to the no-sterile hypothesis is $2\Delta LLH = 4.94$, corresponding to a p-value of 8% (2DOF). Simi-larly, Analysis II was found to have a best fit point at $\sin^2(\theta_{34}) = 0.40$, $\sin^2(\theta_{24}) = 0.006$, with a $2\Delta LLH = 1.74$ corresponding to a p-value of 19% (1DOF).

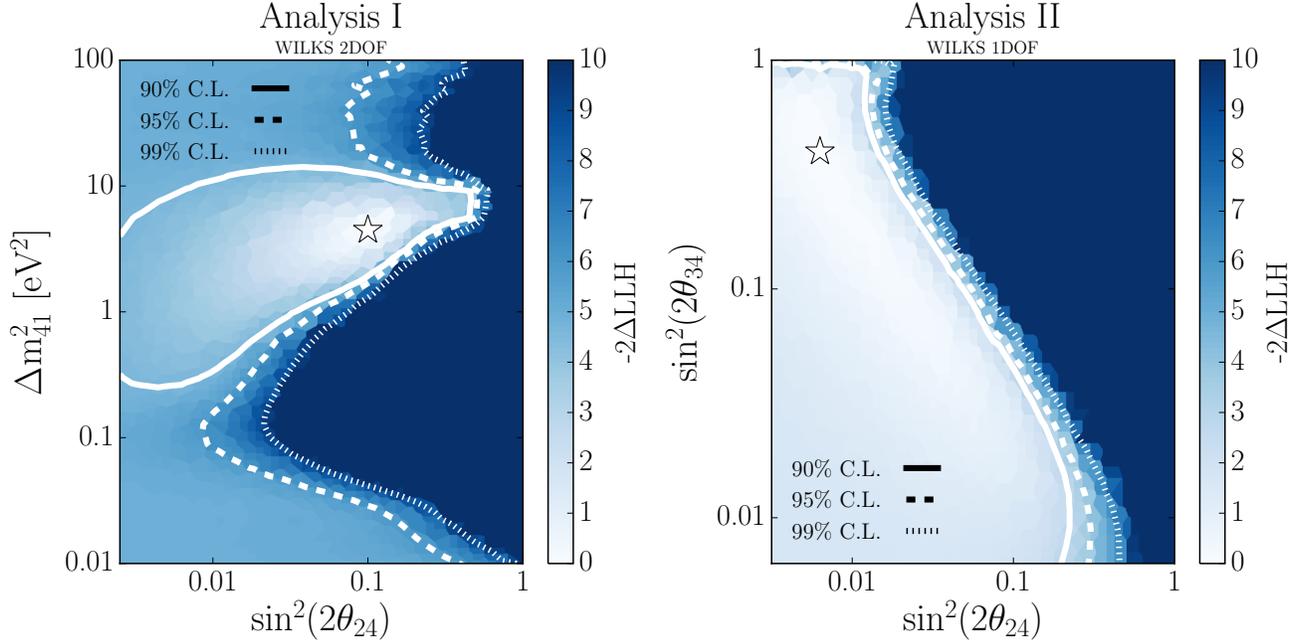

Figure 8-10: The result of Analysis I (left) and Analysis II (right). The best fit points are marked with the star and found to be $\Delta m_{41}^2 = 4.47\,\text{eV}^2$ and $\sin^2(\theta_{24}) = 0.10$ for Analysis I and $\sin^2(\theta_{34}) = 0.40$, $\sin^2(\theta_{24}) = 0.006$ for Analysis II. The 90%, 95%, and 99% CL contours are drawn assuming Wilks' theorem. The color-axis shows the distribution of the TS relative to the best fit point in both analyses.

The best fit values for the model hypothesis are shown in Table 8.3, where the $\pm 1\sigma$ uncertainties are reported as the width of the posterior distribution for each nuisance parameter. The 2D statistical pull distribution of the result compared to the measured best fit point is shown in Fig. 8-11. As described in Sec. 8.2, the distribution of the statistical pulls is consistent with the expectation.

Each IceCube season used in this analysis, IC86.2011 to IC86.2018, was then independently fit using the standard software, albeit with the normalization on the atmospheric and astrophysical neutrino flux scaled according to the relative livetimes given in Table 5.4. The $2\Delta LLH$ between the season-dependent best-fit point (BF) and the result of Analysis I and Analysis II (Fig. 8-10) are presented in Table 8.4 and 8.5, along with the BF compared to the no-sterile hypothesis

| Parameter | Best-fit ±1σ |
|---|---|
| **Physics Mixing Parameters** | |
| $\Delta m^2_{41}$ | $4.47^{+3.53}_{-2.08}$ eV$^2$ |
| $\sin^2(\theta_{24})$ | $0.10^{+0.10}_{-0.07}$ |
| $\sin^2(\theta_{34})$ | 0.0 |
| **Detector parameters** | |
| DOM Efficiency | $0.961 \pm 0.005$ |
| Ice Gradient 0 | $-0.15 \pm 0.25$ |
| Ice Gradient 1 | $0.36 \pm 0.53$ |
| Hole Ice | $-3.44 \pm 0.44$ |
| **Conventional Flux parameters** | |
| Normalization ($\Phi_{\text{conv.}}$) | $1.19 \pm 0.05$ |
| Spectral shift ($\Delta\gamma_{\text{conv.}}$) | $0.068 \pm 0.012$ |
| Atm. Density | $-0.16 \pm 0.71$ |
| Barr WM | $-0.02 \pm 0.28$ |
| Barr WP | $0.00 \pm 0.28$ |
| Barr YM | $-0.06 \pm 0.24$ |
| Barr YP | $-0.10 \pm 0.15$ |
| Barr ZM | $-0.00 \pm 0.11$ |
| Barr ZP | $0.01 \pm 0.09$ |
| **Astrophysical Flux parameters** | |
| Normalization ($\Phi_{\text{astro.}}$) | $0.95 \pm 0.21$ |
| Spectral shift ($\Delta\gamma_{\text{astro.}}$) | $0.11 \pm 0.19$ |
| **Cross sections** | |
| Cross section $\sigma_{\nu_\mu}$ | $1.00 \pm 0.03$ |
| Cross section $\sigma_{\overline{\nu}_\mu}$ | $1.003 \pm 0.075$ |
| Hadronic energy loss $\sigma_{KA}$ | $-0.35 \pm 0.93$ |

| Parameter | Best-fit ±1σ |
|---|---|
| **Physics Mixing Parameters** | |
| $\Delta m^2_{41}$ | $> 10$ eV$^2$ |
| $\sin^2(\theta_{24})$ | $0.006^{+0.004}_{-0.006}$ |
| $\sin^2(\theta_{34})$ | $0.40^{+0.47}_{-0.33}$ |
| **Detector parameters** | |
| DOM Efficiency | $0.965 \pm 0.005$ |
| Ice Gradient 0 | $0.05 \pm 0.24$ |
| Ice Gradient 1 | $0.89 \pm 0.54$ |
| Hole Ice | $-3.23 \pm 0.44$ |
| **Conventional Flux parameters** | |
| Normalization ($\Phi_{\text{conv.}}$) | $1.11 \pm 0.05$ |
| Spectral shift ($\Delta\gamma_{\text{conv.}}$) | $0.066 \pm 0.012$ |
| Atm. Density | $-0.17 \pm 0.68$ |
| Barr WM | $0.00 \pm 0.29$ |
| Barr WP | $0.01 \pm 0.29$ |
| Barr YM | $-0.03 \pm 0.25$ |
| Barr YP | $-0.05 \pm 0.15$ |
| Barr ZM | $-0.00 \pm 0.11$ |
| Barr ZP | $0.016 \pm 0.089$ |
| **Astrophysical Flux parameters** | |
| Normalization ($\Phi_{\text{astro.}}$) | $0.80 \pm 0.21$ |
| Spectral shift ($\Delta\gamma_{\text{astro.}}$) | $-0.06 \pm 0.21$ |
| **Cross sections** | |
| Cross section $\sigma_{\nu_\mu}$ | $1.000 \pm 0.03$ |
| Cross section $\sigma_{\overline{\nu}_\mu}$ | $1.004 \pm 0.074$ |
| Kaon Energy Loss $\sigma_{KA}$ | $-0.06 \pm 0.90$ |

Table 8.3: The measured model parameters for Analysis I (left) and Analysis II (right) at their respective best fit points. The reported ±1σ uncertainties on each of the 18 nuisance parameters are derived from the calculated standard deviations of the posterior distributions shown in Fig. 8-17. A description of the priors on each nuisance parameter can be found in Table 8.1.

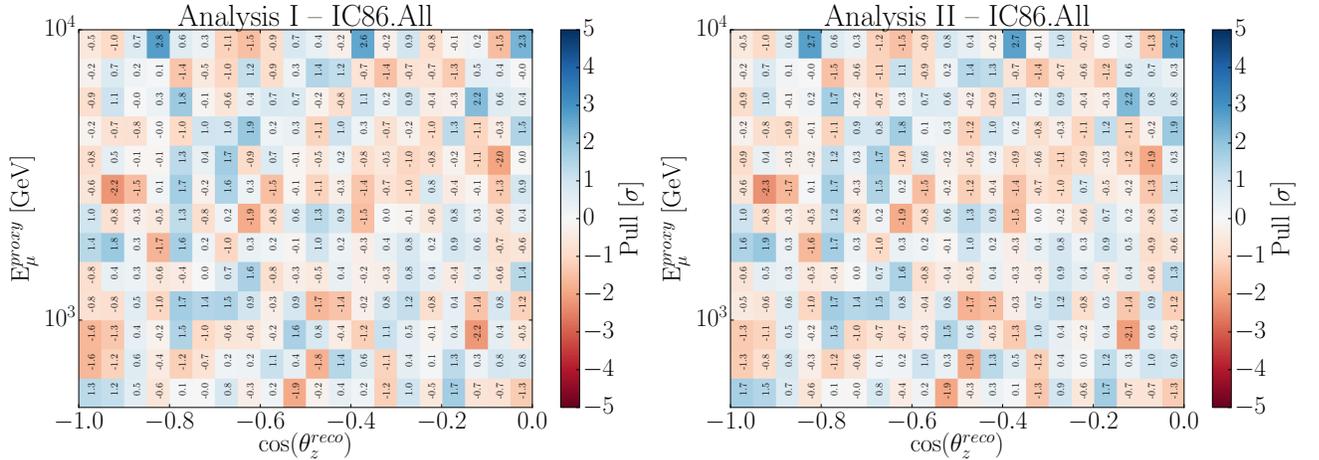

Figure 8-11: The measured statistical pull distribution in reconstructed energy and $\cos(\theta_z)$ for Analysis I (left) and Analysis II (right) at their respective best fit points.
.

(Null). The full TS distribution for each season is shown in Figs. 8-12 and 8-13.

We find the largest difference between the best fit point of the individual seasons and the result of either analysis to be with IC86.2012. The $2\Delta LLH$ between these two points is found to be 3.03 (3.54) for Analysis I (Analysis II), corresponding to a p-value of 42% (13%). With respect to the null hypothesis, the largest difference relative to the individual seasons is also found with IC86.2012. Here, we find a $2\Delta LLH$ of 7.41 (1.88) for Analysis I (Analysis II), corresponding to a p-value of 4% (42%). Nevertheless, given the number of statistically independent data sets, the observation of the result from IC86.2012 is not significant.

| Season | Best fit point $[\sin^2(\theta_{24}), \Delta m_{41}^2]$ | $2\Delta LLH_{\text{Result}-\text{BF}}$ | $2\Delta LLH_{\text{Null}-\text{BF}}$ |
|---|---|---|---|
| IC86.2011-2018 | [0.10, 4.47 eV$^2$] | 0.00 | 4.94 |
| IC86.2011 | [0.14, 5.62 eV$^2$] | 0.12 | 2.83 |
| IC86.2012 | [0.18, 2.00 eV$^2$] | 3.03 | 7.41 |
| IC86.2013 | [0.08, 4.47 eV$^2$] | 0.10 | 1.34 |
| IC86.2014 | [0.50, 3.98 eV$^2$] | 1.92 | 3.47 |
| IC86.2015 | [0.09, 3.16 eV$^2$] | 0.09 | 0.49 |
| IC86.2016 | [0.05, 3.16 eV$^2$] | 0.09 | 0.96 |
| IC86.2017 | [0.89, 0.01 eV$^2$] | 1.31 | 0.92 |
| IC86.2018 | [0.04, 1.26 eV$^2$] | 1.58 | 1.87 |

Table 8.4: The best fit points found when fitting each IceCube season independently for Analysis I, and a TS comparison between the seasonal best fit point (BF) to the best fit point measured when fitting all years (Result) and the null hypothesis (Null).

| Season | Best fit point $[\sin^2(\theta_{24}), \sin^2(\theta_{34})]$ | $2\Delta LLH_{\text{Result}-\text{BF}}$ | $2\Delta LLH_{\text{Null}-\text{BF}}$ |
|---|---|---|---|
| IC86.2011-2018 | [0.006, 0.40] | 0.00 | 1.74 |
| IC86.2011 | [0.00, 1.00] | 0.63 | 0.20 |
| IC86.2012 | [0.03, 1.00] | 3.54 | 1.88 |
| IC86.2013 | [0.01, 0.63] | 0.09 | 0.62 |
| IC86.2014 | [0.01, 0.45] | 0.18 | 1.07 |
| IC86.2015 | [0.00, 0.00] | 0.47 | 0.17 |
| IC86.2016 | [0.13, 0.03] | 0.27 | 0.83 |
| IC86.2017 | [0.00, 0.71] | 0.05 | 0.17 |
| IC86.2018 | [0.02, 0.20] | 0.10 | 0.57 |

Table 8.5: The best fit points found when fitting each IceCube season independently for Analysis II, and a TS comparison between the seasonal best fit point (BF) to the best fit point measured when fitting all years (Result) and the null hypothesis (Null).

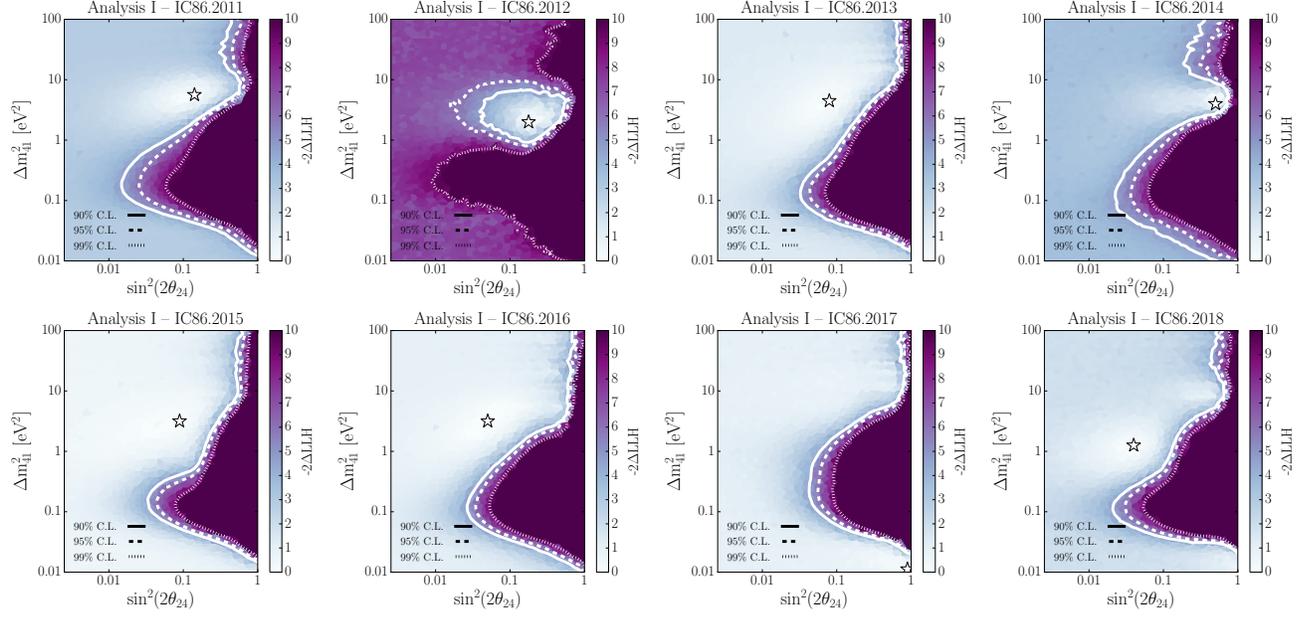

Figure 8-12: The distribution of the TS for Analysis I throughout the physics parameter space for each IceCube season fitted independently with the standard software accounting for the livetime differences.

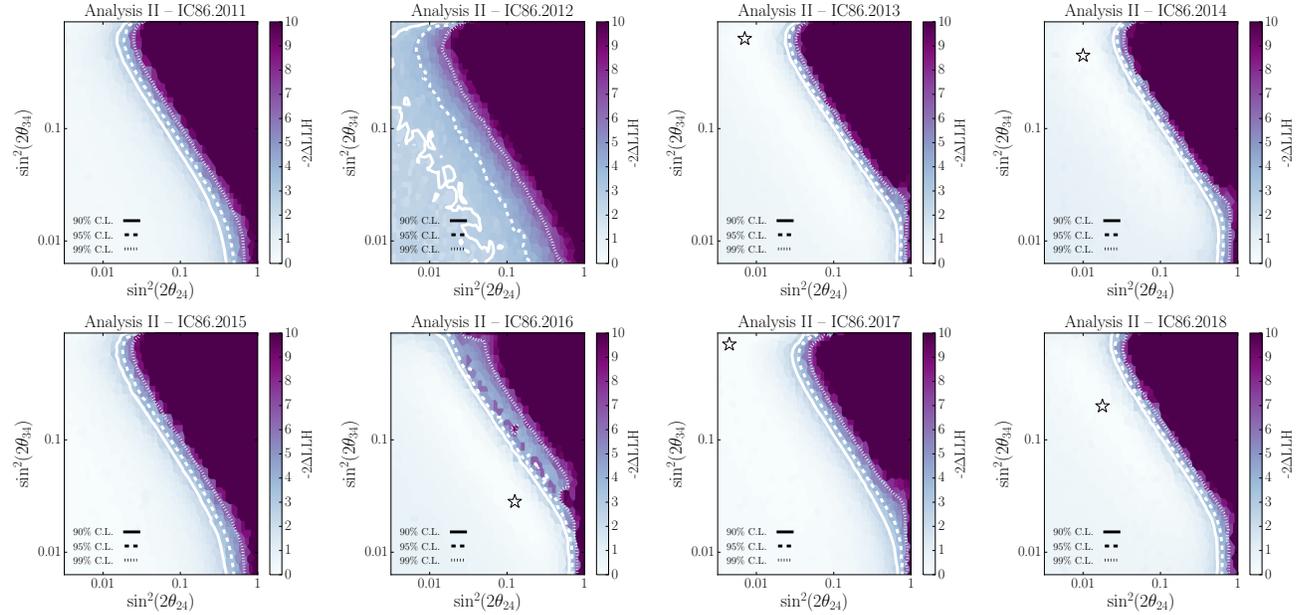

Figure 8-13: The distribution of the TS for Analysis II throughout the physics parameter space for each IceCube season fitted independently with the standard software accounting for the livetime differences.

### 8.3.1 Discussion

**Comparison to the calculated sensitivity**

The sensitivity was reported in Sec. 8.1.2 for both analyses. However, since we observe a closed contour at the 90% CL, the method used to define the Brazil bands means a comparison at this level would not be particularly meaningful. At the 99% CL, we can superimpose the results onto the corresponding Brazil bands. This is shown in Fig. 8-14. We find that the limit placed on Analysis II is slightly lower than the sensitivity along the majority of the parameter space explored, except for in the region below $\sin^2(\theta_{34}) \sim 0.01$. The 99% CL limit placed by Analysis I below $\Delta m_{41}^2$ is slightly stronger than the sensitivity. Above approximately $\Delta m_{41}^2 = 2\,\text{eV}^2$ the limit extends beyond the observed $2\sigma$ Brazil band due to the depth of the best fit point distorting the LLH profile.

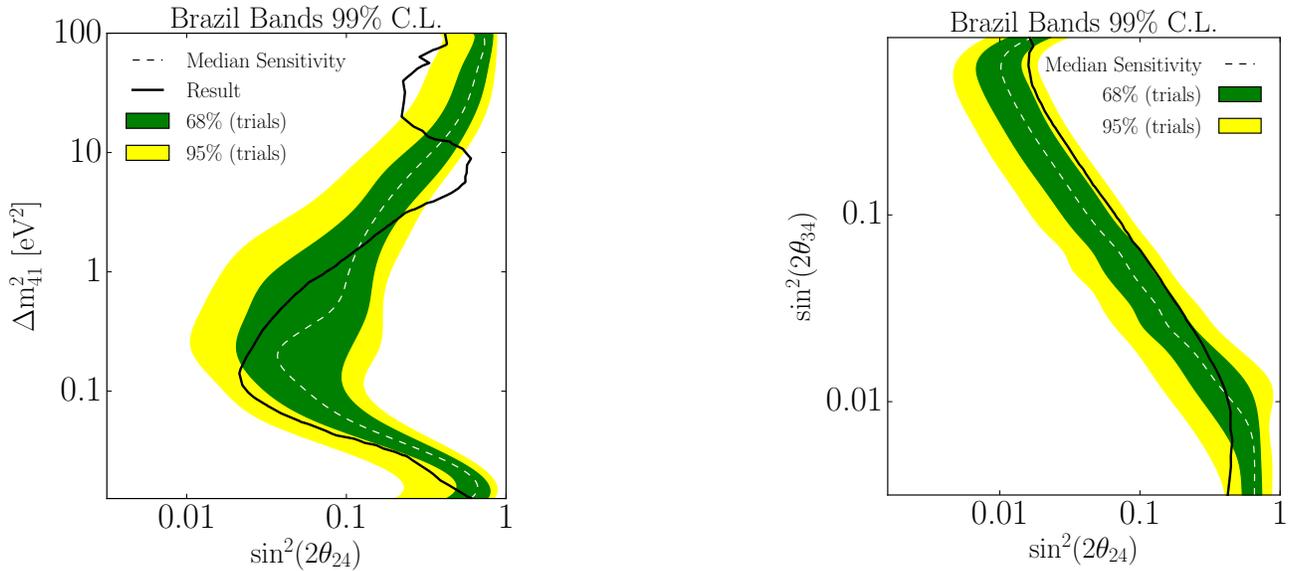

Figure 8-14: The 99% CL Brazil Bands for Analysis I (left) and Analysis II (right) with the observed limit in black. The yellow band corresponds to the 95% spread, while the green to the 68%. The median sensitivity is shown as a dashed white line.
.

**Statistical coverage check at best fit points**

Wilks' theorem verification over the entire physics parameter space is computationally intensive and will be performed at a later date, however, near the best fit points we can verify that this assumption holds. Fig. 8-15 shows a preliminary TS distribution (blue histogram) of the best fit points compared to null for Analysis I (left) and Analysis II (right). The Wilks' 90% and 99% critical values (see Table 6.1) are shown as the vertical red dashed line, and the measured crossing is shown as the green band. The green band represents the $1\sigma$ uncertainty in the measured crossing value given the finite number of pseudo-experiments tested. The cumulative distribution for both the expected distribution and the TS distribution is also shown. In both analyses, we find that the TS distribution follows a $\chi^2$ distribution with 2DOF for Analysis I and 1DOF for Analysis II (black line).

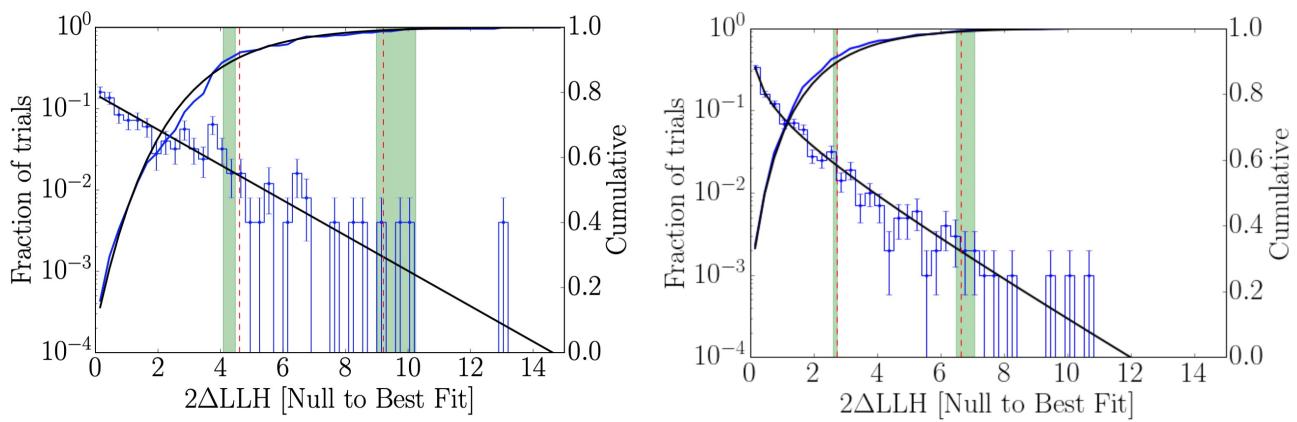

Figure 8-15: The preliminary coverage checks near the best fit points for Analysis I (left) and Analysis II (right). The Wilk's theorem 90% and 99% critical values are shown in dashed red, while the uncertainty in the measured critical values are shown in green.

**Null-like best fit locations**

Using the 2,000 null realizations from the Brazil bands in Sec. 8.1.2, the distribution of best fit points throughout the parameter space were determined. These are shown in blue in Fig. 8-16. The statistical fluctuations tend to populate best fit points around the edge of the sensitivity. The distribution of best fit points for Analysis I shows a slight clustering at large values of $\Delta m^2_{41}$, above $\sim 10\,\text{eV}^2$. It was found in Ref. [204] that the fast oscillations in this region average out pulling the normalization downward with very little signal shape (see for example, Fig. 3.4.6 of Ref. [204]). Statistical fluctuations prefer this area because of this. The observed best fit points for both analyses are shown as the white stars and are found in regions consistent with statistical fluctuations of the null hypothesis.

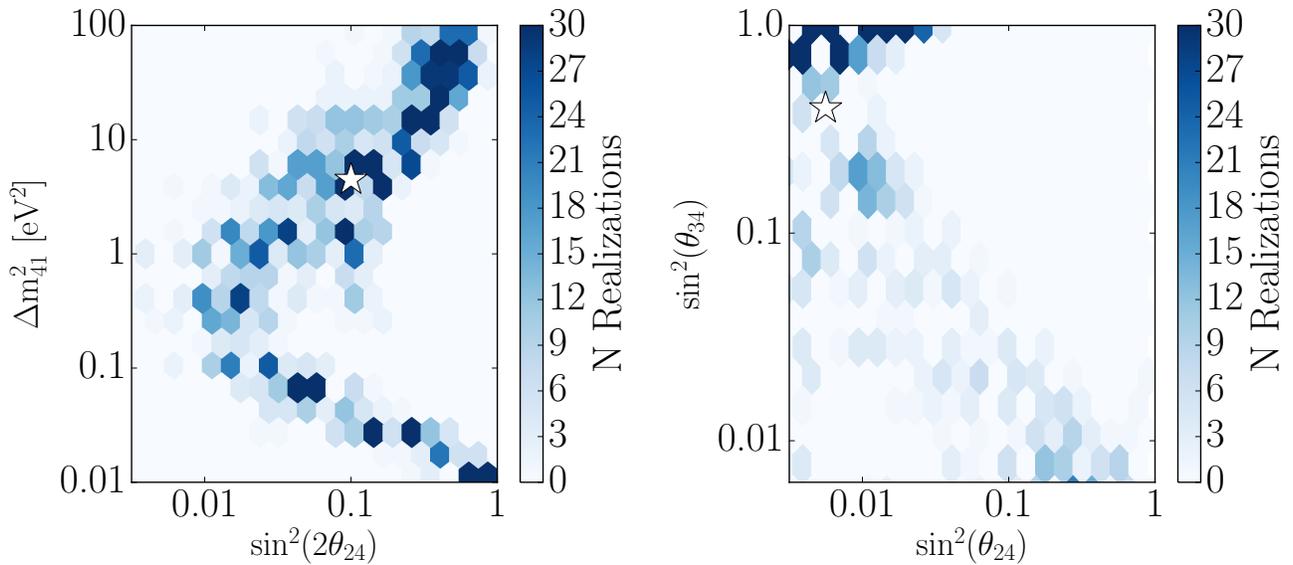

Figure 8-16: The distribution of the best fit points given 2,000 null realizations (blue) throughout the physics parameter space. Also shown is the location of the analyses best fit points, as white stars.

**Nuisance parameters posterior distribution**

This analysis included 18 nuisance parameters relating to the conventional neutrino flux, astrophysical neutrino flux, cross sections, and detector. Table 8.3 shows the minimized values at the best fit points for both analyses. Each nuisance parameter includes a Gaussian prior and central value defined in Table 8.1. The posterior distribution of each nuisance parameter at the best fit point for Analysis I and Analysis II is shown in Fig. 8-17, in the grey and blue histogram, respectively. Both analyses appear to prefer similar systematic pulls. The largest difference observed is between the measured conventional atmospheric neutrino normalization, where they are within 8% of each other corresponding to approximately $1.1\sigma$, given the posterior width. It is also noted that the posterior width of the neutrino-nucleon cross section is identical to the prior width, indicating that we have not significant sensitivity to this systematic uncertainty.

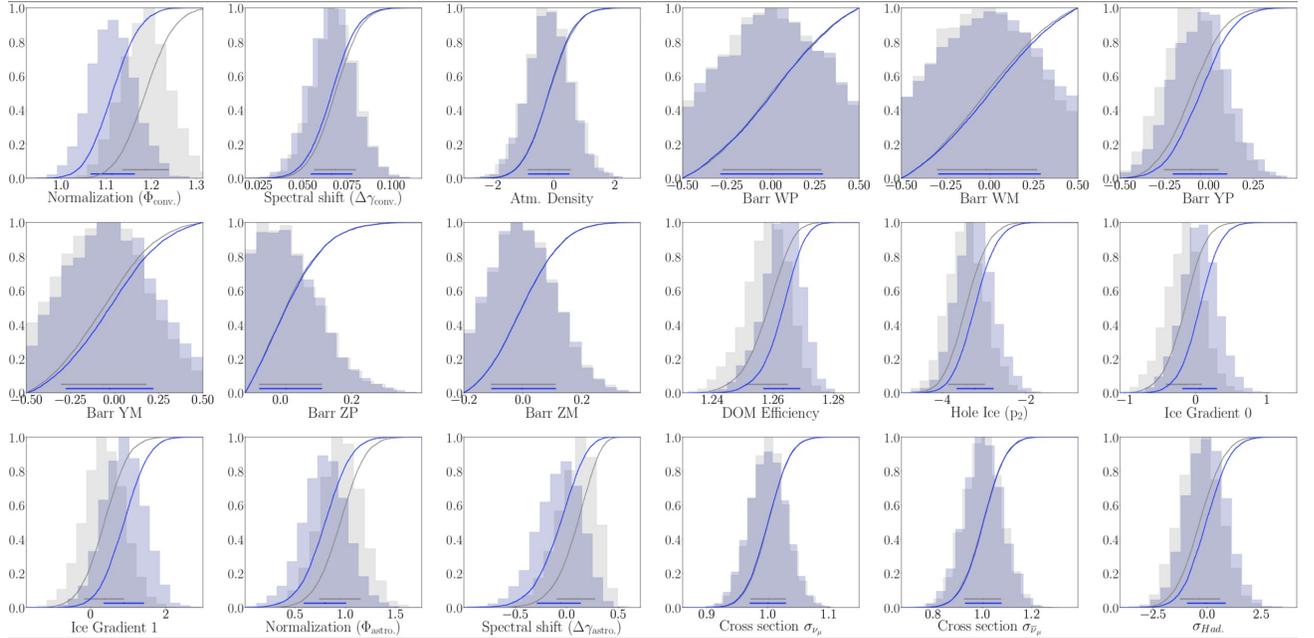

Figure 8-17: The best fit point posterior distributions for each of the nuisance parameters included in Analysis I (grey) and Analysis II (blue). Each subplot shows the cumulative distribution as well as the standard deviation.

**Nuisance parameter correlations at the best fit point**

Fig. 8-18 shows the correlations between each of the 18 nuisance parameters at the best fit point of Analysis I (Analysis II is not shown, but was found to be largely the same). Correlations between subset of nuisance parameters are observable. For example, we find the conventional flux normalization to be anti-correlated with the cosmic ray spectral index as well as the atmospheric density; the DOM efficiency to be highly correlated with the ice properties; and the astrophysical normalization to be correlated with the astrophysical spectral index. The observed anti-correlation between the atmospheric spectral index and the ice gradient 1 is likely to be accidental. The ice gradients were derived using flasher data, completely independent of the cosmic ray flux. They are observed to have similar effects on the energy distribution though, see for example Fig. 7-4 (left) and Fig. 7-13 (right).

The nuisance parameters minimized values throughout the physics parameter space are shown in Fig. 8-19 and 8-20. Each subplot represents a different nuisance parameter, and shows how it pulls in the presence of a sterile neutrino signal.

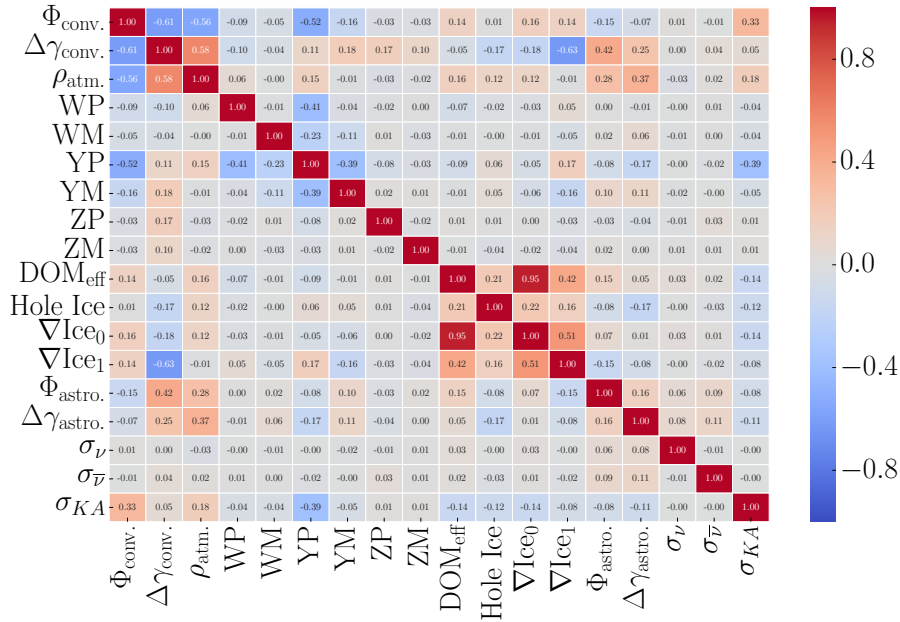

Figure 8-18: The correlation between nuisance parameters at the best fit point of Analysis I.

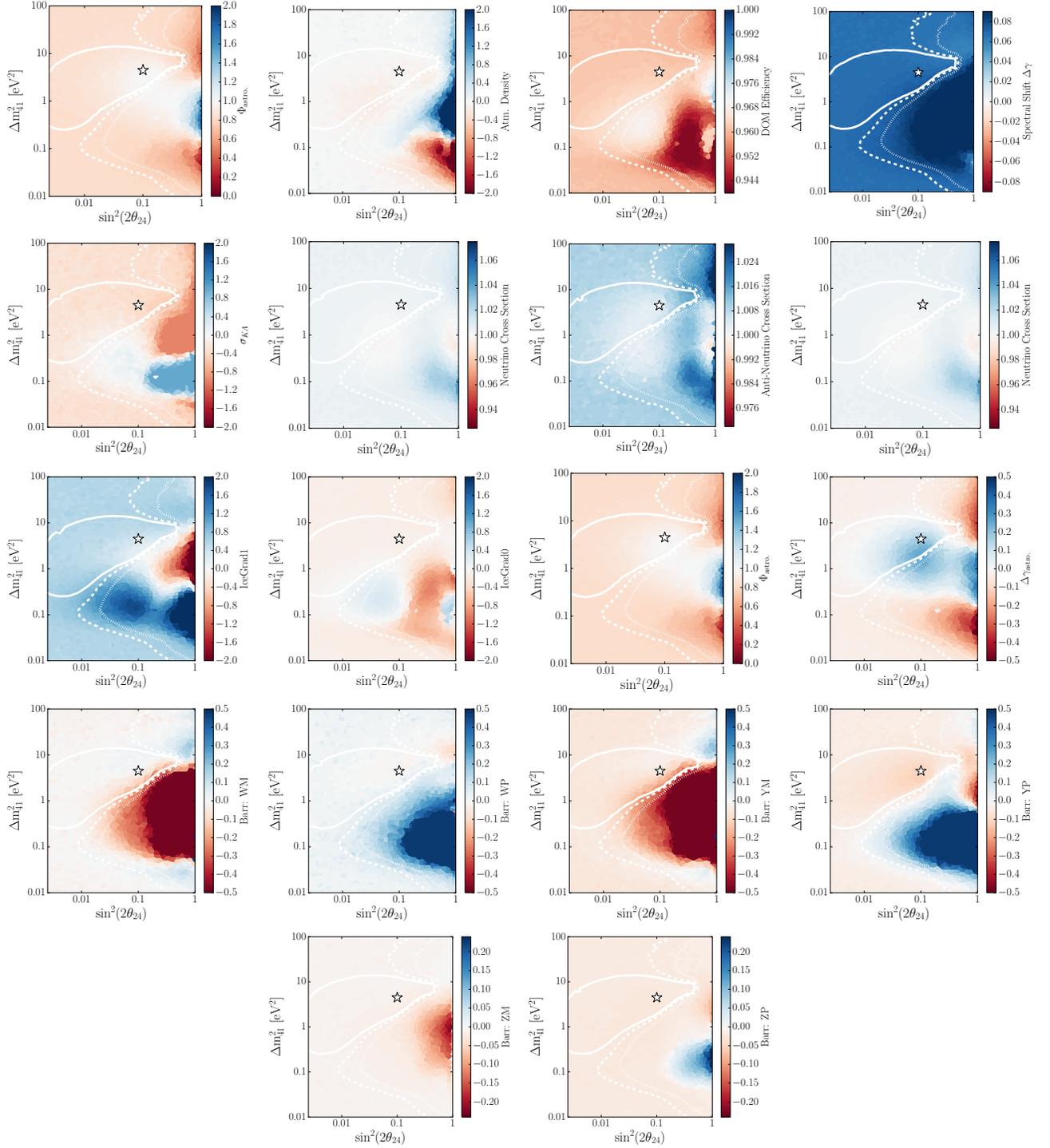

Figure 8-19: The minimized value of each nuisance parameter throughout the physics parameter space of Analysis I. Each subplot is for a separate nuisance parameter. The color scale is set such that the limits are approximately $\pm 2\sigma$ (see Table 8.1 for actual priors).

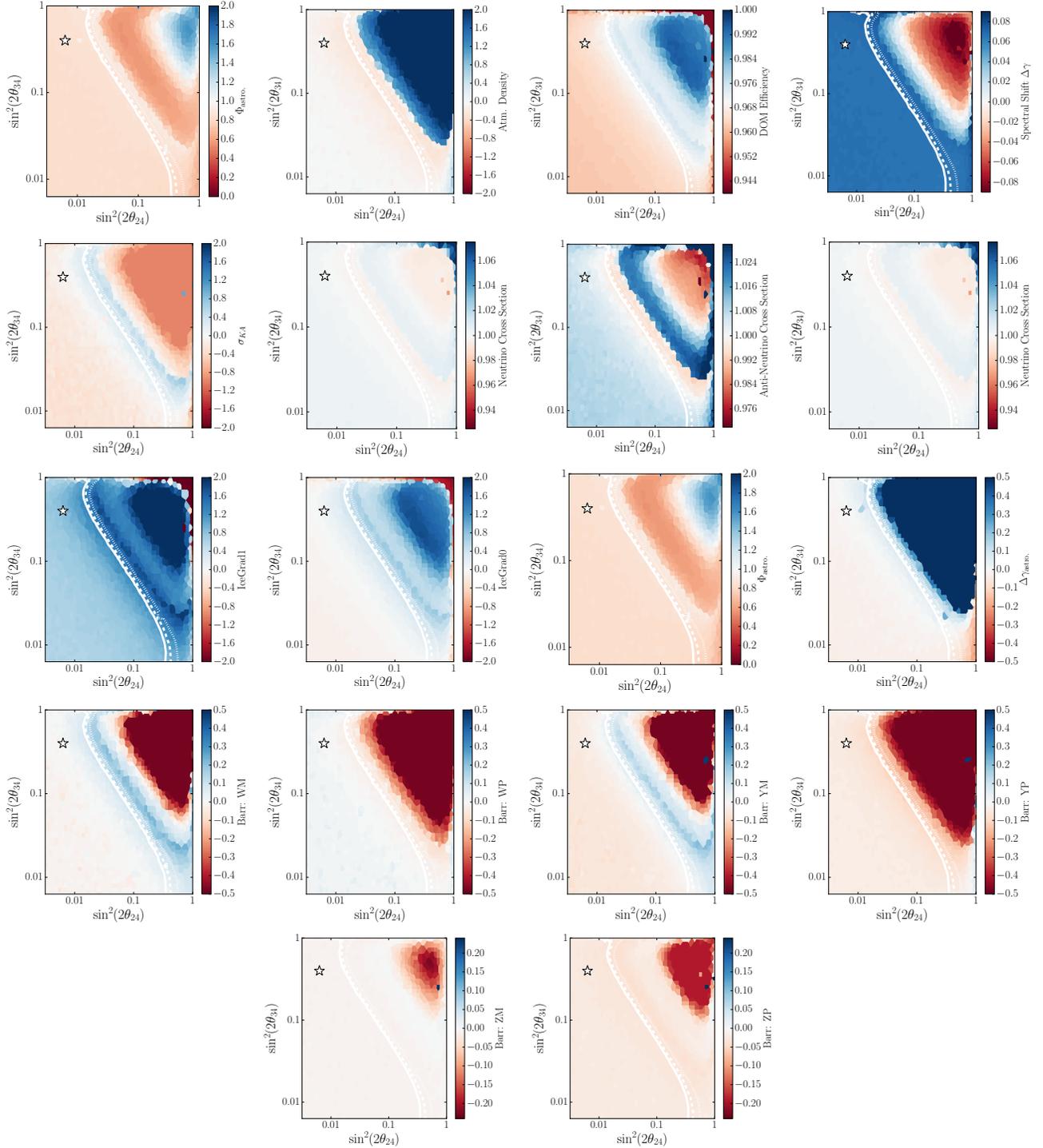

Figure 8-20: The minimized value of each nuisance parameter throughout the physics parameter space of Analysis II. Each subplot is for a separate nuisance parameter. The color scale is set such that the limits are approximately $\pm 2\sigma$ (see Table 8.1 for actual priors).

**The best fit hypothesis shape**

Fig. 8-21 shows a shape-only comparison between the best fit point expectation and null hypothesis expectation in terms of percentage points (top) and statistical pulls (bottom).

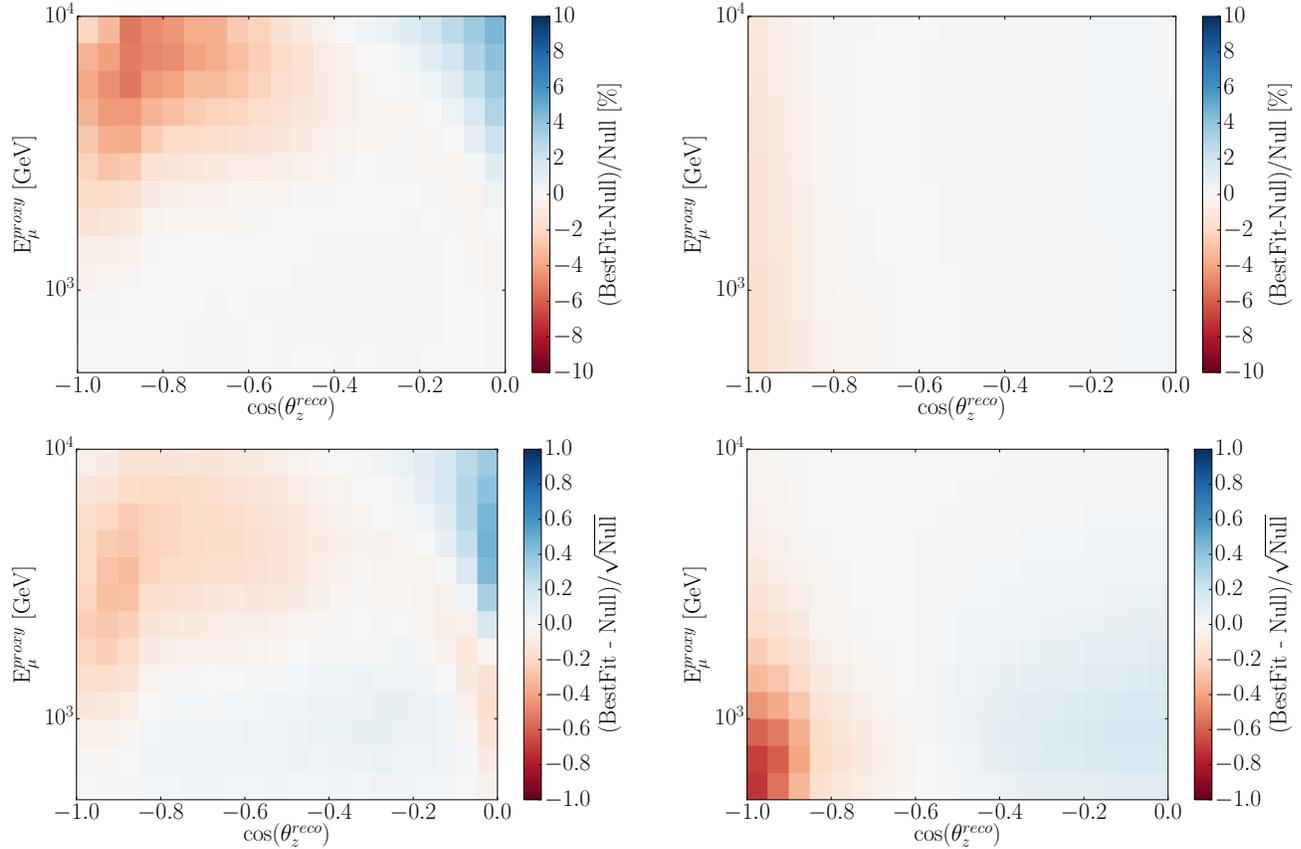

Figure 8-21: Comparison between the best fit point and null for Analysis I (left) and Analysis II (right). Top: Represented in terms of percentage points. Bottom: Represented in terms of statistical pulls.

**N-1 results**

Fig. 8-22 shows the impact on the result after removing various components of the analysis. The solid (dashed) lines in these figures show the 90% CL (99% CL) and the stars represent the best fit location. The top four plots have all been calculated with each nuisance parameter set to the best fit value reported in Table 8.3. The top two plots show the result of the analyses after of removing a single nuisance parameter. The middle plots show the impact of removing a full group of nuisance parameters. Finally, the bottom two figures show the impact of removing a single season from the analysis.

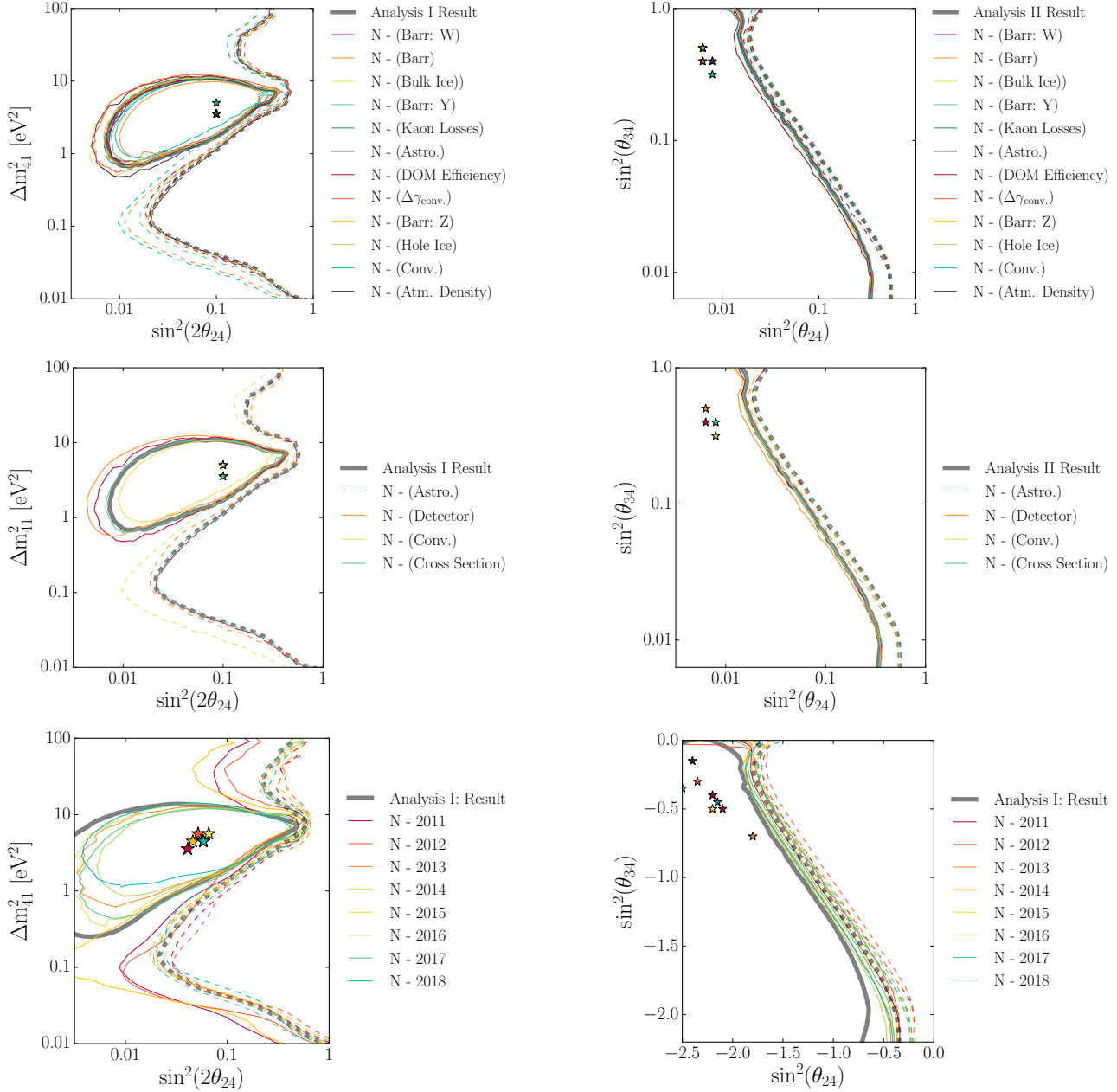

Figure 8-22: The result of the N-1 tests, in which a single part of the analysis was removed for Analysis I (left) and Analysis II (right). Top: The result calculated assuming all but one of the systematic uncertainties. Middle: The result calculated after removing a full group of systematic uncertainties. Bottom: The result after removing a single season from the analysis.
.

**Comparison to previous IceCube measurements**

Currently, there exists three published results from IceCube for a 3+1 sterile neutrino model in the comparable space to Analysis I. The results of these searches are shown in Fig. 8-23 (left) for the 90% CL (solid lines) and 99% CL (dashed lines). The three year DeepCore result [155] (shown in blue) placed a limit on the sterile neutrino parameters above $\Delta m_{41}^2 \approx 1\,\text{eV}^2$ and $\sin^2(\theta_{24}) = 0.39$ at the at 90% CL. The IC86 high energy sterile neutrino search (shown in grey) excluded the region from approximately $0.1\,\text{eV}^2 < \Delta m_{41}^2 < 2.0\,\text{eV}^2$ above $\sin^2(\theta_{24}) = 0.1$, extending out to approximately $\sin^2(\theta_{24}) = 0.016$ and $\Delta m_{41}^2 = 0.27\,\text{eV}^2$. Alongside the publication of the IC86 result, an independent measurement using the IC59 configuration (59 active IceCube strings) reported the limit shown in red at the 99% CL.

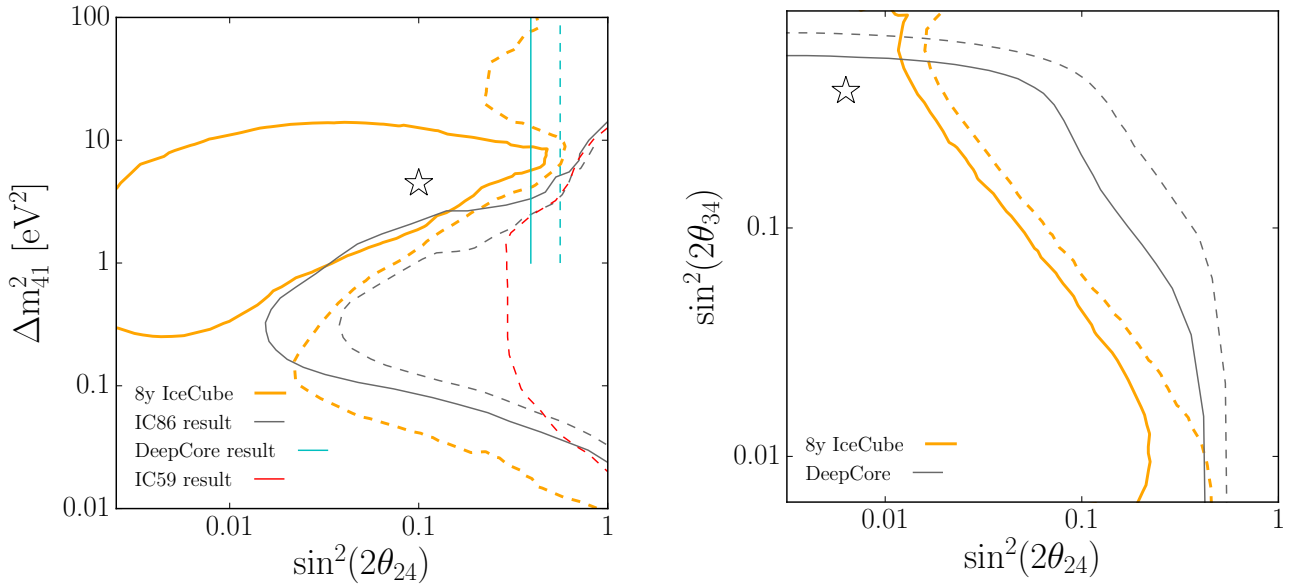

Figure 8-23: The 90% CL (solid) and 99% CL (dashed) for all IceCube published results. Left: Analysis I (orange), compared to: IC86 high energy sterile neutrino analysis result in grey, IC59 result in red, and the low energy DeepCore result in blue. Right: The result of Analysis II (orange) compared to the low energy DeepCore result (grey).

The result of Analysis I (in orange) is shown to be in agreement with the previous IceCube measurements. In particular, the 99% CL limit on the sterile neutrino mixing parameters is increased at all points in the physics parameter space. In the region below $\Delta m_{41}^2 = 0.1\,\text{eV}^2$, the limit is improved by nearly a factor of seven, largely due to the improved statistics at low energies. Although we observe a closed 90% CL contour, it is found to be in agreement with the limits placed from all previous IceCube results.

The three year DeepCore sterile analysis has also placed limits in the comparable space to Analysis II. This is shown in grey in Fig. 8-23. We find that the result of Analysis II improves the limit on the sterile neutrino mixing parameters below approximately $\sin^2(\theta_{34}) = 0.4$. Here, the limit is shown to increase by factor ranging from two to approximately five.

**Comparison to world data**

We now shift the focus of the result to a comparison of the worlds data and global best fits. The result of Analysis I at the 90% and 99% CL is shown compared to the published data from MINOS [227–230], CCFR [80], DeepCore, Super-Kamiokande [76], MiniBooNE-SciBooNE [78, 231], and CDHS [80] in Fig. 8-24. At the 90% CL (Fig. 8-24 left), much of the allowed parameter space is not excluded by other experiments. The best fit point is found to be in a region excluded by MiniBooNE-SciBooNE, however the TS distribution in that region is relatively flat. The 99% CL limits shown on the right side of Fig. 8-24 indicate that this result provides some of the worlds strongest limits in a large portion of the physics parameter space. In particular, we find the worlds strongest limit between $0.04\,\text{eV}^2 < \Delta m_{41}^2 < 1.0\,\text{eV}^2$ when compared to the experiments shown.

The Analysis I results compared to two global best fit calculations: from Diaz et al in blue [73], and its predecessor from Collin et al [81] in green are shown in Fig. 8-25 (left). We find that the 99% CL limit excludes part of the allowed region from Ref. [73], and the lower island from Ref. [81]. Similarly, we show the global best fit regions from Giunti et al [232] compared to the results of Analysis I in Fig. 8-25 (right).

The equivalent comparison to world data for Analysis II is shown in Fig. 8-26. Here, data is compared at the 90% CL (left) and 99% CL (right) to the results above $\Delta m_{41}^2 = 10\,\text{eV}^2$ from Super-Kamiokande [76] and DeepCore [155]. This analysis provides world leading limits in the region $\Delta m^2_{41} > 10\,\text{eV}^2$ from approximately $0.024 < \sin^2(\theta_{34}) < 0.54$ and $0.012 < \sin^2(\theta_{24}) < 0.16$.

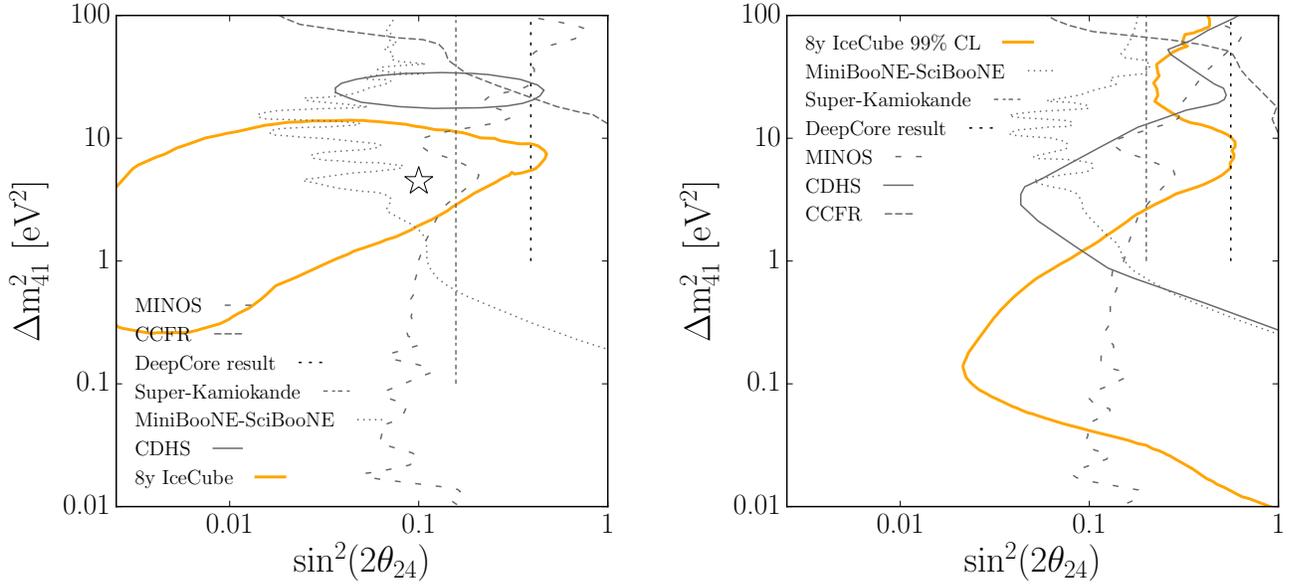

Figure 8-24: The result of Analysis I compared to world data: MINOS [227–230], DeepCore, CCFR [80], MiniBooNE-ScibooNE [78, 231], CDHS [80] and Super-Kamiokande [76]. Left: A comparison at the 90% CL. Right: A comparison at the 99% CL.

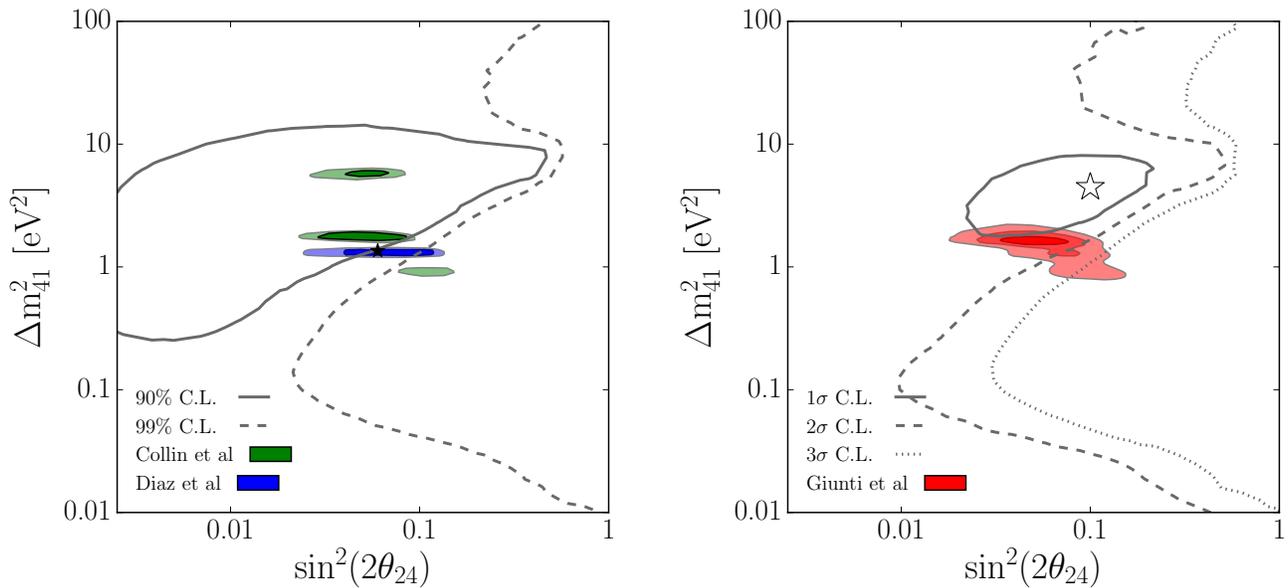

Figure 8-25: Left: the final result for Analysis I, shown at the 90% and 99% CL along with the global best fit regions from Collin et all [81] in green and Diaz et al [73] in blue. Right: the result of Analysis I shown at $1\sigma$, $2\sigma$, and $3\sigma$ compared to the global allow regions from Giunti et al [232].

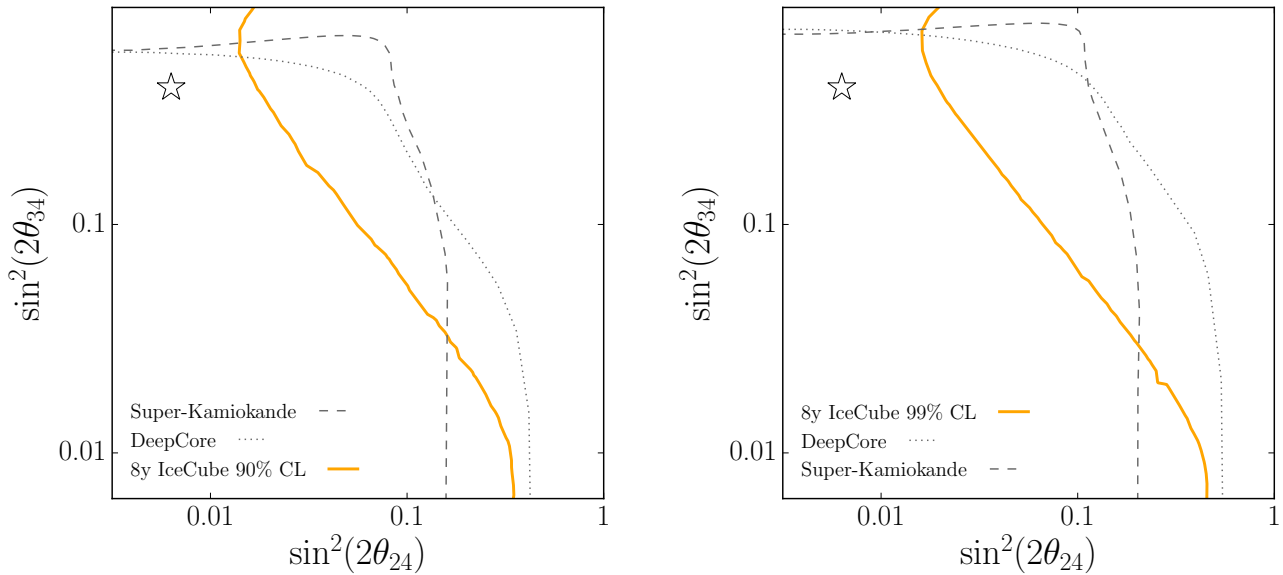

Figure 8-26: The result of Analysis II compared to world data: Super-Kamiokande [76] and DeepCore [155]. Left: A comparison at the 90% CL. Right: A comparison at the 99% CL.

THIS PAGE INTENTIONALLY LEFT BLANK

# Chapter 9

# Conclusion

The pursuit of neutrino oscillation measurements has led to some intriguing anomalous short baseline results which are often interpreted as a sterile neutrino. The IceCube Neutrino Observatory is in a unique position to search of signs of sterile neutrinos using different physical phenomena at vastly different energies and baselines using matter effects.

This thesis has reported results on two searches for sterile neutrinos using 8 years of data from the IceCube Neutrino Observatory. As with most studies on large collaborations, the results presented here have relied upon and benefited greatly from the work of others. My specific contributions to these results are summarized in Appendix. A.

In conclusion of this thesis, no evidence for sterile neutrinos was observed. The first search found a best fit sterile mixing parameters at $\Delta m_{41}^2 = 4.47\,\text{eV}^2$ and $\sin^2(\theta_{24}) = 0.10$. This is an interesting area of the physics parameter space given the worlds data and global best fit regions for the sterile hypothesis, however, we find the result compatible with the no-sterile hypothesis at 8%. The 99% CL limits indicates that this result provides some of the worlds strongest limits on the sterile neutrino mixing parameters between $\Delta m_{41}^2$ and $\sin^2(\theta_{24})$ in the range between $0.04\text{eV}^2 < \Delta m_{41}^2 < 1.0\text{eV}^2$.

The second analysis found the best fit mixing parameters of $\sin^2(\theta_{34}) = 0.40$, $\sin^2(\theta_{24}) = 0.006$,

in the oscillation averaged out region ($\Delta m_{41}^2 > 10\,\text{eV}^2$), with a $2\Delta LLH = 1.74$ relative to the no-sterile hypothesis, corresponding to a p-value of 19%. The limits found here are some of the worlds strongest on the sterile neutrino mixing parameters $\sin^2(\theta_{24})$ and $\sin^2(\theta_{34})$, particularly at values of $\sin^2(\theta_{34})$ below 0.4.

This work has explored important parameter space within the picture of 3+1 models. The results are expected to have a notable effect on the solutions found by global fits to 3+1 models, because the explored parameters space overlaps with allowed regions. This motivates future work in the ongoing sterile neutrino search program. My view of how this work can be carried forward is summarized in Appendix B.

# Appendix A

# Specific contributions

The following represents a summary of the activities that I was particularly involved with throughout the course of my graduate work at MITs. I will begin by detailing the specific contributions to this analysis, then briefly discuss other projects that I was involved with that did not make it into this thesis.

- **Analysis organizer**: My largest contribution to the analysis was that of "analysis organizer". By this, I mean that I was responsible for verifying all updates to software, calculating sensitivities, running pseudo-experiments test, investigating systematic uncertainties, implementing the MC into the analysis chain, testing the minimizer and systematic effects, and reporting to the IceCube Oscillations working group.

- **SPE Templates**: This work is largely my own and I am the primary author of the paper found in Appendix F. I developed the methods used to extract single photoelectrons from the detector as well as the fitting algorithm (with special thanks and help from M. Rongen and the IceCube Calibration working group). This work lead to significant improvements in the low-level data/MC agreement and will be particularly useful for up-coming low-energy analyses. An additional outcome of the SPE Templates was the precise measurement of the discriminator setting. This was also shown to improve data/MC agreement. The result of the SPE templates represents also a precision measurement of the gain setting on each PMT. This is useful for calibration and will be used in the future for data-reprocessing.

- **MC production**: I was responsible for producing the MC used in this analysis (all except the cosmic ray muon background simulation). This was a significant endeavor that took multiple years to complete and is the largest muon neutrino dataset in IceCube. Along the way, many bugs were identified and resolved, as well as new processing techniques implemented.

- **Data reprocessing**: Since this analysis required consistent data from IC86.2011 to IC86.2018, the full IceCube dataset was required to be reprocessed with the same event filters. This too took several years to complete. My contribution to this effort was the measurement of the gain on each PMT in order to re-calibrate the extracted charge. The result of this work is internally known as Pass2.

- **CosmicWatch outreach program**: During my graduate work, I developed an outreach project called CosmicWatch (further information found below). While working at WiPAC, I used this program to teach IceCube interns, high school students, and college professors, about cosmic ray physics and electronics.

- **Master clock update**: While working at the South Pole in late 2018, I was partially responsible for the replacement of the power supplies powering the DOMs and replacing the master clock unit. This work was performed primarily with T. Bendfelt.

- **Working on bulk ice systematic**: The MultiSim/SnoStorm bulk ice method was largely the work of the UTA group. The MC processing for the method and implementation into the processing chain, along with aiding in the development of the method was my contribution. The method used here is subsequently in the process of being implemented on other systematics and has tremendous potential for improving the description of systematic uncertainties in IceCube.

- **Event selection**: The design of the Diamond filter and updates to the Golden filter were also my responsibilities. The output, the Platinum event selection, significantly improves the purity of the $\nu_\mu$ selection, as well as doubled to the total number of extracted events. This work was subsequently implemented into other IceCube analyses.

- **Updated event generator**: This is the first analysis to use LeptonInjector. Understanding any differences in the event distribution compared to previous event generators fell to me. From this, we found multiple undesired features and bugs associated with other generators, and verified the performance of LeptonInjector.

Aside from the work presented in this thesis and with IceCube, I've been particularly involved with three other projects. The paragraphs below outline these projects along with my specific contributions.

The CosmicWatch Desktop Muon Detector is a MIT and Polish National Centre for Nuclear Research (NCBJ) based undergraduate-level physics project that incorporates various aspects of electronics-shop technical development. The detector was designed to be low-power and extremely portable, which opens up a wide range of physics for students to explore. I developed this program during my first few years at MIT and it has since significantly grown in popularity. I've written an article in PhysicsToday [233], which reached over 14,000 shares prior to being recently reset, and was highlighted on MIT News [234]. Since then I have found additional articles written about the project on Symmetry Magazine [235], Science Daily [236], Geek.com [237], Interesting Engineer [238], Next Big Future [239] and many others, as well as hitting the front page of Reddit. I was the primary author of two publications relating to CosmicWatch (see Refs. [1] and [240]) and it was also the subject of my Masters thesis [4]. This project has directly helped over 200 students and teachers build their own detectors, as well as provided online support to hundreds of others.

The ISOtope Decay-At-Rest (IsoDAR) experiment is designed to provide a unique search for short baseline nuebar oscillations. By measuring nuebar disappearance over an L/E of approximately 0.6-7.0 m/MeV with a kiloton class detector like KamLAND, we can conclusively test the current global allowed regions for a 3+1 sterile neutrino hypothesis. IsoDAR expands on several key technologies to make this measurement possible. These include the development of a high-current $H2^+$ ion source (MIST-1), an investigation into using a radio-frequency quadrupole (RFQ) as a buncher/pre-accelerator for the axial injetion into a cyclotron both of which I played an important part in. I was primary responsible for the simulation, design, and manufacturing of MIST-1. This led to the publication of Ref. [241], in which I was the primary author. The work on the beam-line tests also resulted in the paper Ref. [242], in which I was one of the primary authors, along with D. Winklehner. Other than this, I was also involved with the upcoming beam injection strategy using an RFQ. This is described in Ref. [243].

The KPipe publication in Phys. Rev. D [244] was the result of research in the design of a new neutrino detector to be located at the J-PARC Materials and Life Science Experimental Facility's (MLF) spallation neutron source. The MLF represents the world's most intense source of charged kaon decay-at-rest monoenergetic muon neutrinos. This is a very well understood neutrino source, which allowed us to suggest a novel detector configuration together with the implementation of a light collection system based on new silicon photomultiplier technology. I was one of the primary contributing authors to this paper and was responsible for the sensitivity and design of the detector.

# Appendix B

# OUTLOOK

## B.1 The global picture

Given the presented results in this thesis, it is my view that in order to resolve the 3+1 sterile neutrino model tension and elucidate the nature of the short baseline anomalies, we must look towards a decisive experiment capable of probing the global allowed regions to high significance. Ideally, this measurement should be statistically limited with well-understood systematic uncertainties. While near-to-far ratios between different detectors are often used for probing the active neutrino oscillations, the congested and unclear situation with the 3+1 sterile neutrino model motivates future experiments to map out the oscillation wave within a single detector (i.e. observe the disappearance and re-appearance of the neutrino state). This is particularly important for testing alternative hypotheses to the 3+1 model, such as the 3+2 model or sterile neutrino with decay model.

One such example is an experiment capable of this measurement is the IsoDAR experiment [245]. Using a well understood $\bar{\nu}_e$ source, near a kiloton-scale detector like KamLAND, this experiment aims to explore the global best fit point to >$5\sigma$ in just 4 months of operation and and map out the oscillation wave the single detector (see Fig. B-1).

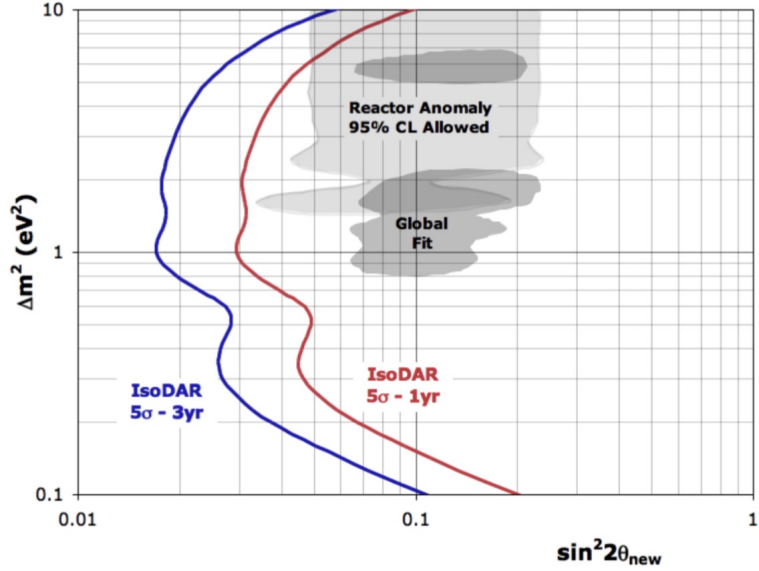

Figure B-1: The projected sensitivity of the IsoDAR experiment.

## B.2 Prospects for future work

The work presented in this thesis is a step in an ongoing thanks to the collaborators at MIT and University of Texas, Arlington. The final verification of Wilks' theorem throughout the full physics parameter space at the 90% and 99% CL is currently underway and we are nearing the finalization the Bayesian-equivalent analysis.

This thesis uses nearly the entire current data set of the 86-string configuration of IceCube. Significant statistical updates will have to wait for new data. The expected sensitivity over the mixing parameters $\Delta m^2_{41}$ and $\sin^2(\theta_{24})$, comparing 8 years of data to 20 years of data is shown in Fig. B-2. While there is a noticeable improvement, we expect the next generation upgrade to IceCube, Gen-2 [246], to dramatically increase the high-energy $\bar{\nu}_\mu$CC sample, providing a significant increase in the sensitivity to eV-scale sterile neutrinos. Until then, several recommended avenues of systematic exploration are:

1. Extend the cosmic ray spectral index into a composition dependent systematic to account for rigidity cutoffs.

2. Move to the unified hole ice description to allow for cross comparison between IceCube measurements.

3. Fit for the prompt atmospheric neutrino flux.

It would be interesting to expand the high energy cut to beyond the current limit of 9,976 GeV. Raising the energy limit may provide extra sensitivity in the region of the best fit point. Removing the high energy cut all together would allow for a sensitive astrophysical neutrino flux measurement and potentially for a prompt atmospheric neutrino measurement.

The scan over $\sin^2(\theta_{34})$ and $\sin^2(\theta_{24})$ could be improved by an extension to lower energies, combining the results presented here with an upgrade to the low-energy Deep-Core analysis [155]. As shown in Fig. 8.3 for Ref. [247], statistical improvements could yield significant improvements to the sensitivity to lower $\sin^2(\theta_{34})$. Extending this search into the mass dependent region, $0.1\,\text{eV}^2 < \Delta m_{41}^2 < 10\,\text{eV}^2$, is also extremely interesting. The analysis presented in this thesis has placed a conservative limit on the sterile mixing parameters, however the sensitivity increases dramatically when extending the parameter search into 3-dimensions, as shown in Fig. B-3

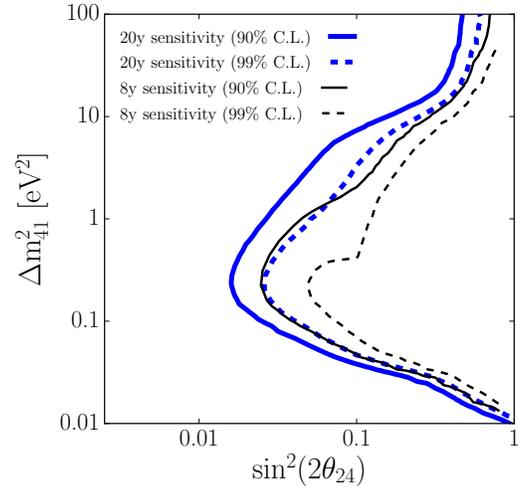

Figure B-2: The expected sensitivity change assuming 20 years of IceCube data (without Gen2 data).

The framework used this analysis opens up a plethora of possible BSM searches, many of which are currently underway. These searches include:

1. A search for sterile neutrinos + decay, where the $\nu_4$ state decays to invisible BSM particles.

2. A search for non-standard interactions.

3. The equivalent search presented here, in the 1+3 sterile neutrino mass ordering.

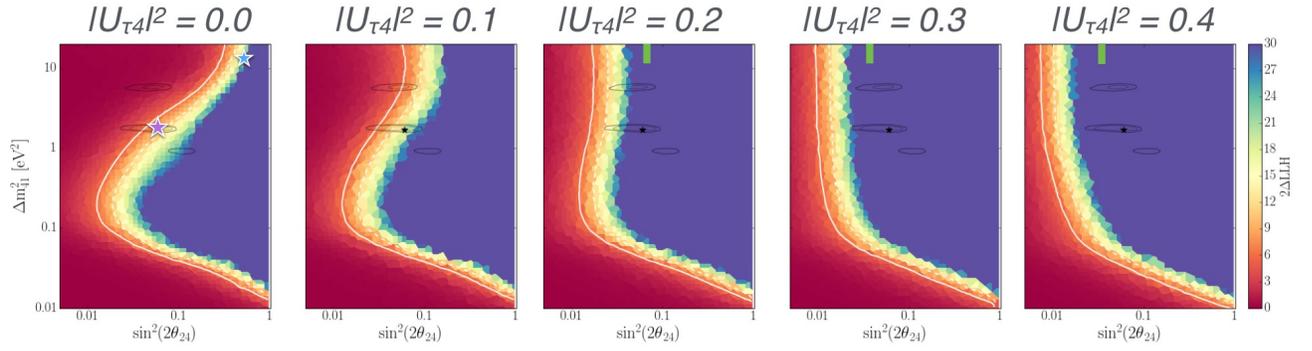

Figure B-3: The sensitivity in terms of $\Delta m_{41}^2$ and $\sin^2(\theta_{24})$ when activating a non-zero $\theta_{34}$ component. The star and black contours indicate the location of the global best fit point from Ref. [81].

# Appendix C

# Further pre-unblinding checks

This section will describe the series of checks performed on the blind data in order to asses any potential issues prior to unblinding. These tests will focus on the results using Asimov datasets to assess coverage of systematic uncertainties and ensure proper performance of the minimizer. The term "standard software" will be used to describe the software used for the result presented in this thesis.

## C.1  Testing the minimizer

The likelihood minimization in this analysis uses the Limited-Memory BFGS-B minimizer (L-BFGS-B) [195]. This is an optimization algorithm that uses quasi-Newton method that approximate the Broyden-Fletcher-Goldfarb-Shanno (BFGS) algorithm with a limited amount of computer memory. The modification "-B" refers to the implementation of box constraints. Prior to reporting any plot found in this work, it was necessary to verify that the minimizer was indeed finding the true minimum in the complicated, multidimensional likelihood space.

## C.1.1 Sterile hypothesis minimization (Inject/Recover signal)

For this test, we inject an Asimov signal somewhere in the physics parameter space and allow the minimizer to recover it with the standard set of software. We find that 399 out of the 400 injected signals are recovered at the exact same sterile hypothesis point in which they were injected (see Fig. C-1). For the single case where this tests does not recover the exact value, it found the minimum in the adjacent sterile hypothesis point, located ∼0.0001 units in $2\Delta LLH$ away from the true point (i.e. in an area that is essentially flat in LLH values). This test concludes that we are able to recover the injected signal hypothesis over the full space. The scan was performed at a resolution of steps of 0.05 in log-space for the physics parameter variables (high resolution scan).

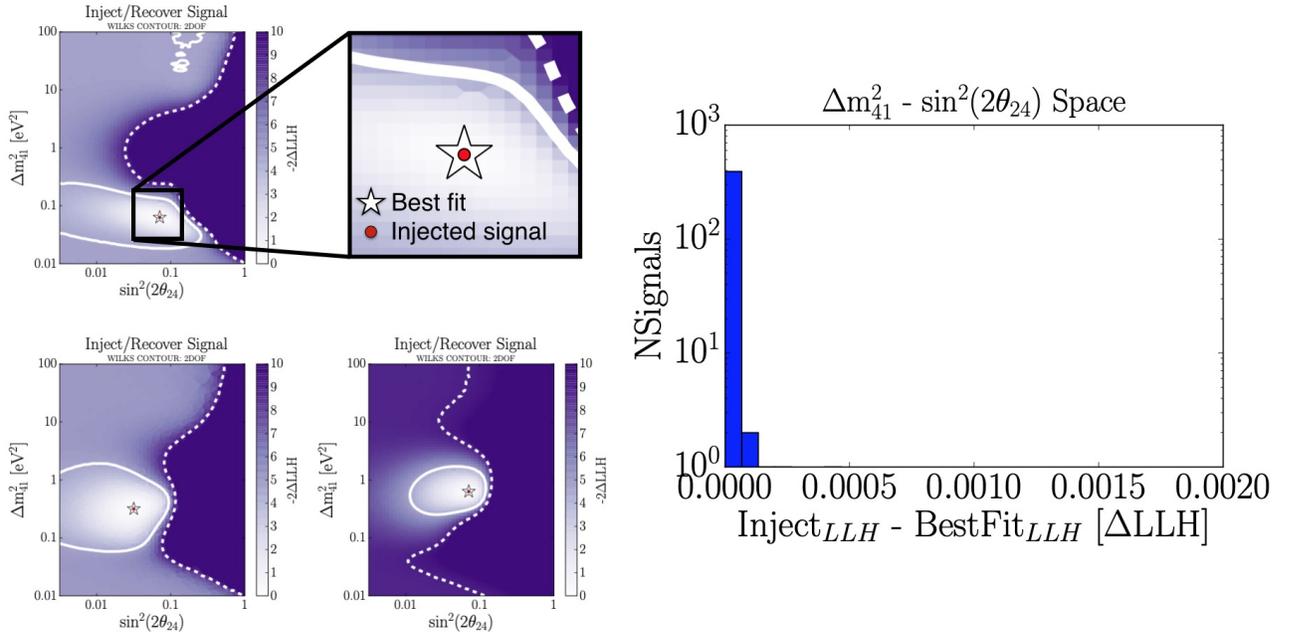

Figure C-1: Left: Thee example of the inject recover signal tests. The signal was injected at the red dot, while the recovered minima was found at the star. Right: The difference in terms of the ΔLLH between the best fit point and the injected flux point.

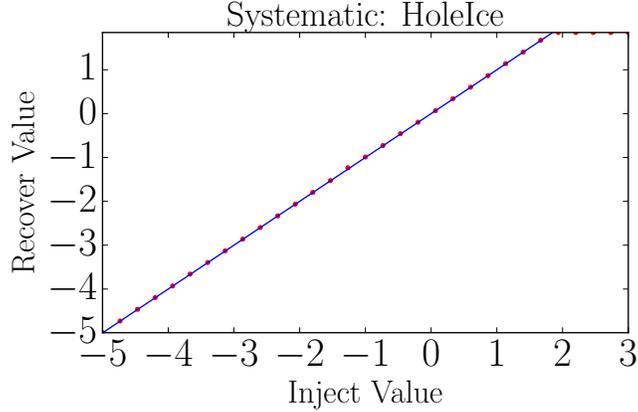

Figure C-2: The inject recover systematic test for the hole ice nuisance parameter. The hole ice has a hard boundary at $-5$ and $2$.

## C.1.2 Nuisance parameter minimization (Inject/Recover systematics)

This test injects the null hypothesis as the signal with the central value of one of the nuisance parameters offset from the default value. We then move the center of the prior of that nuisance parameter to the offset value, and minimize over all nuisance parameters. This is performed on an Asimov dataset and the minimizer should recover the offset value (to within the set precision of the minimizer). The plot shown in Fig. C-2 is one such test for the forward hole ice $p_2$ value (arbitrary selected). The injected (x-axis) and recovered (y-axis) value is shown as a red dot. The blue line shows the ideal case, where the injected value equals the recovered value. As can be seen from this figure, we are properly able to recover all the injected hole ice $p_2$ values over the full space. The conclusion of this plot is the same for all the other nuisance parameters: the minimizer is able to recover an injected signal generated by an offset nuisance parameter. Note: for nuisance parameters with correlations – the astrophysical neutrino flux normalization and spectral index, and bulk ice ice gradients – the correlation was removed prior to performing this test.

## C.2 Cosmic ray muon background contamination

The CORSIKA simulation sets described in introduction to Sec. 5 had an equivalent livetime of less than 30 days. Given that this analysis has such a high level of background rejection, describing the expected distribution of the cosmic ray muon background is computationally intensive and severely limited by the available computational resources. Since CORSIKA starts with the cosmic ray primaries rather than in-ice muons, much of the computational time is spent on simulating the trajectories of low energy particles that will never reach the detector. Similar to the LeptonInjector method described in Sec. 4.2.1, we can decouple the air shower simulation from the muon propagation. Upon generating bundles of cosmic ray muons from some parameterization of the muon flux in-ice, we can then re-weight the muon bundles according to initial conditions, such as the cosmic ray model, hadronic interaction model, and atmospheric temperature profile. This method is used in the toolkit MuonGun [248, 249]. A comparison between the two predicted event distributions is shown in dark blue in Fig. C-3. It is shown that the predicted event rate for MuonGun and CORSIKA, at analysis level, is $0.101\,\mu$Hz and $0.075\,\mu$Hz respectively.

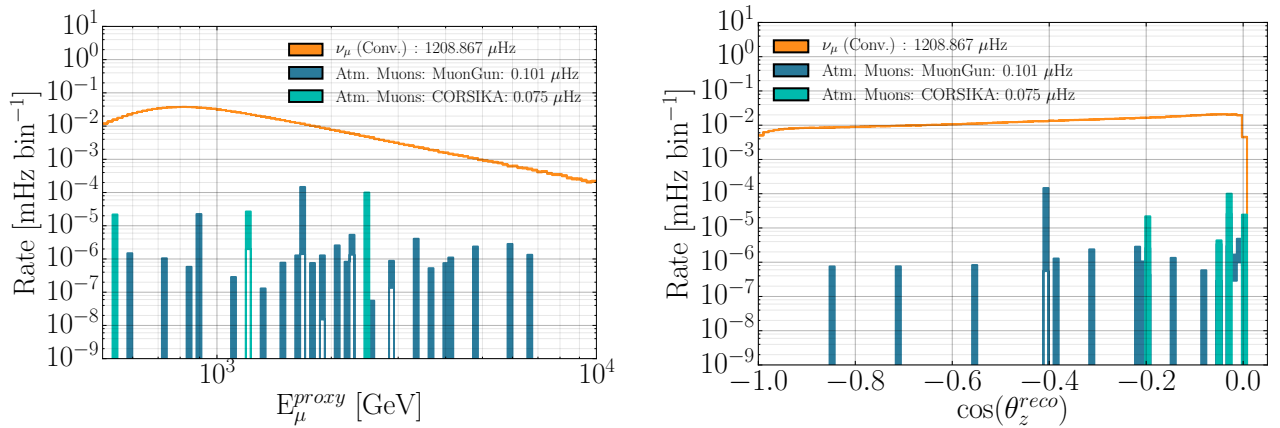

Figure C-3: Comparison between predicted cosmic ray muon background rates from MuonGun and CORSIKA.

Upon unblinding the IC86.2011 high energy sterile neutrino analysis, it was noted that there was ~1% level of mis-reconstructed muon neutrino events due to a coincident cosmic ray muon. This analysis has implemented several additional cuts, along with a new event splitter (see Sec. 5).

# Appendix D

# Results of the first unblinding

On August 1$^{\text{st}}$ 2019, the best fit points for both analyses were revealed. Analysis II was found to be consistent will the null hypothesis at 44%. Analysis I, however, found a best fit point at $\Delta m^2_{41} = 8.9\,\text{eV}^2$ and $\sin^2(\theta_{24}) = 0.50$ with a $2\Delta LLH$ relative to the null of 7.01, corresponding to a p-value of 3%. This chapter will focus on describing the post-unblinding tests performed on Analysis I. The shape represented in terms of statistical pulls between the Analysis I best fit point and the null hypothesis is shown in Fig. D-1 (left). The shape difference between the two hypotheses is shown in Fig. D-1 (right).

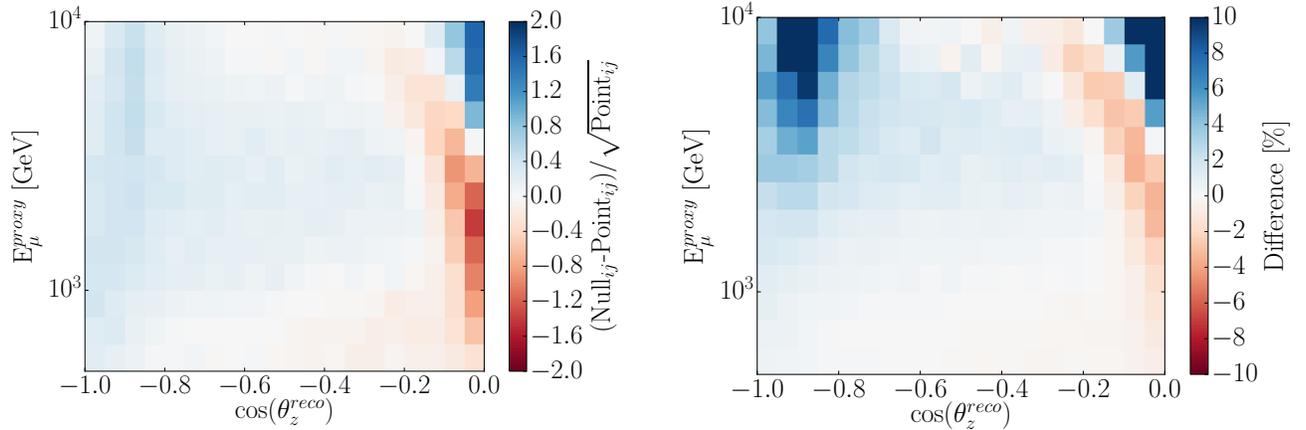

Figure D-1: Left: The shape of the best fit point of the fully blind analysis compared to the null hypothesis, represented in terms of statistical pulls. Right: Percent difference between the best fit point and the null hypothesis.

The low energy DeepCore analysis [155] excluded the values of $\sin^2(\theta_{24}) < 0.39$ at 90% CL and $\sin^2(\theta_{24}) < 0.55$ at 99% CL (for $\Delta m_{41}^2 > 1\,\text{eV}^2$). The Analysis I best fit point was correspondingly excluded by the DeepCore analysis at greater than $2\sigma$, which was flagged as an area of concern in our post-unblinding checks.

It was found that the top reconstructed energy bins contained two $>3\sigma$ statistical pulls (see Fig. D-3), both when compared to the best fit hypothesis and the null hypothesis. The observation of greater than two $3\sigma$ pulls was determined not to be significant, however, both of them arising in the same energy range was unexpected and prompted concern about the modeling of the high energy systematic uncertainties. The pre-unblinding 1D energy projection tests also revealed a slight excess in the high energy event rate (see Fig. D-5).

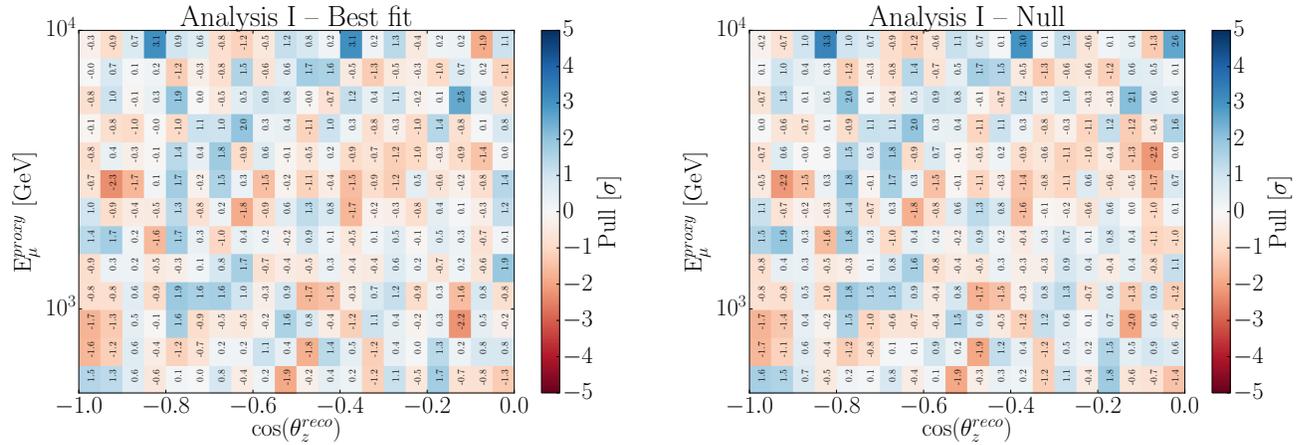

Figure D-2: The statistical pull distribution comparing the data to the best fit point (left) and to the null hypothesis (right).

The difference in the bin-wise $\chi^2$ between the null hypothesis and data, and the best fit point and data is shown in Fig. D-3. This plot indicates the regions of the reconstructed parameter space which prefer the null hypothesis (red) to the best fit point hypothesis (blue). It was found that a single high energy bin (top right bin of Fig. D-3) accounted for $\sim57\%$ of the total difference in the $\Delta\chi^2$ between the two hypotheses. The best fit point appeared to be largely driven by this single bin rather than the shape over the full parameter space. A handscan of 2,000 events reconstructed near the horizon found no evidence of misreconstructions due to cosmic ray muon backgrounds leaking into the sample.

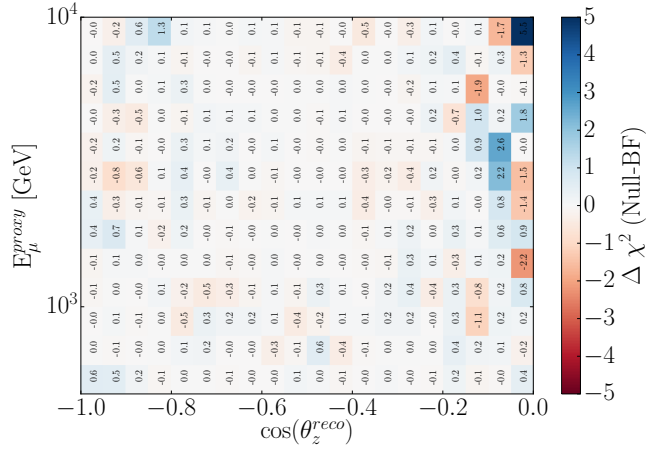

Figure D-3: The $\Delta\chi^2$ comparing the null hypothesis and best fit hypothesis to data.

At this stage, we were confident that the result obtained was not due to a sterile neutrino signal and likely due to an unmodeled high energy systematic or potentially a systematic effect near the horizon. The data was re-blinded and we began pursuing various investigations. The remainder of this section will describe the result of these investigations with an emphasis on the post-unblinding updates to the analysis.

The first update to the analysis was with respect to the Earth density profile description. Since the shape of the best fit point was found to primarily be driven by events near the horizon, we checked that the simulation was properly modeling the South Pole ice sheet. The Earth density profile was described in terms of 200 discrete steps from the center of the Earth to the surface. This meant that every step represented ∼30 km of matter. We updated the description of the density profile to include 3 km of ice rather than 30 km, and created a sharp transition in the bedrock. This update modified the event rate as shown in Fig. D-5 (left).

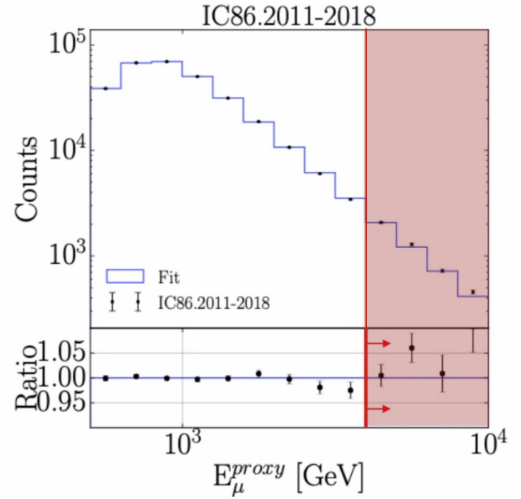

Figure D-4: The 1D energy projection of the pre-unblinding test.

A bug was then identified, in which the software used for generating the fluxes was linked to an outdated and incompatible version of the nuSQuIDS code. After updating the software and reprocessing the fluxes to the final level, a significant change in the observed high-energy $\bar{\nu}_\mu$ flux was observed. This update only affected the $\bar{\nu}_\mu$ rate. The full impact on the reconstructed energy and $\cos(\theta_z)$ is shown in Fig. D-5 (right).

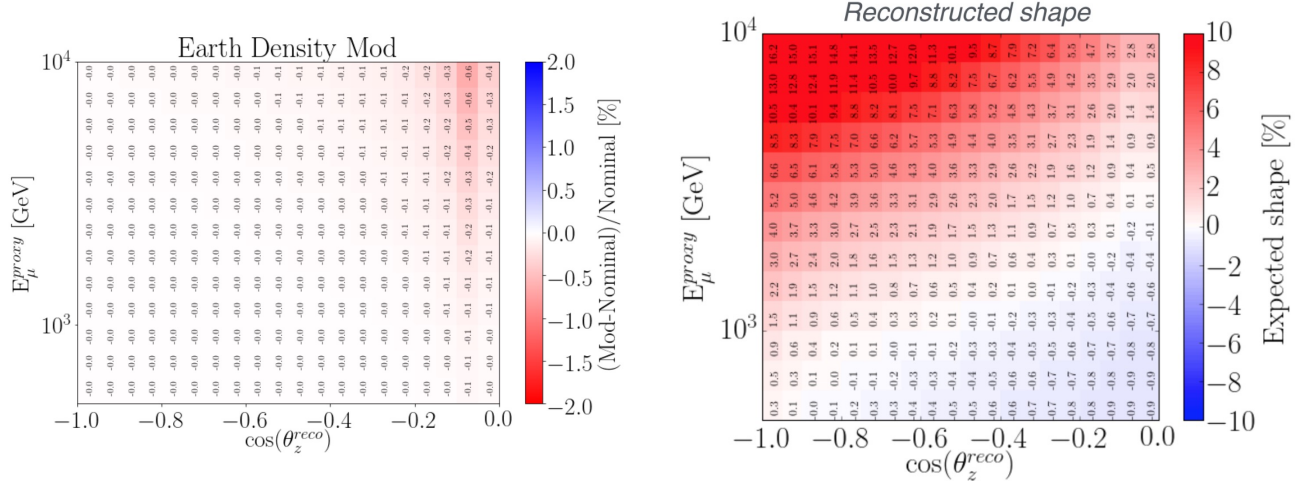

Figure D-5: Left: The shape in reconstructed energy and $\cos(\theta_z)$ upon updating the description of the ice-bedrock transition in the PREM Earth density profile. Right: The shape generated by updating to the compatible version of nuSQuIDS.

We originally found that the astrophysical flux in the energy range of this analysis to be negligible. Based on the the diffuse astrophysical neutrino measurement, IC79 [186], we concluded that we should expect a $\sim$3.7% contamination in our highest energy bins. However, as our analysis developed, the IC79 measurement was superseded but several other measurements [213, 216], all of which reported a significantly larger astrophysical flux. Upon re-examination, we concluded that this flux was significant and should have been included in the analysis. A description of this update can be found in Sec. 7.3. Along these lines, we also included the BERSS prompt atmospheric flux to the flux prediction (Sec. 7.2).

Prior to unblinding, we determined that the neutrino-nucleon cross section uncertainty was negligible. Since a change in neutrino-nucleon cross-section will preferentially impact the highest energy neutrinos, we re-investigated this assumption. Although the conclusion of this remained the same, we introduced the systematic uncertainty described in Sec. 7.4.2 into the analysis. We were also prompted to investigate the energy loss in matter of the kaons propagating through the atmosphere. This was found to have a larger impact on the analysis than the neutrino-nucleon cross section uncertainty, although still having a minimal impact. The resulting implementation of this systematic uncertainty is described in Sec. 7.4.1.

Beyond the updates described above, we investigated the impact of various models on the analysis (central cosmic ray, hadronic interaction, and temperature profile models). These are all described in Appendix E. We also re-assessed our assumptions on various other systematic uncertainties not included as nuisance parameters in the analysis. These are reported in Appendix E.1. We also re-assessed the level of cosmic ray muon contamination by processing an independent set of simulation using MuonGun, a description of which can be cound in Appendix C. The predicted background rate was found to be in agreement with that of CORSIKA ($\sim$0.006 - 0.008% cosmic ray muon contamination).

In summary, the post-unblinding updates were:

1. An update to the PREM Earth density model to include 3 km of ice.

2. The introduction of an astrophysical neutrino flux systematic uncertainty.

3. The introduction of an atmospheric prompt neutrino flux.

4. Update to nuSQuIDS and compatible software.

5. The inclusion of a cross section uncertainty for both the neutrinos and anti-neutrinos.

6. The inclusion of an uncertainty of the kaon energy loss in matter.

Of the items listed above, the bug fix and introduction of the astrophysical neutrino flux had a significant impact on the result. The inclusion of the prompt neutrino flux, cross section uncertainty, update to the PREM model, and uncertainty on the kaon energy loss in matter, were all found to be marginal improvements. The results presented in the main body of this thesis includes all these updates.

THIS PAGE INTENTIONALLY LEFT BLANK

# Appendix E

# MODEL SPECIFIC TESTS

This section describes tests in which we inject specific models (cosmic ray, hadronic interaction, air temperature profiles, ect.), then scan the physics parameter space to assess the impact. We can consider the bottom left of each physics parameter plot in this section to be null-like. That is, the TS value near the bottom left of each plot is considered to be the null TS. The TS of the null indicates the level at which the null is rejected. The color scale will be set such that if it saturates in the null-like region, then the null is said to be rejected at $1\sigma$.

### E.0.1 Cosmic ray flux and hadronic interaction models

As discussed in Sec. 7, we do not include discrete hadronic interaction and cosmic ray models; rather, we have included a set of flux uncertainties that models the actual uncertainty in the atmospheric neutrino production. These are the Barr parameters, cosmic ray spectral index, normalization, and atmospheric density (Sec. 7.2). This description of atmospheric neutrino flux covers the range of discrete models, as shown in Fig. E-1. Here, the energy, $\cos(\theta_z)$, and neutrino to antineutrino ratio are shown for various cosmic ray and hadronic interaction models, along with the $1\sigma$ span of the Barr parameters, conventional normalization, and cosmic ray spectral index.

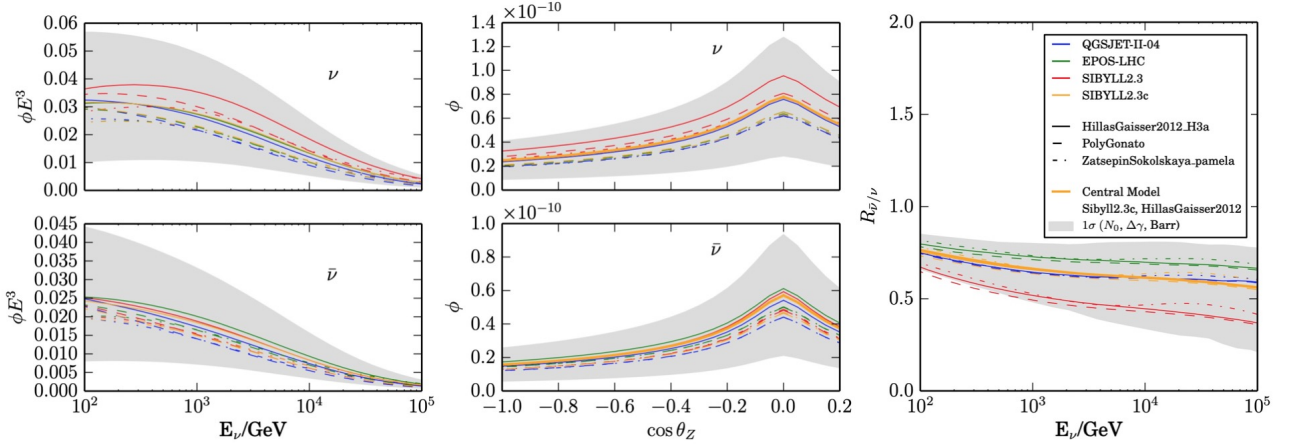

Figure E-1: The true neutrino energy distribution (left) averaged over $\cos(\theta_z)$ at 900 GeV (middle), and neutrino to antineutrino ratio (right), for various cosmic ray and hadronic interaction models, along with the $1\sigma$ span of the Barr parameters, conventional normalization, and cosmic ray spectral index. Plots from M. Moulai.

A natural question to ask is whether the discrete models are covered by the set of nuisance parameters implemented in this analysis. Specifically, we are interested in the question whether or not nature realized a particular model, would that produce a sterile signal in the analysis. We can test this by creating a null signal with a different model, and fitting back with the standard software.

We will examine the impact of using four cosmic ray models: HillasGaisser2012 (H3a) [201], Zatsepin-Sokolskaya/PAMELA [102, 103], Polygonato [104, 105], and Global Spline Fit (GSF) [106]; with the combination of two viable hadronic interaction models: QGSJET-II-04 [210] and SIBYLL2.3C [211]. The percent differences between the hadronic interaction and cosmic ray models to the central model (HillasGaisser2012 (H3a) + SIBYLL 2.3 RC1) is shown in Fig. E-2.

We now inject every combination listed above as a separate null hypothesis and fit back using our standard software. The TS distribution for each combination is shown in Fig. E-3 and E-4. We find that the largest impact occurs with the combination of Zatsepin-Sokolskaya and QGSJET-II-04. It is shown to produce a signal strength equivalent to approximately $0.75\sigma$ in Analysis I. All other combinations are found to be sub-$0.25\sigma$.

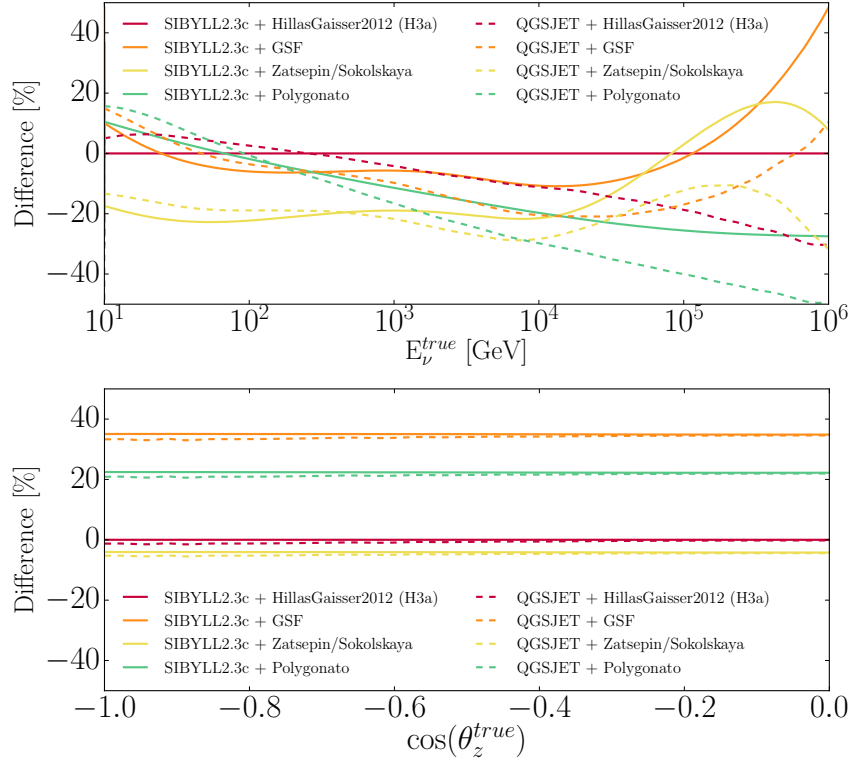

Figure E-2: A comparison between various hadronic interaction and cosmic ray models to the central model used in this analysis (HillasGaiser2012 H3A with SIBYLL2.3c). Top: In terms of the true neutrino energy. Bottom: In terms of the true neutrino $\cos(\theta_z)$.

### E.0.2  Astrophysical neutrino flux models

The prior uncertainty on the astrophysical neutrino flux normalization and spectral index were chosen such that it is able to span the previous IceCube astrophysical neutrino measurements (Sec. 7.3). Nevertheless, it is important to investigate whether a reasonable change in the true astrophysical neutrino spectrum would produce a signal in the physics parameter space. For this test, we create an Asimov dataset with the null hypothesis generated with a different astrophysical neutrino flux model. In this case, we use the IceCube measurements from HESE 7.5 year, Multi-year Cascade, and the Diffuse 6 year central values [213, 215, 216]. We then fit back with the standard software. The results for both analyses are shown in Fig. E-5 and E-6. As expected from the choice in the description of the prior, in all cases, no signal greater than $0.25\sigma$ is generated.

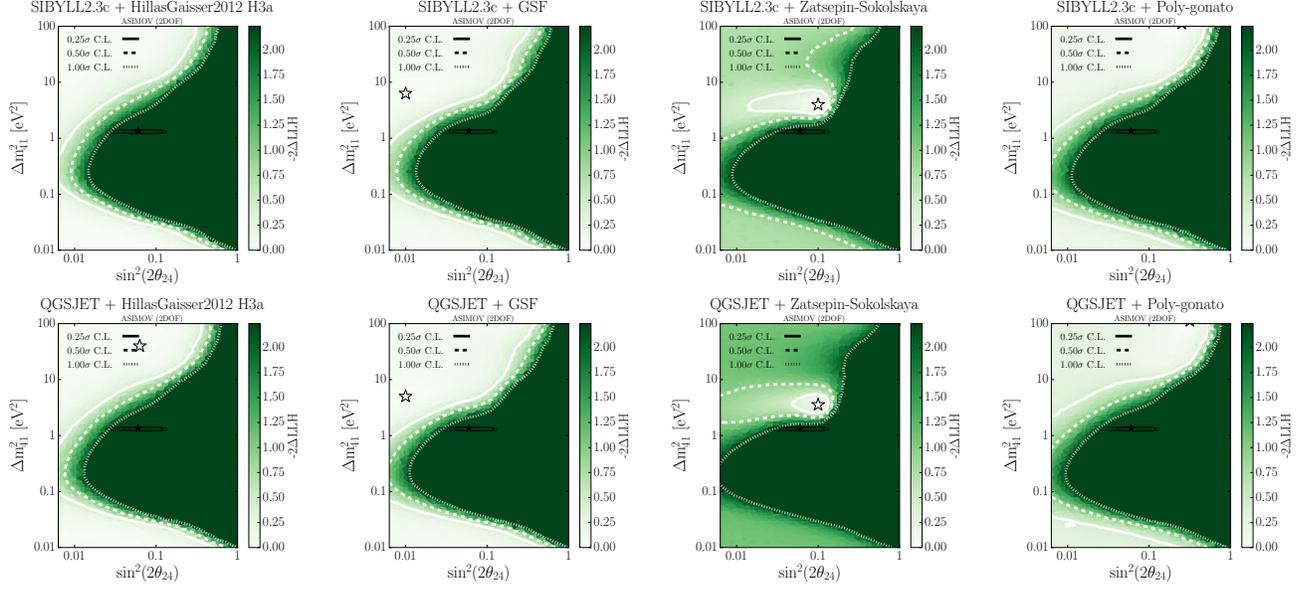

Figure E-3: The distribution of the TS for Analysis I throughout the physics parameter space for various injected signals generated with a different cosmic ray and hadronic interaction model listed in the title. The top left plot corresponds to our standard central model and is therefore also equivalent to the Asimov sensitivity.

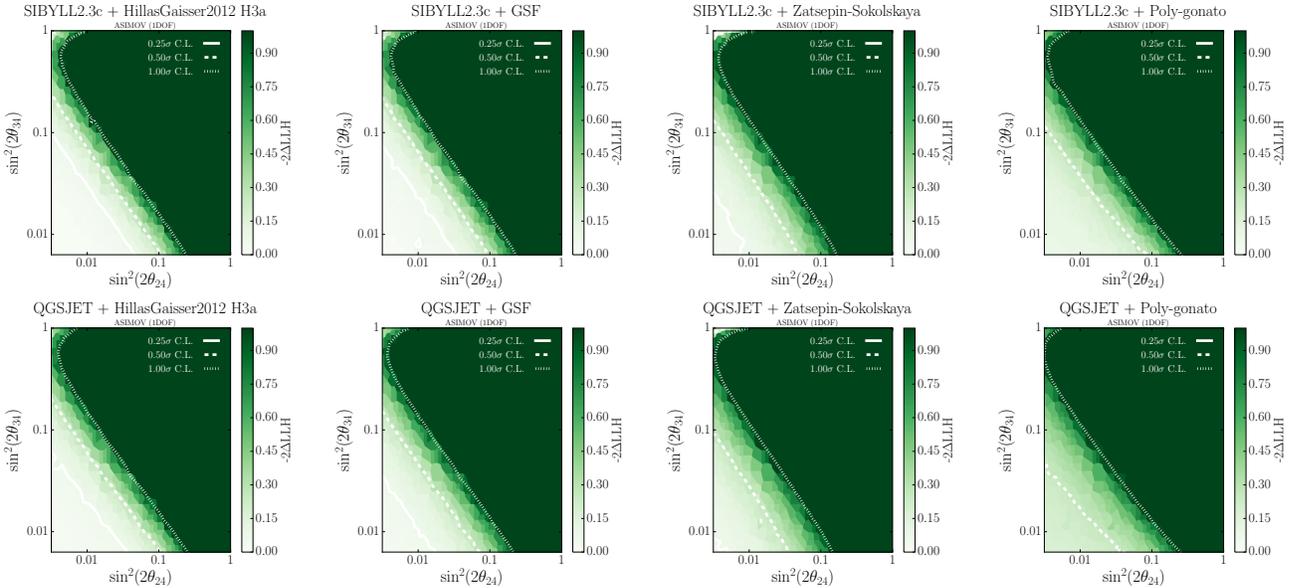

Figure E-4: The distribution of the TS for Analysis II throughout the physics parameter space for various injected signals generated with a different cosmic ray and hadronic interaction models listed in the title. The top left plot corresponds to our standard central model and is therefore also equivalent to the Asimov sensitivity.

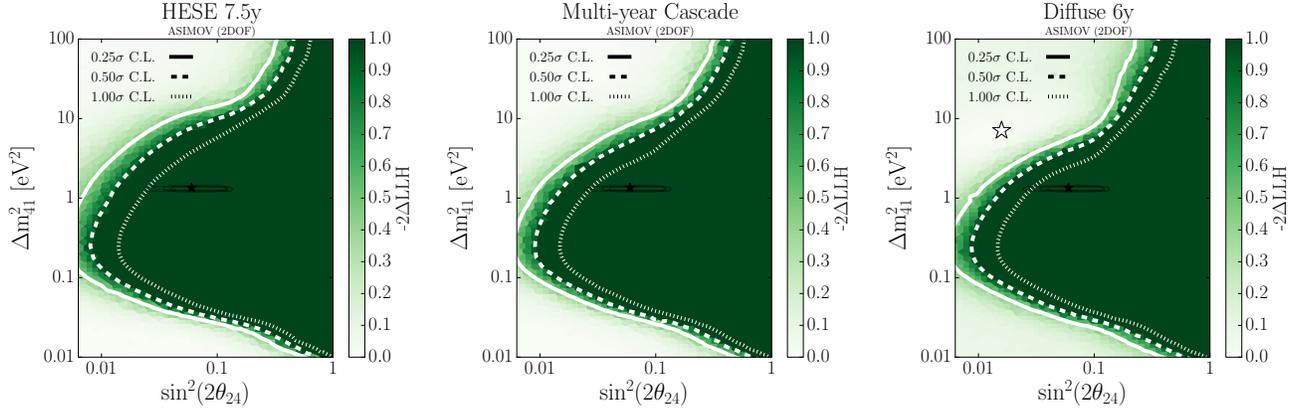

Figure E-5: Inject an astrophysical flux model and scan over the full parameter space.

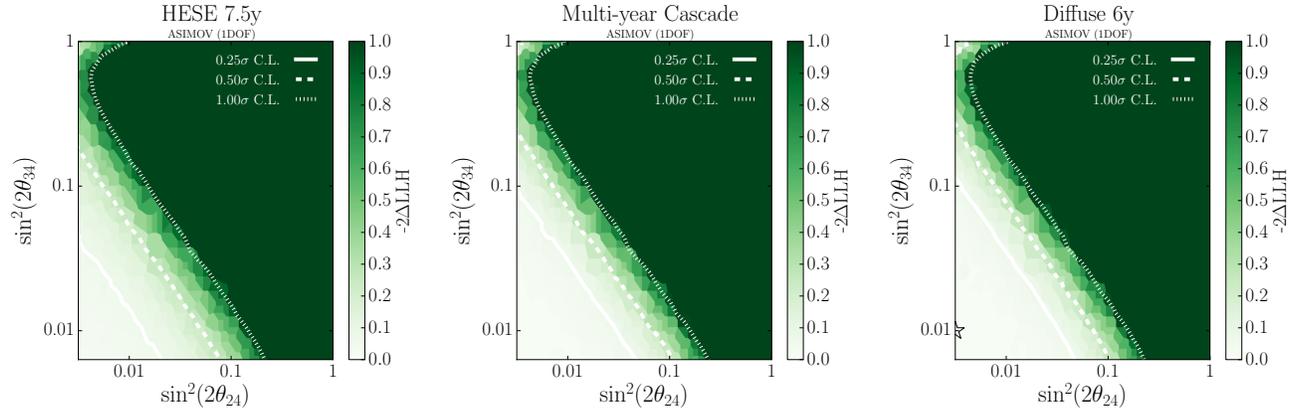

Figure E-6: Inject an astrophysical flux model listed in the title and scan over the full parameter space.

### E.0.3 Atmospheric density profile

We have chosen to model the atmospheric density off of the temperature profile measurements by the AIRS satellite (see Sec. 7.2.4 for a description). The fluxes are calculated separately for each month the summed together such that they match the monthly livetime averages of the IceCube data. Fig. E-7 (left) shows the shape of the monthly averaged AIRS predicted neutrino flux in true quantities compared to the 1976 US Standard temperature profile. The majority of the observed rate+shape shown here is due to the calculated flux difference between the two models, rather than the change in the flux due to the monthly livetime differences. In fact, the plot on the right shows the shape associated with no accounting for the monthly livetime differences (note the scale).

Although the US Standard description is disfavored for several reasons, we can still check to see whether or not a null hypothesis generated with the US Standard atmosphere would manifest itself as a sterile signal in the analysis. For this test, we generate the null hypothesis with the US Standard temperature profile and fit back using the standard software. The result is shown in Fig. E-8. In the case of Analysis I, a signal with a strength of approximately $0.75\sigma$ develops, whereas Analysis II finds a signal just over $1\sigma$.

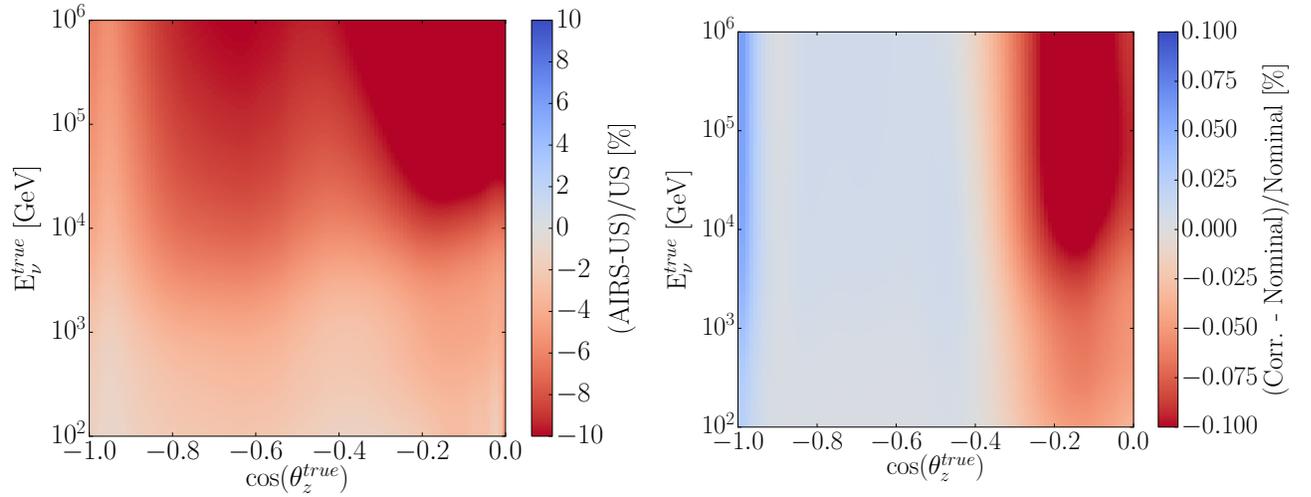

Figure E-7: Left: The shape difference between in true neutrino energy and zenith comparing the produced flux assuming the monthly averaged AIRS temperature profile to the 1976 US Standard. Right: The shape associated with only accounting for the livetime differences in the analysis versus assuming 100% monthly livetime.)

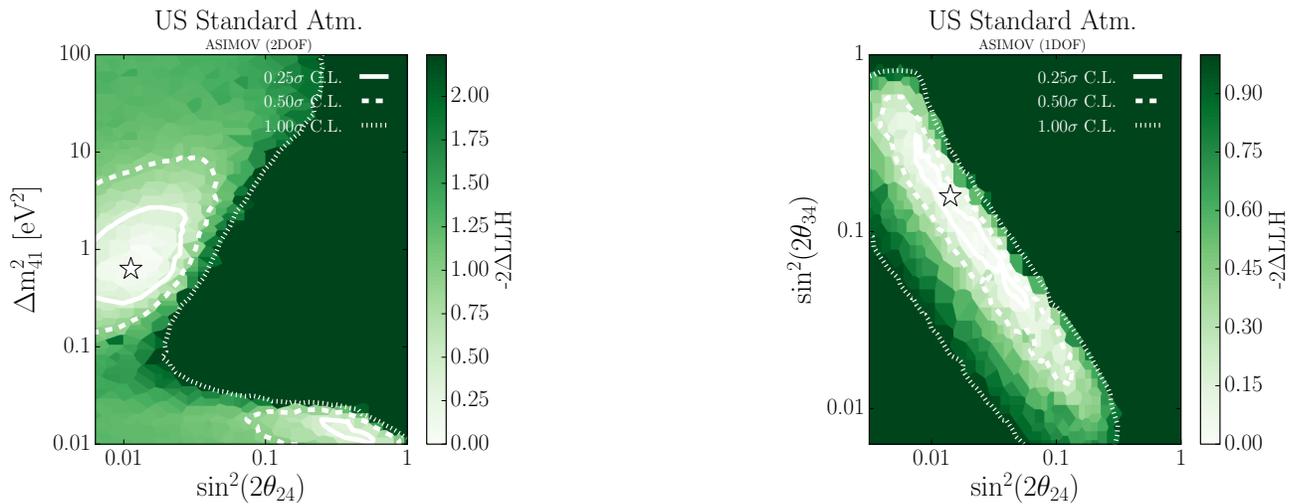

Figure E-8: Inject the US Standard temperature profile and fit back using the standard software.

### E.0.4 Kaon energy loss

Another similar check is to investigate the impact of changing the total cross section for the interaction of the secondary particles created in cosmic ray interaction. Since the charged mesons produced in the collision may either decay (producing our conventional neutrino flux) or interact again with oxygen and nitrogen in the atmosphere, a change in the interaction cross section will lead to a change in the neutrino flux. Fig. E-9 shows the impact of injecting a signal with $\pm 22.5\%$ the kaon total cross section ($3\sigma$ deviation from the central value), then fitting back using our standard software (with the kaon loss systematic, $\sigma_{\text{KA}}$, removed). We find that, in both cases, no significant signal is generated. This is primarily due to the Barr parameters absorbing the signal.

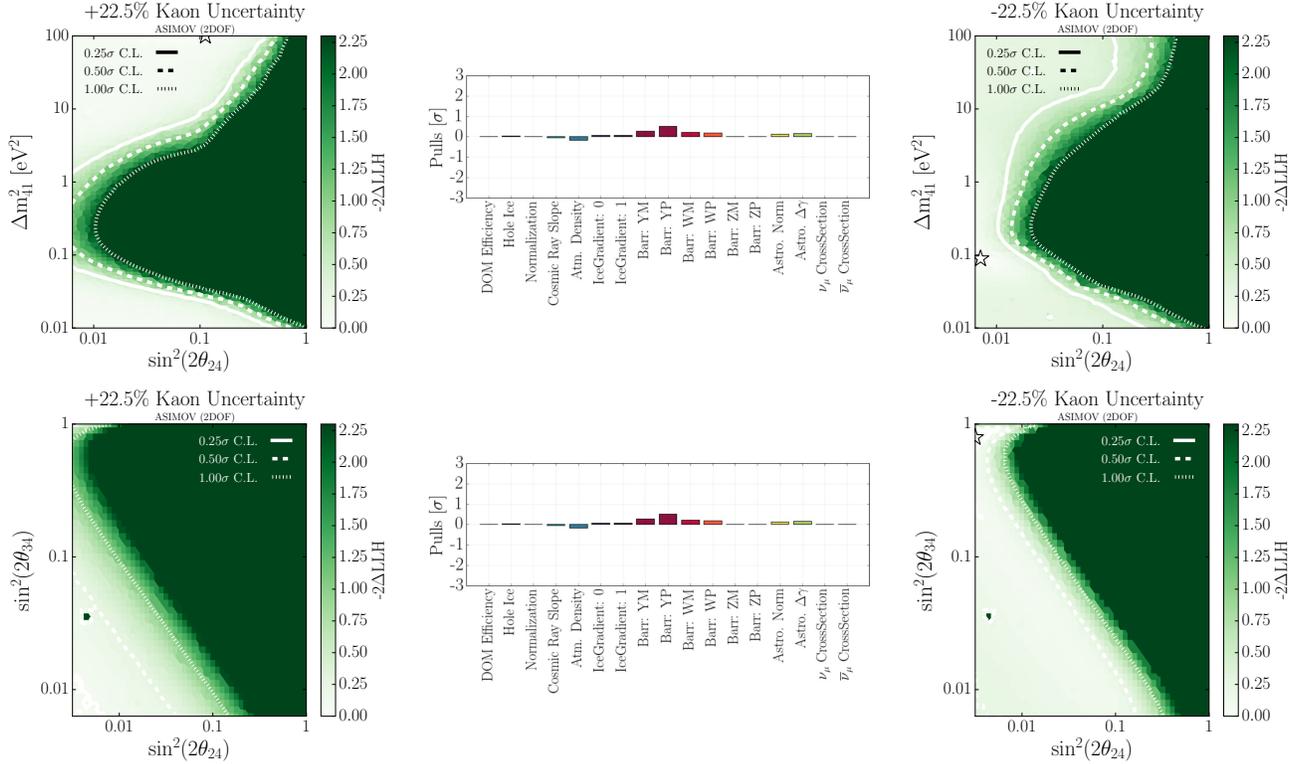

Figure E-9: Injecting a null signal with the kaon total cross section increased and decreased by 22.5% then fitting back without using the kaon energy loss systematic. Top: Analysis I. Bottom: Analysis II. The middle plot shows the systematic pulls at the best fit point.

# E.1 Additional systematic uncertainties investigated

The following subsections describe several investigations into other systematic uncertainties which were found to have a negligible impact on the analysis and were therefore not included as a nuisance parameter.

## E.1.1 Earth density model

The interior of Earth's density profile has been studied extensively using earthquake-generated waves and is a mature science [250]. The upper mantle and crust are also well-understood and are commonly probed using artificial explosions. In both cases, the velocity distribution of compression (P) and shear (S) waves are used to determine the density and transition point between layers. The Earth's internal structure closely resembles oblate ellipsoids, symmetric about the rotation axis. It is therefore reasonable to describe the Earth's density profile radially from the core to the surface. The transition between density regions is also well known. The uncertainty reported in the inner to outer core transition region is now known to within 10 km, with one of the more recent measurements quoting an uncertinaty of 0.6 km [251]. The shape of each of the Earth's layers is believed to be very close to an oblate ellipsoid due to hydrostatic flattening from the Earth's rotation and gravitational potential. The flattening along the longitudinal access is expected to be approximately $1/400^{th}$ that along the latitudinal axis [252]. This corresponds a deformation of 3 km in the inner core.

We use the Preliminary Reference Earth Model [176] (PREM) to describe the 1-D density profile of the Earth. The PREM model was designed to be compatible with variety of different data sets, including surface wave dispersion observations, travel time data for a number of body-wave phases, and basic astronomical data (Earth's radius, mass, and moment of inertia). All current Earth models have values that are reasonably close to PREM; the largest differences are in the description of the upper mantle [253]. In this analysis, the PREM model is described in terms of 204 discrete density steps from the center of the Earth to the surface. In-between steps, the density is interpolated using an Akima spline. Near the surface, we include a layer of ice, with a density 1.02 g/cm$^3$, 3 km thick and a step transition into bedrock just below 3 km. Several other viable earth density models [177–179] are shown in the left side of Fig. E-10.

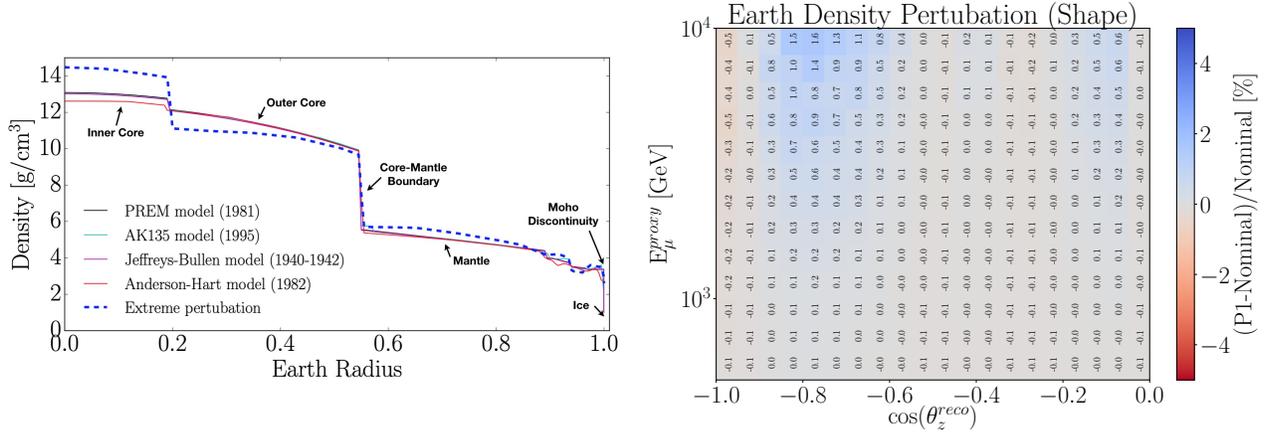

Figure E-10: Left: various Earth density profile models. Right: the shape of the extreme perturbation.

Ref. [204] investigated physically allowed polynomial perturbations of the PREM model, while conserving the mass and moment of inertia of the Earth. An extreme example perturbation is shown in the left side of Fig. E-10 (dashed blue). The corresponding shape introduced by exchanging this extreme model with the PREM model is shown in on the right (similar changes are also observed at various sterile hypothesese). Based on the minimial shape introduced, we conclude that this uncertainty is negligible and we do not include it as a nuisance parameter.

We also investigated the impact of the location of the density transition zones. Perturbations on the location of the transitions of O(100) times larger than the actual uncertainty of these zones is shown in Fig. E-11.

### E.1.2 Pion related Barr parameters

The largest uncertainty in the energy range of interest of this analysis for the pion production is described by the HP and HM Barr parameters. In Fig. E-12, the shape at analysis level when perturbing these uncertainties by $1\sigma$ are shown. The neutrino flux from the pion progenitors are sub-dominant to the kaon mesons. The impact on the analysis space is observed to be minimal.

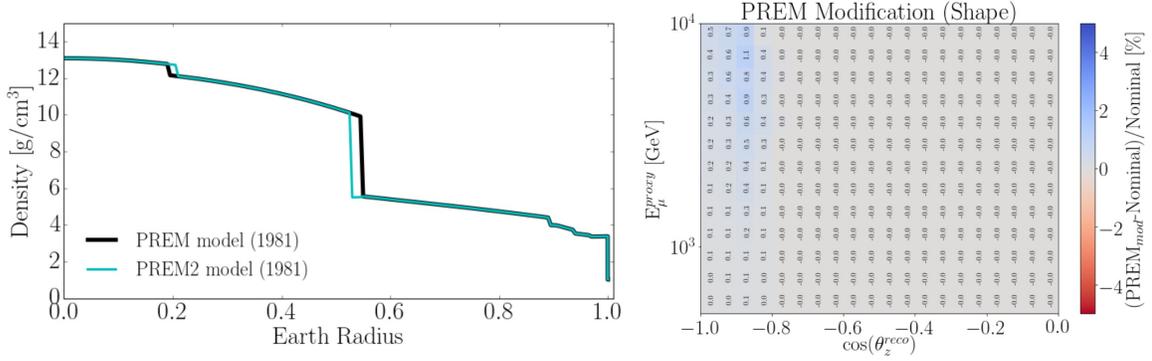

Figure E-11: The shape generated when perturbing the transition zones by several hundred kilometers.

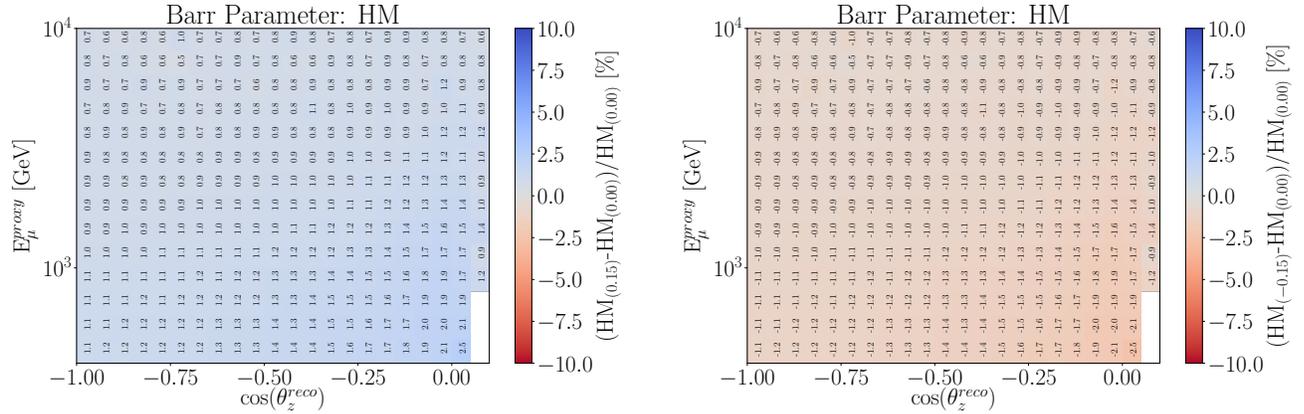

Figure E-12: The shape of the $H^{\pm}$ Barr parameters.

### E.1.3   Astrophysical neutrino to antineutrino ratio

We assume a neutrino to antineutrino ratio of 1:1. This is a safe assumption since IceCube does not have sensitivity to distinguishing between $\nu_\mu$CC and $\overline{\nu}_\mu$CC interactions. We verified this by investigating the impact on the final reconstructed energy and $\cos(\theta_z)$ distribution by increasing and decreasing the relative ratio by 50%. The result of a 50% (150%) decrease (increase) in the neutrino contribution compared to the antineutrinos is shown in Fig. E-13.

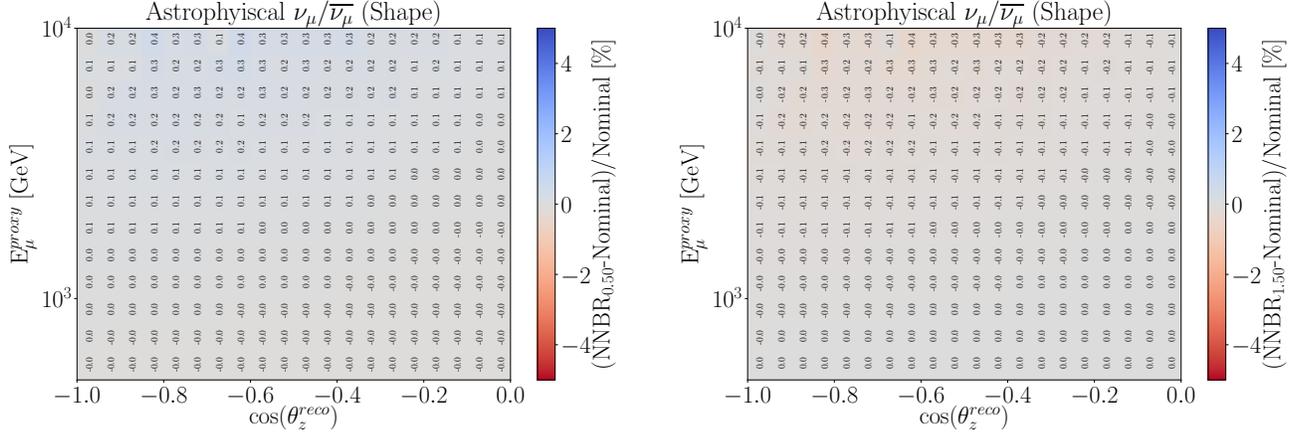

Figure E-13: The shape generated when using an astrophysical neutrino to antineutrino ratio of 0.5 (left) and 1.5 (right).

### E.1.4 Prompt atmospheric flux

This analysis does not fit for the prompt atmospheric neutrino component and instead inserts the BERSS model into the flux hypothesis. We can however show that even if there is an excess prompt component, it does not create a signal in our physics space. Injecting a null signal with 10 times the BERSS prompt prediction then fitting back using our standard software is shown not to produce a signal (Fig. E-14). The prompt contribution is nearly fully absorbed by the astrophysical nuisance parameters in both analyses, as shown by the pull distributions, as shown by the pulls of the astrophysical neutrino flux.

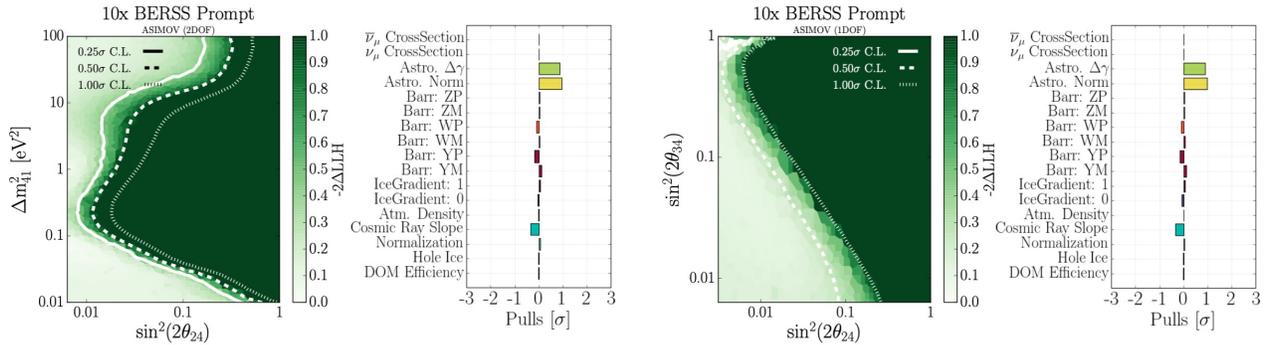

Figure E-14: The effect of injecting a 10 times BERSS prompt neutrino flux and fitting back with the standard software. The effect on Analysis I is shown on the left and Analysis II on the right.

THIS PAGE INTENTIONALLY LEFT BLANK

# Appendix F

# SPE Templates

## F.1 Introduction

The IceCube Neutrino Observatory [126,127] is a cubic-kilometer-sized array of 5,160 photomultiplier tubes (PMTs) buried in the Antarctic ice sheet, designed to observe high-energy neutrinos interacting with the ice [128]. In 2011, the IceCube Collaboration completed the installation of 86 vertical *strings* of PMT modules, eight of which were arranged in a denser configuration known as the DeepCore sub-array [129]. Each string in IceCube contains 60 digital optical modules (DOMs), which contain a single PMT each, as well as all required electronics [130]. The primary 78 strings (excluding DeepCore) are spaced 125 m apart in a hexagonal grid, with the DOMs extending from 1450 m to 2450 m below the surface of the ice sheet. The additional DeepCore strings (79-86) are positioned between the centermost strings in the detector, reducing the horizontal DOM-to-DOM distance in this region to between 42 m and 72 m. The lower 50 DOMs on these strings are located in the deepest 350 m of the detector near the clearest ice [145], while the upper ten provide a cosmic ray veto extending down from 1900 m to 2000 m below the surface. Beyond the in-ice detectors, there exists a surface array, IceTop, consisting of 81 stations located just above the in-ice IceCube strings. The PMTs located in IceTop operate at a lower gain and are not subject to this analysis.

Each DOM consists of a 0.5"-thick spherical glass pressure vessel that houses a single down-facing 10" PMT from Hamamatsu Photonics. The PMT is coupled to the glass housing with optical gel and is surrounded by a wire mesh to reduce the effect of the Earth's ambient magnetic field. The glass housing is transparent to wavelengths 350 nm and above [133].

Of the 5,160 DOMs, 4,762 house a R7081-02 Hamamatsu Photonics PMT, sensitive to wavelengths ranging from 300 nm to 650 nm, with peak quantum efficiency of 25% near 390 nm. These are classified as Standard Quantum Efficiency (Standard QE) DOMs. The remaining 398 DOMs are equipped with the Hamamatsu R7081-02MOD PMTs, which, having a peak quantum efficiency of 34% near 390 nm (36% higher efficiency than the Standard QE DOMs), are classified as High Quantum Efficiency (HQE) DOMs [129]. These DOMs are primarily located in DeepCore and on strings 36 and 43, as shown in the left side of Fig. F-1.

The R7081-02 and R7081-02MOD PMTs have 10 dynode stages and are operated with a nominal gain of $10^7$ and high voltage ranging from approximately $1215 \pm 83$ V and $1309 \pm 72$ V, respectively. A typical amplified single photoelectron generates a $5.2 \pm 0.3$ mV peak voltage after digitization with a full width half maximum of $13 \pm 1$ ns. The PMTs operate with the anodes at high voltage, so the signal is AC coupled to the amplifiers (front-end amplifiers). There are two versions of AC coupling in the detectors, referred to as the *new* and *old toroids*, both of which use custom-designed wideband bifilar wound 1:1 toroidal transformers[1]. The locations of DOMs with the different versions of AC-coupling are shown on the right side of Fig. F-1. The DOMs with the old toroids were designed with an impedance of $43 \, \Omega$, while the new toroids are $50 \, \Omega$ [134]. All HQE DOMs are instrumented with the new toroids.

IceCube relies on two observables per DOM to reconstruct events: the total number of detected photons and their timing distribution. Both the timing and the number of photons are extracted from the digitized waveforms. This is accomplished by deconvolving the digitized waveforms [137] into a series of scaled single photoelectron pulses (so-called pulse series), and the integral of the individual pulses (divided by the load resistance) defines the observed charge. It will often be expressed in units of PE, or photoelectrons, which further divides the measured charge by the charge of a single electron times the nominal gain.

---

[1]The toroidal transformer effectively acts as a high-pass filter with good signal fidelity at high frequencies and offers a higher level of reliability than capacitive coupling. Conventional AC-coupling high-voltage ceramic capacitors can also produce undesirable noise from leakage currents and are impractical given the signal droop and undershoot requirements [133].

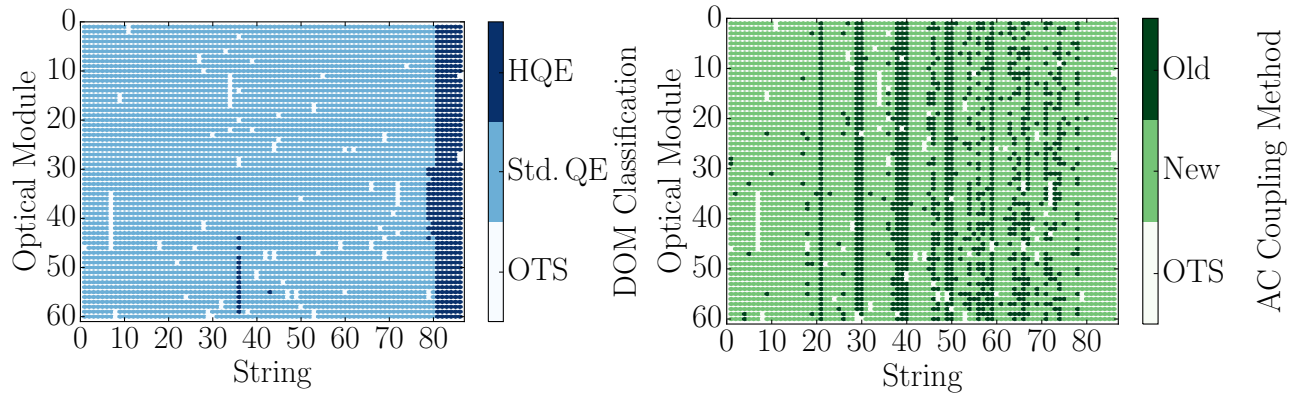

Figure F-1: Left: A mapping of the HQE (dark blue) and Standard QE DOMs (light blue). Right: The version of AC coupling, old toroids (dark green) and new toroids (light green). DOMs that have been removed from service (OTS) are shown in white.

When one or more photons produce a voltage at the anode sufficient to trigger the onboard discriminator the signal acquisition process is triggered. The discriminator threshold is set to approximately 1.2 mV, or equivalently to ∼0.23 PE, via a digital-to-analog converter (DAC) to approximately. The signal is presented to four parallel channels for digitization. Three channels pass through a 75 ns delay loop in order to capture the leading edge of the triggering pulse, and are then subject to different levels of amplification prior to being digitized at 300 million samples per second (MSPS) for 128 samples using a 10-bit Analog Transient Waveform Digitizer (ATWD). The high-gain channel has a nominal amplification of 16 and is most suitable for single photon detection. Two ATWD chips are present on the DOM Mainboard (MB) and alternate digitization between waveforms to remove dead time associated with the readout. The signal to the fourth parallel channel is first shaped and amplified, then fed into a 10-bit fast analog-to-digital converter (fADC) operating at a sampling rate of 40 MSPS. Further detail regarding the description of the DOM electronics can be found in Refs. [130, 254].

This article discusses a method for determining the in-situ individual PMT single-photoelectron charge distributions, which can be used to improve calibration and the overall detector de-scription in Monte Carlo (MC) simulation. The SPE charge distribution refers to the charge probability density function of an individual PMT generated by the amplification of a pure sample of single photoelectrons. The measured shape of the SPE charge distributions is shown to be useful for examining hardware differences and long term stability of the detector. This was recently made possible with the development of two pieces of software:

1. A specially-designed unbiased pulse selection developed to reduce the multiple photoelectron (MPE) contamination while accounting for other physical phenomena (e.g. late pulses, afterpulses, pre-pulses, and baseline shifts) and software-related effects (e.g. pulse splitting). This is further described in Sec. F.2.1.

2. A fitting procedure developed to separate the remaining MPE contamination from the SPE charge distribution by deconvolving the measured charged distribution. This is further described in Sec. F.2.3.

By using in-situ data to determine the SPE charge distributions, we accurately represent the individual PMT response as a function of time, environmental conditions, software version and hardware differences, and realistic photocathode illumination conditions. This is beneficial since it also allows us to inspect the stability and long-term behavior of the individual DOMs, verify previous calibration, and correlate features with specific DOM hardware.

### F.1.1 Single-photoelectron charge distributions

Ideally, a single photon produces a single photoelectron, which is then amplified by a known amount, and the measured charge corresponds to 1 PE. However, there are many physical processes that create structure in the measured charge distributions. For example:

- **Statistical fluctuation due to cascade multiplication** [255]. At every stage of dynode amplification, the number of of emitted electrons that make it to the next dynode is randomly distributed. This in turn causes a smearing in the measured charge after the gain stage of the PMT.

- **Photoelectron trajectory**. Some electrons may deviate from the favorable trajectory, reducing the number of secondaries produced at a dynode or the efficiency to collect them on the following dynode. This can occur at any stage, but it has the largest effect on the multiplication at the first dynode [256]. The trajectory of a photoelectron striking the first dynode will depend on many things, including where on the photocathode it was emitted, the uniformity of the electric field, the size and shape of the dynodes [255], and the ambient magnetic field [257, 258].

- **Late or delayed pulses**. A photoelectron can elastically or inelastically scatter off the first dynode. The scattered electron can then be re-accelerated to the dynode, creating a second pulse. The difference in time between the initial pulse and the re-accelerated pulse in the R7081-02 PMT was previously measured to be up to 70 ns [133, 259]. The two sub-pulses have lower charges but the sum of the two may approximate the original charge. Collecting either the initial pulse or the late pulse will result in the charge being reconstructed in the low-PE region [260].

- **Afterpulses**. When photoelectrons or the secondary electrons produced during the electron cascade gain is sufficient energy to ionize residual gas in the PMT, the positively charged ionized gas will be accelerated in the electric field towards the photocathode. Upon impact with the photocathode, electrons can be released from the photocathode, creating what is called an afterpulse. For the R7081-02 PMTs, the timescale for afterpulses was measured to occur from 0.3 to 11 $\mu$s after the initial pulse, with the first prominent afterpulse peak occurring at approximately 600 ns [133]. The spread in the afterpulse time depends on the position of photocathode, the charge-to-mass ratio of the ion produced, and the electric potential distribution [261], whereas the size of the afterpulse is related to the momentum and species of the ionized gas and composition of the photocathode [262].

- **Pre-pulses**. If an incident photon passes through the photocathode without interaction and strikes one of the dynodes, it can eject an electron that is only amplified by the subsequent stages, resulting in a lower measured charge (lower by a factor of approximately 10). For the IceCube PMTs, the prepulses have been found to arrive approximately 30 ns before the signal from other photoelectrons from the photocathode [133].

- **MPE contamination**. When multiple photoelectrons arrive at the first dynodes within several nanoseconds of each other, they can be reconstructed by the software as a single MPE pulse.

- **Dark noise**. Photoelectron emission, not initiated from an external event, can be attributed to thermionic emission from the low work function photocathode and the dynodes, Cherenkov radiations initiated from radioactive decay within the DOM, and field emission from the electrodes. It is shown in Fig. 28 of Ref. [263] that the dark noise preferentially populates the low-charge region.

- **Electronic noise**. This refers to the fluctuations in the analog-to-digital converters (ATWDs and FADC) and ringing that arises from the electronics.

Beyond the physical phenomena above that modify the measured charge distribution, there is also a lower limit on the smallest charge that can be extracted. For IceCube, the discriminator only triggers for peak voltages above the threshold and subsequent pulses in the readout window are subject to a threshold defined in the software. The software threshold was set conservatively to avoid extracting pulses that originated from electronic noise. This threshold can be modified to gain access to lower charge pulses and will be discussed in Sec. F.2.2.

The standard SPE charge distribution used for all DOMs in IceCube, known as the TA0003 distribution [133], models the above effects as the sum of an exponential plus a Gaussian. The TA0003 distribution represents the average SPE charge distribution extracted from a lab measurement of 118 Hamamatsu R7081-02 PMTs. The measurement was performed in a -32°C freezer using a pulsed UV LED centered along the axis of the PMT, directly in front of the photocathode.

Recently, IceCube has made several lab measurements of the SPE charge distribution of R7081-02 PMTs using single photons from high speed laser pulses. The in-time charge distribution from the laser pulses was found to that the in-time charge distribution includes a steeply falling low-charge component below the discriminator threshold. To account for this, a new functional form including a second exponential was introduced. This form of the charge distribution $f(q)_{\text{SPE}} = \text{Exp}_1 + \text{Exp}_2 + \text{Gaussian}$, is referred to as the *SPE charge template* in this article. Explicitly, it is:

$$f(q)_{\text{SPE}} = \text{E}_1 e^{-q/\text{w}_1} + \text{E}_2 e^{-1/\text{w}_2} + \text{N} e^{-\frac{(q-\mu)^2}{2\sigma^2}}, \tag{F.1}$$

where $q$ represents the measured charge; $\text{E}_1$, $\text{E}_2$, and N represent normalization factors of each component; $\text{w}_1$ and $\text{w}_2$ are the exponential decay widths; and $\mu$, $\sigma$ are the Gaussian mean and width, respectively. This is the assumed functional shape of the SPE charge distributions, and the components of Eq. F.1 are determined in this article for all in-ice DOMs. IceCube has chosen to defines 1 PE as the location of the Gaussian mean ($\mu$) and calibrates the gain of the individual PMTs prior to the start of each season to meet this definition. Any bias in the total observed charge can be absorbed into an efficiency term, such as the quantum efficiency. This is valid since the linearity between the total charge collected and the number of incident photons is satisfied up to $\sim 2\,\text{V}$ [134], or approximately 375 PE. That is, the average charge collected from N photons is N times the average charge of the SPE charge distribution, and the average charge of the SPE charge distribution is a set fraction of the Gaussian mean.

### F.1.2  IceCube datasets and software definitions

The amount of observed light depends on the local properties of the ice [145]. Short term climate variations from volcanoes and longer-term variations from atmospheric dust affect the optical properties of the ice, producing nearly horizontal layers. This layered structure affects how much light the DOMs observes, and, with it, the trigger rate. The largest contribution to the IceCube trigger rate comes from downward-going muons produced in cosmic ray-induced showers [264]. Cosmic ray muons stopping in the detector cause the individual trigger rate to decrease at lower depths.

If a DOM and its nearest or next-to-nearest neighbor observe a discriminator threshold crossing within a set time window, a *Hard Local Coincidence* (HLC) is initiated, and the corresponding waveforms are sampled 128 times and read out on the three ATWD channels. Thermionic emission induced dark noise can be present in the readout, however it is suppressed at lower temperatures and is unlikely to trigger an HLC event.

After waveform digitization, there is a correction applied to remove the measured DC baseline offset. Distortions to the waveform, such as from droop and undershoot [133], introduced by the toroidal transformer AC coupling are compensated for in software during waveform calibration by adding the expected temperature-dependent reaction voltage of the distortion to the calibrated waveform. If the undershoot voltage drops below 0 ADC counts, the ADC values are zeroed and then compensated for once the waveform is above the minimum ADC input. For each version of the AC coupling, scaled single photoelectron pulse shapes are then fit to the digitized waveforms using software referred to as "WaveDeform" (waveform unfolding process), which determines the individual pulse time and charges and populates a pulse series.

The pulse series used in this analysis come from two datasets provided by IceCube:

1. The **MinBias dataset.** This dataset records the full waveform readout of randomly-triggered HLC events, collecting on average 1:1000 events. The largest contribution to this dataset comes from downward-going muons produced in cosmic-ray-induced showers. The average event is approximately 26 PE distributed over an average of 16 triggered DOMs. The full waveform of these events allows us to extract the raw information about the individual pulses. This will be used to measure the individual PMT charge distributions.

2. The **BeaconLaunch dataset.** This dataset is populated with waveforms readout at random time intervals (forced-triggered). It is typically used to monitor the individual DOM baseline and includes the full ATWD-window waveform readout. Since this dataset is forced-triggered, the majority of these waveforms represent electronic noise with minimal contamination from the accidental coincidence pulse that makes it into the readout window. This dataset will be used to examine the noise contribution to the charge distri-butions.

When using this dataset, the weight of every pulse is multiplied by a factor of 28.4 to account for the livetime difference between the MinBias dataset and the BeaconLaunch dataset. Weight, in this context, refers to the number of photons in the MinBias dataset proportional to one photon in the BeaconLaunch dataset for which both datasets have the same equivalent livetime.

This analysis uses the full MinBias and BeaconLaunch datasets from IceCube seasons 2011 to 2016 [265], subsequently referred to as IC86.2011 to IC86.2016. Seasons in IceCube typically start in May of the labeled year and end approximately one year later. Calibration is performed before the start of each season.

## F.2 Extracting the SPE charge templates

### F.2.1 Single photoelectron pulse selection

The pulse selection is the method used to extract candidate, unbiased, single photoelectron pulses from high-gain ATWD channel while minimizing the MPE contamination. It avoids collecting afterpulses, rejects late pulses from the trigger, accounts for the discriminator threshold, reduces the effect of signal droop and baseline undershoot, and gives sufficient statistics to perform a season-to-season measurement. An illustrative diagram of the pulse selection is shown in the left side of Fig. F-2, while a description of the procedure is detailed below.

We restrict the pulse selection to only extract information from waveforms in which the trigger pulse does not exceed $10\,\mathrm{mV}$ ($\sim 2\,\mathrm{PE}$) and no subsequent part of the waveform exceeds $20\,\mathrm{mV}$ ($\sim 4\,\mathrm{PE}$). This reduces the effect of the baseline undershoot due to the AC coupling or other artifacts from large pulses.

In order to trigger a DOM, the input to the front-end amplifiers must exceed the discriminator threshold. To avoid the selection bias of the discriminator trigger, we ignore the trigger pulse as well as the entire first 100 ns of the time window. Ignoring the first 100 ns has the added benefit of also removing late pulses that could be attributed to the triggering pulse. To ensure we are not accepting afterpulses into the selection, we also enforce the constraint that the pulse of interest (POI) is within the first 375 ns of the ATWD time window. This also allows us to examine the waveform up to 50 ns after the POI. In the vicinity of the POI, we ensure that WaveDeform did not reconstruct any pulses up to 50 ns prior to the POI, or 100 to 150 ns after the POI (the light gray region of Fig. F-2 (left)). This latter constraint is to reduce the probability of accidentally splitting a late pulse in the summation window.

If a pulse is reconstructed between 100 and 375 ns after the start of the waveform and the voltage criteria are met, it is accepted as a candidate photoelectron and several checks are performed on the waveform prior to and after the pulse. The first check is to ensure that the waveform is near the baseline just before the rising edge of the POI. This is accomplished by ensuring that the waveform does not exceed 1 mV, 50 to 20 ns prior to the POI, and eliminates cases where the POI is a late pulse. We also ensure the waveform returns to the baseline by checking that no ADC measurement exceeds 1 mV, 100 to 150 ns after the POI. These constraints are illustrated as the horizontal red dotted lines and black arrows in the left side of Fig. F-2.

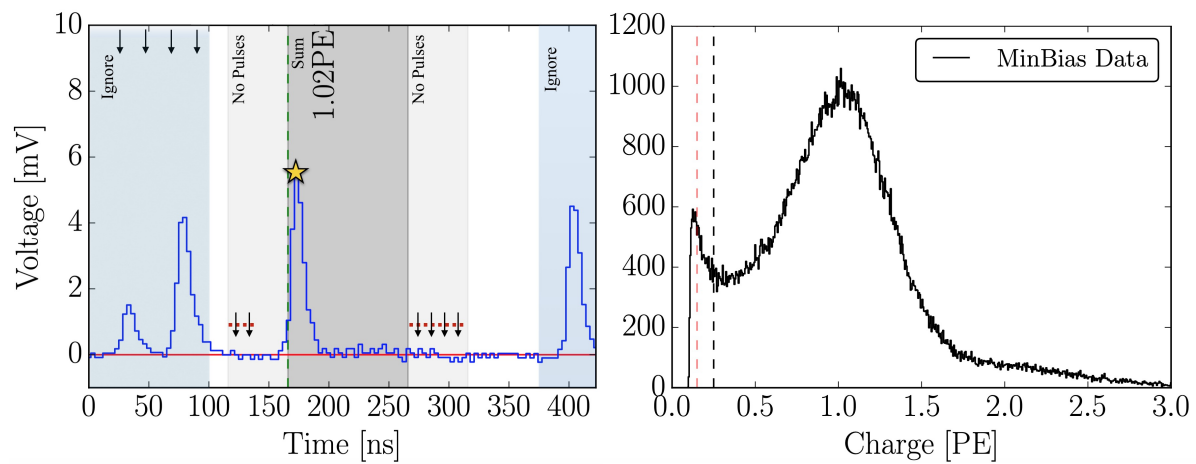

Figure F-2: Left: An illustrative diagram of the pulse selection criteria for selecting a high-purity and unbiased sample of single photoelectrons. The digitized ATWD waveform is shown in blue and the baseline is shown as a solid red line. The pulse of interest is identified with a yellow star. This example waveform was triggered by a small pulse at 25 ns (recall that the delay board allows us to examine the waveform just prior to the trigger pulse), followed by a potential late pulse at 70 ns. At 400 ns, we see a pulse in the region susceptible to afterpulses. Waveform voltage checks are illustrated with arrows, and various time windows described in the text are drawn with semi-opaque regions. The POI is reported to have a charge of 1.02 PE, given by WaveDeform, and would pass the pulse selection criteria. Right: The collected charges from string 1, optical module 1 (DOM 1,1), from the MinBias dataset collected from IC86.2011 to IC86.2016 that pass the pulse selection. For visual purposes, red and black vertical dashed lines are included at 0.15 PE and 0.25 PE.

If all the above criteria are met, we sum the reconstructed charges from the POI time, given by WaveDeform, to +100 ns (the dark gray area in Fig. F-2 (left)). This ensures that any nearby pulses are either fully separated or fully added. WaveDeform may occasionally split an SPE pulse into multiple smaller pulses, therefore it is always critical to perform a summation of the charge within a window. The 100 ns summation also means that the pulse selection will occasionally accept MPE events.

### F.2.2 Characterizing the low-charge region

Fig. F-2 (right) shows the charge distributions of the selected pulses that pass the single photoelectron pulse selection for string 1, optical module 1, DOM(1,1). In the low-charge region (below 0.25 PE), we see a second threshold at approximately 0.13 PE. This is threshold defined in the software that comes from a gradient-related termination condition in WaveDeform. The threshold was set to avoid electronic noise being interpreted as PMT pulses and contaminating the low-charge region.

The steeply falling component of the region from 0.13 PE to 0.25 PE is in agreement with the in-time laser tests mentioned in Sec. F.1.1 and emphasizes the importance of collecting data below the discriminator threshold. This section will assess the noise contribution to this region and examine the effect on the charge distribution and noise contribution by lowering the WaveDeform threshold.

Fig. F-3 (left) shows the charge distributions for the MinBias (black) and the BeaconLaunch (red) datasets using the default settings of WaveDeform. As mentioned in Sec. F.1.2, occa-

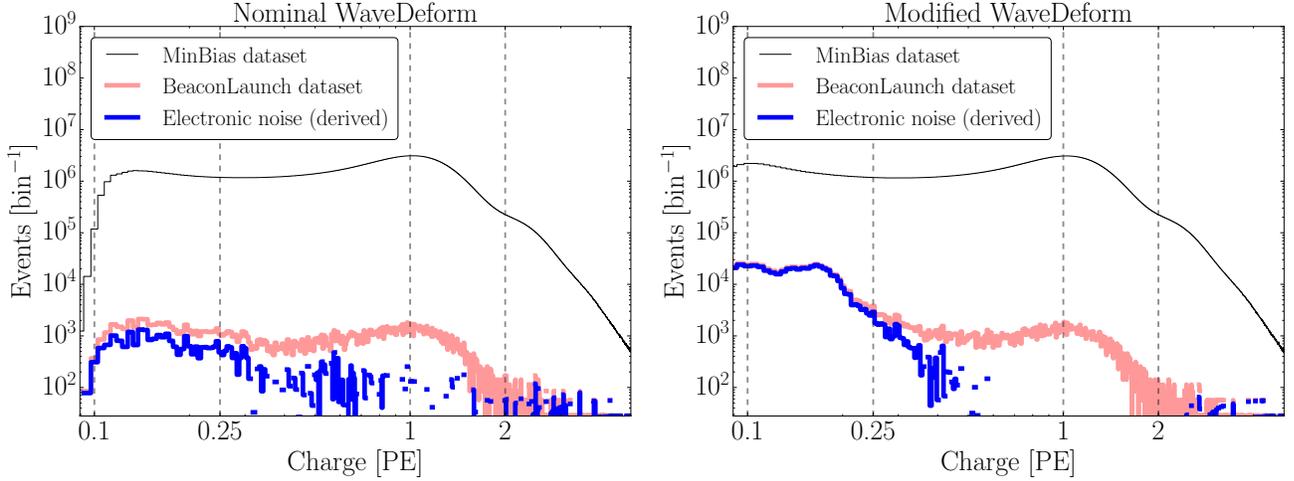

Figure F-3: The cumulative charge distributions of all DOMs for the MinBias (M) and BeaconLaunch (B) datasets. The blue histogram shows the derived contribution from electronic noise. This was found by subtracting the normalized MinBias dataset from the BeaconLaunch dataset (B - M×(B$|_{1PE}$/M$|_{1PE}$)). Left: The charge distributions for the standard WaveDeform settings. Right: The charge distributions for the modified WaveDeform settings.

sionally a photoelectron will be coincident with the forced BeaconLaunch time window. These charges populate a SPE charge distribution. Subtracting the shape of the MinBias charge distribution from the BeaconLaunch dataset yields an estimate of the amount of electronic noise contamination (blue). The bin with the lowest signal-to-noise ratio (SNR) above 0.1 PE was found to have a SNR of 744.7. The SNR for the full distribution was found to be $1.98 \times 10^5$. Fig. F-3 (right) shows the same data after lowering the WaveDeform threshold, and is found to have SNR of 57.9 in the bin with the largest contamination and the total SNR was found to be $0.69 \times 10^5$.

The modified WaveDeform datasets show a minimal increase in the contribution of noise to the low-charge region. From this, we are able to extract charge information down to approximately 0.10 PE and improve the overall description of the charge distribution below the discriminator. This will help constrain the values defining $Exp_1$.

## F.2.3 Fitting procedure

Fitting software is used to determine the components of Eq. F.1 from the measured charge distribution that includes the MPE contamination. The fit assumes that there is a negligible three-PE contribution, which is justified by the lack of statistics in the 3 PE region as well as the significant rate difference between the 1 PE and 2 PE region, as shown in Fig. F-2 (right). The 2 PE charge distribution is assumed to be the SPE charge distribution convolved with itself [266].

The exponential components of Eq. F.1 represent poorly amplified photoelectrons, and we do not allow it to extend beyond the high-charge region of the Gaussian component. In particular, we include a constraint on the the parameter $w_2$ to ensure that it falls off with the Gaussian component:

$$w_2 < \frac{\mu + 2\sigma}{4 - \ln(N/E_2)}. \tag{F.2}$$

This equation was found by setting the $\text{Exp}_2$ to be $\exp^{-2}$ that of the Gaussian component at two sigma (the $\text{Exp}_1$ is neglected from this equation since it falls off in the low-charge region). Eq. F.2 is used as a constraint during the fit to the charge distributions.

Pulses that fall below the WaveDeform threshold and are not reconstructed contribute to an effective efficiency of the individual DOMs. This analysis assumes the same shape of the steeply falling exponential component ($\text{Exp}_1$) for all DOMs in the detector to avoid large fluctuations in the individual DOM efficiencies. The modified WaveDeform data will strictly be used to determine the $\text{Exp}_1$ component. Specifically, using the modified WaveDeform, we background-subtract the BeaconLaunch distribution from the MinBias data, fit the resulting distribution to determine the components of Eq. F.2, and use only the measured shape and normalization of $\text{Exp}_1$ in all subsequent unmodified WaveDeform fits.

As described in Sec. F.1.1, the Gaussian mean ($\mu$) is used to determine the gain setting for each PMT. Therefore, it is particularly important that the fit quality in this region accurately describes the data. While fitting to the full charge distribution improves the overall fit agreement, the mismatch between the chosen functional form (Eq. F.1) and a true SPE charge distribution can cause the Gaussian component to pull away from its ideal location. To compensate for this, the fitting algorithm prioritizes fitting to the data around the Gaussian mean. This is accomplished by first fitting to the full distribution to get an estimate of the Gaussian mean location. Then, the statistical uncertainty is reduced in the region $\pm 0.15$ PE around the original estimated Gaussian mean, and the distribution is re-fitted.

Upon fitting the MinBias data with the predetermined values for $\text{Exp}_1$, the residual of each fit is calculated by measuring the percentage difference between the fit and the data. The average residual is then used as a global scaling factor for all SPE charge templates to account for the difference between the chosen model (Eq. F.2) and the actual data.

### F.2.4 SPE charge template fit results

Using the background-subtracted modified WaveDeform dataset, the $\text{Exp}_1$ component was determined by fitting the distribution from 0.1 PE to 3.5 PE. The result of the fit yielded $E_1 = 6.9 \pm 1.5$ and $w_1 = 0.032 \pm 0.002$ PE. The shape of $\text{Exp}_1$ is then used to describe the low-PE charge region for all subsequent fits.

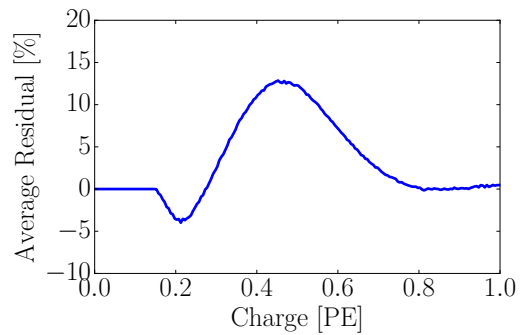

Figure F-4: The measured average residual of the SPE charge templates fit.

Using the MinBias dataset with the measured values of $\text{Exp}_1$, the SPE charge templates are extracted for every DOM, separately for each IceCube season from IC86.2011 to IC86.2016. The fit range for $\text{Exp}_2$ and the Gaussian components is selected to be between 0.15 PE and 3.5 PE. An average fit was also performed on the cumulative charge distribution, in which all the data for a given DOM was summed together (labeled as "AVG").

All the DOMs with "failed fits" are not included in this analysis. A DOM is classified as having a failed fit if it does not pass one of the validity checks on the data requirements (e.g. the number of valid pulses) or goodness of fit. Between 107 and 111 DOMs over the seasons considered have been removed from service and represent the majority of the filed fits. The remaining 6 DOMs that failed the AVG fits are known to have various issues. In the IceCube MC simulation chain, these DOMs are assigned the average SPE charge template.

We can divide the DOMs into subset of hardware differences: the HQE DOMs with the new toroids, the Standard QE DOMs with the new toroids, and the Standard QE DOMs with the old toroids. The mean value and standard error of the IC86.AVG fit parameters, excluding $Exp_1$, for the subset of hardware differences are listed in Table F.1. The average residual for all DOMs from 0 to 1 PE is shown in Fig. F-4.

| Hardware Configuration | $Exp_2$ Amp. ($E_2$) | $Exp_2$ Width ($w_2$) | Gaus. Amp. (N) | Gaus. Mean ($\mu$) | Gaus. Width ($\sigma$) |
|---|---|---|---|---|---|
| HQE / New Toroid | $0.644 \pm 0.003$ | $0.405 \pm 0.003$ | $0.715 \pm 0.002$ | $1.0202 \pm 0.0010$ | $0.311 \pm 0.001$ |
| Std. QE / New Toroids | $0.566 \pm 0.001$ | $0.403 \pm 0.001$ | $0.751 \pm 0.001$ | $1.0238 \pm 0.0004$ | $0.316 \pm 0.001$ |
| Std. QE / Old Toroids | $0.525 \pm 0.002$ | $0.420 \pm 0.002$ | $0.813 \pm 0.002$ | $1.0074 \pm 0.0007$ | $0.294 \pm 0.001$ |

Table F.1: The average values and standard error of each fit parameter for the subset of hardware configurations listed in the first column.

An example fit is shown in Fig. F-5 for the cumulative MinBias charge distribution for DOM (1,1). The collected charge distribution is shown in the black histogram, while the fit to the data is shown as the black line. The extracted SPE charge template from the fit is shown in blue. Both the fit and extracted SPE charge template have been scaled by the average residual shown in Fig. F-4.

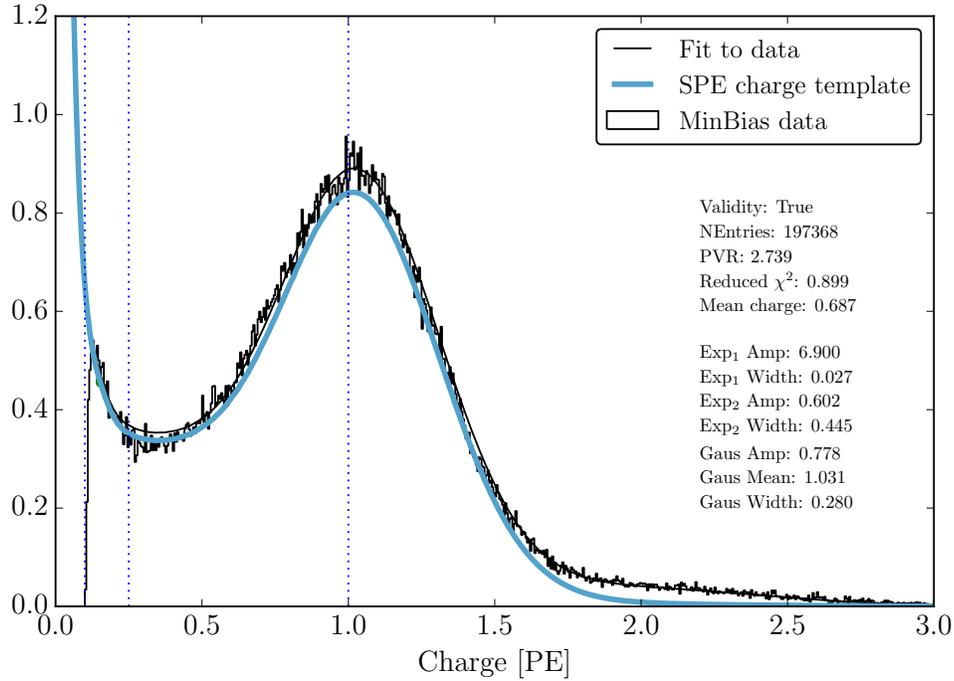

Figure F-5: An example fit for DOM(1,1) using the MinBias dataset including data from seasons IC86.2011 to IC86.2016. The result of the fit is shown as a solid black line and the extracted SPE charge template from the fit is shown in blue. For both the fit and the SPE charge template, the curves include the correction from the average residual shown in Fig. F-4. Repeated figure (Fig. 4-4) for clarity.

## F.3 Discussion

### F.3.1 Correlations between fit parameters and DOM hardware differences

It is evident from the data in Table F.1 that the average shape of the SPE charge templates is correlated with the DOM hardware. These differences can also be seen in the measured peak-to-valley ratios and mean charge of the SPE charge template (see Fig. F-6). When we examine the subset of DOMs instrumented with the new toroids, the average HQE DOM were found to have a $13.8 \pm 0.6\%$ larger $E_2$ component and $4.77 \pm 0.03\%$ smaller Gaussian amplitude. Consequently, the average HQE peak-to-valley ratio is measured to be $2.322 \pm 0.013$, corresponding to $12.12 \pm 0.06\%$ lower than the average Standard QE DOMs. Also, interestingly, the mean charge of the average HQE DOM was found to be $3.34 \pm 0.01\%$ lower than that of the Standard QE DOMs. IceCube compensates for the change in the mean measured charge in simulation, by increasing the HQE DOM efficiency by the equivalent amount. This ensures that the total amount of charge collected by the HQE DOMs remains the same prior to, and after, inserting the SPE charge templates into simulation.

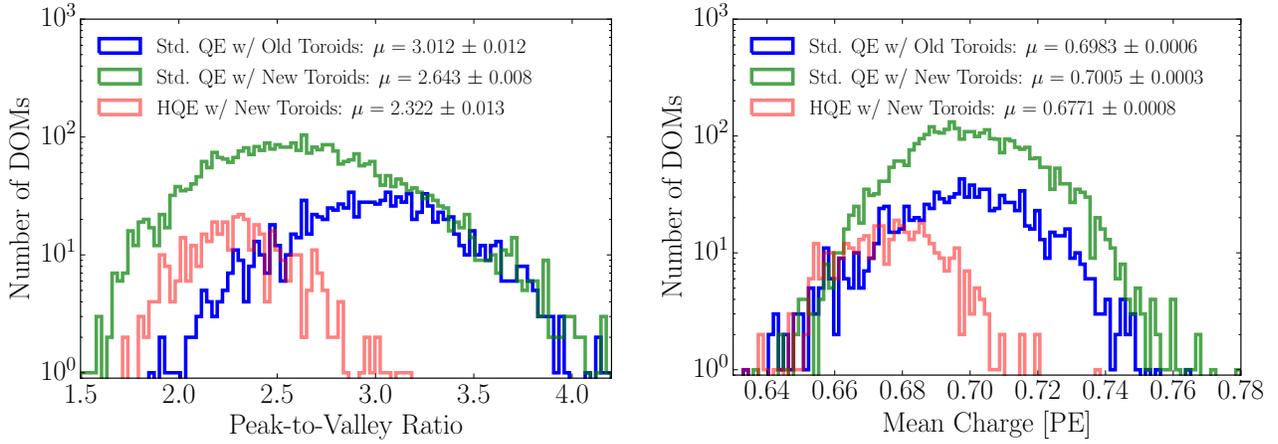

Figure F-6: Comparison between the R7081-02MOD HQE DOMs and standard R7081-02 DOMs. Left: The peak-to-valley ratio for the two subsets of quantum efficiencies. Right: The mean charge of the individual DOM SPE charge templates.

Similarly, using only the subset of Standard QE DOMs, the SPE charge templates comparing the method of AC coupling were found to have measurably different shapes. The average Gaussian amplitude and width for the DOMs instrumented with the old toroids were found to be $8.31 \pm 0.01\%$ and $-6.80 \pm 0.03\%$, respectively. With these differences, we find a peak-to-valley ratio of $2.643 \pm 0.008$ for the new toroid DOMs and $3.012 \pm 0.012$ for the old toroid DOMs. The average Gaussian mean of the fit for the DOMs with the old toroids was also found to be $1.6 \pm 0.1\%$ lower than those with the new toroids. This corresponds proportionally to a change in the expected gain. The mean charge, however, between these two hardware configurations remains very similar ($-0.346 \pm 0.001\%$).

Although the DOMs instrumented with the old toroids were deployed into the ice earlier than those with the new toroids, the differences above are still noted when examining individual deployment years; therefore, the shape differences are not attributed to the change in the DOM behavior over time. However, the DOMs with the old toroids were the first PMTs to be manu-factured by Hamamatsu. A gradual change over time of the fit parameters was observed when ordering the PMTs according to their PMT serial number. This is compelling evidence that the observed differences between the new and old toroids is due to a change in the production procedure rather than version of AC coupling.

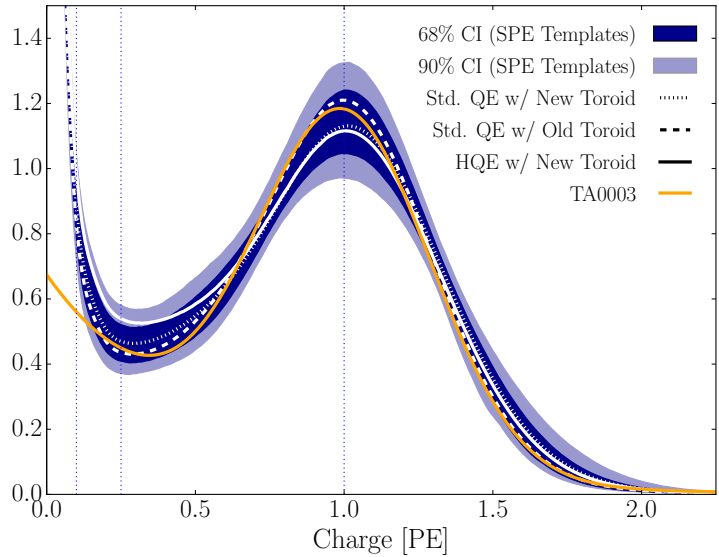

Figure F-7: The inner (outer) dark blue region shows the 68% (90%) confidence interval defined by the measured spread in the extracted SPE charge templates of all DOMs in the detector. Superimposed are the average SPE charge templates for the variety of hardware configurations shown in white. The TA0003 distribution, for comparison, is shown in orange. All curves have been normalized such that the area above 0.25 PE is the same.

Fig. F-7 illustrates the average shape differences in the extracted SPE charge templates between the HQE DOM with the new toroids (solid white line), Standard QE with the new toroids (dotted white line), Standard QE with the old toroids (dashed white line), compared to the spread in the measured SPE charge templates for all DOMs in the detector (dark blue contours). The figure also shows how the TA0003 distribution compares to this recent measurement. All curves in this figure have been normalized such that the area above 0.25 PE is the same. The observable shape differences from the TA0003 are attributed to a better control of the low-charge region, the difference in functional form (described in Section F.1.1), and the fact that the SPE charge templates were generated using a realistic photocathode illumination.

## F.3.2 Fitting parameters variation over time

The SPE charge templates were extracted for each IceCube season independently to investigate the time dependence of the fit parameters. For every DOM in the detector, the change over time of each fit parameter (excluding $Exp_1$) was calculated. Fig. F-8 shows the change in a given fit parameter, relative to the mean value, per year. The measured distribution was found to be consistent with statistically scrambling the yearly measurements. The average of each fit parameters are found to deviate less than 0.1%, which is in agreement with the stability checks performed in Ref. [134]. This observation holds for the individual subset of DOMs with different hardware configurations as well.

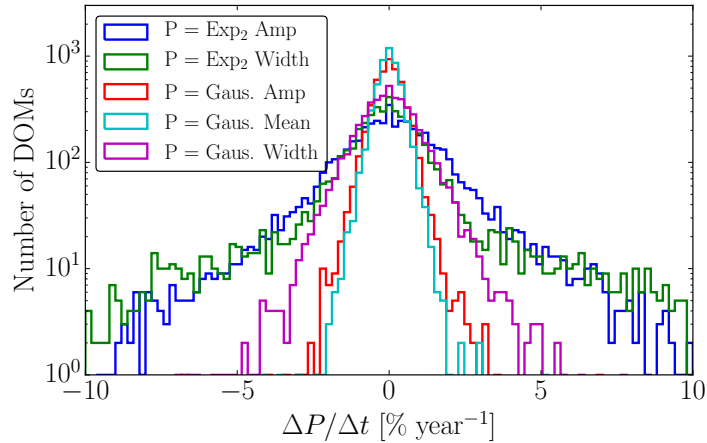

Figure F-8: The change in the individual DOM fitted parameters over time, represented as percentage deviation from the mean fit parameter value.

## F.3.3 Quantifying observable changes when modifying the PMT charge distributions

Changing the assumed gain response in simulation has different implications depending on the typical illumination level present in different analyses. These differences are outlined in the following discussion.

The PMT response is described by a combination of a "bare" efficiency, $\eta_0$, and a normalized charge response function, $f(q)$. The bare efficiency represents the fraction of arriving photons that result in any nonzero charge response, including those below the discriminator threshold. The normalization condition is:

$$\int_0^\infty f(q)dq = 1. \tag{F.3}$$

Generally, $f(q)$ and $\eta_0$ have to be adjusted together to maintain agreement with a quantity known from lab or in-ice measurements, such as the predicted number of pulses above threshold for a dim source.

**Dim source measurements**  Where light levels are low enough, the low occupancy ensures that sub-discriminator pulses do not contribute to any observed charge as they do not satisfy the trigger threshold. Given some independent way of knowing the number of arriving photons, a lab or in-ice measurement determines the trigger fraction above threshold $\eta_{0.25}$ and/or the average charge over threshold $Q_{0.25}$, either of which can be used to constrain the model as follows:

$$\eta_{0.25} = \eta_0 \int_{0.25 q_{pk}}^\infty f(q)\mathrm{d}q \tag{F.4}$$

$$Q_{0.25} = \eta_0 \int_{0.25 q_{pk}}^\infty q f(q)\mathrm{d}q \tag{F.5}$$

Here, the discriminator threshold is assumed to be 0.25 times the peak position $q_{pk}$. It is also useful to multiply observed charges by $q_{pk}$, since we set each PMT gain by such a reference, and then a measurement constraint would be stated in terms of $Q_{0.25}/q_{pk}$.

**Semi-bright source measurements**  For semi-bright sources, pulses that arrive after the readout time window is opened are not subject to the the discriminator threshold. WaveDeform introduces a software termination condition at $\sim 0.13\,\mathrm{PE}$ (described at the end of Section F.2.1).

The average charge of an individual pulse that arrives within the time window is:

$$Q_{0.10} = \eta_0 \int_{0.10 q_{pk}}^{\infty} q f(q) \mathrm{d}q \tag{F.6}$$

**Bright source measurements** For light levels that are large, the trigger is satisfied regardless of the response to individual photons, and the total charge per arriving photon therefore includes contributions below both the discriminator and the WaveDeform thresholds:

$$Q_0 = \eta_0 \int_0^{\infty} q f(q) \mathrm{d}q \tag{F.7}$$

As such, the total charge is directly proportional to the average charge of the SPE charge template, having a strong dependence on $\mathrm{Exp}_1$.

**Model comparison**

A natural question to ask is whether or not a change in $f(q)$ would cause observable changes in the bright-to-dim ratios. When the charge distribution model is changed in a way that preserves agreement with the measured $\eta_{0.25}$ or $Q_{0.25}/q_{pk}$, i.e. $\eta_0$ is adjusted properly for changes in $f(q)$, the physical effect can be summarized by the change in the bright-to-dim ratios $Q_0/Q_{0.25}$, and $Q_0/Q_{0.10}$. Conveniently, these ratios depend only on the shape of $f(q)$. Table F.2 compares these ratios in terms of the TA0003 charge distribution and the SPE charge templates described here. It is shown that there are sub-percent level differences in the physically-observable bright-to-dim ratios.

| Model | Detector | $Q_0/Q_{0.25}$ | $Q_0/Q_{0.10}$ | $\eta_{0.25}/Q_{0.25}$ |
|---|---|---|---|---|
| TA0003 | All DOMs | 1.017 | 1.0031 | 1.05 |
| SPE charge templates | HQE + New Toroids | 1.021±0.002 | 1.0041±0.0004 | 1.05±0.02 |
|  | Std. QE + New Toroids | 1.018±0.002 | 1.0035±0.0005 | 1.03±0.02 |
|  | Std. QE + Old Toroids | 1.017±0.002 | 1.0033±0.0005 | 1.05±0.02 |

Table F.2: The distribution in bright-to-dim ratios for the previous charge distribution (TA0003) and the individual DOM SPE charge templates for the IceCube and DeepCore detectors.

### F.3.4 SPE charge templates for calibration

The gain setting on each PMT is calibrated prior to the beginning of each season such that the Gaussian mean of the charge distribution corresponds to a gain of $10^7$, or equivalently 1 PE. This gain calibration method, run directly on the DOMs, uses waveform integration for charge determination instead of WaveDeform unfolding, resulting in a small systematic shift in gain. This systematic shift was determined for every PMT, and was found to be on average $2.00 \pm 0.03\%$ with a standard deviation of 3.54%, corresponding to an overestimation of the measured charge in the detector.

The correction to the systematic shift in the measured charge can be implemented retroactively by dividing the reported charge from WaveDeform by the corresponding offset for a given DOM. Alternatively, we can account for this by simply inserting SPE charge templates, measured in this analysis, into simulation such that the corresponding systematic shift is also modelled in simulation. This will be performed in the following subsection.

### F.3.5 SPE charge templates in simulation

To model the IceCube instrument, we must implement the PMT response in simulation. The IceCube MC simulation chain assigns a charge to every photoelectron generated at the surface of the photocathode. The charge is determined by sampling from a normalized charge distribution probability density function (PDF). A comparison to data between describing the charge distribution PDF using the SPE charge templates and the TA0003 distribution follows.

Two simulation sets consisting of the same events were processed through the IceCube Monte Carlo simulation chain to the final analysis level of an update to the IC86.2011 sterile neutrino analysis [123]. Here, the events that pass the cuts are >99.9% upward-going (a trajectory oriented upwards relative to the horizon) secondary muons produced by charged current muon neutrino/antineutrino interactions. The muon reconstructed energy range of this event selection is between approximately 500 GeV and 10 TeV.

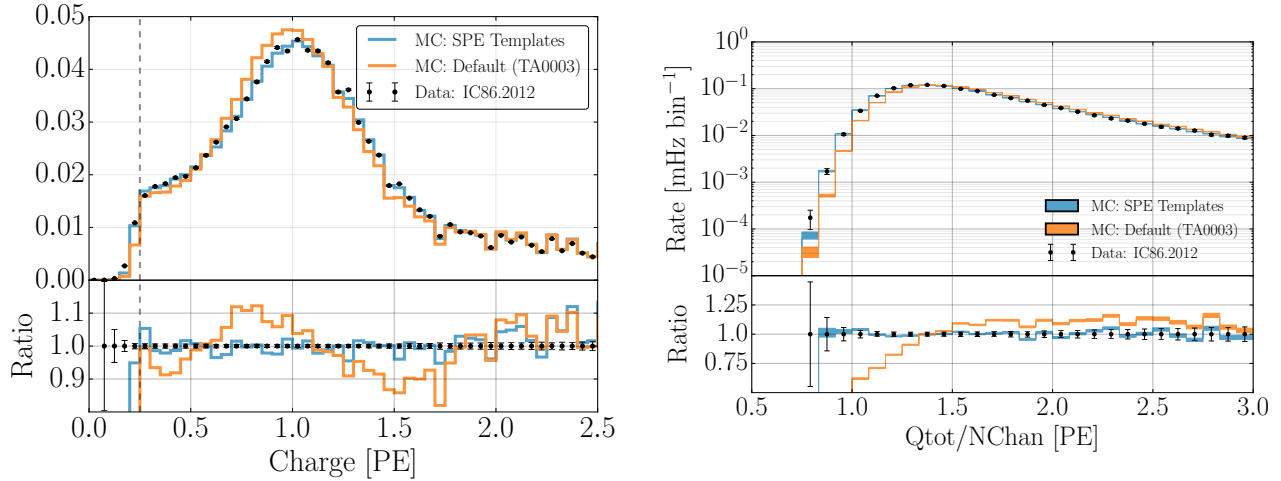

Figure F-9: A comparison between the SPE charge templates (blue) and the TA0003 (orange) model for describing the SPE charge distribution in Monte Carlo. The simulation is compared to the 2012 IceCube season. The data is shown in black. Left: The total measured charge per DOM, per event at analysis level. Right: The distribution of the total measured charge of an event divided by the number of DOMs that participated in the event.

Fig. F-9 (left) shows the distribution of the total measured charge during each event per DOM (data points). The simulation set using the TA0003 charge distribution is shown in orange, and that using the SPE charge templates is shown in blue. The data is shown for the full IC86.2012 season but is statistically equivalent to any of the other seasons. Fig. F-9 (right) shows the distribution of the total measured charge of an event divided by the number of channels (NChan), or DOMs, that participated in the event. Both plots in Fig. F-9 have been normalized such that the area under the histograms is the same.

The SPE charge templates clearly improve the overall MC description of these two low-level variables. This update may be useful for analyses that rely on low-occupancy events (low-energy or dim events) in which average charge per channels is below $1.5\,\text{PE}$, and will be investigated further within IceCube.

## F.4   Conclusion

This article outlines the procedure used to extract the SPE charge templates for all in-ice DOMs in the IceCube detector using in-situ data from IC86.2011 to IC86.2016. The result of this measurement was shown to be useful for improving the overall data/MC agreement as well as calibration of the individual PMTs. It also prompted a comparison between the shape of the SPE charge templates for a variety of hardware configurations and time dependent correlations.

The subset of HQE DOMs were found to have a smaller peak-to-valley ratio relative to the Standard QE DOMs, as well as an overall $3.34 \pm 0.01\%$ lower mean charge. It was also found that the DOMs instrumented with the old toroids used for AC coupling (the first PMTs to be manufactured by Hamamatsu) had narrower and larger Gaussian component corresponding resulting in an increased peak-to-valley ratio of $14.0 \pm 0.6\%$. This was found to be likely due to a change in the manufacturing over time rather than the actual AC coupling method. No significant time dependence in any of the fitted parameters associated with the SPE charge templates over the investigated seasons was observed. A reassessment of the PMT gain settings found a systematic bias of $2.00 \pm 0.03\%$ with a standard deviation of $3.54\%$.

The SPE charge templates were inserted into the MC simulation and the results were compared to the default TA0003 distribution. A significant improvement in the description of the variables total charge per DOM and total charge over the number of channels was shown. Analyses which rely on low-light occupancy measurements, may benefit from this update. As shown in the bright-to-dim ratios, the average mean charge for various light levels will not be affected by this update.

# Reference sheet

In a 3+1 sterile neutrino oscillation framework:

$$\sin^2(2\theta_{ee}) = 4|U_{e4}|^2(1-|U_{e4}|^2) = \sin^2(2\theta_{14})$$

$$\sin^2(2\theta_{\mu\mu}) = 4|U_{\mu 4}|^2(1-|U_{\mu 4}|^2) = 4\cos^2\theta_{14}\sin^2\theta_{24}(1-\cos^2\theta_{14}\sin^2\theta_{24})$$

$$\sin^2(2\theta_{\tau\tau}) = 4|U_{\tau 4}|^2(1-|U_{\tau 4}|^2) = 4\cos^2\theta_{14}\cos^2\theta_{24}\sin^2\theta_{34}(1-\cos^2\theta_{14}\cos^2\theta_{24}\sin^2\theta_{34})$$

$$\sin^2(2\theta_{\mu e}) = 4|U_{\mu 4}|^2|U_{e4}|^2 = \sin^2\theta_{14}\sin^2\theta_{24}$$

$$\sin^2(2\theta_{\mu\tau}) = \sin^2 2\theta_{24}\cos^4\theta_{14}\sin^2\theta_{34}$$

$$\sin^2(2\theta_{e\tau}) = \sin^2 2\theta_{14}\cos^2\theta_{24}\sin^2\theta_{34}$$

$$P_{\nu_e \to \nu_e} = 1 - 4(1-|U_{e4}|)^2(|U_{e4}|^2)\sin^2(1.27\Delta m_{41}^2 L/E)$$

$$P_{\nu_\mu \to \nu_\mu} = 1 - 4(1-|U_{\mu 4}|)^2(|U_{\mu 4}|^2)\sin^2(1.27\Delta m_{41}^2 L/E)$$

$$P_{\nu_\mu \to \nu_e} = 4|U_{e4}|^2|U_{\mu 4}|^2\sin^2(1.27\Delta m_{41}^2 L/E)$$

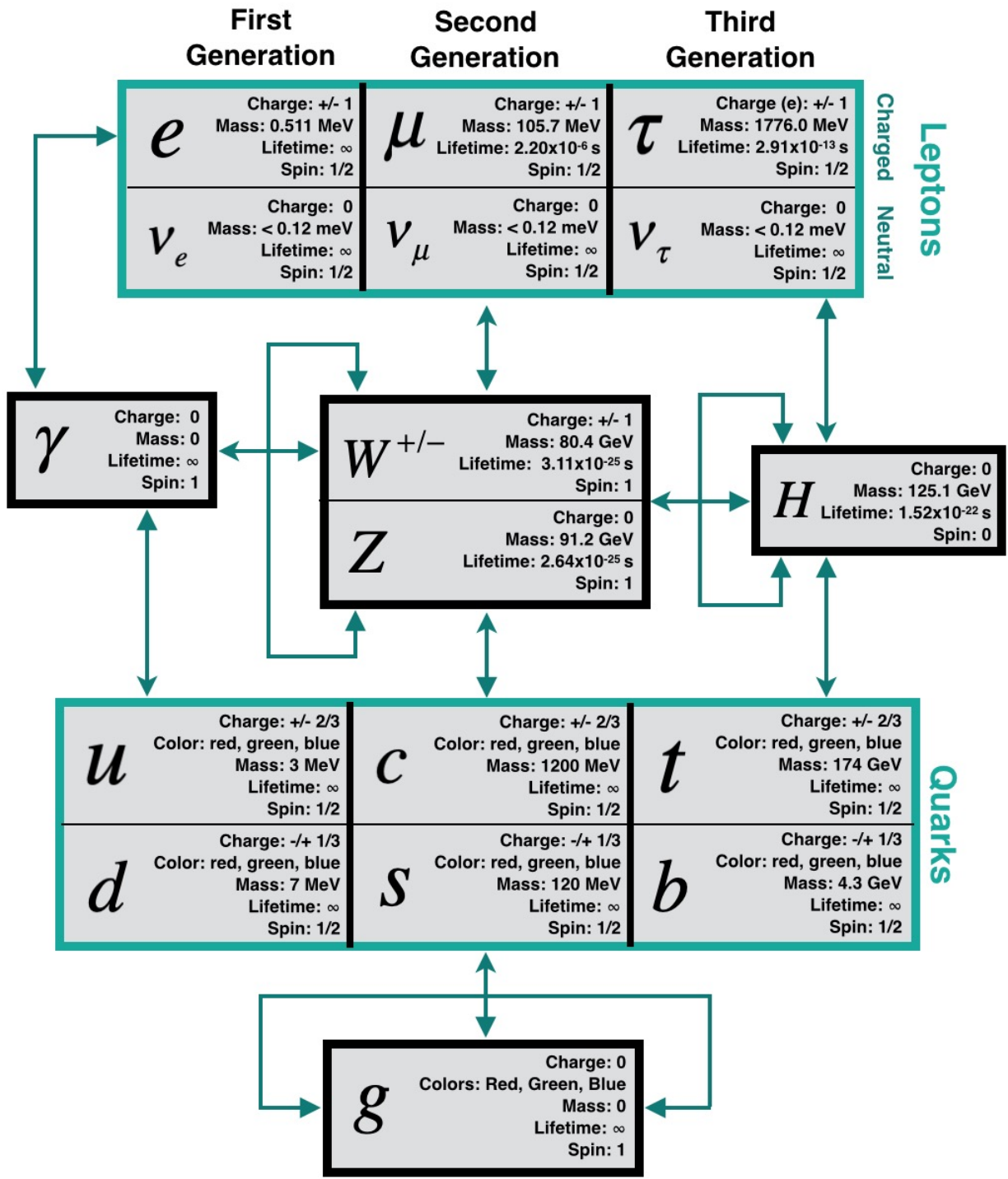

Figure F-10: The Standard Model of Particle Physics.

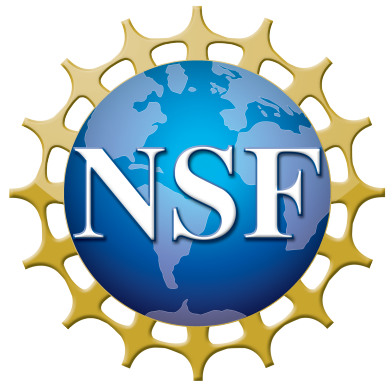

The work in this thesis was supported by the National Science Foundation.

THIS PAGE INTENTIONALLY LEFT BLANK

# Bibliography

[1] S. N. Axani, K. Frankiewicz, and J. M. Conrad, "The CosmicWatch Desktop Muon Detector: A self-contained, pocket sized particle detector," *Journal of Instrumentation*, vol. 13, no. 03, p. P03019, 2018.

[2] S. N. Axani, "Cosmicwatch repository." https://github.com/spenceraxani/CosmicWatch-Desktop-Muon-Detector-v2, accessed: February 2017.

[3] CosmicWatch. www.cosmicwatch.lns.mit.edu, accessed: February 2017.

[4] S. N. Axani, "The physics behind the cosmicwatch desktop muon detectors," *arXiv preprint arXiv:1908.00146*, 2019.

[5] C. Giunti and C. W. Kim, *Fundamentals of Neutrino Physics and Astrophysics*. 2007.

[6] J. Chadwick, "Intensity distribution in the magnetic spectrum of the $\beta$ rays of radium (B + C)," *Verhandl. Dtsc. Phys. Ges.*, vol. 16, p. 383, 1914.

[7] H. Kruse and A. McGu, "Detection of the free neutrino: Confirmation," *Science*, vol. 124, no. 3212, pp. 103–104, 1956.

[8] W. Pauli, C. P. Enz, and K. von Meyenn, *Writings on physics and philosophy*. Springer, 1994.

[9] G. Danby *et al.*, "Observation of high-energy neutrino reactions and the existence of two kinds of neutrinos," *Physical Review Letters*, vol. 9, no. 1, p. 36, 1962.

[10] M. L. Perl *et al.*, "Evidence for anomalous lepton production in e+- e- annihilation," *Physical Review Letters*, vol. 35, no. 22, p. 1489, 1975.

[11] K. Kodama *et al.*, "Observation of tau neutrino interactions," *Physics Letters B*, vol. 504, no. 3, pp. 218–224, 2001.

[12] B. Pontecorvo, "Inverse beta processes and nonconservation of lepton charge," *Zhur. Eksptl'. i Teoret. Fiz.*, vol. 34, 1958.

[13] Z. Maki, M. Nakagawa, and S. Sakata, "Remarks on the unified model of elementary particles," *Progress of Theoretical Physics*, vol. 28, no. 5, pp. 870–880, 1962.

[14] L. Wolfenstein, "Neutrino oscillations in matter," *Physical Review D*, vol. 17, no. 9, p. 2369, 1978.

[15] K. Hirata *et al.*, "Observation of a small atmospheric $\nu_\mu/\nu_e$ ratio in Kamiokande," *Physics Letters B*, vol. 280, no. 1-2, pp. 146–152, 1992.

[16] D. Casper *et al.*, "Measurement of atmospheric neutrino composition with the IMB-3 detector," *Physical Review Letters*, vol. 66, no. 20, p. 2561, 1991.

[17] T. Haines *et al.*, "Calculation of atmospheric neutrino-induced backgrounds in a nucleon-decay search," *Physical review letters*, vol. 57, no. 16, 1986.

[18] K. Hirata *et al.*, "Observation of a neutrino burst from the supernova sn1987a," *Physical Review Letters*, vol. 58, no. 14, p. 1490, 1987.

[19] R. Bionta *et al.*, "Observation of a neutrino burst in coincidence with supernova 1987A in the Large Magellanic Cloud," in *Neutrinos And Other Matters: Selected Works of Frederick Reines*, pp. 340–342, World Scientific, 1991.

[20] G. Ewan, H. Evans, and H. Lee, "Sudbury neutrino observatory proposal," tech. rep., Queen's Univ., 1987.

[21] M. Koshiba, "32 Kton water Cerenkov detector (JACK)," tech. rep., 1984.

[22] M. Koshiba, "22-kton Water Cherenkov Detector (JACK)," tech. rep., 1985.

[23] Y. Suzuki, "The Super-Kamiokande experiment," *Eur. Phys. J. C*, vol. 79, no. 298, 2019.

[24] Y. Fukuda *et al.*, "Evidence for oscillation of atmospheric neutrinos," *Physical Review Letters*, vol. 81, no. 8, p. 1562, 1998.

[25] M. Ambrosio *et al.*, "Measurement of the atmospheric neutrino induced upgoing muon flux using macro," *Physics Letters B*, vol. 434, no. 3-4, pp. 451–457, 1998.

[26] A. McDonald, "Evidence for neutrino oscillations: Solar and reactor neutrinos," *arXiv preprint nucl-ex/0412005*, 2004.

[27] LSND Collaboration, "A proposal to search for neutrino oscillations with high sensitivity in the appearance channels $\nu_\mu \to \nu_e$ and $\overline{\nu}_\mu \to \overline{\nu}_e$," 1989.

[28] C. Athanassopoulos *et al.*, "Evidence for $\nu_\mu \to \nu_e$ Oscillations from the LSND Experiment at the Los Alamos Meson Physics Facility," *Physical Review Letters*, vol. 77, no. 15, p. 3082, 1996.

[29] A. Aguilar-Arevalo *et al.*, "Search for Electron Neutrino Appearance at the $\Delta m^2 \sim 1$ eV$^2$ Scale," *Physical review letters*, vol. 98, no. 23, p. 231801, 2007.

[30] A. Aguilar-Arevalo *et al.*, "Improved search for $\nu\ \mu$âĘŠ $\nu$ e oscillations in the miniboone experiment," *Physical review letters*, vol. 110, no. 16, p. 161801, 2013.

[31] G. Mention, M. Fechner, T. Lasserre, T. A. Mueller, D. Lhuillier, M. Cribier, and A. Letourneau, "Reactor antineutrino anomaly," *Physical Review D*, vol. 83, no. 7, p. 073006, 2011.

[32] W. Hampel *et al.*, "Final results of the Cr-51 neutrino source experiments in GALLEX," *Phys. Lett.*, vol. B420, pp. 114–126, 1998.

[33] W. Hampel *et al.*, "Gallex solar neutrino observations: Results for gallex iv," *Physics Letters B*, vol. 447, no. 1-2, pp. 127–133, 1999.

[34] J. Abdurashitov *et al.*, "The Russian-American gallium experiment (SAGE) Cr neutrino source measurement," *Physical Review Letters*, vol. 77, no. 23, p. 4708, 1996.

[35] J. Abdurashitov *et al.*, "Measurement of the response of a gallium metal solar neutrino experiment to neutrinos from a cr-51 source," *Physical Review C*, vol. 59, no. 4, p. 2246, 1999.

[36] S. Schael *et al.*, "Precision electroweak measurements on the Z resonance," *Phys.Rept.*, vol. 427, no. 257, p. 454, 2006.

[37] M. Goldhaber, L. Grodzins, and A. Sunyar, "Helicity of neutrinos," *Physical Review*, vol. 109, no. 3, p. 1015, 1958.

[38] C.-S. Wu, E. Ambler, R. Hayward, D. Hoppes, and R. P. Hudson, "Experimental test of parity conservation in beta decay," *Physical review*, vol. 105, no. 4, p. 1413, 1957.

[39] T.-D. Lee and C.-N. Yang, "Question of parity conservation in weak interactions," *Physical Review*, vol. 104, no. 1, p. 254, 1956.

[40] Lee and Yang, "1957 nobel prize," Dec 2018.

[41] Q. R. Ahmad *et al.*, "Direct evidence for neutrino flavor transformation from neutral-current interactions in the Sudbury Neutrino Observatory," *Physical review letters*, vol. 89, no. 1, p. 011301, 2002.

[42] Q. Ahmad *et al.*, "Measurement of day and night neutrino energy spectra at SNO and constraints on neutrino mixing parameters," *Physical Review Letters*, vol. 89, no. 1, p. 011302, 2002.

[43] S. Ahmed *et al.*, "Measurement of the Total Active $^8$B Solar Neutrino Flux at the Sudbury Neutrino Observatory with Enhanced Neutral Current Sensitivity," *Physical review letters*, vol. 92, no. 18, p. 181301, 2004.

[44] D. Griffiths, *Introduction to elementary particles*. John Wiley & Sons, 2008.

[45] M. Kobayashi and T. Maskawa, "CP-violation in the renormalizable theory of weak interaction," *Progress of Theoretical Physics*, vol. 49, no. 2, pp. 652–657, 1973.

[46] B. Pontecorvo, "Mesonium and anti-mesonium," *Sov. Phys. JETP*, vol. 6, p. 429, 1957. [Zh. Eksp. Teor. Fiz.33,549(1957)].

[47] I. Esteban, M. Gonzalez-Garcia, M. Maltoni, I. Martinez-Soler, and T. Schwetz, "Updated fit to three neutrino mixing: exploring the accelerator-reactor complementarity," *Journal of High Energy Physics*, vol. 2017, no. 1, p. 87, 2017.

[48] S. Mikheyev and A. Y. Smirnov, "Resonant amplification of $\nu$ oscillations in matter and solar-neutrino spectroscopy," *Il Nuovo Cimento C*, vol. 9, no. 1, pp. 17–26, 1986.

[49] Y. Fukuda *et al.*, "Measurements of the solar neutrino flux from Super-Kamiokande's first 300 days," *Physical Review Letters*, vol. 81, no. 6, p. 1158, 1998.

[50] K. A. Olive *et al.*, "Review of Particle Physics," *Chin. Phys.*, vol. C38, p. 090001, 2014.

[51] D. O. Caldwell, *Current aspects of neutrino physics*. Springer Science & Business Media, 2013.

[52] R. N. Mohapatra and P. B. Pal, *Massive neutrinos in physics and astrophysics*, vol. 72. World scientific, 2004.

[53] E. K. Akhmedov, "Parametric resonance in neutrino oscillations in matter," *Pramana*, vol. 54, no. 1, pp. 47–63, 2000.

[54] C. Athanassopoulos *et al.*, "Results on $\nu_\mu \to \nu_e$ Neutrino Oscillations from the LSND Experiment," *Physical Review Letters*, vol. 81, no. 9, p. 1774, 1998.

[55] A. Aguilar *et al.*, "Evidence for neutrino oscillations from the observation of $\nu_e$ appearance in a $\nu_\mu$ beam," *Physical Review D*, vol. 64, no. 11, p. 112007, 2001.

[56] R. Burman and P. Plischke, "Neutrino fluxes from a high-intensity spallation neutron facility," *NIM-A*, vol. 398, no. 2-3, pp. 147–156, 1997.

[57] A. Aguilar-Arevalo *et al.*, "Significant excess of electronlike events in the MiniBooNE short-baseline neutrino experiment," *Physical review letters*, vol. 121, no. 22, p. 221801, 2018.

[58] S. Gariazzo, C. Giunti, M. Laveder, and Y. Li, "Updated global 3+ 1 analysis of short-baseline neutrino oscillations," *Journal of High Energy Physics*, vol. 2017, no. 6, p. 135, 2017.

[59] J. Bahcall, "An introduction to solar neutrino research," *arXiv preprint hep-ph/9711358*, 1997.

[60] J. Abdurashitov *et al.*, "Measurement of the response of a Ga solar neutrino experiment to neutrinos from a Ar-37 source," *Physical Review C*, vol. 73, no. 4, p. 045805, 2006.

[61] M. A. Acero, C. Giunti, and M. Laveder, "Limits on $\nu_e$ and $\nu_e$ disappearance from Gallium and reactor experiments," *Physical Review D*, vol. 78, no. 7, p. 073009, 2008.

[62] K. N. Abazajian *et al.*, "Light sterile neutrinos: a white paper," *arXiv preprint arXiv:1204.5379*, 2012.

[63] D. Karlen, "The number of light neutrino types from collider experiments," 2004.

[64] C. A. Argüelles, *New physics with atmospheric neutrinos*. PhD thesis, The University of Wisconsin-Madison, 2015.

[65] B. Armbruster *et al.*, "Upper limits for neutrino oscillations $\nu_\mu \to \nu_e$ from muon decay at rest," *Physical Review D*, vol. 65, no. 11, p. 112001, 2002.

[66] P. Astier *et al.*, "Final nomad results on muon-neutrino to tau-neutrino and electron-neutrino to tau-neutrino oscillations including a new search for tau-neutrino appearance using hadronic tau decays," 2001.

[67] N. Agafonova *et al.*, "Search for $\nu_\mu \to \nu_e$ oscillations with the OPERA experiment in the CNGS beam," *Journal of High Energy Physics*, vol. 2013, no. 7, p. 4, 2013.

[68] M. Antonello *et al.*, "Experimental search for the âĂIJLSND anomalyâĂİ with the ICARUS detector in the CNGS neutrino beam," *The European Physical Journal C*, vol. 73, no. 3, p. 2345, 2013.

[69] B. Achkar *et al.*, "Search for neutrino oscillations at 15, 40 and 95 meters from a nuclear power reactor at Bugey," *Nuclear Physics B*, vol. 434, no. 3, pp. 503–532, 1995.

[70] J. Ashenfelter *et al.*, "Measurement of the Antineutrino Spectrum from 235U Fission at HFIR with PROSPECT," *arXiv preprint arXiv:1812.10877*, 2018.

[71] H. Kwon *et al.*, "Search for neutrino oscillations at a fission reactor," *Physical Review D*, vol. 24, no. 5, p. 1097, 1981.

[72] H. Almazán *et al.*, "Sterile Neutrino Constraints from the STEREO Experiment with 66 Days of Reactor-On Data," *Physical review letters*, vol. 121, no. 16, p. 161801, 2018.

[73] A. Diaz, C. Arguelles, G. Collin, J. Conrad, and M. Shaevitz, "Where are we with light sterile neutrinos?," *arXiv:1906.00045 [hep-ex]*, 2019.

[74] I. Alekseev *et al.*, "Search for sterile neutrinos at the DANSS experiment," *Physics Letters B*, vol. 787, pp. 56–63, 2018.

[75] Y. Ko *et al.*, "Sterile neutrino search at the NEOS experiment," *Physical review letters*, vol. 118, no. 12, p. 121802, 2017.

[76] K. Abe *et al.*, "Limits on sterile neutrino mixing using atmospheric neutrinos in Super-Kamiokande," *Physical Review D*, vol. 91, no. 5, p. 052019, 2015.

[77] P. Adamson *et al.*, "Study of muon neutrino disappearance using the Fermilab Main Injector neutrino beam," *Physical Review D*, vol. 77, no. 7, p. 072002, 2008.

[78] K. Mahn *et al.*, "Dual baseline search for muon neutrino disappearance at $0.5\,\text{eV}^2 < \Delta\text{m}^2 < 40\,\text{eV}^2$," *Physical Review D*, vol. 85, no. 3, p. 032007, 2012.

[79] F. Dydak *et al.*, "A Search for Muon-neutrino Oscillations in the $\Delta\text{m}^2$ Range 0.3-eV$^2$ to 90-eV$^2$," *Phys. Lett.*, vol. 134, no. CERN-EP/83-152, p. 281, 1983.

[80] I. Stockdale *et al.*, "Limits on Muon-Neutrino Oscillations in the Mass Range $30 < \Delta\text{m}^2 < 1000\,\text{eV}^2/\text{c}\,4$," *Physical Review Letters*, vol. 52, no. 16, p. 1384, 1984.

[81] G. Collin, C. Argüelles, J. Conrad, and M. Shaevitz, "Sterile neutrino fits to short baseline data," *Nuclear Physics B*, vol. 908, pp. 354–365, 2016.

[82] A. Esmaili, F. Halzen, and O. Peres, "Constraining sterile neutrinos with AMANDA and IceCube atmospheric neutrino data," *Journal of Cosmology and Astroparticle Physics*, vol. 2012, no. 11, p. 041, 2012.

[83] E. K. Akhmedov, "On neutrino oscillations in a nonhomogeneous medium," *Soviet Journal of Nuclear Physics (English Translation)*, vol. 47, no. 2, pp. 301–302, 1988.

[84] P. Krastev and A. Y. Smirnov, "Parametric effects in neutrino oscillations," *Physics Letters B*, vol. 226, no. 3-4, pp. 341–346, 1989.

[85] Q. Liu, S. Mikheyev, and A. Y. Smirnov, "Parametric resonance in oscillations of atmospheric neutrinos?," *Physics Letters B*, vol. 440, no. 3-4, pp. 319–326, 1998.

[86] S. Mikheev and A. Y. Smirnov, "Resonance amplification of oscillations in matter and spectroscopy of solar neutrinos," *Yadernaya Fizika*, vol. 42, no. 6, pp. 1441–1448, 1985.

[87] S. Petcov, "Diffractive-like (or parametric-resonance-like?) enhancement of the earth (day-night) effect for solar neutrinos crossing the earth core," *Physics Letters B*, vol. 434, no. 3-4, pp. 321–332, 1998.

[88] M. Chizhov and S. Petcov, "New conditions for a total neutrino conversion in a medium," *Physical Review Letters*, vol. 83, no. 6, p. 1096, 1999.

[89] M. Ambrosio *et al.*, "Matter effects in upward-going muons and sterile neutrino oscillations," *Physics Letters B*, vol. 517, no. 1-2, pp. 59–66, 2001.

[90] Q. Y. Liu and A. Y. Smirnov, "Neutrino mass spectrum with $\nu\mu \to \nu s$ oscillations of atmospheric neutrinos," *Nuclear Physics B*, vol. 524, no. 3, pp. 505–523, 1998.

[91] M. Lindner, W. Rodejohann, and X.-J. Xu, "Sterile neutrinos in the light of IceCube," *Journal of High Energy Physics*, vol. 2016, no. 1, p. 124, 2016.

[92] M. Aartsen *et al.*, "Observation of high-energy astrophysical neutrinos in three years of IceCube data," *Physical review letters*, vol. 113, no. 10, p. 101101, 2014.

[93] J. A. Formaggio and G. Zeller, "From eV to EeV: Neutrino cross sections across energy scales," *Reviews of Modern Physics*, vol. 84, no. 3, p. 1307, 2012.

[94] E. Pinat, *The IceCube Neutrino Observatory: search for extended sources of neutrinos and preliminary study of a communication protocol for its future upgrade*. PhD thesis, Universite libre de Bruxelles, 2017.

[95] M. G. Aartsen, M. Ackermann, J. Adams, J. Aguilar, M. Ahlers, M. Ahrens, I. Al Samarai, D. Altmann, K. Andeen, T. Anderson, *et al.*, "Measurement of the multi-tev neutrino interaction cross-section with icecube using earth absorption," *Nature*, vol. 551, no. 7682, p. 596, 2017.

[96] K. O. et al, "COSMIC RAYS - Particle Data Group," *NIM-A*, vol. 38, no. 090001, p. 6, 2014.

[97] T. K. Gaisser, R. Engel, and E. Resconi, *Cosmic rays and particle physics*. Cambridge University Press, 2016.

[98] K. Greisen, "End to the cosmic-ray spectrum?," *Physical Review Letters*, vol. 16, no. 17, p. 748, 1966.

[99] G. T. Zatsepin and V. A. Kuzmin, "Upper limit of the spectrum of cosmic rays," *JETP Letters*, vol. 4, no. 3, pp. 78–80, 1966.

[100] T. K. Gaisser, T. Stanev, and S. Tilav, "Cosmic ray energy spectrum from measurements of air showers," *Frontiers of Physics*, vol. 8, no. 6, pp. 748–758, 2013.

[101] T. K. Gaisser, "Spectrum of cosmic-ray nucleons, kaon production, and the atmospheric muon charge ratio," *Astroparticle Physics*, vol. 35, no. 12, pp. 801–806, 2012.

[102] V. Zatsepin and N. V. Sokolskaya, "Three component model of cosmic ray spectra from 10 GeV to 100 PeV," *Astronomy & Astrophysics*, vol. 458, no. 1, pp. 1–5, 2006.

[103] O. Adriani *et al.*, "PAMELA measurements of cosmic-ray proton and helium spectra," *Science*, vol. 332, no. 6025, pp. 69–72, 2011.

[104] J. R. Hoerandel, "On the knee in the energy spectrum of cosmic rays," *Astroparticle Physics*, vol. 19, no. 2, pp. 193–220, 2003.

[105] S. V. Ter-Antonyan and L. Haroyan, "About EAS size spectra and primary energy spectra in the knee region," *arXiv preprint hep-ex/0003006*, 2000.

[106] H. Dembinski, R. Engel, A. Fedynitch, T. Gaisser, F. Riehn, and T. Stanev, "Data-driven model of the cosmic-ray flux and mass composition from 10 GeV to $10^{11}$ GeV," *arXiv preprint arXiv:1711.11432*, 2017.

[107] M. Fukugita and T. Yanagida, *Physics of Neutrinos: and Application to Astrophysics*. Springer Science & Business Media, 2013.

[108] F. Halzen and A. D. Martin, "Quarks and leptons," 1984.

[109] F. Riehn, H. P. Dembinski, R. Engel, A. Fedynitch, T. K. Gaisser, and T. Stanev, "The hadronic interaction model SIBYLL 2.3c and Feynman scaling," *arXiv preprint arXiv:1709.07227*, 2017.

[110] E.-J. Ahn, R. Engel, T. K. Gaisser, P. Lipari, and T. Stanev, "Cosmic ray interaction event generator SIBYLL 2.1," *Physical Review D*, vol. 80, no. 9, p. 094003, 2009.

[111] T. K. Gaisser and M. Honda, "Flux of atmospheric neutrinos," *Annual Review of Nuclear and Particle Science*, vol. 52, no. 1, pp. 153–199, 2002.

[112] A. Fedynitch, R. Engel, T. K. Gaisser, F. Riehn, and T. Stanev, "Calculation of conventional and prompt lepton fluxes at very high energy," in *EPJ Web of Conferences*, vol. 99, p. 08001, EDP Sciences, 2015.

[113] R. Enberg, M. H. Reno, and I. Sarcevic, "Prompt neutrino fluxes from atmospheric charm," *Physical Review D*, vol. 78, no. 4, p. 043005, 2008.

[114] A. Bhattacharya, R. Enberg, M. H. Reno, I. Sarcevic, and A. Stasto, "Perturbative charm production and the prompt atmospheric neutrino flux in light of RHIC and LHC," *Journal of High Energy Physics*, vol. 2015, no. 6, p. 110, 2015.

[115] M. Aartsen *et al.*, "Neutrino emission from the direction of the blazar TXS 0506+ 056 prior to the IceCube-170922A alert," *Science*, vol. 361, no. 6398, pp. 147–151, 2018.

[116] M. Aartsen *et al.*, "Measurement of Atmospheric Neutrino Oscillations at 6–56 GeV with IceCube DeepCore," *Physical review letters*, vol. 120, no. 7, p. 071801, 2018.

[117] P. Adamson *et al.*, "Measurement of the neutrino mixing angle $\theta_{23}$ in NOvA," *Physical review letters*, vol. 118, no. 15, p. 151802, 2017.

[118] P. Adamson *et al.*, "Measurement of neutrino and antineutrino oscillations using beam and atmospheric data in MINOS," *Physical Review Letters*, vol. 110, no. 25, p. 251801, 2013.

[119] K. Abe *et al.*, "Combined analysis of neutrino and antineutrino oscillations at T2K," *Physical review letters*, vol. 118, no. 15, p. 151801, 2017.

[120] M. Aartsen *et al.*, "First search for dark matter annihilations in the Earth with the IceCube detector," *The European Physical Journal C*, vol. 77, no. 2, p. 82, 2017.

[121] M. Aartsen *et al.*, "South Pole glacial climate reconstruction from multi-borehole laser particulate stratigraphy," *Journal of Glaciology*, vol. 59, no. 218, pp. 1117–1128, 2013.

[122] M. Aartsen *et al.*, "Neutrino interferometry for high-precision tests of Lorentz symmetry with IceCube," *Nature Physics*, vol. 14, p. 961âĂŞ966, 2017.

[123] M. Aartsen *et al.*, "Searches for sterile neutrinos with the IceCube detector," *Physical review letters*, vol. 117, no. 7, p. 071801, 2016.

[124] S. Fukuda *et al.*, "The Super-Kamiokande detector," *NIM-A*, vol. 501, no. 2-3, pp. 418–462, 2003.

[125] S. Fukuda *et al.*, "Determination of solar neutrino oscillation parameters using 1496 days of Super-Kamiokande-I data," *Physics Letters B*, vol. 539, no. 3-4, pp. 179–187, 2002.

[126] J. Ahrens *et al.*, "IceCube preliminary design document." http://www.icecube.wisc.edu/science/publications/pdd, 2001.

[127] A. Achterberg *et al.*, "First year performance of the IceCube neutrino telescope," *Astroparticle Physics*, vol. 26, no. 3, pp. 155–173, 2006.

[128] I. Collaboration, "Evidence for high-energy extraterrestrial neutrinos at the IceCube detector," *Science*, vol. 342, no. 6161, p. 1242856, 2013.

[129] R. Abbasi *et al.*, "The design and performance of IceCube DeepCore," *Astroparticle physics*, vol. 35, no. 10, pp. 615–624, 2012.

[130] R. Abbasi *et al.*, "The IceCube data acquisition system: Signal capture, digitization, and timestamping," *NIM-A*, vol. 601, no. 3, pp. 294–316, 2009.

[131] Hamamatsu, "R7081 datasheet," 2003.

[132] R. Abbasi, Y. Abdou, T. Abu-Zayyad, J. Adams, J. Aguilar, M. Ahlers, K. Andeen, J. Auffenberg, X. Bai, M. Baker, *et al.*, "Calibration and characterization of the icecube photomultiplier tube," *NIM-A*, vol. 618, no. 1-3, pp. 139–152, 2010.

[133] R. Abbasi, Y. Abdou, T. Abu-Zayyad, J. Adams, J. Aguilar, M. Ahlers, K. Andeen, J. Auffenberg, X. Bai, M. Baker, *et al.*, "Calibration and characterization of the icecube photomultiplier tube," *NIM-A*, vol. 618, no. 1-3, pp. 139–152, 2010.

[134] M. Aartsen *et al.*, "The IceCube Neutrino Observatory: Instrumentation and Online Systems," *JINST*, vol. 12, no. 03, p. P03012, 2017.

[135] I. Collaboartion, "Icecube muon filter for 2010 pole season," 2010.

[136] M. Rongen, "Measuring the optical properties of IceCube drill holes," in *EPJ Web of Conferences*, vol. 116, p. 06011, EDP Sciences, 2016.

[137] N. Whitehorn *et al.*, "Wavedeform." `http://code.icecube.wisc.edu/svn/projects/wavedeform/`, 2019.

[138] S. Axani, "SPE Templates Paper: v5.5." `https://docushare.wipac.wisc.edu/dsweb/Get/Document-87343/SPE_Paper_v5.5.pdf`, 2019.

[139] S. Axani, "SPE Templates Technote: Extracting and fitting procedure." `https://docushare.wipac.wisc.edu/dsweb/Get/Document-85572/SPE_Templates_Technote_v2.pdf`, 2019.

[140] S. Axani *et al.*, "In-situ calibration of the single-photoelectron charge response of the IceCube photomultipliers." `https://wiki.icecube.wisc.edu/index.php/SPE_Templates_Paper`, 2019.

[141] P. B. Price, K. Woschnagg, and D. Chirkin, "Age vs depth of glacial ice at South Pole," *Geophysical Research Letters*, vol. 27, no. 14, pp. 2129–2132, 2000.

[142] H. Shoji and C. C. Langway Jr, "Air hydrate inclusions in fresh ice core," *Nature*, vol. 298, no. 5874, p. 548, 1982.

[143] P. Askebjer *et al.*, "Optical properties of the South Pole ice at depths between 0.8 and 1 kilometer," *Science*, vol. 267, no. 5201, pp. 1147–1150, 1995.

[144] N. Bramall, R. Bay, K. Woschnagg, R. Rohde, and P. Price, "A deep high-resolution optical log of dust, ash, and stratigraphy in South Pole glacial ice," *Geophysical research letters*, vol. 32, no. 21, 2005.

[145] M. Aartsen *et al.*, "Measurement of South Pole ice transparency with the IceCube LED calibration system," *NIM-A*, vol. 711, pp. 73–89, 2013.

[146] P. B. Price *et al.*, "Temperature profile for glacial ice at the South Pole: Implications for life in a nearby subglacial lake," *Proceedings of the National Academy of Sciences*, vol. 99, no. 12, pp. 7844–7847, 2002.

[147] M. Aartsen *et al.*, "The IceCube neutrino observatory part VI: Ice properties, reconstruction and future developments," *arXiv preprint arXiv:1309.7010*, 2013.

[148] D. Chirkin, "Evidence of optical anisotropy of the South Pole ice," *ICRC2013*, 2013.

[149] D. Chirkin and W. Rhode, "Muon monte carlo: A high-precision tool for muon propagation through matter," tech. rep., 2004.

[150] W. R. Leo, *Techniques for nuclear and particle physics experiments: a how-to approach*. Springer Science & Business Media, 2012.

[151] C. Grupen and B. Shwartz, *Particle detectors*, vol. 26. Cambridge university press, 2008.

[152] K. Nakamura and P. D. Group, "Review of particle physics," *Journal of Physics G: Nuclear and Particle Physics*, vol. 37, no. 7A, p. 075021, 2010.

[153] J. Van Santen, *Neutrino Interactions in IceCube above 1 TeV*. PhD thesis, UW-Madison, 2014.

[154] D. Xu, *Search for Astrophysical Tau Neutrinos in Three Years of IceCube Data*. PhD thesis, Alabama U., 2015.

[155] M. Aartsen *et al.*, "Search for sterile neutrino mixing using three years of IceCube Deep-Core data," *Physical Review D*, vol. 95, no. 11, p. 112002, 2017.

[156] B. J. P. Jones, "IceCube Sterile Neutrino Searches," *Very Large Volume Neutrino Telescopes (VLVnT-2018)*, vol. 207, no. 04005, p. 6, 2019.

[157] R. Pordes *et al.*, "The open science grid," in *Journal of Physics: Conference Series*, vol. 78, p. 012057, IOP Publishing, 2007.

[158] I. Sfiligoi, D. C. Bradley, B. Holzman, P. Mhashilkar, S. Padhi, and F. Wurthwein, "The pilot way to grid resources using glideinWMS," in *2009 WRI World congress on computer science and information engineering*, vol. 2, pp. 428–432, IEEE, 2009.

[159] K. Hoshina, "Neutrino Generator." http://code.icecube.wisc.edu/svn/projects/neutrino-generator/, 2019.

[160] C. Arguelles and C. Weaver, "Lepton Injector." http://code.icecube.wisc.edu/svn/sandbox/cweaver/LeptonInjector/, 2019.

[161] C. Arguelles, B. Jones, A. Schneider, and C. Weaver, "LeptonWeighter." http://code.icecube.wisc.edu/svn/sandbox/aschneider/lepton_weighter/, 2019.

[162] C. Arguelles, J. Salvado, and C. Weaver, "MuonInjector." http://code.icecube.wisc.edu/svn/sandbox/cweaver/MuonInjector/, 2015.

[163] J.-H. Koehne *et al.*, "PROPOSAL: A tool for propagation of charged leptons," *Computer Physics Communications*, vol. 184, no. 9, pp. 2070–2090, 2013.

[164] C. Kopper, "CLSim." http://code.icecube.wisc.edu/svn/projects/clsim/, 2019.

[165] P. Liu, "A new phase function approximating to mie scattering for radiative transport equations," *Physics in medicine & biology*, vol. 39, no. 6, p. 1025, 1994.

[166] D. Chirkin *et al.*, "Photon tracking with GPUs in IceCube," *NIM-A*, vol. 725, pp. 141–143, 2013.

[167] C. Weaver, "PMTResponseSimulator.." `http://code.icecube.wisc.edu/svn/projects/DOMLauncher/trunk/private/DOMLauncher/PMTResponseSimulator.cxx?p=122325`.

[168] T. Feusels, "PMT saturation at low gain," 2019.

[169] M. J. Larson, *Simulation and identification of non-poissonian noise triggers in the IceCube neutrino detector*. PhD thesis, University of Alabama Libraries, 2013.

[170] M. Aartsen *et al.*, "Energy reconstruction methods in the IceCube neutrino telescope," *JINST*, vol. 9, pp. 1748–0221, 2014.

[171] A. Fedynitch, J. B. Tjus, and P. Desiati, "Influence of hadronic interaction models and the cosmic ray spectrum on the high energy atmospheric muon and neutrino flux," *Physical Review D*, vol. 86, no. 11, p. 114024, 2012.

[172] Jet Propulsion Laboratory , "AIRS/AMSU/HSB Version 6 Level 3 Product User Guide," *Version 1.2*, November 2014.

[173] U. S. Atmosphere, "National oceanic and atmospheric administration," *National Aeronautics and Space Administration, United States Air Force, Washington, DC*, 1976.

[174] C. Arguelles, J. Salvado, and C. Weaver, "$\nu$-SQuIDS," 2015.

[175] C. Arguelles, J. Salvado, and C. Weaver, "SQuIDS," 2015.

[176] A. M. Dziewonski and D. L. Anderson, "Preliminary reference Earth model," *Physics of the earth and planetary interiors*, vol. 25, no. 4, pp. 297–356, 1981.

[177] B. L. Kennett, E. Engdahl, and R. Buland, "Constraints on seismic velocities in the Earth from traveltimes," *Geophysical Journal International*, vol. 122, no. 1, pp. 108–124, 1995.

[178] K. E. Bullen, K. E. Bullen, and B. A. Bolt, *An introduction to the theory of seismology.* Cambridge university press, 1985.

[179] L. Volgyesi and M. Moser, "The inner structure of the earth," *Periodica Polytechnica Chemical Engineering*, vol. 26, 01 1982.

[180] M. Aartsen et al., "Characterization of the atmospheric muon flux in IceCube," *Astroparticle physics*, vol. 78, pp. 1–27, 2016.

[181] D. Heck, G. Schatz, J. Knapp, T. Thouw, and J. Capdevielle, "CORSIKA: A Monte Carlo code to simulate extensive air showers," tech. rep., 1998.

[182] D. Heck and T. Pierog, "Extensive air shower simulation with CORSIKA: A users guide," *Forschungszentrum Karlsruhe, Institut für Kernphysik*, 2000.

[183] C. Weaver, "Topologicalsplittersphere." `http://code.icecube.wisc.edu/svn/projects/TopologicalSplitter/branches/TopologicalSplitterSphere/`, 2015.

[184] C. Weaver, "Evidence for Astrophysical Muon Neutrinos from the Northern Sky," 2015.

[185] T. Neunhöffer, "Estimating the angular resolution of tracks in neutrino telescopes based on a likelihood analysis," *Astroparticle Physics*, vol. 25, no. 3, pp. 220–225, 2006.

[186] M. Aartsen et al., "Evidence for astrophysical muon neutrinos from the northern sky with IceCube," *Physical review letters*, vol. 115, no. 8, p. 081102, 2015.

[187] C. Weaver, "Diffuse event selection." `https://icecube.wisc.edu/~cweaver/DiffuseAnalysis/EventSelection/index.html`, Dec 2018.

[188] J. Tatar, "Truncated energy." `http://software.icecube.wisc.edu/documentation/projects/truncated_energy/index.html`, 2019.

[189] T. Gaisser, "Seasonal variation of atmospheric neutrinos in IceCube," 2013.

[190] C. A. A. Delgado, J. Salvado, and C. N. Weaver, "A simple quantum integro-differential solver (squids)," *Computer Physics Communications*, vol. 196, pp. 569–591, 2015.

[191] A. Cooper-Sarkar, P. Mertsch, and S. Sarkar, "The high energy neutrino cross-section in the Standard Model and its uncertainty," *Journal of High Energy Physics*, vol. 2011, no. 8, p. 42, 2011.

[192] J. Ahrens *et al.*, "Muon track reconstruction and data selection techniques in AMANDA," *NIM-A*, vol. 524, no. 1-3, pp. 169–194, 2004.

[193] C. Argüelles *et al.*, "GolemFit." https://github.com/IceCubeOpenSource/GolemFit/tree/SterilizeSuperOptimized, 2019.

[194] S. A. Murphy and A. W. Van der Vaart, "On profile likelihood," *Journal of the American Statistical Association*, vol. 95, no. 450, pp. 449–465, 2000.

[195] Wiki, "Limited-memory bfgs." https://en.wikipedia.org/wiki/Limited-memory_BFGS, 2019.

[196] T. Y. Carlos A. Argüelles, Austin Schneider, "A binned likelihood for stochastic models," *arXiv:1901.04645*, 2019.

[197] S. S. Wilks, "The large-sample distribution of the likelihood ratio for testing composite hypotheses," *The Annals of Mathematical Statistics*, vol. 9, no. 1, pp. 60–62, 1938.

[198] N. Whitehorn, J. van Santen, and S. Lafebre, "Penalized splines for smooth representation of high-dimensional Monte Carlo datasets," *Computer Physics Communications*, vol. 184, no. 9, pp. 2214–2220, 2013.

[199] M. Aartsen *et al.*, "Efficient propagation of systematic uncertainties from calibration to analysis with the SnowStorm method in IceCube," *arXiv preprint arXiv:1909.01530*, 2019.

[200] S. Fiedlschuster, "The Effect of Hole Ice on the Propagation and Detection of Light in IceCube," *arXiv preprint arXiv:1904.08422*, 2019.

[201] A. M. Hillas, "Cosmic rays: Recent progress and some current questions," *arXiv preprint astro-ph/0607109*, 2006.

[202] S. Ostapchenko, "Monte Carlo treatment of hadronic interactions in enhanced Pomeron scheme: QGSJET-II model," *Physical Review D*, vol. 83, no. 1, p. 014018, 2011.

[203] G. Barr, S. Robbins, T. Gaisser, and T. Stanev, "Uncertainties in atmospheric neutrino fluxes," *Physical Review D*, vol. 74, no. 9, p. 094009, 2006.

[204] B. J. Jones, *Sterile neutrinos in cold climates*. PhD thesis, MIT, 2015.

[205] M. Honda, T. Kajita, K. Kasahara, S. Midorikawa, and T. Sanuki, "Calculation of atmospheric neutrino flux using the interaction model calibrated with atmospheric muon data," *Physical Review D*, vol. 75, no. 4, p. 043006, 2007.

[206] Y. Yoon *et al.*, "Proton and helium spectra from the CREAM-III flight," *The Astrophysical Journal*, vol. 839, no. 1, p. 5, 2017.

[207] R. Alfaro *et al.*, "All-particle cosmic ray energy spectrum measured by the HAWC experiment from 10 to 500 TeV," *Physical Review D*, vol. 96, no. 12, p. 122001, 2017.

[208] M. Iacovacci *et al.*, "Cosmic Ray physics with ARGO-YBJ," *Nuclear Physics B-Proceedings Supplements*, vol. 239, pp. 157–162, 2013.

[209] A. Karelin *et al.*, "The Proton and Helium cosmic ray spectra from 50 GeV to 15 TeV," *Astrophysics and Space Sciences Transactions*, vol. 7, no. 2, pp. 235–238, 2011.

[210] C. Lattes, Y. Fujimoto, and S. Hasegawa, "Hadronic interactions of high energy cosmic-ray observed by emulsion chambers," *Physics Reports*, vol. 65, no. 3, pp. 151–229, 1980.

[211] R. Fletcher, T. Gaisser, P. Lipari, and T. Stanev, "SIBYLL: An event generator for simulation of high energy cosmic ray cascades," *Physical Review D*, vol. 50, no. 9, p. 5710, 1994.

[212] G. Barr, T. Gaisser, and T. Stanev, "Flux of atmospheric neutrinos," *Physical Review D*, vol. 39, no. 11, p. 3532, 1989.

[213] M. Aartsen *et al.*, "Observation and Characterization of a Cosmic Muon Neutrino Flux from the Northern Hemisphere using six years of IceCube data," *The Astrophysical Journal*, vol. 833, no. 1, p. 3, 2016.

[214] A. Schneider, "Characterization of the Astrophysical Diffuse Neutrino Flux with IceCube High-Energy Starting Events," *arXiv preprint arXiv:1907.11266*, 2019.

[215] M. Chianese, G. Miele, and S. Morisi, "Interpreting IceCube 6-year HESE data as an evidence for hundred TeV decaying Dark Matter," *Physics Letters B*, vol. 773, pp. 591–595, 2017.

[216] J. Stachurska, "IceCube High Energy Starting Events at 7.5 Years–New Measurements of Flux and Flavor," *arXiv preprint arXiv:1905.04237*, 2019.

[217] M. T. et al. (Particle Data Group), "Data files and plots of cross-sections and related quantities in the 2018 Review of Particle Physics," *Phys. Rev. D 98, 030001*, 2018.

[218] R. Glauber, "Cross sections in deuterium at high energies," *Physical Review*, vol. 100, no. 1, p. 242, 1955.

[219] R. Glauber and G. Matthiae, "High-energy scattering of protons by nuclei," *Nuclear Physics B*, vol. 21, no. 2, pp. 135–157, 1970.

[220] V. Gribov, "A reggeon diagram technique," *Sov. Phys. JETP*, vol. 26, no. 414-422, p. 27, 1968.

[221] V. Khachatryan *et al.*, "Measurement of the inelastic cross section in proton-lead collisions at $\sqrt{s} = 5.02$ TeV," *Physics Letters B*, vol. 759, pp. 641–662, 2016.

[222] A. Toia *et al.*, "Bulk properties of Pb–Pb collisions at TeV measured by ALICE," *Journal of Physics G: Nuclear and Particle Physics*, vol. 38, no. 12, p. 124007, 2011.

[223] P. Abreu *et al.*, "Measurement of the proton-air cross section at $\sqrt{s} = 57$ TeV with the Pierre Auger Observatory," *Physical Review Letters*, vol. 109, no. 6, p. 062002, 2012.

[224] H. Dembinski, R. Ulrich, and T. Pierog, "Future proton-oxygen beam collisions at the LHC for air shower physics," in *36th International Cosmic Ray Conference (ICRC2019)*, vol. 36, 2019.

[225] F. Halzen, K. Igi, M. Ishida, and C. S. Kim, "Total hadronic cross sections and $\pi \pm \pi +$ scattering," *Physical Review D*, vol. 85, no. 7, p. 074020, 2012.

[226] B. Allbrooke *et al.*, "Measurement of the inelastic proton-proton cross section at $\sqrt{s}=13$ TeV with the ATLAS detector at the LHC," *Physical review letters*, vol. 117, p. 182002, 2016.

[227] P. Adamson *et al.*, "Active to sterile neutrino mixing limits from neutral-current interactions in MINOS," *Physical review letters*, vol. 107, no. 1, p. 011802, 2011.

[228] P. Adamson *et al.*, "Improved measurement of muon antineutrino disappearance in MINOS," *Physical review letters*, vol. 108, no. 19, p. 191801, 2012.

[229] P. Adamson *et al.*, "Search for the disappearance of muon antineutrinos in the NuMI neutrino beam," *Physical Review D*, vol. 84, no. 7, p. 071103, 2011.

[230] P. Adamson *et al.*, "Search for sterile neutrinos mixing with muon neutrinos in MINOS," *Physical review letters*, vol. 117, no. 15, p. 151803, 2016.

[231] G. Cheng *et al.*, "Dual baseline search for muon antineutrino disappearance at 0.1 eV$^2$< $\Delta$ m$^2$< 100 eV$^2$," *Physical Review D*, vol. 86, no. 5, p. 052009, 2012.

[232] C. Giunti, "Light sterile neutrinos: status and perspectives," *Nuclear Physics B*, vol. 908, pp. 336–353, 2016.

[233] Physics Today, "An easy-to-build desktop muon detector." `https://physicstoday.scitation.org/do/10.1063/PT.6.1.20170614a/full/`, 2017.

[234] MIT News, "Physicists design $100 handheld muon detector." `http://news.mit.edu/2017/handheld-muon-detector-1121/`, 2017.

[235] Symmetry Magazine, "The $100 muon detector." `https://www.symmetrymagazine.org/article/the-100-muon-detector`, 2016.

[236] Science Daily, "Physicists design $100 handheld cosmic ray muon detector." `https://www.sciencedaily.com/releases/2017/11/171120174502.htmr`, 2017.

[237] Geek.com, "Beat Boredom By Building $100 DIY Muon Detector." `https://www.geek.com/tech/beat-boredom-by-building-100-diy-muon-detector-1724031/`, 2017.

[238] Interesting Engineer, "You Can Make Your Own $100 Particle Detector Thanks to MIT Engineers." https://interestingengineering.com/you-can-make-your-own-100-particle-detector-thanks-to-mit-engineers, 2017.

[239] Next Big Future, "$100 pocket size muon detector." https://www.nextbigfuture.com/2017/11/100-pocket-size-muon-detector.html, 2016.

[240] S. N. Axani, J. M. Conrad, and C. Kirby, "The Desktop Muon Detector: A simple, physics-motivated machine-and electronics-shop project for university students," *American Journal of Physics*, vol. 85, no. 12, pp. 948–958, 2017.

[241] S. Axani, D. Winklehner, J. Alonso, and J. Conrad, "A high intensity $H2^+$ multicusp ion source for the isotope decay-at-rest experiment, IsoDAR," *Review of Scientific Instruments*, vol. 87, no. 2, p. 02B704, 2016.

[242] J. Alonso, S. Axani, L. Calabretta, D. Campo, L. Celona, J. M. Conrad, A. Day, G. Castro, F. Labrecque, and D. Winklehner, "The IsoDAR high intensity H2+ transport and injection tests," *Journal of Instrumentation*, vol. 10, no. 10, p. T10003, 2015.

[243] D. Winklehner, R. Hamm, J. Alonso, J. Conrad, and S. Axani, "Preliminary design of a RFQ direct injection scheme for the IsoDAR high intensity H2+ cyclotron," *Review of Scientific Instruments*, vol. 87, no. 2, p. 02B929, 2016.

[244] S. Axani, G. Collin, J. Conrad, M. Shaevitz, J. Spitz, and T. Wongjirad, "Decisive disappearance search at high $\Delta$ m$^2$ with monoenergetic muon neutrinos," *Physical Review D*, vol. 92, no. 9, p. 092010, 2015.

[245] IsoDAR, "Isodar website." https://www.nevis.columbia.edu/daedalus/exp/isodar.html, 2019.

[246] I. Collaboration, "IceCube-Gen2," 2019.

[247] A. Terliuk, *Measurement of atmospheric neutrino oscillations and search for sterile neutrino mixing with IceCube DeepCore*. PhD thesis, Humboldt-Universität zu Berlin, 2018.

[248] IceCube, "Muongun software," 2019.

[249] Y. Becherini, A. Margiotta, M. Sioli, and M. Spurio, "A parameterisation of single and multiple muons in the deep water or ice," *Astroparticle Physics*, vol. 25, no. 1, pp. 1–13, 2006.

[250] T. Lay and T. C. Wallace, *Modern global seismology*, vol. 58. Elsevier, 1995.

[251] E. Engdahl, E. A. Flinn, and R. P. Massé, "Differential PKiKP travel times and the radius of the inner core," *Geophysical Journal International*, vol. 39, no. 3, pp. 457–463, 1974.

[252] C. Denis, Y. Rogister, M. Amalvict, C. Delire, A. I. Denis, and G. Munhoven, "Hydrostatic flattening, core structure, and translational mode of the inner core," *Physics of the earth and planetary interiors*, vol. 99, no. 3-4, pp. 195–206, 1997.

[253] P. M. Shearer, *Introduction to seismology*. Cambridge university press, 2019.

[254] R. Stokstad, "Design and performance of the icecube electronics," 2005.

[255] Hamamatsu, "Basics and applications," Third Edition.

[256] Hamamatsu, "Handbook, chapter 4."

[257] J. Brack *et al.*, "Characterization of the Hamamatsu R11780 12 in. photomultiplier tube," *NIM-A*, vol. 712, pp. 162–173, 2013.

[258] E. Calvo *et al.*, "Characterization of large-area photomultipliers under low magnetic fields: Design and performance of the magnetic shielding for the Double Chooz neutrino experiment," *NIM-A*, vol. 621, no. 1-3, pp. 222–230, 2010.

[259] F. Kaether and C. Langbrandtner, "Transit time and charge correlations of single photo-electron events in R7081 photomultiplier tubes," *Journal of Instrumentation*, vol. 7, no. 09, p. P09002, 2012.

[260] B. Lubsandorzhiev, P. Pokhil, R. Vasiljev, and A. Wright, "Studies of prepulses and late pulses in the 8" electron tubes series of photomultipliers," *NIM-A*, vol. 442, no. 1-3, pp. 452–458, 2000.

[261] K. o. Ma, "Time and amplitude of afterpulse measured with a large size photomultiplier tube," *NIM-A*, vol. 629, no. 1, pp. 93–100, 2011.

[262] S. Torre, T. Antonioli, and P. Benetti, "Study of afterpulse effects in photomultipliers," *Review of scientific instruments*, vol. 54, no. 12, pp. 1777–1780, 1983.

[263] Hamamatsu, "Photomultiplier tubes: Construction and Operating Characteristics," URL: `https://www.hamamatsu.com/resources/pdf/etd/PMT_TPMZ0002E.pdf`, 2016.

[264] M. Aartsen *et al.*, "Characterization of the atmospheric muon flux in IceCube," *Astroparticle physics*, vol. 78, pp. 1–27, 2016.

[265] M. Aartsen *et al.*, "Search for steady point-like sources in the astrophysical muon neutrino flux with 8 years of IceCube data," *The European Physical Journal C*, vol. 79, no. 3, p. 234, 2019.

[266] R. Dossi, A. Ianni, G. Ranucci, and O. J. Smirnov, "Methods for precise photoelectron counting with photomultipliers," *NIM-A*, vol. 451, no. 3, pp. 623–637, 2000.